\documentclass[11pt,DIV11]{scrbook} %DIV11 := divide page width and height by 11 and then upper most und 2 lower most stripes are margin and for double sided prints the inner most stripe (+bindung) and the two outer most stripes are margin :)

\usepackage[latin1]{inputenc}
\usepackage{appendix}
\usepackage{float,amsmath,amstext,amsthm,amssymb,amsfonts,slashed,epigraph}
\usepackage[nouppercase]{scrpage2}      % Seitenlayout nach Koma-Script
\usepackage{makeidx}
\usepackage{hyperref}
\usepackage{graphicx}
\usepackage{caption}

    \clearscrheadfoot
    \ohead[\pagemark]{\pagemark}
    
    \small
    \automark[section]{chapter}
    \normalsize
    \setheadsepline{0.5pt}
\declareslashed{}{\not}{-0.4}{0}{\int}

\makeindex

\makeatletter
\newcommand{\contraction}[5][1ex]{%
  \mathchoice
    {\contraction@\displaystyle{#2}{#3}{#4}{#5}{#1}}%
    {\contraction@\textstyle{#2}{#3}{#4}{#5}{#1}}%
    {\contraction@\scriptstyle{#2}{#3}{#4}{#5}{#1}}%
    {\contraction@\scriptscriptstyle{#2}{#3}{#4}{#5}{#1}}}%
\newcommand{\contraction@}[6]{%
  \setbox0=\hbox{$#1#2$}%
  \setbox2=\hbox{$#1#3$}%
  \setbox4=\hbox{$#1#4$}%
  \setbox6=\hbox{$#1#5$}%
  \dimen0=\wd2%
  \advance\dimen0 by \wd6%
  \divide\dimen0 by 2%
  \advance\dimen0 by \wd4%
  \vbox{%
    \hbox to 0pt{%
      \kern \wd0%
      \kern 0.5\wd2%
      \contraction@@{\dimen0}{#6}%
      \hss}%
    \vskip 0.9ex%  how far above the line starts
    \vskip\ht2}}

\newcommand{\contraction@@}[3][0.05em]{%
  \hbox{%
    \vrule width #1 height 0pt depth #3%
    \vrule width #2 height 0pt depth #1%
    \vrule width #1 height 0pt depth #3%
    \relax}}
\makeatother
\newcounter{saveeqn}

\begin{document}
  \begin{titlepage}
  \vspace*{0cm}
  \begin{center}
  \Huge
   \textsc{Collective Phenomena\\ in the Non-Equilibrium\\ Quark-Gluon Plasma}\\
  \vspace{2cm}
   Dissertation \\
    \LARGE
   zur Erlangung des Doktorgrades\\
   der Naturwissenschaften\\
  \vspace{2cm}
   vorgelegt beim Fachbereich Physik\\
   der Johann Wolfgang Goethe-Universit\"at\\
   in Frankfurt am Main\\
  \vspace{2cm}
  \LARGE
  von\\
  \LARGE
  Bj\"orn Peter Schenke\\
  \LARGE
  \vspace{0.2cm}
  aus Iserlohn\\
  \vspace{1cm}
  Frankfurt am Main 2008\\
  (D30)
  \vspace{5cm}
  \end{center}
  \large
  \vfill
\newpage
~\\
\vspace{10cm}
~\\
vom Fachbereich Physik (13) der Johann Wolfgang Goethe-Universit\"at\\
als Dissertation angenommen.\\
\vspace{6cm}\\
Dekan: Prof. Dr. D. H. Rischke\\
\vspace{0.5cm}\\
Gutachter: Prof. Dr. C. Greiner, JProf. Dr. A. Dumitru\\
\vspace{0.5cm}\\
Datum der Disputation: 03.07.2008
\end{titlepage}

%  \listoffigures
  \chapter*{Zusammenfassung\markboth{Zusammenfassung}{Zusammenfassung}}
\section*{\"Ubersicht}
In dieser Arbeit werden die Eigenschaften und die dynamische
Entwicklung des Quark-Gluon-Plasmas, wie es in
Schwer\-ionen-Kollisionen erzeugt wird, behandelt. Insbesondere
wird untersucht welche Rolle instabile kollektive Moden bei der
Iso\-tropi\-sierung und Ther\-ma\-li\-sie\-rung des Systems spielen k\"onnen.
Dazu wird zun\"achst der Einfluss von Kollisionen zwi\-schen Teilchen im
Medium auf die Wachstumsraten dieser Instabilit\"aten sowohl in
einer Modell\-rechnung als auch in einer dynamischen
\textsc{Wong-Yang-Mills} Realzeit-Gitter-Rechnung bestimmt. Diese
Simulation beinhaltet neben direkten elastischen
Teilchen-Kollisionen auch die Wechselwirkung der Teilchen mit
selbstkonsistent generierten Farb\-fel\-dern. Sie wird weiterhin
benutzt, um dynamisch den Energieverlust und die Aufweitung von
Jets (Teilchen(-schauer) mit hohem Transversalimpuls) zu
betrachten. Insbesondere ist es auf diese Weise m\"oglich, den
Transport\-koeffizienten $\hat{q}$ zu bestimmen, wobei das
Ergebnis weitgehend von der Gittergr\"o\ss e unabh\"angig ist. 
Des Weiteren wird die Jet-Propagation in Nichtgleichgewichts-Systemen untersucht. 
So wird ein anisotropes System in dem eine
Chromo-\textsc{Weibel}-Instabilit\"at auftritt simuliert, und der
Einfluss der dadurch entstehenden gro\ss en Dom\"anen von starken
Farb\-fel\-dern auf die pro\-pa\-gieren\-den Jet-Teilchen bestimmt.
Weiterhin werden fermionische Moden in anisotropen Systemen
untersucht. Dabei wird gezeigt, dass zumindest in der
Hard-Loop-N\"aherung keine instabilen fermionischen Moden
auftreten. Schlie\ss lich wird die M\"oglichkeit studiert, ob
hochenergetische Photonen zur Be\-stimmung der Impulsraum-Anisotropie
des in Schwer\-ionen-Kollisionen erzeugten Systems verwendet
werden k\"onnen. Das Ausma\ss \,der auftretenden Anisotropie ist
entscheidend f\"ur die Rolle, welche Instabilit\"aten f\"ur
Iso\-tropi\-sierung oder gar Ther\-ma\-li\-sie\-rung spielen k\"onnen, denn je
gr\"o\ss er die Anisotropie, desto gr\"o\ss er sind die
Wachstumsraten der instabilen Moden. Daher ist ihre Kenntnis extrem wichtig.

\section*{Einleitung}
Vor bereits \"uber drei\ss ig Jahren wurde vorgeschlagen sich
einen tieferen Einblick in die Natur der starken Wechselwirkung zu
verschaffen, indem man ``hohe Energie oder hohe Nukleonendichten
\"uber ein relativ gro\ss es Volumen'' \cite{Baym:2001in}
verteilt. Erst sp\"ater wurde reali\-siert, dass die asymptotische
Freiheit \cite{Gross:1973id,Politzer:1973fx} der Theorie der
starken Wechselwirkung (QCD) impliziert, dass bei hohen
Energiedichten ein Zustand mit freien Quarks und Gluonen erreicht
werden kann, welcher später Quark-Gluon-Plasma (QGP)
\cite{Shuryak:1977ut,Shuryak:1978ij,Shuryak:1980tp,Kalashnikov:1979dp,Kapusta:1979fh}
genannt wurde. Die einzige M\"oglichkeit, auf der Erde ausgedehnte
Systeme mit entsprechend hohen Energiedichten zu erzeugen, sind
Schwer\-ionen-Kollisionen. Erste systematische Studien fanden am
BEVALAC in Berkeley \cite{Bevalac:2007} statt, gefolgt von
Experimenten am Alternating Gradient Synchrotron (AGS)
\cite{AGS:2007} im Brookhaven National Laboratory (BNL), am CERN
Super Proton Synchrotron (SPS) \cite{SPS:2007} sowie am
Relativistic Heavy Ion Collider (RHIC) \cite{RHIC:2007} am BNL. In
naher Zukunft werden noch h\"ohere Energien am CERN Large Hadron
Collider (LHC) \cite{LHC:2007} untersucht werden, sowie hohe
Netto-Baryonendichten am RHIC und am internationalen Beschleunigerzentrum FAIR (Facility for
Antiproton and Ion Reasearch at GSI, Darmstadt).

Besonders die Experimente am RHIC haben wichtige Ergebnisse \"uber die
Erzeugung und die Eigen\-schaften des QGPs geliefert. Neben vielen
Indizien f\"ur ein Entstehen des QGPs in
Schwer\-ionen\--Kollisionen bei RHIC-Energien, deutet der starke
elliptische Fluss auf eine fr\"uhe Thermalisierung und eine sehr
niedrige Viskosit\"at des Mediums hin. Weiterhin l\"asst die
starke Unterdr\"uckung von Jets auf einen hohen Energieverlust der
Partonen (Quarks und Gluonen) im Medium schlie\ss en.

In dieser Arbeit wird insbesondere die M\"oglichkeit der schnellen
Isotropisierung und Thermalisierung durch Plasma-Instabilit\"aten
n\"aher untersucht. Sie bieten eine Erkl\"arung f\"ur schnelle
Equilibrierung, auch ohne die Annahme eines stark-gekoppelten
QGPs, welche der asymptotischen Freiheit bei hohen Energiedichten
widersprechen w\"urde. Des Weiteren wird die Wechselwirkung von
hochenergetischen Partonen mit dem Medium untersucht. Dabei wird
besonderer Wert auf die Bestimmung des Einflusses der instabilen
Moden in Nichtgleichgewichts\-systemen gelegt.

\section*{Theoretischer Hintergrund}
Zun\"achst werden die kinetischen Gleichungen f\"ur ein System von
Quarks und Gluonen aus den \textsc{Kadanoff-Baym}-Gleichungen
abgeleitet. Dazu wird eine Mean-field N\"aherung sowie eine
Gradienten-Entwicklung herangezogen. Man erh\"alt so eine
effektive Theorie f\"ur die kollektiven Moden des Systems bei der
Skala $gT$, wobei $T$ die Temperatur ist und $g$ die zun\"achst
als infinitesimal klein angenommene Kopplungskonstante. Im Limes
unendlich hoher Energien ist diese Annahme korrekt. Im
kollisionslosen Fall findet man auf diese Weise folgende
\textsc{Vlasov}-Gleichungen f\"ur Gluonen bzw. Quarks
\begin{align}\label{zus:vlasov}
     \left[V\cdot D_X,\delta n^g(\mathbf{k},X)\right]+g V_\mu
    F^{\mu\nu} \partial_{\nu}n^g(K)=0\,,\notag\\
 \left[ V\cdot D_X,\,\delta n^{q/\bar{q}}({{\bf
k}},X)\right]\pm\, g\,V_\mu F^{\mu\nu}(X)\partial_\nu n(K)=0\,,
\end{align}
wobei $D_X=\partial_X+igA_X$ die kovariante Ableitung bez\"uglich
der Koordinate $X$ bezeichnet, $V^\mu$ die Vierergeschwindigkeit
der Teilchen, $n_i$ die Hintergrunds-Verteilungsfunktion der
Gluonen ($i=g$) bzw. Quarks ($i=q/\bar{q}$), sowie $\delta n_i$
die entsprechenden Abweichungen vom statischen (zum Beispiel
Gleichgewichts-) Hintergrund. $F^{\mu\nu}\equiv[D^\mu,D^\nu]/(ig)$
ist der Feldst\"arketensor des Eichfeldes. Die
\textsc{Vlasov}-Gleichungen beschreiben die zeitliche Ent\-wicklung
eines Teilchensystems unter dem Einfluss selbstkonsistent
erzeugter Farbfelder.

L\"ost man die kinetischen Gleichungen f\"ur die
``harten'' Teilchen, d.h. solche mit hohem Impuls, so l\"asst sich
der induzierte Strom allein durch die Freiheitsgrade niedrigen
Impulses darstellen. Dies f\"uhrt letztlich zu einer effektiven
Theorie f\"ur die kollektiven Moden mit Impulsen der Gr\"o\ss
enordnung $gT$. Aus dem induzierten Strom erh\"alt man die
Selbstenergie
\begin{equation}\label{zus:selfenergyresult}
    \Pi^{\mu\nu}_{ab}(K)=g^2 \delta_{ab}\int_{\mathbf{p}} V^{\mu}
    \partial_{\beta}^{(p)}
    f(\mathbf{p})\left(g^{\nu\beta}-\frac{V^{\nu}K^{\beta}}{K\cdot
    V+i\epsilon}\right)\,\text{,}
\end{equation}
das selbe Ergebnis, wie man es mittels der diagrammatischen
Methode in der so ge\-nann\-ten ``hard-thermal-loop''-N\"aherung
erh\"alt, in der alle externen Impulse als klein gegen\"uber den
Loop-Impulsen angenommen werden. Mit Hilfe der Selbstenergie
(\ref{zus:selfenergyresult}) lassen sich die Dispersionsrelationen
der kollektiven Moden bestimmen. Dies wird neben einem isotropen
System auch f\"ur ein System mit einer Impulsraum-Anisotropie
durchgef\"uhrt. Man findet, dass in diesem Falle auch instabile
Moden auftreten, welche der \textsc{Weibel}-Instabilit\"at in
elektro-magnetischen Plasmen entsprechen. Die anisotrope
Verteilung wird in den expliziten Rechnungen durch Strecken bzw.
Stauchen einer isotropen Verteilung erhalten:
\begin{equation}
    f(\mathbf{p})=\mathcal{N}(\xi)\,f_{\text{iso}}\left(\mathbf{p}^2+\xi(\mathbf{p}\cdot\mathbf{\hat{n}})^2\right)\label{zus:anisodist}
\end{equation}
Die Richtung der Anisotropie ist durch $\mathbf{\hat{n}}$ gegeben,
$\xi>-1$ ist ein ver\"anderlicher Anisotropie\-parameter und
$\mathcal{N}(\xi)$ ein Normierungsfaktor. Neben der Ableitung der
Wachstumsraten dieser Instabilit\"aten unter der Annahme der oben
angegebenen Art von Anisotropie, werden die physikalischen
Mechanismen besprochen, die zu einer solchen Instabilit\"at
f\"uhren.

Um numerische Simulationen mittels Testteilchen durchzuf\"uhren,
werden die \textsc{Wong}-Gleichungen abgeleitet, welche, gekoppelt
an die \textsc{Yang-Mills}-Gleichungen f\"ur die Felder
\begin{equation}\label{zus:currentwym}
    D_\mu F^{\mu\nu}=J^\nu=g\sum_i q_i v_i^\nu\delta(\mathbf{x}-\mathbf{x}_i(t))\,,
\end{equation}
die Entwicklung eines mikroskopischen Systems individueller
Teilchen beschreiben. Sie lauten
\begin{align}
    \dot{\mathbf{x}}_i(t)&=\mathbf{v}_i(t)\,, \label{zus:wong1}\\
    \dot{\mathbf{p}}_i(t)&=g q^a_i(t)\left(\mathbf{E}^a(t)+\mathbf{v}_i(t)\times\mathbf{B}^a(t)\right)\,,\label{zus:wong2}\\
    \dot{q}_i(t)&=-igv_i^\mu(t)[A_\mu(t),q_i(t)]\,.\label{zus:wong3}
\end{align}
Besonders interessant ist die letzte Gleichung, die es in der
Elektrodynamik nicht gibt. Sie beschreibt die Rotation der
Farbladung $q$ (Farbvektor) durch die Farbfelder. Es
wird gezeigt, dass die \textsc{Wong}-Gleichungen \"aquivalent zu
den \textsc{Vlasov}-Gleichungen sind, sofern man die
kontinuierliche Teilchenverteilung $n$ durch eine gro\ss e Menge
Testteilchen ersetzt.

Nach der detaillierten Beschreibung der numerischen Simulation
wird erl\"autert, wie bin\"are Kollisionen in die Simulation,
welche zun\"achst nur Teilchen-Feld-Wechsel\-wirkungen
be\-schreibt, aufgenommen werden k\"onnen. Mittels der
stochastischen Methode wird anhand der Wir\-kungs\-quer\-schnitte
f\"ur Kollisionen zwi\-schen zwei Teilchen bestimmt, ob eine
Kollision stattfindet und welcher Impuls ausgetauscht wird.

Da die Wechselwirkung \"uber die Felder sozusagen schon niedrige
Impulsaust\"ausche zwi\-schen Teilchen enth\"alt (ein Teilchen
erzeugt durch seine Farbladung ein Feld, in dem ein anderes
Teilchen abgelenkt wird), werden nur solche Kollisionen
implementiert, deren Impulsaustausch gr\"o\ss er ist als der
Impuls der maximalen Feldmode auf dem Gitter $k^*\sim \pi/a$,
wobei $a$ der Gitterabstand ist.

Physikalisch ist es sinnvoll, $k^*$ so zu w\"ahlen, dass es im
Bereich der Temperatur (f\"ur ein System im Gleichgewicht), welche
die Impulsskala der Teilchen festlegt, liegt, denn oberhalb der
Temperatur ist eine Beschreibung mittels klassischer Felder nicht
mehr anwendbar.

Die Simulation erlaubt das Studium von
Nichtgleichgewichts\-situationen, da die Infrarotdivergenz der
Kollisionsterme nicht k\"unstlich mit Hilfe von Gr\"o\ss en aus
der Gleichgewichtsbeschreibung, wie der \textsc{Debye}-Masse,
reguliert werden m\"ussen.

\section*{Ergebnisse}
\subsection*{Modell-Beschreibung des Effekts von Kollisionen auf Plasma-Instabilit\"aten}
Kollisionen zwischen den ``harten'' Teilchen werden mittels eines
zus\"atzlichen Kollisions\-terms in den Gleichungen
(\ref{zus:vlasov}) eingef\"uhrt:
\begin{equation}
V\cdot \partial_X \delta f^{i}_a(p,X) + g \theta_{i}
V_{\mu}F^{\mu\nu}_a(X)\partial_{\nu}^{(p)}f^{i}(\mathbf{p})=\mathcal{C}^{i}_a(p,X)\label{zus:trans2}\,\text{,}
\end{equation}
mit
\begin{equation}
    \mathcal{C}^{i}_a(p,X)=-\nu\left[f^{i}_a(p,X)-\frac{N^{i}_a(X)}{N^{i}_{\text{eq}}}f^{i}_{\text{eq}}(|\mathbf{p}|)\right]\,\text{.}\label{zus:collision}
\end{equation}
Dieser der Relaxationszeit-N\"aherung \"ahnliche Term beschreibt,
wie die Kollisionen das System innerhalb einer Zeit proportional
zur inversen Kollisionsrate $\nu^{-1}$ ins Gleichgewicht bringen.
Gleichung (\ref{zus:trans2}) gilt im effektiv Abelschen Limes,
welcher gilt, solange die Felder $A$ nicht zu gro\ss \,werden. Daher
$D_X\rightarrow \partial_X$, $F^{\mu\nu}\rightarrow
\partial^\mu A^\nu-\partial^\nu A^\mu$, und die Gleichungen f\"ur die einzelnen
Farbkomponenten entkoppeln. Die $f$ bezeichnen hier die
Komponenten der Teilchenverteilungen, $f_{\text{eq}}$ ist die
Verteilung im Gleichgewicht.

\begin{figure}[htb]
      \begin{center}
        \includegraphics[height=6cm]{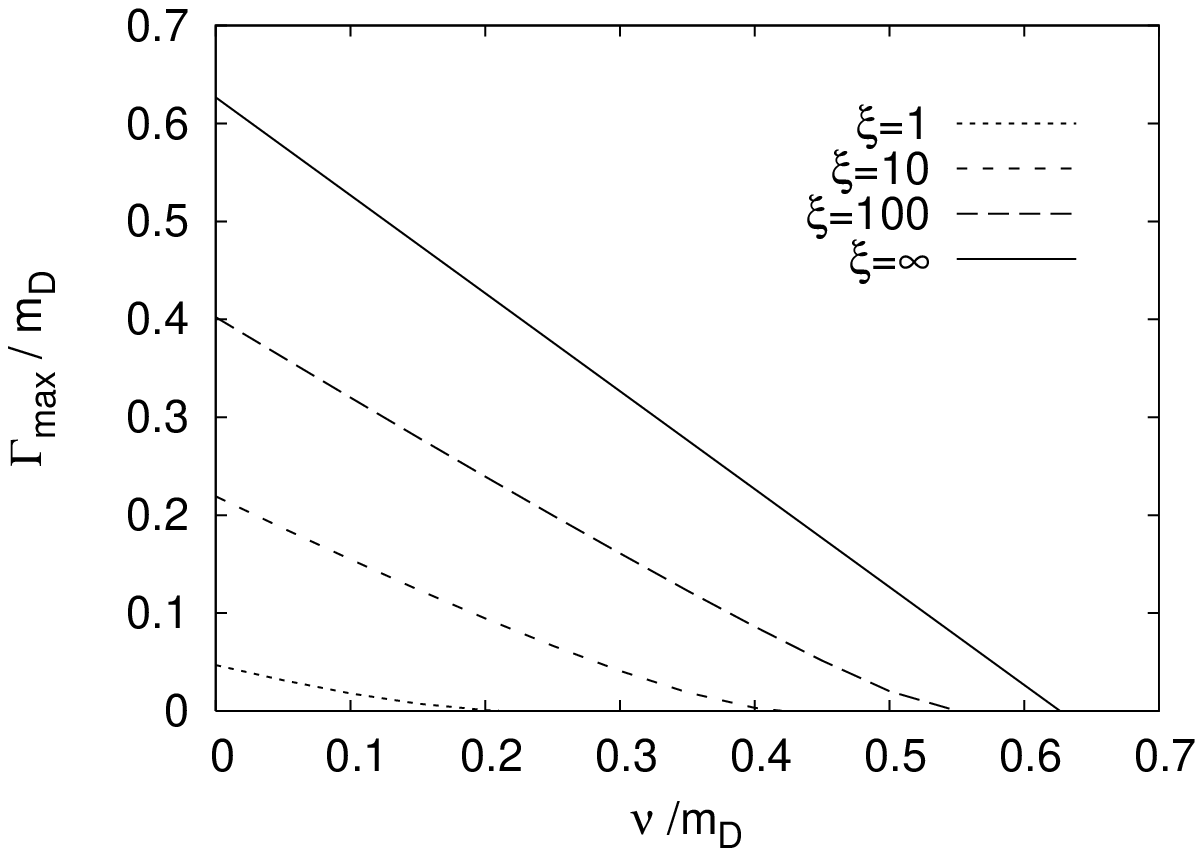}
        \caption{Die maximale Wachstumsrate der Instabilit\"at als Funktion der Kollisionsrate $\nu$.}
        \label{fig:zus:gammamax}
      \end{center}
  \end{figure}

Nun k\"onnen die kinetischen Gleichungen gel\"ost werden
und analog zum kollisionslosen Fall der induzierte Strom und
daraus die Selbstenergie der kollektiven Moden bestimmt werden.
Nachdem der Einfluss des Kollisionsterms auf die stabilen Moden
besprochen wurde, wird die Unterdr\"uckung der Wachstumsrate der
instabilen Moden bestimmt. Fig. \ref{fig:zus:gammamax} zeigt die
Abh\"angigkeit der maximalen Wachstumsrate von der Kollisionsrate
in Einheiten der \mbox{\textsc{Debye}}\--Mas\-se f\"ur unterschiedliche
Anisotropieparameter $\xi$.
Es stellt sich heraus, dass abh\"angig vom Grad der Anisotropie
maximale Kollisionsraten existieren, oberhalb derer keine
Instabilit\"aten mehr auftreten. 

Die Absch\"atzung der
Kollisionsrate $\nu$ wird dadurch erschwert, dass alle bisherigen
Rechnungen strikt perturbativ sind und bei f\"ur
Schwer\-ionen-Kollisionen relevanten Kopplungskonstanten
$\alpha_s\approx 0.2-0.3$ nicht uneingeschr\"ankt verwertbar sind. Eine grobe
Absch\"atzung jedoch f\"uhrt auf $\nu\approx 0.1-0.2\, m_D$, was
f\"ur starke Anisotropien deutlich unterhalb der kritischen
Kollisionsrate liegt, oberhalb derer keine Instabilit\"aten mehr
auftreten.

\subsection*{Wong-Yang-Mills Simulation}
\begin{figure}[hbt]
  \begin{center}
    \includegraphics[width=12cm]{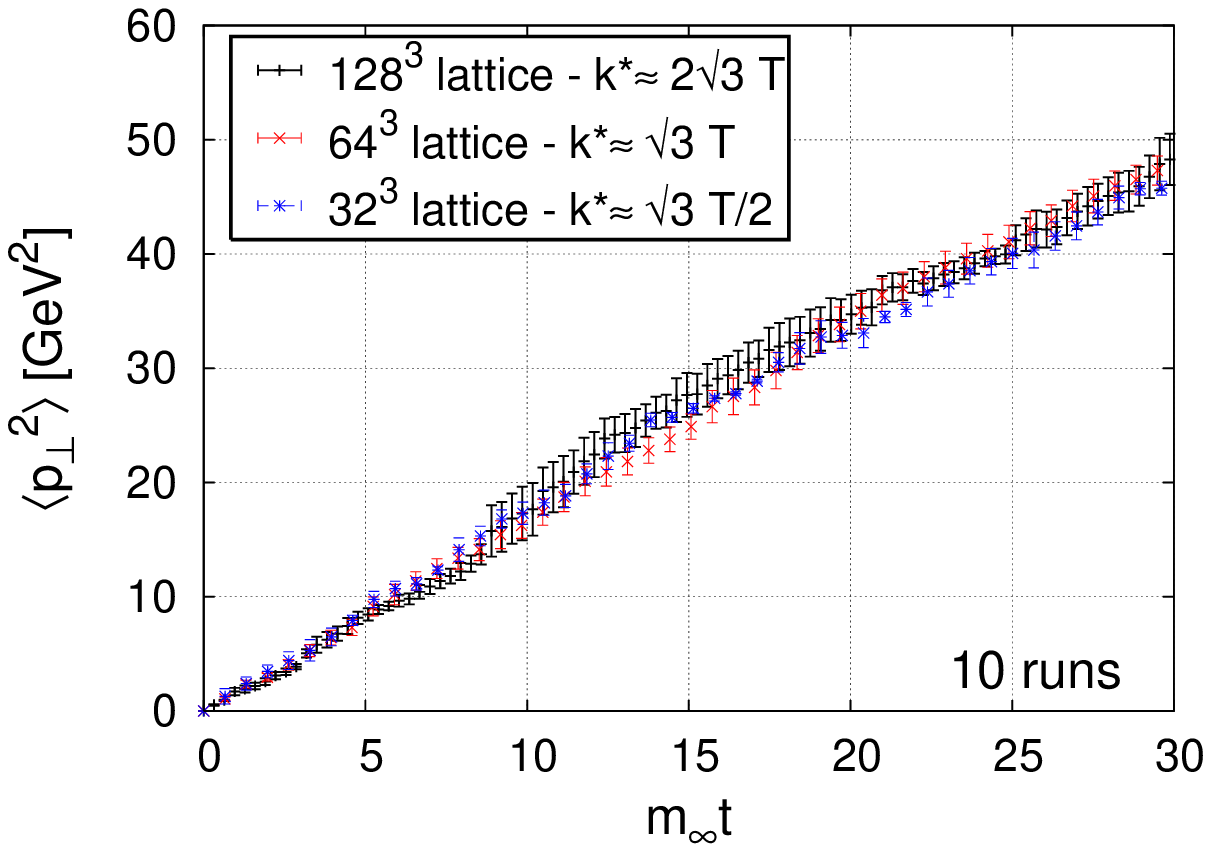}
    \caption{Jet Impulsraum-Diffusion bedingt durch unterschiedliche Anteile Feld-Teilchen Wechselwirkung und direkter bin\"arer Kollisionen. $T=4$ GeV,
    $g=2$, $n_g=10/\text{fm}^3$, $m_\infty=1.4/$fm.}
    \label{fig:zus:ptcoll}
  \end{center}
\end{figure}

Die eingangs bereits beschriebene numerische Simulation eines
nicht-Abelschen Plasmas einschlie\ss lich Feld-Teilchen und
Teilchen-Teilchen-Wechselwirkung erlaubt das Studium von Plasmen
im und weit entfernt vom Gleichgewicht. Im Falle eines isotropen
Plasmas wird die Impulsdiffusion von Teilchen mit hohem Impuls
(Jets) untersucht. Es stellt sich heraus, dass wenn die
Energiedichte und das Spektrum der Felder der Teilchenverteilung
korrekt angepasst wird, weitgehend Gitter-unabh\"angige Ergebnisse f\"ur den
Transportkoeffizienten $\hat{q}$, definiert als
\begin{equation}
    \hat{q}=\frac{1}{\sigma\lambda}\int d^2p_\perp p_\perp^2 \frac{d
    \sigma}{dp_\perp^2}\,,
\end{equation}
erzielt werden. Letztlich ist dies die Steigung des akkumulierten
$\langle p_\perp^2\rangle (t)$, gezeigt in Fig.
\ref{fig:zus:ptcoll}. Je gr\"o\ss er das Gitter, desto gr\"o\ss er
der Anteil der Felder an der Wechselwirkung zwischen den
Jet-Teilchen und dem Medium. Je kleiner das Gitter, desto gr\"o\ss
er der Anteil der direkten Kollisionen. Dennoch ist deutlich zu
sehen, dass bei verschiedenen Gittergr\"o\ss en, also
unterschiedlichen $k^*$, das gleiche Ergebnis f\"ur $\langle
p_\perp^2\rangle (t)$ erhalten wird.

Wie bereits erw\"ahnt k\"onnen auch Nichtgleichgewichtssituationen
simuliert werden. F\"ur den Fall eines Teilchenhintergrundes mit
anisotroper Impulsverteilung treten auch in der Simulation
Plasma-Instabilit\"aten auf. Erlaubt man Kollisionen zwischen den
Teilchen wird eine Verringerung der mittleren Wachstumsrate um
zwischen 10 und 15\% festgestellt. In diesem Fall ist das Ergebnis
abh\"angig von der Wahl von $k^*$. Dieses sollte jedoch wie oben erw\"ahnt
aus physikalischen Gr\"unden in den Bereich des mittleren Impulses der Teilchen gelegt werden. Kollisionen
reduzieren das Anwachsen von Instabilit\"aten also nicht dramatisch, wie
bereits aufgrund von Ergebnissen der Modell-Rechnung geschlossen
werden konn\-te.

\begin{figure}[htb]
  \begin{center}
    \includegraphics[width=12cm]{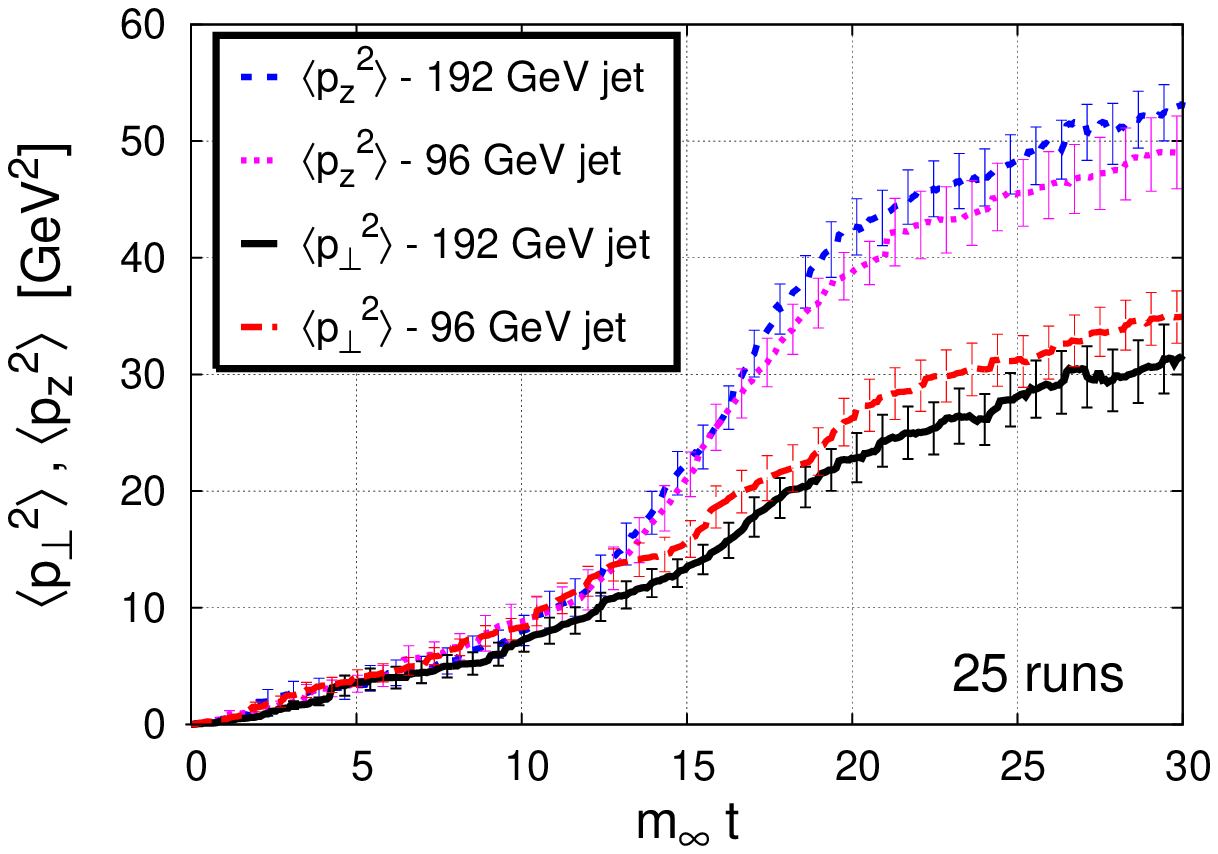}
    \caption{Impulsaufweitung von Jets in die Richtungen senkrecht zu seinem anf\"anglichen Impuls.
      $p_z$ zeigt entlang der Strahlrichtung, $p_\perp$ steht senkrecht zum Strahl.}
    \label{fig:zus:pxyjet192}
  \end{center}
\end{figure}

Betrachtet man nun, wie im isotropen Fall zuvor, Jets und misst
ihren transversalen Impuls, stellt sich heraus dass dieser entlang
der Richtung der Anisotropie $\mathbf{\hat{n}}$ st\"arker anw\"achst als
senkrecht dazu. Im Falle einer Schwer\-ionen-Kollision entspricht
dies einer st\"arkeren Aufweitung des Jets in Richtung der
Strahlachse ($z$) als senkrecht zu dieser. Dies wurde auch
experimentell in Zwei-Teilchen-Korrelationen gefunden. Fig.
\ref{fig:zus:pxyjet192} zeigt, wie w\"ahrend der Zeit in der die
Instabilit\"at anw\"achst, $\langle p_z^2\rangle$ deutlich
st\"arker ansteigt als $\langle p_\perp^2\rangle$.
Der Grund f\"ur diesen Effekt ist das Entstehen ausgedehnter
Dom\"amen von starken Farbfeldern mit $E_z>B_z$ und
$B_\perp>E_\perp$ durch die Instabilit\"at, wodurch eine
st\"arkere Ablenkung in die $z$-Richtung bewirkt wird. Dass dies
geschieht ist nicht trivial, denn im nicht-Abelschen Fall wachsen
im Gegensatz zum Abelschen Fall alle Komponenten des Farbfeldes
an. Der Quotient der Steigungen von $\langle p_z^2\rangle$ und
$\langle p_\perp^2\rangle$ im Bereich der Instabilit\"at betr\"agt
ca. 2.3, was f\"ur die Messgr\"\ss e $\langle \Delta
\eta\rangle/\langle\Delta \phi\rangle\approx 1.5$ ergibt, wobei
$\eta$ die Pseudorapidit\"at und $\phi$ der amzimuthale Winkel
ist. Chromo-\textsc{Weibel}-Instabilit\"aten im anisotropen Plasma
bieten also zumindest qualitativ eine m\"ogliche Erkl\"arung f\"ur die experimentell
bestimmte asymmetrische Aufweitung von Jets in
Schwer\-ionen-Kollisionen.

\subsection*{Fermionische kollektive Moden}
Um zu bestimmen, ob auch instabile fermionische kollektive Moden
zur Isotropisierung des Plasmas beitragen k\"onnen werden diese im
Rahmen der ``hard-loop'' (HL) N\"aherung untersucht. Es wird
sowohl numerisch als auch f\"ur bestimmte F\"alle analytisch mit
Hilfe komplexer Kontur-Integration gezeigt, dass zumindest im
Rahmen der HL-N\"aherung keine fermionischen Instabilit\"aten im
QGP auftreten.

\subsection*{Photonenproduktion als Ma\ss\, f\"ur die Anisotropie des Quark-Gluon-Plasmas}
Wie eingangs erw\"ahnt ist die St\"arke der
Impulsraum-An\-isotropie des Systems entscheidend f\"ur die Rolle,
welche Instabilit\"aten f\"ur Isotropisierung und Thermalisierung
spielen k\"onnen. Zudem gibt es keinen theoretischen Beweis
daf\"ur, dass das System tats\"achlich so schnell isotropisiert,
wie aus der Anwendbarkeit idealer Hydrodynamik auf die
Beschreibung experimenteller Daten geschlossen wurde.

%%%%%%%%%%%%%%%%%%%%%%%%%%%%%%%%%%%%%%%%%%%%%%%%%%%%%%%%%%%%%%%%%%%%%%%%%%%%%%
\begin{figure}[htb]
\begin{center}
\includegraphics[height=8cm]{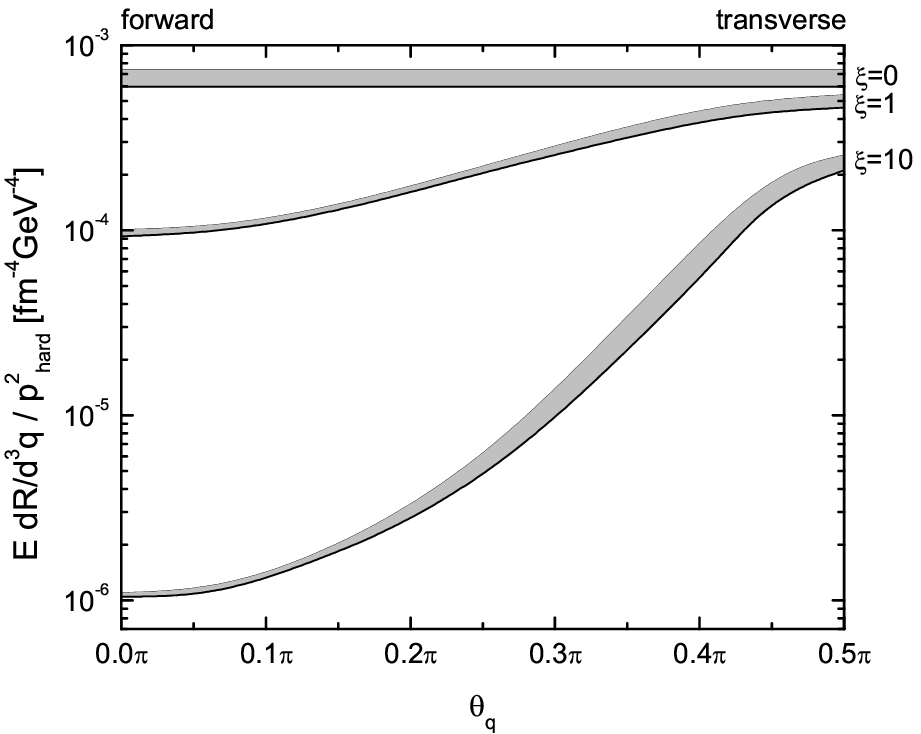}
\end{center}
\vspace{-5mm} \caption{Photonenrate f\"ur $\xi=\{0,1,10\}$ als
Funktion des Winkels zwischen Photon und Strahlachse, $\theta_q$,
f\"ur $E/p_{\rm hard} = 5$ und $\alpha_s=0.3$.}
\label{fig:zus:thetaPlot}
\end{figure}
%%%%%%%%%%%%%%%%%%%%%%%%%%%%%%%%%%%%%%%%%%%%%%%%%%%%%%%%%%%%%%%%%%%%%%%%%%%%%%
Daher ist man gezwungen nach Observablen zu suchen, welche
sensitiv auf fr\"uhe An\-iso\-tropien in den Quark und Gluon
Verteilungsfunktionen sind. Ideale Kandidaten sind
elektromagnetische Observable, wie hochenergetische Dileptonen
oder Photonen. Ihre hohe Energie l\"asst darauf schlie\ss en, dass
sie fr\"uhzeitig in der Kollision entstanden sind. Weiterhin
wechselwirken sie nur elektromagnetisch mit dem Medium, haben
daher eine lange mittlere freie Wegl\"ange und k\"onnen das Medium
nahezu immer ohne weitere Wechselwirkung verlassen. Daher bieten
sie ungest\"orte Informationen \"uber die fr\"uhen Stadien des in
Schwer\-ionen-Kollisionen erzeugten Mediums.

In dieser Arbeit wird die Produktion von Photonen in einem
Quark-Gluon-Plasma mit einer Impuls\-raum-An\-iso\-tropie, wie sie
im fr\"uhen Stadium einer Schwer\-ionen-Kollision aufgrund der
longitudinalen Expansion erwartet wird, berechnet. Dabei werden
sowohl der Mechanismus der \textsc{Compton}-Streuung, $q(\bar{q})
\, g \rightarrow q(\bar{q}) \, \gamma$, als auch der
Quark-Antiquark-An\-nihila\-tion, $q \, \bar{q} \rightarrow g \,
\gamma$, einbezogen.

Es wird gezeigt, dass die \textsc{Braaten-Yuan}-Methode zur
Regulierung der Infrarot-Divergenz der Photonenrate auch im
anisotropen Fall angewandt werden kann. F\"ur kleine
Kopplungskonstanten $g\ll 1$ ist im Bereich des geometrischen
Mittels zwischen der harten Impulsskala $T$ und der Skala der
niedrigen Impulse $gT$ die totale Rate nahezu unabh\"angig von der
Lage der Separation zwischen den beiden Bereichen. F\"ur gr\"o\ss
ere $g$ findet sich dort anstelle eines Plateaus nur ein Minimum.
In diesem Bereich kann die totale Rate berechnet werden. Fig.
\ref{fig:zus:thetaPlot} zeigt die Winkelabh\"angigkeit der Rate in
Plasmen mit unterschiedlichen Anisotropieparametern $\xi$.
%%%%%%%%%%%%%%%%%%%%%%%%%%%%%%%%%%%%%%%%%%%%%%%%%%%%%%%%%%%%%%%%%%%%%%%%%%%%%%
\begin{figure}[htb]
\begin{center}
\vspace{8mm}
\includegraphics[angle=270,width=9cm]{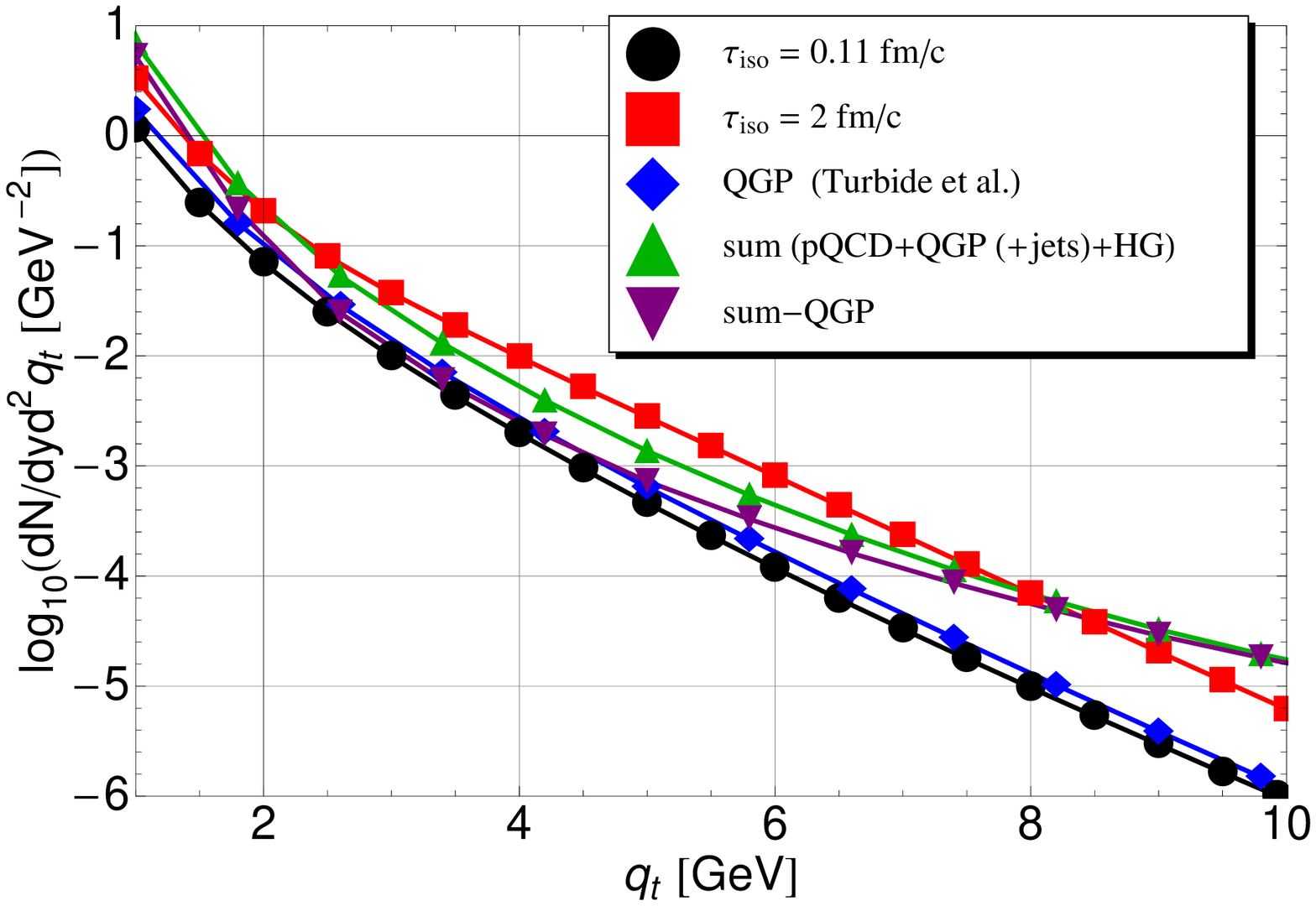}
\end{center}
\vspace{-5mm} \caption{Photonen-Yield bei festen
Anfangsbedingungen. $\sqrt{s}=5.5$ TeV, $T_0=845$ MeV,
$y_\gamma=0$. Die Rechnungen f\"ur $\tau_{\text{iso}}=\tau_0=0.11$
fm/c (reine hydrodynamische Expansion) und $\tau_{\text{iso}}=2$
fm/c werden mit den Ergebnissen von Turbide et al.
\cite{Turbide:2005fk} verglichen, wobei bei diesen das Ergebnis
f\"ur den thermischen QGP-Beitrag (auch mit
$\tau_{\text{iso}}=\tau_0=0.11$ fm/c) zus\"atzlich Bremsstrahlung
und inelastische Paar-Annihilation in f\"uhrender Ordnung
enth\"alt.} \label{fig:zus:turbidecomp1}
\end{figure}
%%%%%%%%%%%%%%%%%%%%%%%%%%%%%%%%%%%%%%%%%%%%%%%%%%%%%%%%%%%%%%%%%%%%%%%%%%%%%%
Man erkennt eine starke Winkelabh\"angigkeit der Rate in
anisotropen Plasmen. Observabel ist aber nur der totale
Ertrag an Photonen, den wir erhalten, wenn wir die Rate \"uber die
gesamte Raum-Zeit-Geschichte der Schwer\-ionen-Kollision
integrieren. Fig. \ref{fig:zus:turbidecomp1} zeigt das Ergebnis
f\"ur den Yield bei instantaner Isotropisierung und einer
Isotropisierungs\-zeit von 2 fm, in der das System vom frei
str\"omenden Zustand in hydrodynamische Expansion \"ubergeht. Zum
Vergleich sind weitere Beitr\"age, auch aus der Hadronengas-Phase
aufgef\"uhrt.

Insgesamt zeigt sich, dass obwohl die starke Abh\"angigkeit der
Rate von der St\"arke der Anisotropie sich nicht in der Gesamtzahl der Photonen
widerspiegelt, ein Effekt in der messbaren Photonen-Ausbeute bestehen
bleibt. Die experimentelle Extraktion der berechneten Unterschiede
im Bereich eines Faktors 2-6 ist aufgrund des starken
Hintergrundes, besonders auch von Photonen aus Pion-Zerf\"allen, ein
schwieriges Unterfangen. Weitere Untersuchungen des Einflusses
einer Impulsraum-Anisotropie auf die anderen Beitr\"age zur
Photonenproduktion k\"onnten weiteren Aufschluss geben und werden
in Zukunft durchgef\"uhrt werden.

  \chapter*{Abstract\markboth{}{}}
\label{abstract} In this work we study the non-equilibrium
dynamics of a quark-gluon plasma, as created in heavy-ion
collisions. We investigate how big of a role plasma instabilities
can play in the isotropization and equilibration of a quark-gluon
plasma. In particular, we determine, among other things, how much
collisions between the particles can reduce the growth rate of
unstable modes. This is done both in a model calculation using the
hard-loop approximation, as well as in a real-time lattice
simulation combining both classical \textsc{Yang-Mills}-fields as
well as inter-particle collisions. The new extended version of the
simulation is also used to investigate jet transport in isotropic
media, leading to a cutoff-independent result for the transport
coefficient $\hat{q}$. The precise determination of such transport
coefficients is essential, since they can provide important
information about the medium created in heavy-ion collisions. In
anisotropic media, the effect of instabilities on jet transport is
studied, leading to a possible explanation for the experimental
observation that high-energy jets traversing the plasma
perpendicular to the beam axis experience much stronger broadening
in rapidity than in azimuth. The investigation of collective modes
in the hard-loop limit is extended to fermionic modes, which are
shown to be all stable. Finally, we study the possibility of using
high energy photon production as a tool to experimentally
determine the anisotropy of the created system. Knowledge of the
degree of local momentum-space anisotropy reached in a heavy-ion
collision is essential for the study of instabilities and their
role for isotropization and thermalization, because their growth
rate depends strongly on the anisotropy.

  \tableofcontents
    \cleardoublepage
      \pagenumbering{arabic}
      \setcounter{page}{1}
      \chapter{Introduction}
\epigraphwidth 250pt \epigraph{Auf diesem beweglichen Erdball ist
doch nur in der wahren Liebe, der Wohlt\"atigkeit und den
Wissenschaften die einzige Freude und Ruhe.}{\emph{an Charlotte
von Stein, 6. Dezember 1781}\\
Johann Wolfgang von Goethe (1749-1832)}
\label{introduction}
\section{Nuclear matter under extreme conditions}
\begin{figure}[htb]
    \begin{center}
        \includegraphics[width=8.5cm]{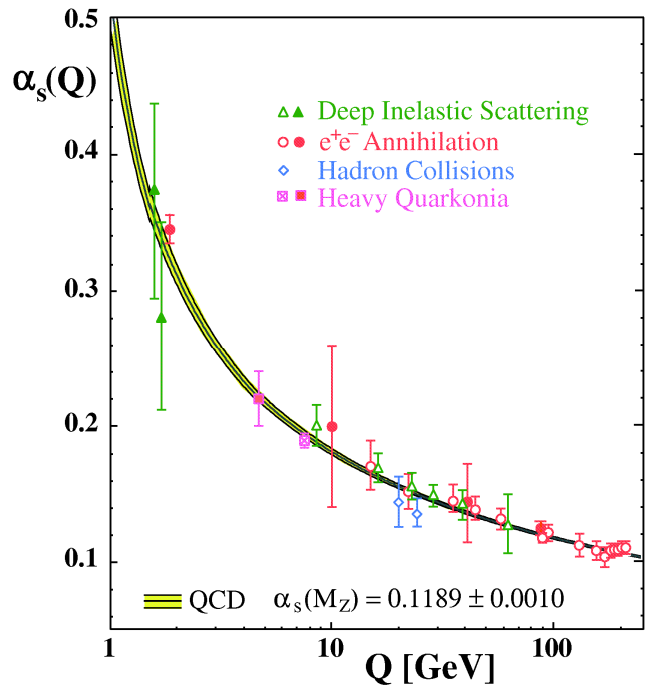}
        \caption{Summary of measurements of $\alpha_s(Q)$ as a function of the
respective energy scale $Q$. Open symbols indicate (resummed)
next-to-leading-order (NLO), and filled symbols next-to-NLO (NNLO)
QCD calculations used in the respective analysis. The curves are
the QCD predictions for the combined world average value of
$\alpha_s(M_{Z^0}) = 0.1189 \pm 0.0010$ \cite{Bethke:2006ac} (the
value of the strong coupling constant at the $Z^0$-boson mass $M_{Z^0}=91.1876 \pm 0.0021$ GeV \cite{Yao:2006px}), in
4-loop approximation and using 3-loop threshold matching at the
heavy quark pole masses $M_c = 1.5$~GeV and $M_b = 4.7$~GeV.
Figure taken from \cite{Bethke:2006ac}. Please see
\cite{Bethke:2006ac} for the references to the respective
experiments.} \label{fig:asq-2006}
    \end{center}
\end{figure}

Over thirty years ago (1974) \cite{Baym:2001in}, it was suggested
to explore new phenomena ``by distributing high energy or high
nucleon density over a relatively large volume'', to temporarily
restore broken symmetries of the physical vacuum and possibly
create novel states of nuclear matter \cite{Lee:1974ma}. At this
point, the idea of quark matter as the ultimate state of nuclear
matter at high energy density had not taken hold. However,
concurrently \textsc{Collins} and \textsc{Perry} and others
\cite{Collins:1974ky,Baym:1976yu,Freedman:1976ub,Chapline:1976gy}
realized that the asymptotic freedom of the theory of strong
interactions, quantum chromodynamics (QCD), shown in 1973 by
\textsc{Gross}, \textsc{Wilczek} and \textsc{Politzer}
\cite{Gross:1973id,Politzer:1973fx} \footnote{Gross, Politzer and
Wilczek won the 2004 Nobel Prize in physics ``for the
discovery of asymptotic freedom in the theory of the strong
interaction''\cite{Nobel:2004}}, implies the existence of an
ultra-dense form of matter with deconfined quarks and gluons,
later called the quark-gluon plasma (QGP)
\cite{Shuryak:1977ut,Shuryak:1978ij,Shuryak:1980tp,Kalashnikov:1979dp,Kapusta:1979fh}.
Fig. \ref{fig:asq-2006} shows the experimental verification of the
running of $\alpha_s$ and of asymptotic freedom in excellent
agreement with the predictions from QCD.

The possibility of accelerating and colliding uranium ions in the
CERN \footnote{\emph{European Organization for Nuclear Research},
acronym derived from the name of the council formed to set up the
laboratory: \emph{Conseil Europ\'{e}en pour la Recherche
Nucl\'{e}aire} (CERN)} Intersecting Storage Rings (ISR) to create
states of high energy density was contemplated already in the late
1960's. In the following years, due to the arising expectations
described above, heavy-ion physics was moved to the forefront as a
research tool. Heavy-ion experiments have sometimes been referred to as
little Big-Bangs in the laboratory because the produced conditions
are also expected to have existed for a brief time shortly after
the Big Bang, with temperatures exceeding 200 MeV (about $2 \times
10^ {12}$ K, which is about 100,000 times the temperature of the
core of the sun \cite{Nasa:2007}) for the first 10 $\mu$s in the
history of our universe. Systematic studies with heavy-ions
started with experiments at the BEVALAC \cite{Bevalac:2007}
\footnote{The references indicated after the facilities' names are
brief descriptions of the respective facility.} at Berkeley,
followed by the Alternating Gradient Synchrotron (AGS)
\cite{AGS:2007} at Brookhaven National Laboratory (BNL) and the
CERN Super Proton Synchrotron (SPS) \cite{SPS:2007}, culminating
at present at the Relativistic Heavy Ion Collider (RHIC)
\cite{RHIC:2007} at BNL. In the near future, even higher energies
will be explored at the CERN Large Hadron Collider (LHC)
\cite{LHC:2007}, and large net baryon densities will be studied at
RHIC and, later, at GSI-FAIR \footnote{Gesellschaft f\"ur
Schwerionenforschung, Darmstadt, Germany. Facility for Antiproton
and Ion Research at GSI.} \cite{FAIR:2007}. At RHIC, a vast data
base \cite{RHICdata:2007} on $p+p$, $D+Au$, and $Au+Au$ at a
center of mass energy $\sqrt{s}=20-200$ AGeV (GeV per nucleon) has
been harvested. Energies at the LHC will reach up to
$\sqrt{s}=5.5$ TeV for $Pb+Pb$ collisions.

To estimate when the transition to the QGP takes place, one can
calculate the pressure depending on the degrees of freedom in a
hadron gas and in the QGP using the bag model. When the pressures
become equal, a phase transition occurs. The degrees of freedom
for a gas of massless pions are 3, for a QGP with 2 active light
flavors ($u$ and $d$) they are 16 for the gluons and 24 for the
quarks and antiquarks together. Equating the resulting pressures
leads to a transition temperature $T_c\approx 160$ MeV at zero
quark chemical potential. This corresponds to an energy density of
$\varepsilon \approx 1$ GeV fm$^{-3}$, roughly 7.5 times that of
normal nuclear matter ($\varepsilon_0\approx 0.135$ GeV
fm$^{-3}$). To obtain more reliable information about the equation
of state of QCD matter or to find out in which temperature domain
the above rough estimate applies, one must turn to exact
calculations of the energy density. At present, this requires
massive computer simulations of QCD discretized on a lattice. Over
the past few years, increasingly precise lattice calculations of
thermal QCD, extrapolated to the continuum and thermodynamic
limits and to small quark masses, have become available
\cite{Allton:2003vx,Karsch:2000ps,Karsch:2001cy,Bernard:1996cs,Gupta:2003be,Fodor:2004nz,Csikor:2004me,Aoki:2006br,Cheng:2006qk}.
\begin{figure}[htb]
    \begin{center}
        \includegraphics[width=10.5cm]{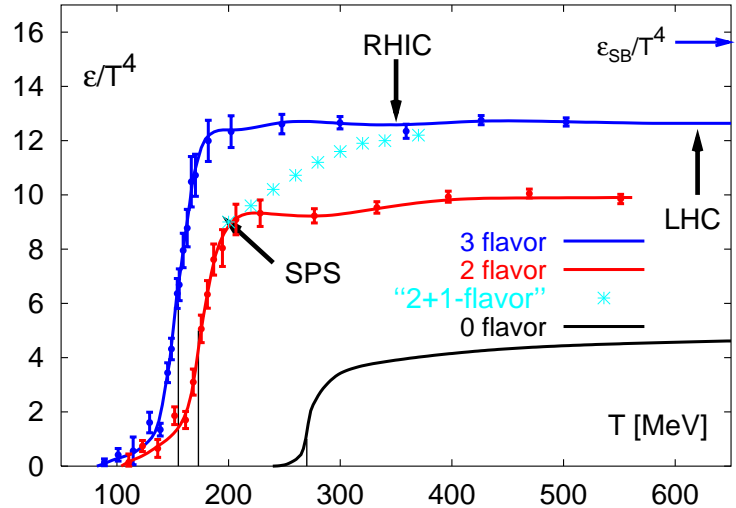}
        \caption{The energy density as a function of temperature scaled by $T^4$
from lattice QCD \cite{Allton:2003vx,Karsch:2000ps,Karsch:2001cy},
taken from \cite{Gyulassy:2004zy}. Various number of species of
quarks are considered, including the realistic case of 2+1
flavors. An estimate of the typical temperature reached at SPS and
RHIC, and estimated for LHC is included in the figure.}
\label{fig:gm01_fig1a}
    \end{center}
\end{figure}
\index{Lattice QCD} Such computations show that there is a rapid
rise of the energy density $\varepsilon(T)$ of matter when the
temperature reaches $T\approx T_c\sim 160$ MeV, about the same
temperature found for the phase transition using the rough
estimate. The energy density changes about an order of magnitude
in a narrow range of temperatures $\Delta T \sim 10$-$20$ MeV as
can be seen from Fig. \ref{fig:gm01_fig1a}. Since the energy
density, pressure and entropy are all roughly proportional to the
number of degrees of freedom, one can understand this rapid rise
in the energy density over a narrow range of temperature as a
change in the degrees of freedom between the confined and
deconfined states. The system above $T_c$ is called a plasma
because the degrees of freedom carry the non-Abelian analog of
charge, the so-called color-charge. Note that the transition
between the confined hadronic state and the deconfined QGP may or
may not be a phase transition in the strict statistical mechanical
sense. Strictly speaking, a phase transition requires a
mathematical discontinuity in the energy density or one of its
derivatives in the infinite volume limit. The QGP transition may
in fact be a ``cross over'', or rapid change, as is suggested by
numerical computations and a number of theoretical arguments.
Nevertheless, the change as measured in numerical computation is
very abrupt as seen in Fig. \ref{fig:gm01_fig1a}. Note that the
lattice calculations are performed for matter with an equal number
of baryons and anti-baryons, i.e., for vanishing baryo-chemical
potential $\mu_b$. The ab-initio evaluation of the phase boundary
in the $(T, \mu_b)$-plane poses major numerical difficulties,
related to the \textsc{Fermi-Dirac}-statistics of the quarks
(fermion-sign problem). Only recently new methods have been
developed to investigate the region of finite $\mu_b$
\cite{Allton:2003vx,Fodor:2001au,deForcrand:2002ci}. To derive the
full structure of the phase diagram one has to rely on a
combination of information from several models. The bag-model (see
\cite{BraunMunzinger:1996mq}) predicts that the critical
temperature decreases with increasing $\mu_b$. It describes a
first-order phase transition for all chemical potentials by
construction. There are also indications from various QCD-inspired
model studies, mainly the \textsc{Nambu-Jona-Lasinio} (NJL) model,
that the (chiral) phase transition is indeed a first-order one.
Since the lattice results discussed above indicate a cross over at
small $\mu_b$, this would imply the existence of a critical
endpoint in the phase diagram (see \cite{Philipsen:2007rm} for
details).
\begin{figure}[htb]
    \begin{center}
        \includegraphics[width=12.5cm]{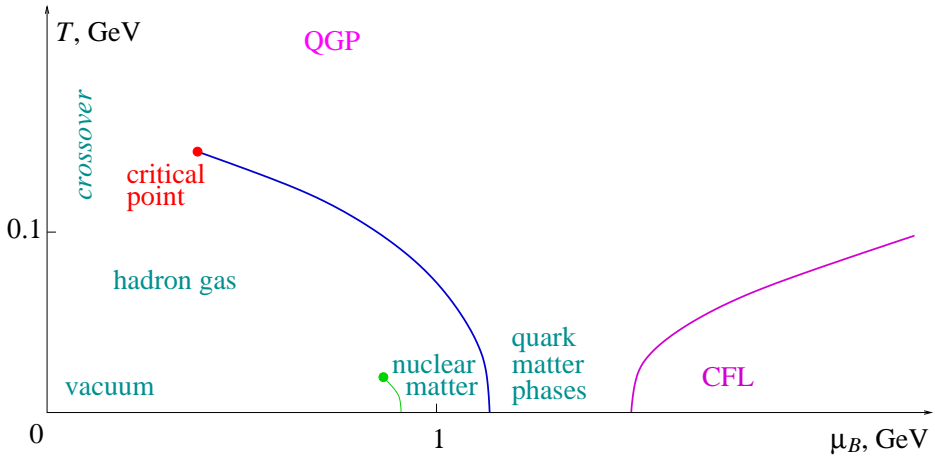}
        \caption{Semi-quantitative QCD-phase-diagram. Figure taken from \cite{Stephanov:2007fk}.}
\label{fig:pdqcd-v6}
    \end{center}
\end{figure}
Fig. \ref{fig:pdqcd-v6} shows a contemporary sketch of the phase
diagram of strongly-interacting matter, including
color-super-conducting phases at high $\mu_b$ (particularly the
color-flavor-locked (CFL) phase, shown to be present for QCD with
3 quark flavors; see \cite{Alford:2006wn} for more details). For a
recent overview of the properties of the phase-diagram of strongly
interacting matter, see \cite{BraunMunzinger:2008tz}.

\section{Important results from RHIC}
Due to the many insights into the nature of the quark-gluon plasma
the first phase of RHIC has become a great success. In its first
six runs (2000-2006), RHIC has collected data on four different
collision systems ($Au+Au$, $d+Au$, $Cu+Cu$, and $p+p$) at a
variety of energies, ranging from nucleon-nucleon center-of-mass
energies of 19.6 to 200 GeV. The largest data samples were
collected at the highest energy of 200 GeV, where the accelerator
has achieved sustained operation at four times the design
luminosity. The ability to study proton-proton, deuteron-nucleus,
and nucleus-nucleus collisions at identical center-of-mass
energies with the same detectors allows for systematic control of
the measurements to large extent. The experiments also measure the
impact parameter (distance of closest approach) and the
orientation of the reaction plane in each event. This detailed
categorization of the collision geometry provides a wealth of
differential observables, which have proven to be essential for a
precise, quantitative study of the plasma. The results obtained by
the four RHIC experiments (BRAHMS, PHENIX, PHOBOS, and STAR),
summarized in
\cite{Arsene:2004fa,Back:2004je,Adams:2005dq,Adcox:2004mh} are in
remarkable quantitative agreement with each other. The initial set
of heavy-ion results from RHIC has provided evidence for the
creation of a new state of thermalized matter at an unprecedented
energy density of (30 - 100)$\varepsilon_0$, which exhibits almost
ideal hydrodynamic behavior. Important results from the RHIC
experiments include \cite{Muller:2006ee,Muller:2007rs}
\begin{itemize}

\item chemical (flavor) and thermal equilibration of all observed hadrons
including multi-strange baryons; only the reaction $gg \rightarrow \bar{s}s$ is known
to achieve this on the time-scale of the nuclear reaction

\item strong elliptic flow, indicating early thermalization (at times on the order of
1 fm/c) and a very low viscosity of the produced medium

\item collective flow patterns described well by the recombination model, related to independently flowing valence quarks,
not hadrons

\item strong jet quenching, implying a very large parton energy loss in the
medium and a high color opacity of the produced matter

\item strong suppression of open heavy flavor mesons at high transverse
momentum, implying a large energy loss of heavy (c and b) quarks in
the medium

\item direct photon emission at high transverse momentum that remains
unaffected by the medium

\item charmonium suppression effects, likely due to the screening of color charges, that are similar to those observed at
the lower energies of the CERN SPS.
\end{itemize}

In the following we will concentrate mainly on the second and
fourth point, because this work deals particularly with the
theoretical explanation of the apparent fast isotropization and/or
thermalization, and the observed energy loss and broadening of
jets.

\section{Thermalization of the QGP and plasma instabilities}
\index{Equilibration} \index{Isotropization} The matter created in
relativistic heavy-ion collisions at RHIC manifests a strongly
collective hydrodynamic behavior \cite{Heinz:2005ja} which is
particularly evident in studies of the so-called elliptic flow
\cite{Retiere:2004wa}. When the heavy nuclei do not collide
head-on, the initial energy density is not azimuthally isotropic.
As flow follows the energy density gradient, it will be stronger
along the short overlap direction (so-called in-plane) than along
the long overlap direction (out-of-plane), leading to an azimuthal
momentum anisotropy of the particle emission. This anisotropy is
quantified by decomposing the particle azimuthal momentum
distribution in \textsc{Fourier} coefficients such as
\cite{Poskanzer:1998yz}
\begin{equation}
\frac{dN}{dp_\perp d\varphi} = \frac{dN}{dp_\perp} (1 + 2v_1\cos(\varphi) + 2v_2\cos(2\varphi) + 2v_4\cos(4\varphi) + ...) \,,
\end{equation}
where $v_2$ is the elliptic flow coefficient.\index{Elliptic flow}
For ideal hydrodynamic models to apply, it is assumed that the
stress-energy tensor is isotropic in momentum space, since having a
local momentum-space anisotropy requires the inclusion of shear
viscosity. Furthermore the assumption of local thermal equilibrium
is required, because an equilibrium equation of state (EOS) is
used. Therefore, the success of ideal hydrodynamic models in
describing the measured elliptic flow implies that the
equilibration time of the system is as short as 1 fm/c
\cite{Heinz:2004pj}. Note, however, that it has also been argued
\cite{Arnold:2004ti} that the hydrodynamic collective behavior
does not require local thermodynamic equilibrium but merely an
isotropic momentum distribution of liquid components, meaning that
the pressure in $T_{ij}=p\delta_{ij}$ does not have to be the
equilibrium pressure. Thus, the above mentioned estimate of 1 fm/c
would apply to the isotropization rather than to the
equilibration time.

Such a fast isotropization or even equilibration can be explained
assuming that the quark-gluon plasma is strongly-coupled
\cite{Shuryak:2004kh}. However, it can not be excluded that due to the
high-energy density at the early stage of the collision, when the
elliptic flow is generated \cite{Sorge:1998mk}, the plasma is
weakly-coupled because of asymptotic freedom. Thus, the question
arises whether the weakly-coupled plasma can be equilibrated or at least isotropized
within 1 fm/c.

Models that assume parton-parton collisions to be responsible
for the thermalization of weakly-coupled plasmas lead to a
longer equilibration time. Calculations performed
within the ``bottom-up'' thermalization scenario
\cite{Baier:2000sb}, where binary and $2\rightarrow 3$
collision processes are taken into account, give an equilibration
time of at least 2.6 fm/c \cite{Baier:2002bt}. 
Numerical parton-cascade simulations also including $2\leftrightarrow 3$ collisions
lead to a value of $\sim 1$ fm/c for the equilibration time \cite{Xu:2004mz}.
To thermalize the system one needs either a few hard collisions of momentum transfer
of order of the characteristic parton momentum, usually denoted by
$p_{\text{hard}}$ or $T$ as the temperature of an equilibrium
system, or many collisions of smaller momentum transfer.
For small coupling $g$, the orders of $g$ are used to classify the different time scales
in the system. When only regarding binary collisions, the inverse equilibration time is of
order $g^4 \text{ln}(g)T$ (with $g$ being the QCD coupling
constant) \cite{Arnold:1998cy}.

However, it has been proven recently that the previous
calculations of the isotropization and equilibration times had
overlooked an important aspect of non-equilibrium gauge field
dynamics, namely the possibility of plasma instabilities
\cite{Mrowczynski:1993qm,Mrowczynski:1994xv,Arnold:2004ti}. These
unstable collective modes cause the early stage QGP to have
non-perturbative occupation numbers for soft fields. These
non-perturbatively large field amplitudes ($f \sim 1/\alpha_s$,
with $\alpha_s=g^2/4\pi$) mean that even at small values of the
strong coupling constant the system can be strongly-interacting due to
strong possible coherent particle-field interactions. The possibility of generating
strongly-coupled systems from a system which does not necessarily
have a strong coupling constant is familiar from studies of
conventional QED plasmas where the electromagnetic coupling is
very small and still the system can be strongly-coupled via
collective modes.

One of the chief obstacles to thermalization in ultra-relativistic
heavy-ion collisions is the intrinsic expansion of the produced matter.
If the matter expands too quickly then there will not be
time enough for its constituents to interact before flying apart
and therefore the system will not reach thermal equilibrium.
In a heavy-ion collision the
longitudinal expansion (along the beam-line) is the most relevant,
because at early times it is much faster than the radial
expansion. At weak coupling this longitudinal
expansion causes the system to quickly become much ``colder'' in
the longitudinal than in the transverse (radial) direction,
$\langle p_L\rangle \ll \langle p_\perp \rangle$.

The question is now how long it would take for interactions to
restore isotropy in the $p_\perp-p_L$ plane. In the bottom-up
scenario \cite{Baier:2000sb} isotropy is reached through hard
collisions between the high-momentum modes (particles) which
interact via an isotropically screened gauge interaction
(\textsc{Debye}-screening). In deriving the results, it was assumed that the underlying
soft gauge modes responsible for the screening were the same in an
anisotropic plasma as in an isotropic one. In fact, this turns out
to be incorrect and in anisotropic plasmas the most important
collective mode corresponds to an instability to transverse
magnetic field fluctuations
\cite{Mrowczynski:1994xv,Arnold:2004ti}.

Equilibration is sped up by these instabilities, as growth of the
unstable modes is associated with the system's isotropization. This is due to the fact that
the unstable modes generate large longitudinal pressure, which compensates for the decreased longitudinal pressure due to expansion.
The characteristic inverse time of instability development is roughly
of order $gT$ for a sufficiently anisotropic momentum distribution
\cite{Mrowczynski:1994xv,Arnold:2004ti,Randrup:2003cw,Romatschke:2003ms,Arnold:2003rq,Rebhan:2004ur}.
Thus, the instabilities are much faster than the collisions in the
weak coupling regime (recall that the inverse equilibration was determined to be of
order $g^4 \text{ln}(g)T$ for binary collisions).

Note again that the isotropization should be clearly distinguished
from the equilibration. The instabilities driven isotropization is
a mean-field reversible phenomenon which is not accompanied by
entropy production \cite{Mrowczynski:1994xv,Dumitru:2005gp}.
Therefore, the collisions, which are responsible for the
dissipation, are needed to reach the equilibrium state of maximal
entropy. The instabilities contribute to the equilibration
indirectly, shaping the parton momentum distributions.

To understand the full dynamics, in this work we study the
interplay between collisions and instabilities for the realistic
situation in heavy-ion collisions, when the coupling is not
arbitrarily small, both in an analytic model calculation as well
as in a detailed microscopic simulation including binary hard
momentum exchange collisions as well as particle field
interactions.

\section{Hard observables}
\begin{figure}[htb]
    \begin{center}
        \includegraphics[width=10.5cm]{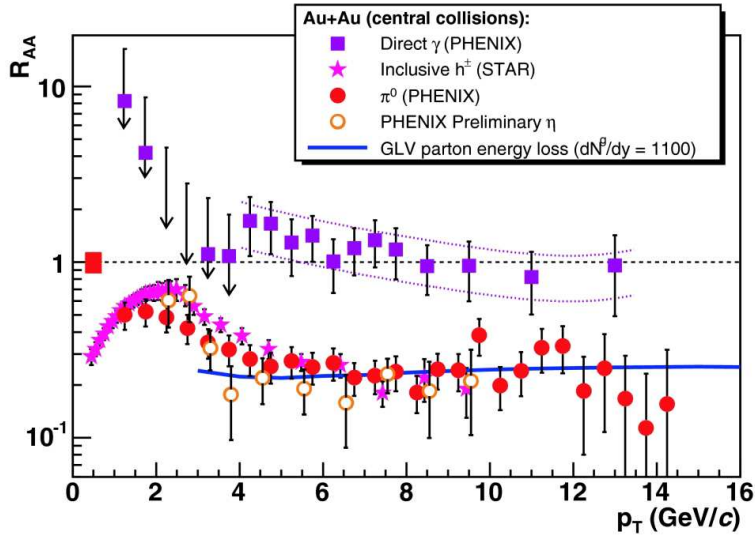}
        \caption{$R_{AA}(p_\perp)$ measured in central $Au+Au$ at $\sqrt{s}=200$ GeV for:
         direct photons \cite{Adler:2005ig}, inclusive charged hadrons \cite{Adams:2003kv}, $\pi^0$,
         and $\eta$ \cite{Busching:2005yy} compared to theoretical predictions for parton energy
         loss in a dense medium with $dN_g/dy = 1100$ \cite{Vitev:2002pf}. The
         shaded band at $R_{AA} = 1$ represents the overall fractional uncertainty
         of the data. The baseline $p+p$ reference of the direct
         $\gamma$ $Au+Au$ data is a NLO calculation whose uncertainties are
         indicated by the dotted lines around the points \cite{Adler:2005ig}. Figure taken from \cite{d'Enterria:2005cs}.}
\label{fig:raa}
    \end{center}
\end{figure}

\index{Jets} Via the hard scattering of incoming quarks and gluons
and their subsequent fragmentation into directionally aligned
hadrons, so-called jets are created. The rates for jet production
and other hard-scattering processes grow rapidly with increasing
collision energy, which was a primary motivation for constructing
RHIC with high center-of-mass energy. The investment paid off with
the discovery of jet quenching and its development as a
quantitative tomographic probe of the QGP. Jet quenching means the
suppression of high transverse momentum hadrons, such as $\pi^0$
and $\eta$ mesons in central $Au+Au$ collisions compared with
expectations from measurements in $p+p$ collisions. Whereas pions
and $\eta$ mesons show the same amount of suppression at high
$p_\perp$, where quark fragmentation is the dominant production
mechanism, direct photons were found to be
unsuppressed\footnote{Note that when using the measured $p+p$
direct-photon data as reference, direct photons are also
suppressed at high $p_\perp$ (preliminary data presented at
QM2006, see e.g. \cite{Csanad:2007fj}). If $R_{AA}^\pi =
R_{AA}^\gamma$, the whole concept of energy loss changes, since
then the modification is more likely to be an initial state effect
due to the modification of the parton distribution functions.}.
This indicates that the suppression is a final-state effect
related to the absorption (energy loss) of energetic partons in
the medium. A quantitative measure of jet quenching, shown in Fig.
\ref{fig:raa} as a function of $p_\perp$, is the nuclear
modification factor $R_{AA}$, the measured yield of hadrons
relative to the expected yield from proton-proton reactions scaled
by the ratio of the incident parton flux of two gold nuclei to
that of two protons. In contrast to the direct-photon data, a
suppression by a factor of 5 is observed for the hadrons.

Additional evidence for the color opacity of the medium is seen in studies
of the angular correlation of the radiation associated with a high-$p_\perp$ trigger
particle \cite{Adler:2002tq,Adler:2005ee}. In $p+p$ and $d+Au$ collisions, a hard recoiling hadron
frequently occurs at 180 degrees in azimuth to the trigger, showing the
back-to-back nature of jets in leading order QCD. In contrast to this, central
$Au+Au$ collisions show a strong suppression of such recoils, accompanied
by an enhancement and broadening of low-$p_\perp$ particle production. Detailed
analyses indicate that the response of the medium to the passage of an
energetic parton may be of a characteristic hydrodynamical nature: the
energy lost by a high-energy parton may re-appear as a collective Mach cone \cite{Baumgardt:1975qv,Stoecker:2007su,Baeuchle:2007qw}.

Furthermore, recent measurements of di-hadron correlations provide
evidence for an asymmetric broadening of jet profiles in the plane
of pseudo-rapidity ($\eta$) and azimuthal angle ($\phi$). The
interaction with the medium causes a much stronger broadening in
$\eta$ than in $\phi$ \cite{Putschke:2007mi,Adams:2004pa}, the so-called ``ridge''. The detailed origin of this effect is yet
unknown, and in this work we study the influence of plasma
instabilities on jet broadening that can lead to an explanation.

\section{Electromagnetic probes}
Photons, as well as dileptons, interact only electromagnetically
with the surrounding matter during the evolution of the plasma and
subsequent hadronic phases. Due to the smallness of the
electromagnetic coupling they have a very long mean-free-path and
can leave the reaction region without further interaction. This
makes them a valuable tool for gaining undistorted information on
the early stages of a heavy-ion reaction. For this reason
isotropic thermal photon production
\cite{Shuryak:1978ij,Kajantie:1981wg,Halzen:1981kz,Kajantie:1982nj,Sinha:1983jm,
Hwa:1985xg,Staadt:1985uc,Neubert:1989hu,Kapusta:1991qp,Redlich:1992fr,Aurenche:1998nw,
Aurenche:1999tq,Arnold:2001ba,Arnold:2001ms,Turbide:2003si} as
well as dilepton production
\cite{we90,gk91,ra96,br97,le98,Cooper:1998hu,Rapp:1999ej,Renk:2002md,Ruppert:2005id,Schenke:2005ry,Schenke:2006uh}
has been studied extensively.

Low-mass dileptons, for example, may provide information on chiral
symmetry restoration via the measurement of the $\rho$ meson and
its medium modifications. It is expected to become degenerate with
the $a_1$ meson, i.e., the two masses should approach each other
when chiral symmetry is restored
\cite{Brown:1991kk,Hatsuda:1991ez}. This is complicated by the
fact that the $a_1$ decays solely into hadrons, however, the
$\rho$-meson (as well as others, like the $\omega$- and $\phi$- or
$J/\Psi$-meson) has a finite branching ratio for decays into
dileptons and thus may be investigated using $e^+-e^-$ or
$\mu^+-\mu^-$ measurements (see e.g. \cite{ce95,Arnaldi:2006jq}).

Coming back to the isotropization and thermalization of the QGP,
it would be useful to have an experimental observable, which is
sensitive to a momentum-space anisotropy in the quark and gluon
distribution functions. One such observable is the high-energy
photon production. In this work we will calculate the photon
production from an anisotropic QGP and discuss under which
circumstances it may provide a means to estimate the evolving
anisotropy of the system. Knowledge of the degree of local
momentum-space anisotropy reached in a heavy-ion collision is
essential for the study of instabilities and their role for
isotropization and thermalization, as will be discussed in detail
in Chapter \ref{instabilities}.

\section{Outline of this work}
This work is organized as follows: In Chapter \ref{kintheory}, we
introduce the kinetic theory for QCD plasmas and present a
derivation of the non-Abelian \textsc{Vlasov} equations for quarks
and gluons. In Chapter \ref{HTL}, we show how by solving the
kinetic equations an effective theory for the soft plasma modes,
the so-called hard-thermal-loop effective theory, can be obtained.
We extend the discussion to anisotropic plasmas in Chapter
\ref{anisotropy}, and discuss in detail the appearing unstable
modes and the physical processes leading to them. In addition, we
present recent development, mostly of the numerical simulations of
instabilities in non-Abelian plasmas.

In Chapter
\ref{collisionsmodel}, we investigate the question whether collisions,
which can become important for a coupling that is not arbitrarily small,
reduce the growth of the unstable modes.
Therefore we present a model
for the inclusion of collisions to the non-Abelian
\textsc{Vlasov} equations, and study
their effect on instability growth.
Without collisions the system will never thermalize
because the pure \textsc{Vlasov}-equation is time reversible.
Hence, to achieve fast thermalization the instability
and the collisions are needed, and it is necessary to understand
the dynamics including both processes.

This study is then extended to
a microscopic computation, a so-called \textsc{Wong-Yang-Mills}
simulation, in which we included the possibility of hard binary
scatterings among particles in addition to the particle-field
interactions. The used techniques and implementation of the
collision term are discussed, together with the investigation of
jet transport in isotropic media in great detail in Chapter
\ref{chap:wong}, whereas the simulation of instabilities including
collisions is covered in Chapter \ref{wyminstabilities}. Here we
also present results on the influence of instabilities on jet
transport and a potential explanation for the observed asymmetric
broadening of jets along the longitudinal and transverse direction
in heavy-ion collisions.

We complete the discussion of collective modes in
Chapter \ref{fermionicmodes}, where we present studies of the
fermionic modes and show that in the hard-loop approximation no
unstable modes appear. Finally to experimentally determine how strong the parton momentum-space
anisotropy becomes in a heavy-ion collision, and by that how much
potential there is for instabilities, we study the photon
production from an anisotropic QGP and determine to what extent
the resulting yield can provide information on the system's anisotropy
in Chapter \ref{chap:photons}.

      \chapter{Kinetic theory for hot QCD plasmas}
\epigraphwidth 250pt \epigraph{Wenn man sich nur bewegt, andere in Bewegung bringt, so f\"ugt sich gar manches sch\"on und gut.}{\emph{an Friedrich Heinrich Jacobi, 18. Oktober 1784}\\
Johann Wolfgang von Goethe (1749-1832)} \label{kintheory} Kinetic
theory, as derived from quantum field theory, is a powerful tool
to construct effective theories for the soft modes of a QCD
plasma, which then can be treated non-perturbatively
\cite{Elze:1989un,Blaizot:1993zk,Blaizot:1993be,Kelly:1994dh,Blaizot:2001nr}.
Therefore, kinetic theory is an ideal tool to investigate plasma
instabilities.

For now, we assume the temperature $T$ to be high enough such that
$g(T)\ll 1$ and a weak coupling expansion can be performed. The
plasma constituents, i.e., the quarks and gluons, have typical
momenta $k\sim T$ and take part in collective excitations, which
typically develop on a space-time scale $\lambda\sim 1/(gT)$ (The
mean particle distance is $\bar{r}\sim1/T$). Those excitations are
similar to the familiar charge oscillations of the electromagnetic
plasma \cite{Lifshitz81}, and can indeed be described by simple
kinetic equations of the \textsc{Vlasov} type. By formally solving
these equations for the hard particles, one can express the
induced current in terms of the soft gauge fields and thus obtain
an effective \textsc{Yang-Mills} equation, which involves the soft
fields alone. At the softer scale $g^2T$ the non-Abelian plasmas
enter a non-perturbative regime where the coupling constant is
small but the field strengths are large such that perturbation
theory breaks down due to non-linear effects \cite{Linde:1980ts}.

Following \cite{Blaizot:2001nr}, in this chapter we will derive
the collisionless kinetic equations from the
\textsc{Kadanoff-Baym} equations in the mean-field approximation
using a gradient expansion. This leads to an effective theory for
the collective modes at the scale $gT$. We will also derive the
response function of a QCD plasma and the induced current.

\section{Non-Abelian plasmas}
For sufficiently weak gauge fields $A_a^{\mu}$ the linear response
approximation is valid even for a non-Abelian theory as QCD and
the induced current \index{Induced current} is of the form
\begin{equation}\label{linapp}
    J_{\text{ind}}^{\mu\,a}(x)=\int d^4y\,\Pi_{ab}^{\mu\nu}(x,y)A_{\nu}^b(y)\,,
\end{equation}
where the polarization tensor $\Pi_{ab}^{\mu\nu}$ is diagonal in
color $\Pi_{ab}^{\mu\nu}(x,y)=\delta_{ab}\Pi^{\mu\nu}(x,y)$ and
receives contributions from all colored particles. Considering the
scale $gT$, at which collective motion first appears, to leading
order in $g$ the color channels are decoupled and individually
conserved: $\partial_{\mu} J_{\text{ind}}^{\mu\,a}=0$. Note,
however, that in QCD the fields have to be much weaker than in QED
for the linear approximation to hold, because
$J_{\text{ind}}^{\mu\,a}$ needs to be covariantly conserved, i.e.,
\begin{equation}
    [D_{\mu},J_{\text{ind}}^{\mu}]=0\,,
\end{equation}
with the covariant derivative $D_\mu=\partial_\mu+igA_\mu$, and a
commutator of the color generators.
$J_{\text{ind}}^{\mu}=J_{\text{ind}}^{\mu\,a}T^a$, with the
generator $T^a$ in the appropriate representation for gluons
(adjoint) or quarks (fundamental). The linearized conservation law
$\partial_{\mu}J_{\text{ind}}^{\mu}=0$ is only a good
approximation to the exact law $[D_{\mu},J_{\text{ind}}^{\mu}]=0$
when the fields are so weak that the term $gA_{\mu}$ in the soft
covariant derivative can be neglected over the derivative
$\partial_x\sim gT$. This requires $A\ll T$, a limit in that all
non-linear terms in the \textsc{Yang-Mills} equation
\begin{equation}
    [D_{\nu},F^{\nu\mu}]^{a}(x)=J_{\text{ind}}^{\mu\,a}(x)
\end{equation}
can be neglected and the equation reduces to a set of uncoupled
\textsc{Maxwell} equations. This means that the linear
approximation is only good if the theory is effectively Abelian.
Otherwise, in non-Abelian theory, linear response is not
sufficient: constraints due to gauge symmetry force us to take
into account specific non-linear effects and a more complicated
formalism needs to be worked out. Still, simple kinetic equations
can be obtained in this case also, but in contrast to QED, the
resulting induced current is a non-linear functional of the gauge
fields. It may be expanded in powers of $A^µ$, thus generating the
one-particle irreducible amplitudes of the soft gauge fields:
\begin{equation}
J_\mu^a=\Pi_{\mu\nu}^{ab}A_b^\nu+\frac{1}{2}\Gamma_{\mu\nu\rho}^{abc}A_b^\nu
A_c^\rho+\dots
\end{equation}
The additional terms represent vertex corrections. These
amplitudes are the ``hard thermal loops'' (HTL)
\cite{Pisarski:1988vd,Braaten:1989mz} which define the effective
theory for the soft gauge fields at the scale $gT$. In the
following we present a derivation of the kinetic equations and the
induced current beyond the linear approximation, leading to the
same result as the diagrammatic approach which isolates the
leading order contributions to one-loop diagrams with soft
external lines.

\section{Derivation of the kinetic equations in background field gauge}
\index{Background field gauge} To preserve explicit gauge
covariance with respect to the background fields in the derivation
of the kinetic equations we follow \cite{Blaizot:2001nr} and
introduce the background field gauge
\cite{DeWitt:1967,DeWitt:1975ys,Abbott:1980hw,Hansson:1987um}. We
express the generating functional $Z[J]$ as the following
functional integral in imaginary time:
\begin{equation}\label{genfunc1}
    Z[J]=\int{\cal D} A \, \,{\rm det}\left(\frac{\delta G^a}{\delta \theta^b}\right)
    \exp\left\{-\int {\rm d}^4 x \left(\frac{1}{4}(F_{\mu\nu}^a)^2 +
        \frac{1}{2\lambda}(G^a[A])^2+J_\mu^a A_a^\mu \right)\right\}\,,
\end{equation}
where $G^a[A]$ is the gauge fixing term \index{Gauge fixing} (e.g.
$G^a=\partial^{\mu} A^a_{\mu}$ for covariant gauges or
$G^a=\partial^i A^a_i$ for \textsc{Coulomb} gauges).
$F^a_{\mu\nu}=\partial_\mu A^a_\nu-\partial_\nu A^a_\mu - g
f^{abc} A_\mu^b A_\nu^c$ is the gauge field strength tensor,
$\lambda$ is the free gauge fixing parameter, and $\theta^a(x)$
the parameter of the infinitesimal gauge transformations
\begin{equation}
    A^a_{\mu}\rightarrow A^a_{\mu} -\frac{1}{g}\partial_\mu\theta^a+f^{abc}A_\mu^b\theta^c=A^a_{\mu}-\frac{1}{g}[D_\mu,\theta]^a\,.
\end{equation}
The gauge fixed Lagrangian in Eq. (\ref{genfunc1}) is not gauge
invariant and hence the corresponding equations of motion do not
have simple transformation properties under the gauge
transformations of the external sources or the average fields. In
order to derive equations of motion which do fulfill these
properties, we use the method of the background field gauge
\cite{DeWitt:1967,DeWitt:1975ys,Abbott:1980hw}, in which the gauge
field is split into a classical background field $A_\mu^a$ and a
fluctuating quantum field $a_\mu^a$. The corresponding generating
functional is then defined to be \index{Generating functional}
\begin{equation}\label{genfunc2}
    Z[J,A]=\int{\cal D} a \, \,{\rm det}\left(\frac{\delta G^a}{\delta \theta^b}\right)
    \exp\left\{-\int {\rm d}^4 x \left(\frac{1}{4}(F_{\mu\nu}^a[A+a])^2 +
        \frac{1}{2\lambda}(G^a[a])^2+J_\mu^b a_b^\mu \right)\right\}\,,
\end{equation}
where the new gauge fixing term $G^a$ is chosen to be covariant
under the gauge transformations of the background fields. We show
explicitly in Appendix \ref{appgaugecov} that the \textsc{Coulomb} type
gauge fixing term
\begin{equation}\label{gft}
    G^a\equiv\left[D_i[A],a^i\right]^a=\partial^ia_i^a-gf^{abc}A_i^ba^{i\,c}
\end{equation}
transforms covariantly under the following gauge transformations
of the fields
\begin{align}\label{gt}
    A_\mu & \rightarrow h A_\mu h^{\dag}-\frac{i}{g}h\partial_\mu
    h^{\dag}\notag\,,\\
    a_\mu &\rightarrow h a_\mu h^{\dag}\,,
\end{align}
with
\begin{equation}\label{hotqcd:ah}
h(x)=\exp(i\theta^a(x)T^a)=1+i\theta^a(x)T^a+\mathcal{O}(\theta^2)\,.
\end{equation}
These are accompanied by the transformations
\begin{align}\label{gt+}
    J_\mu \rightarrow h J_\mu h^{\dag}\,,~~ \zeta \rightarrow h \zeta h^{\dag}\, ~\text{and~~} \bar{\zeta} &\rightarrow h \bar{\zeta} h^{\dag}
\end{align}
of the current $J_\mu$ and the anticommuting ghost fields
\index{Ghost fields} in the adjoint representation, $\zeta^a$ and
$\bar{\zeta}^a$, which appear when we rewrite the
\textsc{Faddeev-Popov} determinant as a functional integral:
\begin{equation}
{\rm det}\left(\frac{\delta G^a}{\delta \theta^b}\right)=\,
\int{\cal D}\bar\zeta \,{\cal D}\zeta \,\exp\left\{-\int {\rm d}^4
x \, \bar\zeta^a\Bigl(D_i[A] D^i[A+a]\Bigr)_{ab}\zeta^b\right\}.
\end{equation}
%This determinant involves the variation of $G^a$ in the gauge
%transformation of the total gauge field
%\begin{equation}
%    A^a_{\mu}+a^a_{\mu}\rightarrow
%    A^a_{\mu}+a^a_{\mu}-\frac{1}{g}\partial_\mu\theta^a+f^{abc}(A_\mu^b+a^b_{\mu})\theta^c=A^a_{\mu}+a^a_{\mu}-\frac{1}{g}[D_\mu[A+a],\theta]^a\,,
%\end{equation}
%where $D_\mu[A+a]=\partial_\mu+ig(A_{\mu}+a_{\mu})$ is the
%covariant derivative for the total field.
We thus obtain:
\begin{equation}\label{Zbk1}
Z[J,A]=\int{\cal D}a\,{\cal D}\bar\zeta\, {\cal D}\zeta\,
\,\exp\Bigl\{-S_{FP}[a,\zeta,\bar\zeta;A] - \int {\rm d}^4x\,
J_\mu^b a_b^\mu\Bigr\}\,,
\end{equation}
with the \textsc{Faddeev-Popov} action:
\begin{equation} \label{SFP}
S_{FP}[a,\zeta,\bar\zeta;A]=\int{\rm d}^4 x \biggl\{\frac{1}{4}
\Bigl(F_{\mu\nu}^a [A+a]\Bigr)^2+ \frac{1}{2\lambda}\Bigl(D_i[A]
a^i\Bigr)^2 +\bar\zeta^a\Bigl(D_i[A] D^i[A+a]\Bigr)_{ab}\zeta^b
\biggr\}\,.\nonumber\\\
\end{equation}
The complete action in the generating functional (\ref{Zbk1}) is
invariant with respect to the gauge transformations (\ref{gt}) and
(\ref{gt+}), and so is the generating functional itself, which can
be seen when transforming the integration variables as well. This
symmetry of $Z[J,A]$ guarantees the covariance of the
\textsc{Green} functions under the gauge transformations
(\ref{gt}) and (\ref{gt+}) of the external field and current
\cite{Blaizot:2001nr}, which was the desired property and reason
for the introduction of the background field gauge. After adding
fermionic fields and sources, the full generating functional reads
\begin{equation}\label{Zbk}
Z[j,\eta,\bar\eta,A,\Psi,\bar\Psi]\,=\, \int {\cal D}a {\cal
D}\bar\zeta {\cal D}\zeta
 {\cal D}\bar\psi {\cal D}\psi
\,\exp\biggl\{-\,S_{FP}-\int d^4x\Bigl(j_\mu^b a_b^\mu
+\bar\eta\psi+\bar\psi\eta\Bigr)\biggr\},
% + \bar C\zeta + \bar\zeta C  \nonumber\\
\end{equation}
with
\begin{align}\label{SFP0}
S_{FP}\equiv \int d^4x &\left\{
\frac{1}{4}\left(F_{\mu\nu}^a[A+a]\right)^2+
\left(\bar\Psi+\bar\psi\right)(-i\slashed{D}[A+a])\left(\Psi+\psi\right)\right.\notag\\
&~~+\left. \,\frac{1}{2\lambda}\Bigl(D_i[A] a^i\Bigr)^2 +
\bar\zeta^a\Bigl(D_i[A] D^i[A+a]\Bigr)_{ab}\zeta^b\right\}.
 \nonumber\\
\end{align}
At finite temperature, the gluonic fields to be integrated over
are periodic in imaginary time with period $\beta=1/T$:
$a_\mu(\tau=0) = a_\mu(\tau=\beta)$. The fermionic fields $\psi$,
$\bar\psi$ satisfy antiperiodic boundary conditions (e.g.,
$\psi(\tau=0) =-\psi(\tau=\beta)$), and the ghost fields $\zeta$
and $\bar\zeta$ are again periodic in spite of their
\textsc{Grassmann}ian nature: this is because the
\textsc{Faddeev-Popov} determinant is defined on the space of
periodic gauge fields. The full generating functional (\ref{Zbk})
is invariant under the gauge transformations (\ref{gt}) and
(\ref{gt+}) together with the transformations of the fermionic
fields and sources:
\begin{align}\label{GT3}
 \Psi\,\to\, h\Psi, ~~~~ &\bar\Psi \,\to\, \bar\Psi h^{-1},\nonumber\\
 \eta\,\to\, h\eta, ~~~~ &\bar\eta \,\to\, \bar\eta h^{-1}.
\end{align}
The associated \textsc{Green} functions are covariant under the
same transformations. Finally, the classical fields $A$, $\Psi$,
and $\bar\Psi$ can be identified with the respective average
fields by requiring that
\begin{equation}
\label{consist} \langle a_\mu\rangle  \,=\, \langle\psi\rangle \,=
\,  \langle\bar\psi\rangle \,=\,0\,,
\end{equation}
which are conditions that have a gauge invariant meaning because
the average values of the quantum fields transform covariantly
\cite{Blaizot:2001nr}.

With the full generating functional \index{Generating functional}
(\ref{Zbk}) at hand, we can derive the mean field kinetic
equations. First, we choose to work in \textsc{Coulomb} gauge,
which can be defined by imposing the transversality constraint
\begin{equation}
    D_i[A]a^i=0\,,
\end{equation}
within (\ref{SFP0}). In this gauge the gluon \textsc{Green}
functions are transverse, meaning that
\begin{equation}
    D_x^i[A]G_{i\nu}(x,y)=0\,.
\end{equation}
Only the physical transverse gluons are present in this gauge at
tree level, which is convenient because in other gauges
contributions from present longitudinal gluons and ghosts cancel
in the final result
\cite{Braaten:1989mz,Blaizot:1993zk,Blaizot:1993be}. Hence ghosts
shall be ignored from now on.

Ignoring the ghost fields as motivated above, the equations of
motion following from (\ref{Zbk}) read
\begin{align}\label{Acl1}
\Bigl\langle D_{ab}^\nu[A+a] F_{\nu\mu}^b[A+a]\Bigr \rangle -
g\Bigl\langle(\bar\Psi+ \bar\psi) \gamma_\mu
t^a(\Psi+\psi)\Bigr\rangle &=J_\mu^a(x),\\
\label{psicl1}
i\,\Bigl\langle\slashed{D}[A+a](\Psi+\psi)\Bigr\rangle&=\eta(x),
\end{align}
together with the hermitian conjugate equation for $\bar{\Psi}$.
Note that the first term on the left hand side of Eq. (\ref{Acl1})
is to be understood as
\begin{align}
\Bigl\langle D_{ab}^\nu[A+a] F_{\nu\mu}^b[A+a]\Bigr \rangle
=\Bigl\langle\left(\right.&\left.\partial_\nu\delta_{ab}+ig(A_\nu^c+a_\nu^c)(T^c)_{ab}\right)\notag\\&\times\left(\partial^\nu(A+a)^\mu_b-\partial^\mu(A+a)^\nu_b-g(A+a)^\nu_e(A+a)^\mu_d
f^{edb}\right)\Bigr\rangle,
\end{align}
where $(T^c)_{ab}=-if^{abc}$.
 After imposing the conditions (\ref{consist}) we obtain
the physically relevant equations
\begin{align}
\label{avA1} \left [\, D^\nu,\, F_{\nu\mu}(x)\,\right ]^a \,-\,g
\bar\Psi (x)\gamma_\mu t^a \Psi(x)
&=J_\mu^a(x)+J_\mu^{\text{ind}\, a}(x),\\
\label{avpsi1} i\slashed{D}
\,\Psi(x)&=\eta(x)+\eta^{\text{ind}}(x),
\end{align}
where $D_\mu$ and $F_{\mu\nu}$ are the covariant derivative and
field strength tensor associated with the background field only.
The induced currents on the right hand side follow from the
non-vanishing correlators of two or three fluctuating quantum
fields in Eqs.\,(\ref{Acl1}) and (\ref{psicl1}). In detail, we can
write
\begin{equation}
\label{jindfb} J^{\text{ind}\,\mu}_a(x)\,=\,J^{\mu}_{{\rm
f}\,a}(x)\,+\,J^\mu_{{\rm g}\,a}(x)\,,
\end{equation}
with the quark and gluon contributions
\begin{align}\label{jindf}
J^{\mu\,a}_{{\rm f}}(x)&= g \left\langle \bar\psi (x)
\gamma^\mu t^a \psi (x) \right\rangle,\\
J^{\mu\,a}_{\rm  g}(x)&=
 gf^{abc}\,\Gamma^{\mu\rho\lambda\nu}\left
\langle a^b_\nu\left (D_\lambda a_\rho\right )^c\right\rangle\,
+\,g^2f^{abc}f^{cde}\left\langle a_\nu^b a^\mu_d
a^{\nu}_e\right\rangle,\label{jindgG}
\end{align}
where
$\Gamma_{\mu\rho\lambda\nu}\equiv\,2g_{\mu\rho}g_{\lambda\nu}
-g_{\mu\lambda}g_{\rho\nu}-g_{\mu\nu}g_{\rho\lambda}$.
The induced fermionic source reads
\begin{equation}\label{etaind}
    \eta^{\text{ind}}(x)=g\gamma^\nu t^a\langle a_\nu^a(x)\psi(x)\rangle\,.
\end{equation}
In equilibrium, symmetry causes both the mean fields and the
induced sources to vanish, because the expectation values are then
thermal averages over color singlet states. We now introduce
notations for the appearing two-point functions (The appearing
three-point function in (\ref{jindgG}) is negligible at leading
order because it contains at least two more powers of $g$ than the
other terms). The usual quark and gluon propagators read
\begin{align}\label{nor}
S_{ij}(x,y)&\equiv\langle{\rm T}\psi_i(x)\bar\psi_j(y)\rangle
=-\frac {\delta \langle \psi_i(x)\rangle }{\delta\eta_j(y)}, \nonumber\\
G_{\mu\nu}^{ab}(x,y)&\equiv\langle{\rm T}a_\mu^a(x)a_\nu^b(y)\rangle
=-\frac{\delta\langle a_\mu^a(x)\rangle}{\delta j^\nu_b(y)},
\end{align}
with the time-ordering operator $T$. We shall also define the
following ``abnormal'' propagators:
\begin{align}\label{abn}
K_{i\nu}^b(x,y)&\equiv\langle{\rm T}\psi_i(x)a^b_\nu(y)\rangle
=-\frac{\delta\langle \psi_i(x)\rangle }{\delta j_b^\nu(y)}
=-\frac{\delta \langle a_\nu^b(y)\rangle}{\delta\bar\eta_i(x)},
\nonumber\\
H_{\nu i}^b(x,y)&\equiv\langle{\rm T}
a^b_\nu(x)\bar\psi_i(y)\rangle=-\frac{\delta
\langle \bar\psi(y)\rangle}{\delta
j_b^\nu(x)}=\frac{\delta \langle a_\nu(x)\rangle}{\delta\eta(y)},
\end{align}
which mix fermionic and bosonic degrees of freedom and vanish in equilibrium.

We can express the time-ordered propagators by
\begin{align}\label{analytic}
S_{ij}(x,y) &=\theta(\tau_x-\tau_y)
S^>_{ij}(x,y)-\theta(\tau_y-\tau_x)S^<_{ij}(x,y) \,,\notag\\
G_{\mu\nu}^{ab}(x,y) &=\theta(\tau_x-\tau_y)
G_{\mu\nu}^{>\,ab}(x,y)-\theta(\tau_y-\tau_x)G_{\mu\nu}^{<\,ab}(x,y) \,,
\end{align}
and find analogous expressions for $K$ and $H$.

The induced sources in (\ref{jindf}) - (\ref{etaind}) involve products of fields with equal time arguments and may be expressed by
\begin{align}\label{jf}
J_{{\rm f}}^{\mu\,a}(x)&=g\,{\rm Tr}\left(\gamma^\mu
 t^aS^<(x,x)\right),\nonumber\\
J^{\mu\,a}_{\rm  g}(x)&=i\,g\,\Gamma^{\mu\rho\lambda\nu}\,
{\rm Tr}\,\,T^a\,
 D^x_\lambda\,G_{\rho\nu}^<(x,y) |_{y\to x^+},\nonumber\\
\eta^{\text{ind}}(x)&=g\gamma^\nu t^aK^<_{a\,\nu}(x,x).
\end{align}
To arrive at the second equation it was used that
\begin{align}
    i{\rm Tr}\left(T^a D^x G^<(x,y)\right)&=i N D^x G^{<\,a}(x,y)=i (T^a)_{bc} D^x(G^{<\,a}(x,y)T^a)^{cb}\notag\\
    &=f^{abc}D^x G^{<\,cb}(x,y)=f^{abc}\langle a(y)^b (D a(x))^c\rangle\notag\,,
\end{align}
because $T^aT^a=N$.

\subsection{Mean field approximation}
\index{Mean field approximation} 
Approximations are needed to
develop kinetic equations for the QCD plasma. The first
approximation is a mean field approximation, equivalent to the
one-loop approximation in the diagrammatic approach. 
First, let us write down the general equation of motion for the quark propagator, 
found by differentiating Eq. (\ref{psicl1}) with respect to $\eta(y)$:
\begin{equation}\label{eqmsigma}
 -i\slashed{D}_x S(x,y)-g \gamma^\nu t^a \Psi(x) H_\nu^a (x,y)+ \int d^4z\, \Sigma(x,z) S(z,y)=\delta(x-y)
\end{equation}
with the quark self-energy $\Sigma(x,y)$, defined as
\begin{equation}
 	\int d^4z\, \Sigma(x,z) S(z,y) \equiv \frac{\delta \eta^{\text{ind}}(x)}{\delta \eta(y)}=g \langle T\slashed{a}(x)\psi(x)\bar{\psi}(y)\rangle_c\,.
\end{equation}
The equations of motion for the other propagators are found analogously.
To justify the mean field approximation,
where the hard particles only interact via the soft mean fields, i.e., where we neglect the last term in 
Eq. (\ref{eqmsigma}), let us analyze the magnitude of the different terms in the equations of motion.
For simplicity let us consider the gluon propagator $G(x,y)$ in a
soft color background field $A_a^\mu(x)$. Its \textsc{Wigner}
transform will obey a kinetic equation, which involves a drift
term $(K\cdot\partial_X)G^<$, a mean field term and a collision
term $C(K,X)=-(G^>\Sigma^<-\Sigma^>G^<)$.\footnote{Note, that we
will use capital letters for four-momenta to avoid confusion with
the absolute values of three-momenta, for which we will use lower
case letters.} Both the drift and the collision term vanish in equilibrium
\cite{Blaizot:2001nr}. Considering small fluctuations of the \textsc{Green}
function and self energy
\begin{align}
    G^<(K,X)&=G_{\text{eq}}(K)+\delta G^<(K,X)\notag\\
    \Sigma^<(K,X)&=\Sigma^<_{\text{eq}}(K)+\delta\Sigma^<(K,X)\notag\,,
\end{align}
the drift term becomes $(K\cdot\partial_X)\delta G^<$, while the
collision term reads
\begin{equation}
    C(K,X)=-(\Sigma^<_{\text{eq}}\delta G^>-\Sigma^>_{\text{eq}}\delta G^<)+(\delta\Sigma^> G^<_{\text{eq}}-\delta\Sigma^< G^>_{\text{eq}}))+\dots\,,
\end{equation}
where the dots stand for higher order deviations from equilibrium.
The order $\Sigma_{\text{eq}}(k)\sim g^2T^2$ is fixed by the
physics in equilibrium, such that the order of the drift term,
which is given by the scale $\partial_X$ of inhomogeneities,
determines the relative importance of the two terms. For
$\partial_X\sim gT$ the collision term is suppressed by one order
of $g$ relative to the drift term, whereas for $\partial_X\sim
g^2T$ or less, the collisions are as important as the drift term.
Alterations of this argument due to accidental cancellations are
discussed in \cite{Blaizot:2001nr}. Regarding strict perturbation
theory to leading order for the collective dynamics at the scale
$gT$ the collision terms can be neglected and the mean field
approximation is valid. That is, the hard particles only interact
with the soft mean fields and do not collide among each other. 
In this limit, the relevant equations for the two-point functions
read in \textsc{Coulomb} gauge:
\begin{align}\label{S10}
\slashed{D}_x S^<(x,y)&=ig\gamma^\nu t^a\Psi(x)H^{<\,a}_\nu(x,y)\,,\\
\label{K10} \slashed{D}_x
K^{<\,b}_\nu(x,y)&=-ig t^a\gamma^\mu\Psi(x)G^{<\,ab}_{\mu\nu}(x,y)\,, \\
\label{K11}
\Bigl(g_{\mu\nu}D^2- D_\mu D_\nu +
2igF_{\mu\nu}\Bigr)_y^{ab}
 K^{<\,\nu}_b(x,y)&=-gS^<(x,y)\gamma_\mu t^a\Psi(y)\,,\\
\label{D10}
\Bigl(g_{\mu}^\rho D^2 - D_\mu D^\rho
+2igF_{\mu}^{\,\,\,\rho}\Bigr)_x^{ac}G^{<\,cb}_{\rho\nu}
(x,y)&=g\bar\Psi(x)\gamma_\mu t^a K_\nu^{<\,b}(x,y)
+ g H_\nu^{<\,b}(y,x)\gamma_\mu t^a\Psi(x)\,.
\end{align}
The \textsc{Coulomb} gauge fixing conditions read
\begin{align}
\label{TR} D^i_x G_{i\nu}^<(x,y)=0\,,&{}\qquad\,\,
G_{\mu j}^<(x,y) D^{j \dagger}_y=0\,,\nonumber\\
D^i_x H^{<\,a}_i(x,y)=0\,,&{}\qquad\,\,D^i_y K^{<\,a}_i(x,y)=0\,,
\end{align}
and the initial conditions are to be chosen such that without external sources the system is in equilibrium:
The mean fields vanish and the two-point functions reduce to their equilibrium form, which, to the order of interest, are the free functions.

Under the gauge transformations (\ref{gt}) and (\ref{GT3}) the
equations (\ref{S10}) - (\ref{D10}) transform covariantly. Solving
these equations without further approximations would yield the
induced sources to one-loop order. However, by itself the mean
field approximation is not consistent: Additional powers of $g$
are hidden in the soft non-equilibrium inhomogeneities. These can
be isolated by the gradient expansion.

\subsection{Gradient expansion}
\index{Gradient expansion} To set the stage for the gradient
expansion we first rewrite the two-point functions in terms of
\textsc{Wigner} functions. Let $G_{ab}(x,y)$ be a generic
two-point function. Its \textsc{Wigner} transform is
\begin{equation}
    \mathcal{G}_{ab}(K,X)=\int d^4 s \,
    G_{ab}\left(X+\frac{s}{2},X-\frac{s}{2}\right)\,,
\end{equation}
with
\begin{equation}\label{relcent}
    s^\mu=x^\mu-y^\mu, ~~~~~~~~X^\mu=\frac{x^\mu+y^\mu}{2}\,.
\end{equation}
 The problem with this definition is that unlike
$G_{ab}(x,y)$, which is gauge covariant at $x$ and $y$ separately,
$\mathcal{G}_{ab}(K,X)$, which mixes $x$ and $y$ in its
definition, is not covariant. It is, however, possible to
construct a gauge covariant \textsc{Wigner} function. First we
define
\begin{equation}\label{newg}
    \tilde{G}_{ab}(s,X)=U_{a\bar{a}}\left(X,X+\frac{s}{2}\right)G_{\bar{a}\bar{b}}\left(X+\frac{s}{2},X-\frac{s}{2}\right)U_{\bar{b}b}\left(X-\frac{s}{2},X\right)\,,
\end{equation}
using the non-Abelian transporter or \textsc{Wilson} line
\begin{equation}\label{wilson}
    U(x,y)=P \exp\left(-ig\int_{\mathcal{C}} dz^\mu\,A_\mu(z)\right)\,,
\end{equation}
where $\mathcal{C}$ is an arbitrary path going from $y$ to $x$,
$A_\mu=A_\mu^aT^a$, and the symbol $P$ denotes the path ordering
of the color matrices in the exponential. Under the gauge
transformations of $A_\mu$ the \textsc{Wilson} line transforms as
\begin{equation}
    U(x,y)\rightarrow h(x) U(x,y) h^{\dag}(y)\,,
\end{equation}
such that the function (\ref{newg}) transforms covariantly at $X$ for any given $s$:
\begin{equation}
    \tilde{G}(s,X)\rightarrow h(X) \tilde{G}(s,X)h^{\dag}(X)
\end{equation}
Its \textsc{\textsc{Wigner}} transform
\begin{equation}
    G_{ab}(K,X)=\int d^4s\,e^{i K\cdot s}\tilde{G}_{ab}(s,X)
\end{equation}
transforms covariantly as well: For any given $k$ we have
\begin{equation}
    G(K,X)\rightarrow h(X) G(K,X)h^{\dag}(X)\,,
\end{equation}
and this \textsc{Wigner} function will satisfy a gauge covariant
equation of motion, which makes the physical interpretation more
transparent than one that is not gauge covariant. This way all
two-point functions can be expressed by the corresponding gauge
covariant \textsc{Wigner} functions, where we will not use
different notations, because their arguments clearly identify
them.

The induced sources (\ref{jf}) can now be expressed in terms of
the newly defined \textsc{Wigner} functions:
\begin{align}
\label{jf1} J_{{\rm  f}}^{\mu\,a}(X) &= g\int\frac{{\rm d}^4
k}{(2\pi)^4}
\,{\rm Tr}\,\Bigl(\gamma^\mu t^a S(K,X)\Bigr)\,,\\
\label{jg1} J^{\mu\,a}_{\rm  g}(X)&= g
\,\Gamma^{\mu\rho\lambda\nu} \int\frac{{\rm d}^4k}{(2\pi)^4}\,{\rm
Tr}\,T^a \biggl\{ k_\lambda G_{\rho\nu}(K,X)+\frac{i}{2}\Bigl[
D_\lambda^X,\,G_{\rho\nu}(K,X)\Bigr] \biggr\}\,,\\
\label{eind1} \eta^{\text{ind}}(X)&=g\int\frac{{\rm
d}^4k}{(2\pi)^4}\,
 \gamma^\mu t^a K_\mu^a(K,X)\,.
\end{align}
The path $\mathcal{C}$ is still arbitrary and in fact, the results
will not depend on the exact form of the path, because we need
$U(x,y)$ only in situations where $x$ is close to $y$ as we will
argue below. Choosing $\mathcal{C}$ to be a straight line
connecting $y$ and $x$, the transition from the non-covariant to
the gauge covariant \textsc{Wigner} function corresponds to the
replacement of the canonical momentum $K^\mu=i\partial_s$ by the
kinetic momentum $P^\mu=K^\mu-gA^\mu(X)$ as discussed in
\cite{Heinz:1984yq},\cite{Elze:1986qd} (also see
\cite{Gyulassy:1986da} for details). For soft and rather weak
background fields, $\tilde{G}(s,X)$ stays close to its equilibrium
value, meaning that it is peaked at $s=0$ and vanishes for
$s\gtrsim 1/T$. Over this short range the mean field does not vary
significantly and we may write
\begin{equation}
    g \int_{\mathcal{C}}dz^\mu\,A_\mu(z)\approx g(s\cdot A(X))\,,
\end{equation}
omitting terms which involve at least one soft derivative
$\partial_X A\sim gTA$. For $s\sim 1/T$, $g(s\cdot A) \sim g$,
because $gA\sim gT$, and we can expand the exponent in
(\ref{wilson}) giving
\begin{equation}
    U_{ab}(x,y)\approx \delta_{ab}-ig(s\cdot A_{ab}(X))\,.
\end{equation}
For the off-equilibrium derivation $\delta G=G-G_{\text{eq}}$, this yields
\begin{equation}
    \delta\tilde{G}(s,X)\simeq\delta G(x,y)+ig(s\cdot A(X))G_{\text{eq}}(s)\,,
\end{equation}
and for its \textsc{Wigner} transform
\begin{equation}\label{gaugecovwigner}
    \delta G(K,X)\simeq\delta \mathcal{G}(K,X)+g(A(X)\partial_k)G_{\text{eq}}(K)\,.
\end{equation}
Using
\begin{equation}
    \left.D_x^\mu G(x,y)\right|_{x=y}=\left.\partial_s^\mu \tilde{G}(s,X)\right|_{s=0}\,,
\end{equation}
we can perform a similar simplification on the gluonic current (\ref{jg1}) to get:
\begin{equation}
\label{jb1} J^{\mu\,a}_{\rm  g}(X)=g\int\frac{{\rm
d}^4k}{(2\pi)^4}\,{\rm Tr} \,T^a \Bigl\{- K^\mu \delta
G_{\nu}^{\,\,\,\nu}(K,X) + \delta G^{\mu\nu}(K,X)K_\nu \Bigr\}\,.
\end{equation}
Since the gauge-fixing conditions (\ref{TR}) imply that the gauge
covariant gluon \textsc{Wigner} function is spatially transverse,
we can write
\begin{equation}\label{trans}
    \delta G(K,X)=(\delta_{ij}-K_i K_j)\delta G(K,X)\,.
\end{equation}
The spatial components are the only ones to contribute to the
induced current to leading order in $g$. This is specific to the
\textsc{Coulomb} gauge and shall be shown later. With that and Eq.
(\ref{trans}) we can write \index{Induced current}
\begin{equation}\label{jb12}
J^{\mu\,a}_{\rm  g}(X)=2 g\int\frac{{\rm d}^4k}{(2\pi)^4}\,K^\mu
\,{\rm Tr} \,\Bigl\{T^a \delta G(K,X)\Bigr\}\,.
\end{equation}

We are now ready to perform a gauge covariant gradient expansion
\index{Gradient expansion} and by that extract the terms of
leading order in $g$ in Eqs. (\ref{S10})-(\ref{D10}).

Consider the gluon two-point function $G^{<\,ab}_{\mu\nu}(x,y)$ in
the presence of a soft background field $A_a^\mu$, but without
fermionic fields ($\Psi=\bar{\Psi}=0$). We start with the
following \textsc{Kadanoff-Baym} equations (cf. Eq. (\ref{D10}))
\index{Kadanoff-Baym equations}:
\begin{align}\label{MFEQ}
\left(g_\mu^{\,\,\rho}D^2-D_\mu D^\rho+ 2igF_{\mu}^{\,\,\rho}
\right)_xG^<_{\rho\nu}(x,y)&=0,\nonumber\\
G_{\mu}^{<\,\rho}(x,y)\left(g_{\rho\nu}\bigl(D^\dagger\bigr)^2
- D_\rho^\dagger D_\nu^\dagger + 2ig F_{\rho\nu}\right)_y&=0\,,
\end{align}
and take their difference, which then includes terms like
\begin{equation}
    \label{Xi}
    \Xi(x,y)\equiv D^2_x G(x,y)\,-\,G(x,y) (D^\dag_y)^2\,,
\end{equation}
where $D_x^2\,=\,\partial_x^2+2igA\cdot\partial_x
+ig(\partial\cdot A)-g^2A^2\/$, and \textsc{Minkowski} indices are
omitted to simplify notation. We will concentrate on the quantity
$\Xi$ to illustrate the performed approximations.

After introduction of the relative and center variables
(\ref{relcent}), we typically have $s\sim 1/T$, $\partial_s\sim
T$, $\partial_X\sim gT$, meaning that the system is slowly varying
in space and time with respect to the center variable. We expand
in powers of the soft derivative $\partial_X$ and keep only terms
involving up to one soft derivative. For example
\begin{equation}
    A_\mu\left(X+\frac{s}{2}\right)\approx A_\mu(X)+ \frac{1}{2}(s\cdot \partial_X) A_\mu(X)\,.
\end{equation}
We find after explicit insertion of $D_x^2$ and $(D_y^\dag)^2$:
\begin{align} \label{DIFF1}
\Xi(s,X)&= 2\partial_s\cdot\partial_X G
+2ig\Bigl[A_\mu(X),\partial^\mu_s G\Bigr]
+ig\Bigl\{A_\mu(X),\partial^\mu_X G\Bigr\}
+ig\Bigl\{(s\cdot\partial_X)A_\mu,\partial^\mu_s G\Bigr\}\nonumber\\
&~~~+ig\Bigl\{(\partial_X\cdot A), G\Bigr\} -
g^2\left[A^2(X),G\right]
-\frac{g^2}{2}\left\{(s\cdot\partial_X)A^2,G\right\}\,+\,...\,,
\end{align}
where $[\cdot,\cdot]$ denote commutators and $\{\cdot,\cdot\}$
anticommutators of color matrices.

As done in \cite{Blaizot:2001nr} we use $A\sim T$ and $\delta
G=G-G_{\text{eq}}\sim g G_{\text{eq}}$, with $G_{\text{eq}}\approx
G_0$ in the present approximation. Note that at the momentum scale
$gT$ $A$ is actually of order $\sqrt{g}T$. \footnote{\label{orderofA} To see this,
consider the free propagator of the magnetic gluon in imaginary
time:\begin{equation}
    \langle A(\tau,x)A(0)\rangle=T\sum_n\int\frac{{\rm
    d}^3k}{(2\pi)^3}e^{-i\omega_n\tau+i\mathbf{k}\cdot\mathbf{x}}\frac{1}{k^2+\omega_n^2}\notag
\end{equation}
By letting $\tau\rightarrow 0$ and $x\rightarrow 0$ and keeping
only the contribution of the static modes ($\omega_n=0$) one
obtains for momenta of order $gT$: \begin{equation} \langle
A^2\rangle\simeq T\int\frac{{\rm d}^3k}{k^2}\sim g T^2\,,\notag
\end{equation}
so that $|A|=\sqrt{\langle A^2 \rangle}\sim \sqrt{g}T$.} However,
for now we assume it to be larger in order to include all
non-linear effects in the theory. We will later use the fact that
$A\sim\sqrt{g}T$ to linearize the theory.

Keeping only terms of leading order in $g$, (\ref{DIFF1})
simplifies to
\begin{equation}
\label{DIFF2} \Xi(s,X)\approx 2(\partial_s\cdot\partial_X) \delta
G +2ig\left[A_\mu, \partial^\mu_s \delta G\right]
+2ig(s\cdot\partial_X)A_\mu\,(\partial^\mu_s G_0) +2ig (\partial_X
\cdot A) G_0,\,\,
\end{equation}
where all terms on the right hand side are of order $g^2T^2G_0$.
\textsc{Fourier} transformation leads to
\begin{equation}
\Xi(K,X)\,\approx\,2\Bigl[K\cdot D_X,\, \delta {\cal G}(K,X)\Bigr]
+ 2 gK^\mu \Bigl(\partial_X^\nu A_\mu(X)\Bigr)\partial_\nu
G_0(K)\,.
\end{equation}
The ordinary \textsc{Wigner} transform $\delta {\cal G}(K,X)$ can
be expressed by the gauge covariant one using
(\ref{gaugecovwigner}), which finally leads to:
\begin{equation}
\label{KIN} \Xi_{\mu\nu}(K,X)\,\approx \,2\Bigl[K\cdot D_X, \delta
G_{\mu\nu}(K,X)\Bigr] -
 2 gK^\alpha F_{\alpha\beta}(X) \,\partial^\beta G_{\mu\nu}^{(0)}(K)\,,
\end{equation}
where we reintroduced the \textsc{Minkowski} indices.

We recognize here the familiar structure of the \textsc{Vlasov}
equation, generalized to a non-Abelian plasma: Eq.~(\ref{KIN})
involves a (gauge-covariant) drift term $(K\cdot D_X)\delta G$,
together with a ``force term'' proportional to the background
field strength tensor. In fact, this ``force term'' involves the
{equilibrium} distribution function $G_0\equiv G^<_0$, so, in this
respect, it is closer to the { linearized} version of the
\textsc{Vlasov} equation. However, Eq.~(\ref{KIN}) is still
non-linear, because of the presence of the covariant drift
operator $(K\cdot D_X)$, and because the non-Abelian field
strength tensor is itself non-linear.

\section{The non-Abelian Vlasov equations}
\label{derivevlasov} \index{Vlasov equation} We will now derive
the kinetic equations which determine the color current induced by
a soft gauge field $A_a^\mu$, disregarding the fermionic mean
fields. This is done using the expressions for the currents
(\ref{jf1}) and (\ref{jb12}). In order to evaluate these
expressions, we need the equations satisfied by the appearing
quark and gluon \textsc{Wigner} functions $\delta S$ and $\delta
G_{\mu\nu}$ in the presence of the background field $A_a^\mu$.

\subsection{Vlasov equation for gluons}
\index{Vlasov equation!Gluon Vlasov equation} We start with the
determination of the kinetic equation for the gluon
\textsc{Wigner} function. The transverse components $\delta
G_{ij}$ are expected to be dominant, and we concentrate on the
spatial components of Eq. (\ref{MFEQ}):
\begin{align}
\label{MFIJ} D_x^2
G_{ij} - D^x_iD^x_0G_{0j}+ 2igF_{i}^{\,\rho}(x)
G_{\rho j}&=0,\nonumber\\
G_{ij}\Bigl(D_y^\dagger\Bigr)^2 - G_{i0} D_{0\,y}^\dagger
D_{j\,y}^\dagger + 2igG_{i\rho}F^{\rho}_{\,j}(y) &=0,
\end{align}
where we used the gauge fixing conditions (\ref{TR}) for
simplifications. As in the last section we take their difference
and meet
\begin{equation}
\label{GIJ1} D_x^2 G_{ij} -G_{ij}\Bigl(D_y^\dagger\Bigr)^2\,
\longrightarrow \, 2\Bigl[K\cdot D_X, \delta G_{ij}\Bigr]
 - 2 gK^\alpha F_{\alpha\beta}(X) \,\partial^\beta G^{(0)}_{ij}(k).
\end{equation}
It was stated in Eq. (\ref{trans}) that $\delta G_{ij}$ is
transverse. However, the second term on the right hand side of Eq.
(\ref{GIJ1}) involves non-transverse components:
\begin{align}\label{ID1}
\lefteqn{ K^\alpha F_{\alpha\beta}\partial^\beta
G^{(0)}_{ij}(K)\equiv K^\alpha F_{\alpha\beta}\partial^\beta [
(\delta_{ij}-\hat K_i\hat K_j)G_0(K)]}\nonumber\\
& &=(\delta_{ij}-\hat K_i\hat K_j)K^\alpha
F_{\alpha\beta}\partial^\beta G_0 -K^\alpha F_{\alpha
l}\,\frac{K_i\delta_{jl} +K_j\delta_{il}-2\hat K_i\hat K_j K_l}
{{\bf k}^2}\,G_0.\,\,\,
\end{align}
because the derivative acts on both the projector and $G^{0}$. It
can be shown \cite{Blaizot:2001nr} that the non-transverse
components cancel with other terms present in the full difference
equation, such that Eq. (\ref{trans}) holds and $\delta G(K,X)$
satisfies
\begin{equation}\label{VLAS}
\Bigl[K\cdot D_X,\,\delta G(K,X)\Bigr] \,=\,g\,K^\alpha
F_{\alpha\beta}(X)\partial^\beta G_0(K)\,.
\end{equation}
Indeed it follows that the transverse part is dominant 
because $D_X\sim gT$ and $g F_{\alpha\beta}\sim (D_X)^2\sim g^2
T^2$ and $k\sim T$ we have
\begin{equation}
    \delta G \sim (D_X/T) G_0 \sim g G_0 \,,
\end{equation}
whereas $G_{0 j} \sim (g F_{0i}/T^2) G_0 \sim g^2 G_0$ is
suppressed by one order of $g$.

Remember that $G_0\equiv G_0^<(K)= 2 \pi
\left(K_0/|K_0|\right)\delta(K^2) n^g(K)$, which only has support
on the tree-level mass shell $(K^2=0)$. It follows from Eq.
(\ref{VLAS}) that the same holds for $\delta G(K,X)$. Using the
symmetry property
\begin{equation}
    G^{>\,ab}_{\mu\nu}(x,y)=G^{<\,ba}_{\nu\mu}(y,x)\,,
\end{equation}
we can write
\begin{equation}
    \delta G_{ab}(K,X)=2\pi \delta(K^2)\left\{\theta(K_0)\delta
    n^g_{ab}(\mathbf{k},X)+\theta(-K_0)\delta
    n^g_{ba}(-\mathbf{k},X)\right\}\,,
\end{equation}
where $\delta n^g_{ab}(\mathbf{k},X)$ is a density matrix, which
satisfies the following \textsc{Vlasov} equation:
\begin{equation}\label{gluovlasov}
    \left[V\cdot D_X,\delta n^g(\mathbf{k},X)\right]+g V_\mu
    F^{\mu\nu} \partial_{\nu}n^g(K)=0\,,
\end{equation}
with $V^\mu=(1,\mathbf{k}/k)$. The $\delta n^g$ have the same
color structure as the fields, which is $\delta f^g_a T^a$ assuming that they are small fluctuations.
Otherwise more complicated structures are possible.
The components $\delta f^g_a(\mathbf{k},X)$ transform like a
color vector in the adjoint representation. They are found by the
projections
\begin{align}
    \delta f^{g}_a(p,X) &= \frac{1}{N_c} \mathrm{Tr}\left[T_a n^{g}(p,X)\right]\text{\,,}
\end{align}
while the scalar coefficient in the color neutral background field
$n^{g}(\mathbf{p})=f^{g}(\mathbf{p})\mathcal{I}$ is given by
\begin{align}
    f^{g}(\mathbf{p}) &= \frac{1}{N_c^2-1} \mathrm{Tr}\left[n^{g}(p,X)\right]\text{\,.}
\end{align}
In terms of the density matrix $\delta n^g$ the induced gluonic
current (\ref{jb12}) reads
\begin{equation}\label{jb1A} J_{{\rm g}}^{\mu\,a}(X)\,=\,2g\int\frac{{\rm
d}^3k}{(2\pi)^3}\,V^\mu \,{\rm Tr}\Bigl(T^a \delta n^g({\bf
k},X)\Bigr)\,=\,2g N_{{\rm{c}}} \int\frac{{\rm
d}^3k}{(2\pi)^3}\,V^\mu\,\delta f^{g}_a({\bf k},X) \,.
\end{equation}
Equation (\ref{gluovlasov}) is covariant under the gauge
transformation of the background field and independent of the
gauge-fixing for the quantum fields, as proven in
\cite{Blaizot:1993zk,Blaizot:1993be}.

\subsection{Vlasov equation for quarks}
\index{Vlasov equation!Quark Vlasov equation} For the evaluation
of Eq. (\ref{jf1}) we need the kinetic equation for the quark
\textsc{Wigner} function $S\equiv S^<$. We start with Eq.
(\ref{S10}) for $\Psi=0$,
\begin{equation}
    \slashed{D}_x S(x,y)=0, ~~~~~~~  S(x,y)\slashed{D}^\dag_y=0\,.
\end{equation}
Using
\begin{align}
    \slashed{D} \slashed{D} &= \gamma^\mu \gamma^\nu D_\mu
    D_\nu \cdot = 2g^{\mu\nu} D_\mu D_\nu \cdot - \gamma^\nu
    \gamma^\mu D_\mu D_\nu \cdot \notag\\
    &= 2 D^2 - 2i\sigma^{\mu\nu} D_\mu D_\nu \cdot - \gamma^\mu \gamma^\nu D_\mu D_\nu
    \cdot\notag\\
    \Leftrightarrow \slashed{D} \slashed{D} &= D^2 - i\sigma^{\mu\nu} D_\mu D_\nu
    \cdot = D^2 - \frac{i}{2}(\sigma^{\mu\nu} D_\mu D_\nu -\sigma^{\mu\nu} D_\nu D_\mu)
    \cdot\notag\\
    \Leftrightarrow \slashed{D} \slashed{D} &= D^2+\frac{g}{2}\sigma^{\mu\nu}F_{\mu\nu}(x)\,,
\end{align}
with
$\sigma^{\mu\nu}=-\sigma^{\nu\mu}=i/2\left[\gamma^\mu,\gamma^\nu\right]$,
and $F_{\mu\nu}=[D_\mu,D_\nu]/(ig)$, we find the following
difference equation:
\begin{equation}\label{SD}
D_x^2
S(x,y)-S(x,y)\bigl(D_y^\dagger)^2+\frac{g}{2}\,\Bigl(\sigma^{\mu\nu}
F_{\mu\nu}(x)S(x,y)-S(x,y)\sigma^{\mu\nu}
F_{\mu\nu}(y)\Bigr)=0.\,\,\,\,
\end{equation}
As before we perform a covariant gradient expansion, which leads
to
\begin{equation}
\label{Vlas0}
 \left [K\cdot D_X,\, \delta S(K,X) \right ]=
g K\cdot F(X)\,\partial_k S_0
-i\,\frac{g}{4}\,F^{\mu\nu}(X)\left[\sigma_{\mu\nu},\,S_0\right]\,,
\end{equation}
where $F(X)=F_\mu^\mu(X)$. Using $S_0\equiv S_0^<(K)=\slashed{K}
\rho_0(K)n(K)$,
$$\partial_k(\slashed{K} \rho_0(K)n(K))=\slashed{K}/k
\rho_0(K)n(K)+ \slashed{K}\partial_k(\rho_0(K)n(K))\,,$$ 
and $\left[\sigma^{\mu\nu},\slashed{K}\right]=2i(K^\nu\gamma^\mu-K^\mu\gamma^\nu) $, we find
\begin{equation}
\label{Vlasquark1}
 \left [K\cdot D_X,\, \delta S(K,X) \right ]= g \slashed{K} K\cdot F(X)\partial_k (\rho_0(K)n(K))\,.
\end{equation}
$\delta S$ is a color matrix of the form $\delta S=\delta S_a
t^a$, with the components $\delta S_a$ transforming as a color
vector in the fundamental representation. It also has the same
spin and mass-shell structure as the free two-point function
$S_0^<$:
\begin{equation}\label{gens1}
 \delta S(K,X)\,=\,\slashed{K}\,2\pi\delta(K^2)
\left\{ \theta(k_0) \delta n^q({\bf k}, X)+ \theta(-k_0) \delta
n^{\bar{q}}(-{\bf k},X)\right\}\,.
\end{equation}
The density matrices
 $\delta n^{q/\bar{q}} ({\bf k},X) \equiv\delta f^{q/\bar{q}}_a({\bf k},X)\,t^a$
 satisfy the following kinetic equation:
\begin{equation}
\label{n} \left[ V\cdot D_X,\,\delta n^{q/\bar{q}}({{\bf
k}},X)\right]\pm\, g\,V_\mu F^{\mu\nu}(X)\partial_\nu n(K)=0\,,
\end{equation}
which is the non-Abelian version of the \textsc{Vlasov} equation
for quarks. The components $\delta f^{q/\bar{q}}_a$ are found by
the projections
\begin{align}
    \delta f^{q/\bar{q}}_a(p,X) &= 2 \mathrm{Tr}\left[t_a n^{q/\bar{q}}(p,X)\right]\text{\,,}
\end{align}
while the scalar coefficient in the color neutral background field
$n^{q/\bar{q}}(\mathbf{p})=f^{q/\bar{q}}(\mathbf{p})I$ is given by
\begin{align}
    f^{q/\bar{q}}(\mathbf{p}) &= \frac{1}{N_c} \mathrm{Tr}\left[n^{q/\bar{q}}(p,X)\right] \text{\,.}
\end{align}
The induced current finally reads \index{Induced current} (cf.
eq.~(\ref{jf1})):
\begin{equation}\label{jf1A}
J_{{\rm f}}^{\mu\,a}(X) =g N_{{\rm f}}\int\frac{{\rm
d}^3k}{(2\pi)^3}\,V^\mu \Bigl( \delta
 f^q_a({\bf k},X)-\delta f^{\bar{q}}_a({\bf k},X)\Bigr)\,.
\end{equation}
Altogether we find for the induced current:
\begin{align}\label{inducedcurrent}
    J^{\mu\,a}(X) &= J_{{\rm g}}^{\mu\,a}(X) + J_{{\rm
    f}}^{\mu\,a}(X)\notag\\
    &=g \int\frac{{\rm d}^3k}{(2\pi)^3}\,V^\mu\,\left\{2 N_{{\rm{c}}} \delta f^g_a({\bf
    k},X)+N_{{\rm f}}\left[\delta f^q_a({\bf k},X)-\delta f^{\bar{q}}_a({\bf
    k},X)\right]\right\}
\end{align}

      \chapter{Effective theory for the soft modes %with momenta of the order $gT$
}
%\epigraphwidth 250pt \epigraph{Begriff ist Summe, Idee Resultat der Erfahrung;\\ jene zu ziehen, wird Verstand, %dieses zu erfassen, Vernunft erfordern.}{\emph{Maximen und Reflexionen}\\
%Johann Wolfgang von Goethe (1749-1832)}
\label{HTL} In order to obtain an effective theory for the soft
collective modes with momenta of the order $gT$ we have to solve
the kinetic equations (\ref{gluovlasov}) and (\ref{n}) for the
hard particles and by that express the current
(\ref{inducedcurrent}) in terms of the soft fields alone. This
leads to an effective \textsc{Yang-Mills} equation for the soft
modes, in which the hard modes do not appear explicitly.
\section{Solving the kinetic equations}
\label{solving} We combine Eqs. (\ref{gluovlasov}) and (\ref{n})
by writing
\begin{equation}\label{HTLvlasov}
[V\cdot D_X,\delta n^{i}(\mathbf{p},X)] + g \theta_{i}
V_{\mu}F^{\mu\nu}(X)\partial_{\nu}^{(p)}n^{i}(\mathbf{p})=0\text{\,,}
\end{equation}
with $i\in\{g,q,\bar{q}\}$, $\theta_{g}=\theta_{q}=1$ and
$\theta_{\bar{q}}=-1$. We remind the reader of the color
structure, which for the color neutral background fields is
$n^{q/\bar{q}}(\mathbf{p})=f^{q/\bar{q}}(\mathbf{p})I$ and
$n^{g}(\mathbf{p})=f^{g}(\mathbf{p})\mathcal{I}$. $I$ and
$\mathcal{I}$ are unit matrices in the fundamental and adjoint
representation, respectively. The induced color fluctuations are
constructed like $\delta n^{q/\bar{q}}(\mathbf{p},X)=\delta
f^{q/\bar{q}}_b(\mathbf{p},X)t^b$ and $\delta n^{g}(\mathbf{p},X)=\delta
f^{g}_b(\mathbf{p},X)T^b$, with the generators $t_b$ and $T_b$ in the
fundamental and adjoint representation, respectively.

The induced current (\ref{inducedcurrent}) reads
\begin{equation}\label{indcurrent}
    J_{\text{ind}\,a}^{i\,\mu}=g \int_{\mathbf{p}} V^{\mu}\left\{2N_c \delta f^{g}_a(\mathbf{p},X)+N_{f}[\delta f^{q}_a(\mathbf{p},X)
    -\delta f^{\bar{q}}_a(\mathbf{p},X)]\right\}\,\text{,}
\end{equation}
where we introduced
\begin{equation}
    \int_{\mathbf{p}} := \int \frac{d^3p}{(2\pi)^3}
\end{equation}
for convenience.

As mentioned before (see footnote on page \pageref{orderofA}) $A\sim\sqrt{g}T$ and when we neglect terms of subleading order in $g$ the theory becomes
effectively Abelian as $D_X\rightarrow \partial_X$ and $F^{\mu\nu}\rightarrow\partial^\mu A^\nu-\partial^\nu A^\mu$.
The color channels decouple and Eq. (\ref{HTLvlasov}) can be written for each color channel separately:
 \begin{equation}\label{HTLvlasov2}
V\cdot \partial_X\delta f^{i}_a(\mathbf{p},X) + g \theta_{i}
V_{\mu}F^{\mu\nu}_a(X)\partial_{\nu}^{(p)}f^{i}(\mathbf{p})=0\text{\,,}
\end{equation}
where $F^{\mu\nu}_a=\partial^\mu A_a^\nu-\partial^\nu A_a^\mu$.
This equation can be solved using a \textsc{Fourier}
transformation. First we write
\begin{align}\label{HTLvlasov3}
(\partial_t + \mathbf{v}\cdot\boldsymbol{\nabla})\delta f^{i}_a(\mathbf{p},X) + g \theta_{i}
V_{\mu}F^{\mu\nu}_a(X)\partial_{\nu}^{(p)}f^{i}(\mathbf{p})&=0\notag\\
\xrightarrow{\text{\textsc{Fourier}}}(-i\,\omega +
i\,\mathbf{k}\cdot\mathbf{v})\delta f^{i}_a(\mathbf{p},K) + g
\theta_{i}
V_{\mu}F^{\mu\nu}_a(K)\partial_{\nu}^{(p)}f^{i}(\mathbf{p})&=0\,.
\end{align}
Solving for $\delta f^{i}_a(\mathbf{p},K)$ we get
\begin{align}\label{HTLvlasov4}
\delta f^{i}_a(\mathbf{p},K) &= - g \theta_{i} \frac{
V_{\mu}F^{\mu\nu}_a(K)\partial_{\nu}^{(p)}f^{i}(\mathbf{p})}{-i\,\omega
+ i\,\mathbf{k}\cdot\mathbf{v}+\epsilon}=
- i\,g \theta_{i} \frac{i\,V_{\mu}(K^\mu A_a^\nu-K^\nu A_a^\mu)\partial_{\nu}^{(p)}f^{i}(\mathbf{p})}{\omega - \,\mathbf{k}\cdot\mathbf{v}+i\epsilon}\notag\\
&=
g \theta_{i} \frac{\left[
(\omega-\mathbf{k}\cdot\mathbf{v}) A_a^\nu - V_\mu K^\nu A_a^\mu\right]\partial_{\nu}^{(p)}f^{i}(\mathbf{p})}{\omega - \,\mathbf{k}\cdot\mathbf{v}+i\epsilon}\notag\\
&=
g \theta_{i}\,\partial_{\nu}^{(p)}f^{i}(\mathbf{p})\left( A_a^\nu + \frac{V_\mu K^\nu A_a^\mu}{\omega - \,\mathbf{k}\cdot\mathbf{v}+i\epsilon}\right)\notag\\
&= g \theta_{i}\,\partial^\beta_{(p)}f^{i}(\mathbf{p})\left(
g_{\nu\beta} + \frac{V_\nu K_\beta }{\omega -
\,\mathbf{k}\cdot\mathbf{v}+i\epsilon}\right)A_a^\nu\,.
\end{align}
The induced current finally reads
\begin{equation}
    J^{\mu}_{\text{ind}\,a}(K)=g^2 \int_{\mathbf{p}} V^{\mu}
    \partial^{\beta}_{(p)}
    f(\mathbf{p})\left(g_{\gamma\beta}-\frac{V_{\gamma}K_{\beta}}{K\cdot
    V+i\epsilon}\right)A^{\gamma}_a(K)+\mathcal{O}(g^3 A^2)
    \,\text{,}
\end{equation}
where we combined the gluon and quark distributions in
\begin{equation}
f(\mathbf{p})=2N_c f^{g}(\mathbf{p})+N_f\left[f^{q}(\mathbf{p})+f^{\bar{g}}(\mathbf{p})\right] \label{f}\text{\,.}
\end{equation}
By functional differentiation
\begin{equation}
    \Pi^{\mu\nu}_{ab}(K)=\frac{\delta J^{\mu}_{\text{ind}\,a}(K)}{\delta
    A_{\nu}^b(K)}\,\text{,}
\end{equation}
we obtain the self energy
\begin{equation}\label{selfenergyresult}
    \Pi^{\mu\nu}_{ab}(K)=g^2 \delta_{ab}\int_{\mathbf{p}} V^{\mu}
    \partial_{\beta}^{(p)}
    f(\mathbf{p})\left(g^{\nu\beta}-\frac{V^{\nu}K^{\beta}}{K\cdot
    V+i\epsilon}\right)\,\text{,}
\end{equation}
the same result that one gets in the diagrammatic approach using
the HTL approximation \index{HTL approximation} and the assumption
that $f$ is symmetric under $\mathbf{p}\rightarrow -\mathbf{p}$
\cite{Mrowczynski:2000ed}. One finds that this tensor is symmetric
$\Pi^{\mu\nu}=\Pi^{\nu\mu}$ and transverse $K_\mu\Pi^{\mu\nu}=0$,
if $f(\mathbf{p})$ vanishes on a two-sphere at infinity:
$\lim_{\mathbf{p}\rightarrow\infty}f(\mathbf{p})=0$. We
demonstrate the transversality for one example:
\begin{align}
K^\mu\Pi_{\mu 0}&=g^2\int_{\mathbf{p}}K^\mu V_\mu \partial_j f(\mathbf{p}) \frac{k^j}{K\cdot V+i\epsilon}\notag\\
        &=g^2\int_{\mathbf{p}} v_j k^j \frac{{\rm d}}{{\rm d}p}f(\mathbf{p})\notag\\
        &=g^2 k\int_{\mathbf{p}} \cos{\theta} \frac{{\rm d}}{{\rm d}p}f(\mathbf{p})\notag\\
        &=0\,.
\end{align}
Due to the transversality not all components of $\Pi^{\mu\nu}$ are independent,
and we can restrict our investigation to the spatial components.
As we have already seen, in the linear approximation the current that
is induced by the fluctuations can be expressed in terms of the
self energy:
\begin{equation}
    J^{\mu}_{\text{ind}}(K)=\Pi^{\mu\nu}(K)A_{\nu}(K)\,\text{.}
\end{equation}
Inserting this into \textsc{Maxwell}'s equation
\begin{equation}
    iK_{\mu}F^{\mu\nu}(K)=J^{\nu}_{\text{ind}}+J^{\nu}_{\text{ext}}\,\text{,}
\end{equation}
we get
\begin{equation}
    \left[K^2g^{\mu\nu}-K^{\mu}K^{\nu}+\Pi^{\mu\nu}(K)\right]A_{\mu}(K)=J^{\nu}_{\text{ext}}(K)\,\text{,}
\end{equation}
with the external current $J^{\nu}_{\text{ext}}$. Using the gauge
covariance of the self energy in the HTL-approximation \index{HTL
approximation} we can write this in terms of a physical electric
field by specifying a certain gauge. Using the temporal axial
gauge, where $A_0=0$, yields
\begin{equation}\label{propdef}
    \left[(k^2-\omega^2)\delta^{ij}-k^ik^j+\Pi^{ij}(K)\right]E^j(K)=[\Delta^{-1}(K)]^{ij}E^{j}(K)=i\omega
    J^{i}_{\text{ext}}(K)\,\text{,}
\end{equation}
and the response of the system to the external source is given by
\begin{equation}
    E^i(K)=i \omega \Delta^{ij}(K)J^{j}_{\text{ext}}(K)\text{\,.}
\end{equation}
The dispersion relations are obtained by finding the poles of the
propagator $\Delta^{ij}(K)$, defined in (\ref{propdef}).
%%%%%%%%%%%%%%%%%%%%%%%%%%%%%%%%%%%%%%%%%%%%%%%%%%%%%%%%%%%%%%%%%%%%%%%%%%%%%%%%%%%%%%%%%%%%%%%%%%%%%%%%%%%%%%%%%
\section{Dispersion relations in the isotropic limit}
\index{Dispersion relations} \index{Isotropic limit} To determine
the dispersion relations for an isotropic system, we assume $f$ to
depend on the length of $\mathbf{p}$ only, and take a closer look
at the spatial components of the self energy:
\begin{equation}
    \Pi^{ij}_{ab}(K)=-g^2 \delta_{ab}\int_{\mathbf{p}} v^{i}
    \partial_{l}^{(p)}
    f(p)\left(\delta^{jl}+\frac{v^{j}k^{l}}{K\cdot
    V+i\epsilon}\right)\,\text{.}
\end{equation}
First, the derivative becomes
\begin{equation}
    \partial_{l}f(p)=\frac{p_l}{p}\frac{{\rm d}f(p)}{{\rm d}p}=v_l\frac{{\rm d}f(p)}{{\rm d}p}\,.
\end{equation}
Then, we can perform the radial and $\phi$ integrations to obtain
\begin{equation}
    \Pi^{ij}(K)=\frac{m_D^2}{2}\int_0^\pi \sin\theta {\rm d}\theta\,v^iv^l\left(\delta^{jl}+\frac{v^jk^l}{K\cdot V+i\epsilon}\right)\,,
\end{equation}
after neglecting the trivial color structure and introducing the
\textsc{Debye} mass \index{Debye mass}
\begin{equation}\label{debyemass}
    m_D^2=-\frac{g^2}{2\pi^2}\int_0^\infty{\rm d}p\,p^2 \frac{{\rm d}f(p)}{{\rm d}p}\xrightarrow{\text{equilibrium}}
    g^2T^2\left(\frac{2N_c+N_f}{6}\right)\,,
\end{equation}
for zero quark chemical potential.
In the isotropic case there is only one distinct spatial direction in the system, the direction of $\mathbf{k}$.
Therefore we can decompose the self energy into components parallel and transverse to $\mathbf{k}$:
\begin{equation}
    \Pi^{ij}=P_T\,\Pi_T+P_L\,\Pi_L\,,
\end{equation}
with the projectors
\begin{equation}\label{projectors}
    P_T=\left(\delta^{ij}-\frac{k^ik^j}{k^2}\right) \text{ and } P_L=\frac{k^ik^j}{k^2}\,.
\end{equation}
They obey the usual projector properties
\begin{align}
    &(P_T)^2=P_T\,% \footnote{$\left((P_T)_{ik}(P_T)_{kj}=(P_T)_{ij}\right)$}
    \text{, } (P_L)^2=P_L\,,\\
    &P_T\,P_L=P_L\,P_T = 0\,.
\end{align}
The scalar self energies are defined by the following projections, for which the integrals can be solved analytically:
\begin{align}\label{isopist}
    \Pi_T&=\frac{1}{2}P_T\Pi^{ij}=\frac{m_D^2}{2}\left[\frac{\omega^2}{k^2}+\omega\frac{(k^2-\omega^2)}{2k^3}\ln\left(\frac{\omega+k}{\omega-k}\right)\right]\,,\\
    \Pi_L&=P_L\Pi^{ij}=m_D^2\frac{\omega^2}{k^2}\left[-1+\frac{\omega}{2k}\ln\left(\frac{\omega+k}{\omega-k}\right)\right]\,.\label{isopisl}
\end{align}
The factor $1/2$ in the definition of $\Pi_T$ is convenient
because there are two transverse modes in three dimensions.
Generally it is a factor of $1/(d-1)$, where $d$ is the number of
dimensions. Without it, the $1/2$ would show up in front of
$\Pi_T$ in Eq. (\ref{dispit}) below.

\begin{figure}[htb]
    \begin{center}
         \includegraphics[width=12cm]{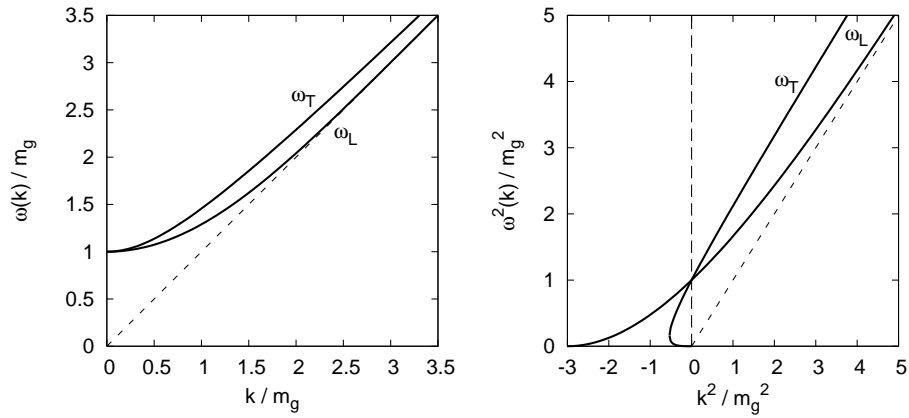}
         \caption{Dispersion laws for transverse and longitudinal gluons (left) on a linear scale in units of $m_g=m_D/\sqrt{3}$, and (right)
             in quadratic scales. Modes with $k^2>0$ are propagating whereas solutions for negative $k^2$ show screening phenomena.}
    \label{fig:gluondispersionboth}
    \end{center}
\end{figure}

Making use of the projectors (\ref{projectors}), we can determine the dispersion relations of the transverse and longitudinal modes separately. By applying $P_T$ to the equation that determines the zeros of the inverse propagator $\Delta^{-1}(K)$ (see Eq. (\ref{propdef})), we get the transverse dispersion relation:
\begin{align}\label{dispit}
    k^2-\omega_T^2+\Pi_T=0\,,
\end{align}
whereas application of $P_L$ leads to the longitudinal one:
\begin{align}\label{dispil}
    \Pi_L-\omega_L^2=0\,.
\end{align}
The solutions to Eqs. (\ref{dispit}) and (\ref{dispil}) are shown
in Fig. \ref{fig:gluondispersionboth}. Taking the limit
$k\rightarrow 0$ we get $\omega\rightarrow 1/3\,m_D^2=m_g^2$,
representing the lowest frequency with which modes can propagate.
We present the dispersion laws in units of $m_g=m_D/\sqrt{3}$ in
linear scales in Fig. \ref{fig:gluondispersionboth}(a) and in
quadratic scales in Fig. \ref{fig:gluondispersionboth}(b) in order
to include both propagating modes and screening phenomena in one
plot.

Propagating modes exist above a common plasma frequency $\omega
> \omega_{\rm{pl}} =\sqrt{m^2_D/3}$, are always above the
light cone ($\omega > k$) and are called normal modes. For large
momenta the transverse mode tends to a mass hyperboloid with
asymptotic mass $m_\infty^2=m_D^2/2$, whereas the longitudinal one
approaches the light-cone with exponentially vanishing residue.
For $\omega<m_g$, $|\mathbf k|$ is the inverse screening length,
which in the static limit ($\omega\rightarrow 0$) vanishes for the
transverse mode (absence of magnetostatic screening) but reaches
the \textsc{Debye} mass, $m_D^2=3\, m_g^2$, for the longitudinal
mode (electrostatic screening). For $0< \omega <
\omega_{\rm{pl}}$, $k$ becomes imaginary~\cite{Weldon:1982aq} and
there are no stable normal modes, however, there is a collective
behavior which corresponds to dynamical screening both in the
electric and magnetic sector as long as $\omega > 0$.

%%%%%%%%%%%%%%%%%%%%%%%%%%%%%%%%%%%%%%%%%%%%%%%%%%%%%%%%%%%%%%%%%%%%%%%%%%%%%%%%%%%%%%%%%%%%%%%%%%%%%%%%%%%%%%%%%

      \chapter{The anisotropic quark-gluon plasma}
\epigraphwidth 250pt \epigraph{Ungleich erscheint im Leben viel,\\
doch bald und unerwartet ist es ausgeglichen.}{\emph{Die nat\"urliche Tochter. Vierter Aufzug.\\ Zweiter Auftritt. Gerichtsrat.}\\
Johann Wolfgang von Goethe (1749-1832)} \label{anisotropy}
\index{Anisotropy} The only system above the QCD-deconfinement
transition one can (probably) study on earth, namely that of a
fireball following an ultra-relativistic heavy-ion collision, is
anisotropic in nature. Therefore, it is interesting to investigate
which effects the presence of an anisotropy has on the dynamics of
the system and which differences one can expect when comparing to
the usually studied isotropic case.

We will consider systems which are anisotropic only in {\em
momentum space}, therefore taking a snapshot of the system at some
short time after the collision and before the system had time to
convert the anisotropy from momentum space to configuration space.
The appearance of local momentum anisotropies can be best
understood when regarding a local cell in which initially the
transverse momentum (transverse to the beam line) is set by the
hard momentum scale $Q_s$, while the transverse momentum drops
like $1/\tau$ (where $\tau$ is the proper time) as particles with
momenta larger than that of the cell leave it. For pure free
streaming only particles with a certain longitudinal momentum
remain in the cell.

The momentum scale $Q_s$ comes about the following way.
The bulk of multiparticle production at central rapidities in A-A collisions
emerges from partons with very small momentum fractions $x$ (at LHC $x\leq 10^{-3}$). 
They interact coherently over distances much larger than the radius of the nucleus, which
potentially makes understanding the properties of the nuclear wave function at small $x$ a difficult task.
However, some features of these wave functions simplify the problem.
First, perturbative QCD predicts that parton distributions grow very rapidly at small $x$.
At sufficiently small $x$, the parton distributions saturate, i.e., when the density of
partons is such that they overlap in the transverse plane, their repulsive interactions
are sufficiently strong to limit further growth to be at most logarithmic \cite{Gribov:1984tu,Mueller:1985wy,Blaizot:1987nc}.
The large parton density per transverse area then provides a scale, $Q_s$.
The physics of small-$x$ parton distributions can be formulated in an effective field theory,
where the saturation scale $Q_s$ appears as the only scale \cite{McLerran:1993ni,McLerran:1993ka,McLerran:1994vd}. The coupling must therefore 
run as a function of this scale, and since $Q_s^2\ll \Lambda_{\text{QCD}}^2$ at RHIC or the LHC, $\alpha_s\ll 1$. Small coupling also ensures that the occupation number is large, $\sim 1/\alpha_s$.
Thus, although the physics of small-$x$ partons is non-perturbative, it can be studied with weak coupling \cite{JalilianMarian:1996xn,Kovchegov:1996ty,JalilianMarian:1997jx,JalilianMarian:1997gr,McLerran:1998nk,Kovner:1999bj,Iancu:2000hn,Iancu:2001ad}. This is analogous to many systems in condensed-matter physics. In particular, the physics of small-$x$ wave functions is similar to that of a spin glass \cite{Gavai:1996vu,JalilianMarian:1997jx,JalilianMarian:1997gr,McLerran:1998nk,Kovner:1999bj,Iancu:2000hn}. Further, since in this state the mean transverse momentum of the partons is of order $Q_s$, and their occupation numbers are large, of order $1/\alpha_s$, as in a condensate, the partons in the nuclear wave function form a so called color glass condensate\index{Color Glass Condensate}(CGC) \cite{Iancu:2001ad,Gavai:1996vu}.

Considering momentum anisotropies alone is certainly not the whole
story. However, the generalization from fully isotropic
distributions already provides great insight into the properties
of the QGP within the fireball. Moreover, a treatment of
configuration space anisotropies would involve a full
non-equilibrium quantum field theoretical description of the
system, which despite recent progress is not yet available.

\begin{figure}[htb]
   \begin{center}
      \includegraphics[width=12cm]{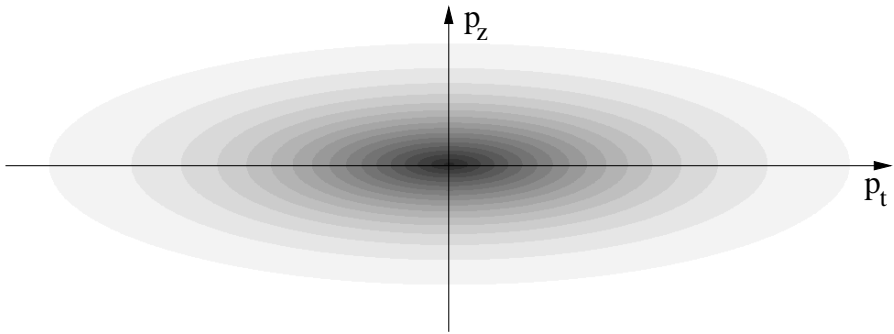}
      \caption{Contour plot of a squeezed \textsc{Fermi-Dirac} distribution with anisotropy
      parameter $\xi=10$. The anisotropy vector ${\bf \hat{n}}$ is taken to be along the $p_z$ direction.}
       \label{fig:squeeze}
   \end{center}
\end{figure}

In this chapter we will analyze the collective modes of
high-temperature QCD in the case when there is an anisotropy in
the momentum-space distribution function. We concentrate on a
class of anisotropic distribution functions, which can be obtained
by stretching or squeezing an isotropic one along a certain
direction, thereby preserving a cylindrical symmetry in momentum
space. This particular class of distribution functions has been
first discussed by \textsc{Romatschke} and \textsc{Strickland} in
\cite{Romatschke:2003ms} and \cite{Romatschke:2004jh}. We write
\begin{equation}
    f(\mathbf{p})=\mathcal{N}(\xi)\,f_{\text{iso}}\left(\mathbf{p}^2+\xi(\mathbf{p}\cdot\mathbf{\hat{n}})^2\right)\,\text{,}\label{anisodist}
\end{equation}
for an arbitrary isotropic distribution function
$f_{\text{iso}}(\mid\!\!\mathbf{p}\!\!\mid)$. The direction of the
anisotropy is given by $\mathbf{\hat{n}}$ and $\xi>-1$ is an
adjustable anisotropy parameter. $\xi>0$ corresponds to a
contraction of the distribution in the $\mathbf{\hat{n}}$
direction, whereas $-1<\xi<0$ represents a stretching of the
distribution in the $\mathbf{\hat{n}}$ direction. The factor
$\mathcal{N}(\xi)$ is used for normalization. For example, when
choosing $\mathcal{N}(\xi)=\sqrt{1+\xi}$, we ensure that the
overall particle number is the same for both the anisotropic and
the isotropic distribution function.\footnote{Note that for the
\textsc{Fermi} distribution this way of normalizing $f$ is not
appropriate, because it violates the \textsc{Pauli} exclusion
principle. The following considerations will not depend on the
particular choice of $\mathcal{N}(\xi)$ qualitatively. When we
turn to the discussion of photon production from an anisotropic
QGP we will set $\mathcal{N}(\xi)=1$, because there the compliance
with the \textsc{Pauli} exclusion principle becomes important -
otherwise the \textsc{Pauli} blocking term may become negative.}
For a certain anisotropy Fig.\,\ref{fig:squeeze} shows how this
procedure deforms the \textsc{Fermi-Dirac} distribution.
One of the most interesting aspects of the anisotropic QGP is the
appearance of unstable gluonic modes, which we discuss in detail in
Sec. \ref{instabilities}.

\section{Tensor decomposition}
\label{tensordecomp} \index{Tensor decomposition} The anisotropy
vector introduces a second special direction and we can no longer
decompose the propagator and the self energy into transverse and
longitudinal parts. However, we can decompose the spatial part of
the self energy into four structure functions using the tensor
basis introduced in \cite{Romatschke:2003ms,Dumitru:2007hy}:
\begin{equation}
    \Pi^{ij}=\alpha A^{ij}+\beta B^{ij}+\gamma C^{ij}+\delta
    D^{ij}\,\text{,}
\end{equation}
\index{Structure functions} where
\begin{align}
    A^{ij}&=\delta^{ij}-k^ik^j/k^2 \notag\\
    B^{ij}&=k^ik^j/k^2 \notag\\
    C^{ij}&=\tilde{n}^i\tilde{n}^j/\tilde{n}^2 \notag\\
    D^{ij}&=k^i\tilde{n}^j+k^j\tilde{n}^i \,\text{,}
\end{align}
with $\tilde{n}^i=A^{ij}n^j$ the part of ${\bf n}$ that is
perpendicular to ${\bf k}$, i.e., $\tilde{{\bf n}}\cdot {\bf
k}=0$. We determine the structure functions by taking the
contractions:
\begin{align}
    k^i\Pi^{ij}k^j&=k^2\beta \notag\\
    \tilde{n}^i\Pi^{ij}k^j&=\tilde{n}^2k^2\delta\notag\\
    \tilde{n}^i\Pi^{ij}\tilde{n}^j&=\tilde{n}^2(\alpha+\gamma)\notag\\
    \text{Tr}\,\Pi^{ij}&=2\alpha+\beta+\gamma\text{\,.}\label{contractions}
\end{align}
The four structure functions $\alpha,\beta,\gamma,\delta$ depend
on the \textsc{Debye} mass $m_D$ \index{Debye mass}, the frequency
and spatial momentum $\omega$ and $k$, the strength of the
anisotropy $\xi$ and the angle between the spatial momentum and
the anisotropy direction
${\bf \hat{k}}\cdot{\bf \hat n}=\cos\theta_n$. %For symmetry reasons one can restrict $0<\theta_n<\frac{\pi}{2}$ in the following.
In the isotropic limit \index{Isotropic limit} ($\xi \rightarrow
0$) the structure functions $\alpha(K,\xi)$ and $\beta(K,\xi)$ reduce to the
isotropic HTL self-energies and $\gamma$ and $\delta$ vanish,
\begin{align}
\alpha(K,0) &=\Pi_T(K) \; , \nonumber \\
\beta(K,0) &=\Pi_L(K) \; , \nonumber \\
\gamma(K,0) &=0 \; , \nonumber \\
\delta(K,0) &=0 \; , \label{isolimit}
\end{align}
where $\Pi_T(K)$ and $\Pi_L(k)$ are given by Eqs.\,(\ref{isopist})
and (\ref{isopisl}).
Note that for finite $\xi$ the analytic structure of
$\alpha,\beta,\gamma,\delta$ is the same as for $\Pi_L,\Pi_T$ in
the isotropic case, namely there is a cut in the complex $\omega$
plane which can be chosen to run along the real $\omega$ axis within
$-k<\omega<k$. For real-valued $\omega$ the structure functions
are complex for all $\omega < k$ (corresponding to the
\textsc{Landau} damping regime) and real for $\omega
>k$ while for imaginary-valued $\omega$ all four structure
functions are real-valued.

In this basis, the inverse of the propagator in
Eq.~(\ref{propdef}) can be written as
\begin{equation}
    \mathbf{\Delta}^{-1}(K)=(k^2-\omega^2+\alpha)\mathbf{A}+(\beta-\omega^2)\mathbf{B}+\gamma
    \mathbf{C}+\delta \mathbf{D}\,\text{,}
\end{equation}
while the propagator itself is given by \cite{Romatschke:2003ms}:
\begin{equation}
    \mathbf{\Delta}(K)=\Delta_A(\mathbf{A}-\mathbf{C})+\Delta_G\left[(k^2-\omega^2+\alpha+\gamma)\mathbf{B}+(\beta-\omega^2)\mathbf{C}-\delta\mathbf{D}\right]\,\text{,}
    \label{propagator}
\end{equation}
with
\begin{align}
    \Delta_A^{-1}(K)&=k^2-\omega^2+\alpha\,\text{,}\label{deltaa}\\
    \Delta_G^{-1}(K)&=(k^2-\omega^2+\alpha+\gamma)(\beta-\omega^2)-k^2\tilde{n}^2\delta^2\text{\,.}\label{deltag}
\end{align}
Furthermore, with the particular choice for the distribution
functions (\ref{anisodist}), the spatial components of the self
energy (\ref{selfenergyresult})
\begin{equation}
    \Pi^{ij}_{ab}(K)=-g^2 \delta_{ab}\int_{\mathbf{p}} v^{i}
    \partial_{l}^{(p)}
    f(\mathbf{p})\left(\delta^{jl}+\frac{v^{j}k^{l}}{K\cdot
    V+i\epsilon}\right)
\end{equation}
can be simplified by performing the change of variables to
$\tilde{p}$:
\begin{equation}
    \tilde{p}^2=p^2\left[1+\xi({\bf v}\cdot{\bf
    \hat{n}})^2\right]\,,
\end{equation}
which allows for integrating out the $\tilde p$-dependence. We
find
\begin{equation}
\Pi^{i j}(K) = m_{D}^2 \int \frac{d \Omega}{4 \pi} v^{i}%
\frac{v^{l}+\xi({\bf v}\cdot{\bf \hat{n}}) n^{l}}{%
(1+\xi({\bf v}\cdot{\bf \hat{n}})^2)^2} \left( \delta^{j
l}+\frac{v^{j} k^{l}}{K\cdot V + i \epsilon}\right) \,,
\end{equation}
where (cf. Eq. (\ref{debyemass}))
\begin{equation}\label{aniso:debyemass}
    m_D^2=-\frac{g^2}{2\pi^2}\int_0^\infty{\rm d}p\,p^2 \frac{{\rm d}f_{\text{iso}}(p)}{{\rm d}p}\,.
\end{equation}

\section{Stable modes} Considering first the stable collective
modes which correspond to poles of the propagator at real-valued
$\omega>k$ one factorizes $\Delta_G^{-1}$ as
\cite{Romatschke:2003ms}:
\begin{equation} \Delta_G^{-1} = (\omega^2 - \Omega_+^2)(\omega^2-\Omega_-^2)
\,,
\end{equation}
where
\begin{equation}
2 \Omega_{\pm}^2 = \bar\Omega^2 \pm \sqrt{\bar\Omega^4- 4 ((\alpha+\gamma+%
k^2)\beta-k^2\tilde n^2\delta^2) } \; , \label{omegapm}
\end{equation}
and
\begin{equation} \bar\Omega^2 = \alpha+\beta+\gamma+k^2 \,.
\end{equation}
 Since the quantity under the square root in (\ref{omegapm})
can be written as $(\alpha-\beta+\gamma+k^2)^2+4k^2\tilde
n^2\delta^2$ (which is always positive for real $\omega>k$), there
are at most two stable modes coming from $\Delta_G$, while the
remaining stable collective mode comes from the zero of
$\Delta_A^{-1}$.

The dispersion relations \index{Dispersion relations} for all of
the collective modes are then given by the solutions to
\begin{align}
\omega^2_\pm &= \Omega_\pm^2(\omega_\pm) \; , \\
\omega^2_\alpha &= k^2 + \alpha(\omega_\alpha) \,.
\end{align}
In the isotropic limit ($\xi\rightarrow 0$) one finds the
correspondence
 $\omega_\alpha = \omega_+ = \omega_T$ and $\omega_- = \omega_L$.
Accordingly, for finite $\xi$, there are three stable
quasi-particle modes with dispersion relations \index{Dispersion
relations} that depend on the angle of propagation with respect to
the anisotropy vector, $\theta_n$. The resulting dispersion
relations for all three modes for the case $\xi=10$ and different
angles $\theta_n$ are shown in Ref.\,\cite{Romatschke:2003ms}.
\index{Dispersion relations}

\section{Unstable modes}
\label{unstable modes} One of the most interesting features of the
anisotropic quark-gluon plasma is the presence of unstable modes
that, instead of being damped, grow exponentially with time.
Mathematically they arise due to poles of the propagator along the
positive imaginary $\omega$ axis. We will discuss the physical
explanation for these so called \textsc{Weibel} instabilities
\index{Instabilities!Weibel instability} in Section
\ref{instabilities}. There are also additional damped modes along
the negative imaginary $\omega$-axis, which are not present in the
isotropic case. Checking for poles at complex (not purely
imaginary) $\omega$ can be done numerically but no poles on the
physical sheet have been found \cite{Romatschke:2003ms}. In
\cite{Romatschke:2004jh} it was shown analytically for certain
special cases that there are no other modes than the ones listed
above. However, there exist solutions on unphysical sheets,
which come very close to the border of the physical sheet for
sufficiently large values of the anisotropy parameter $\xi$ (see
Ref.\,\cite{Romatschke:2004jh}).

For the unstable modes of the system we can write $\omega
\rightarrow i \Gamma$, with $\Gamma$ being the real-valued
solution to the equations
\begin{align}
\Delta_G^{-1} &= (\Gamma^2 + \Omega_+^2)(\Gamma^2 + \Omega_-^2)=0 \;%
,\nonumber\\
\Delta_A^{-1} &= (\Gamma^2+k^2+\alpha)=0.
\end{align}
It turns out that in contrast to the stable modes there is at most
one solution for $\Delta_G^{-1}(\omega=i \Gamma)=0$ since
numerically one finds that $\Omega_+^2>0$ for all $\Gamma>0$,
while $\Delta_A^{-1}(\omega=i \Gamma)=0$ has only solutions for
$\xi>0$. Therefore, the system possesses one or two unstable modes
depending on the sign of the anisotropy parameter
\cite{Romatschke:2003ms}.

\begin{figure}[htb]
   \begin{center}
      \includegraphics[width=10cm]{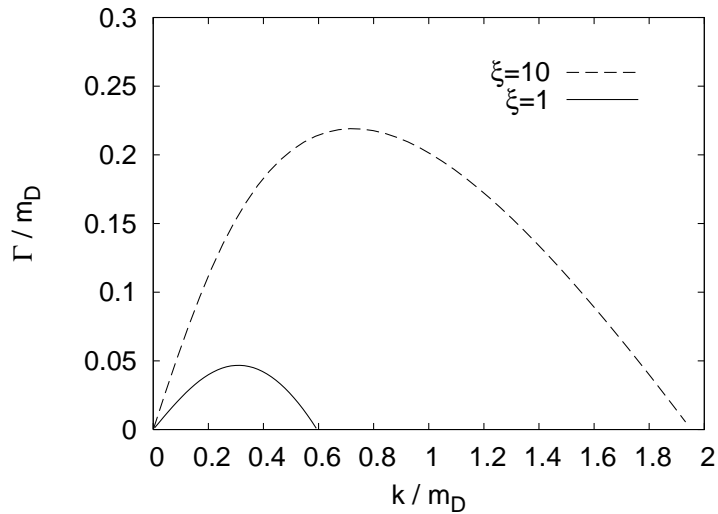}
      \caption{Growth rate of the unstable $\alpha$-mode for the case $\mathbf{k}\parallel {\bf \hat{n}}$ (where this is the only unstable mode)
        and different anisotropy parameters $\xi$. The maximal growth rate as well as the maximal $k$ increase with increasing $\xi$.}
       \label{fig:unstable}
   \end{center}
\end{figure}

We present a plot of the growth rate \index{Instabilities!Growth
rate of} $\Gamma_\alpha$ for the case that $\mathbf{k}\parallel
{\bf \hat{n}}$ ($\theta_n=0$) and different anisotropy parameters
$\xi$ in Fig. \ref{fig:unstable}. For solutions for arbitrary
angles see \cite{Romatschke:2003ms}. It turns out that the growth
rate is the largest for $\theta_n=0$, which we will explain on
physical grounds in Section \ref{instabilities}. One can also see
that the growth rate and the maximal value of $k$ for the unstable
mode become larger with increasing anisotropy.

\section{Discussion of instabilities}
\label{instabilities}
\index{Instabilities} The electron-ion
plasma is known to experience a large variety of instabilities
\cite{Kra73}. They can be either caused by coordinate space
inhomogeneities, in particular by the boundaries of finite systems
(hydrodynamic instabilities) or by non-equilibrium momentum
distributions of plasma particles (kinetic instabilities). Hardly
anything is known about hydrodynamic instabilities of the quark-gluon 
plasma and we will concentrate on the kinetic instabilities,
which are initiated either by charge or current fluctuations. In
the first case, the electric field is parallel to the wave vector
$\mathbf{k}$ ($\mathbf{E}\parallel\mathbf{k}$), while in the
second case the field is perpendicular to $\mathbf{k}$
($\mathbf{E}\perp\mathbf{k}$). This is why the corresponding
instabilities are called longitudinal and transverse,
respectively. Since the electric field plays a crucial role in the
generation of longitudinal modes, they are also called electric,
while the transverse modes are called magnetic.

Soon after the concept of quark-gluon plasma had been established,
the existence of the color kinetic instabilities, fully analogous
to those known in the electrodynamic plasma, was suggested
\cite{Heinz:1985vf,Pokrovsky:1988bm,Pokrovsky:1990sz,Pokrovsky:1990uh,
Mrowczynski:1988dz,Pavlenko:1990as,Pavlenko:1991ih}.

However, in these works a two-stream system, or more generally, a
momentum distribution with more than one maximum, was considered.
Such a distribution is common in the electron-ion plasma, however
it is rather inappropriate for describing the quark-gluon plasma
produced in relativistic heavy-ion collisions where the global as
well as local momentum distribution is expected to be monotonously
decreasing in every direction from the maximum. Electric
instabilities are absent in such a system but, as we have already
seen in Section \ref{unstable modes} and was first demonstrated in
\cite{Mrowczynski:1993qm,Mrowczynski:1994xv}, a magnetic unstable
mode known as the filamentation or \textsc{Weibel} instability
\cite{Weibel:1959} \index{Instabilities!Weibel instability} can possibly
appear since a sufficient condition for its existence is the
anisotropy of the momentum distribution. That these instabilities
really exist in nature is strongly supported by rather recent
experimental observations in plasma physics. Please refer to
Appendix \ref{experiment} for further details.

The filamentation instability was shown
\cite{Mrowczynski:1993qm,Mrowczynski:1994xv} to be possibly
relevant for the quark-gluon plasma produced in relativistic
heavy-ion collisions. The characteristic time of instability
growth is shorter or at least comparable to other time scales of
the parton system evolution. And the instabilities -- usually not
one but several modes are generated -- drive the system towards
isotropy, thus speeding up its equilibration. Since the apparent
very fast equilibration \index{Equilibration} of the QGP is not
understood so far, this aspect has been under strong investigation
in the recent past
\cite{Mrowczynski:2000ed,Birse:2003qp,Randrup:2003cw,Romatschke:2003ms,
Arnold:2003rq,Romatschke:2004jh,Mrowczynski:2004kv,Rebhan:2004ur,Dumitru:2005gp,Arnold:2005vb,Rebhan:2005re}.

The equilibration due to instabilities only happens indirectly,
because the instabilities driven isotropization is a mean-field
reversible process, which does not produce entropy
\cite{Mrowczynski:1994xv,Dumitru:2005gp}. Therefore, the
collisions\index{Collisions}, which are responsible for the
dissipation, are needed to reach the equilibrium state of maximal
entropy. Recently it has been argued \cite{Arnold:2004ti} that the
hydrodynamic collective behavior does not require local
thermodynamic equilibrium but a merely isotropic momentum
distribution of liquid components. This would mean that the
success of hydrodynamic models in describing the data indicates
that not the equilibration time but rather the isotropization time
is of order 1 fm$/$c.

In 2003, \textsc{Mr\'owczy\'nski} and \textsc{Randrup}
\cite{Randrup:2003cw} have performed a phenomenological estimate
of the growth rate of the instabilities for scenarios relevant to
the quark-gluon plasma produced at RHIC or LHC, using however a
different implementation of the anisotropy than
Eq.\,(\ref{anisodist}). They found that the degree of
amplification of the instabilities is not expected to dominate the
dynamics of the quark-gluon plasma, but instead their effect would
be comparable to the contribution from elastic \textsc{Boltzmann}
collisions. However, they also pointed out that if a large number
of unstable modes would be excited then their combined effect on
the overall dynamics could well be significant. In the same year,
\textsc{Arnold, Lenaghan} and \textsc{Moore} \cite{Arnold:2003rq}
investigated the case corresponding to the $\xi\rightarrow \infty$
limit in the prescription described above, arguing that it drastically modifies the
scenario of ``bottom-up thermalization'' advocated by
\textsc{Baier, Mueller, Schiff} and \textsc{Son}
\cite{Baier:2000sb}, which would then have to be replaced by a
different scheme.

%Furthermore, the presence of instabilities will in general
%prohibit the calculation of physical quantities in a perturbative
%framework, since they correspond to unregularized singularities of
%the propagator.

\subsection{Seeds and mechanism of filamentation}
\label{secfilamentation}
\index{Filamentation} An anisotropic system has a natural tendency
to split into current filaments parallel to the direction of
the momentum surplus. These currents are seeds of the
filamentation instability.\index{Instabilities!Weibel instability}
We now explain in terms of elementary physics why the fluctuating
currents, which flow in the direction of the momentum surplus, can
grow in time. To simplify the discussion, which follows
\cite{Mrowczynski:1996vh}, we consider an electromagnetic
anisotropic system. The form of the fluctuating current is chosen
to be\footnote{For now we assume a distribution with the momentum
surplus in the $z$ direction.}
\begin{equation}
 \label{flu-cur}
    {\bf j}(x) = j \: \hat {\bf e}_z \: {\rm cos}(k_x x) \,,
\end{equation} where $\hat {\bf e}_z$
is the unit vector in the $z$ direction. From Eq.~(\ref{flu-cur})
we can infer the existence of current filaments of the thickness
$\pi /\vert k_x\vert$ with the current flowing in the opposite
directions in the neighboring filaments.
\begin{figure}
    \begin{center}
        \includegraphics[width=19.5pc]{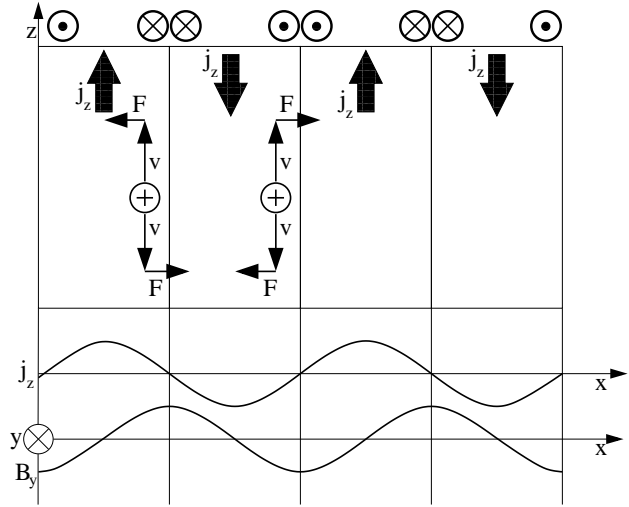}
        \caption{The mechanism that leads to the filamentation instability, see text for a
    description.} \label{fig:mechanism}
    \end{center}
\end{figure}
The magnetic field generated by the current (\ref{flu-cur}) is
given by
\begin{align}
    {\bf B}(x) = \frac{j}{k_x} \: \hat {\bf e}_y \: {\rm sin}(k_x x) \,,
\end{align}
and the \textsc{Lorentz} force acting on the partons, which move
along the $z$ direction, equals
\begin{align}
     {\bf F}(x) = q \: {\bf v} \times {\bf B}(x) = - q \: v_z \: \frac{j}{k_x} \: \hat {\bf e}_x \: {\rm sin}(k_x x) \,,
\end{align}
where $q$ is the electric charge. In Fig.\,\ref{fig:mechanism} it
is shown how the force distributes the partons in such a way that
those, which positively contribute to the current in a given
filament, are focused in the filament center while those, which
negatively contribute, are moved to the neighboring one. Thus,
both the initial current and the magnetic field generated by this
current are growing. The instability is driven by the the energy
transferred from the particles to fields. More specifically, the
kinetic energy related to a motion along the direction of the
momentum surplus is used to generate the magnetic field. A
somewhat different explanation of the mechanism of the
\textsc{Weibel} \index{Instabilities!Weibel
instability}instability is given in \cite{Arnold:2003rq}.

\subsection{Isotropization and Abelianization}
\index{Isotropization} \label{iso-abel} When the instabilities
grow the system becomes more isotropic because the
\textsc{Lorentz} force changes the particles' momenta.
Additionally the growing fields carry momentum directed into
directions with momentum deficit. Let us stick with the assumption
that there is a momentum surplus in the $z$ direction. The
fluctuating current flows in the $z$ direction with the wave
vector pointing in the $x$ direction. Since the magnetic field has
a $y$ component, the \textsc{Lorentz} force, which acts on partons
flying along the $z$ axis, pushes the partons in the $x$ direction
where there is a momentum deficit. Numerical simulations
\cite{Rebhan:2004ur,Dumitru:2005gp,Arnold:2005vb,Rebhan:2005re}
show that growth of the instabilities is indeed accompanied by
the system's fast isotropization.

The system isotropizes not only due to the effect of the
\textsc{Lorentz} force but also due to the momentum carried by the
growing field. When the magnetic and electric fields are oriented
along the $y$ and $z$ axes, respectively, the \textsc{Poynting}
vector points in the direction $x$ that is along the wave vector.
Thus, the momentum carried by the fields is oriented in the
direction of the momentum deficit of particles.

Unstable modes cannot grow to infinity and even in the
electron-ion plasma there are several possible mechanisms which
stop the instability growth \cite{Kato:2005wv}. The actual
mechanism depends on the plasma state as well as on the external
conditions. In the case of the quark-gluon plasma one suspects
that non-Abelian non-linearities can play an important role here.
An elegant argument \cite{Arnold:2004ih} suggests that the
non-linearities do not stabilize the unstable modes because the
system spontaneously chooses an Abelian configuration in the
course of instability development. In Appendix \ref{app:abel} we
explain this argument in detail.

\subsection{Summary of recent numerical simulations}
As mentioned above, the temporal evolution of an anisotropic
quark-gluon plasma has been studied numerically
\cite{Rebhan:2004ur,Dumitru:2005gp,Arnold:2005vb,Rebhan:2005re,Arnold:2005ef,Dumitru:2005hj,Dumitru:2006pz,Romatschke:2006wg}.
The three groups of authors use two very different dynamical
schemes. In the simulations
\cite{Rebhan:2004ur,Arnold:2005vb,Rebhan:2005re,Arnold:2005ef,Romatschke:2006wg}
the dynamics are described by Eq.\,(\ref{HTLvlasov}) together with
the \textsc{Yang-Mills} equation, both in the local
representation, where the quark, anti-quark and gluon
distributions are parameterized by the field
$W^{\mu}(\mathbf{v},X)$ through the relations
\begin{align}
    \delta n^{q}(\mathbf{p},X)=g \frac{\partial f^{q}(\mathbf{p})}{\partial p^\mu}W(\mathbf{v},X)\,,\notag\\
    \delta n^{\bar{q}}(\mathbf{p},X)=-g \frac{\partial f^{\bar{q}}(\mathbf{p})}{\partial p^\mu}W(\mathbf{v},X)\,,\notag\\
    \delta n^{g}(\mathbf{p},X)=g \frac{\partial f^{g}_a(\mathbf{p})}{\partial p^\mu} T^a \, \text{Tr} \left(t^a W(\mathbf{v},X)\right)\,.
\end{align}
Note that all used quantities have been defined below
Eq.\,(\ref{HTLvlasov}). Instead of the three transport equations
(\ref{HTLvlasov}) one has only one equation
\begin{equation}
    v_\mu D^\mu W^\nu(\mathbf{v},X)=-v_\rho F^{\rho\nu}(X)\,,\label{Wfield}
\end{equation}
and the \textsc{Yang-Mills} equation reads
\begin{equation}
    D_\mu F^{\mu\nu}(X)=J^\nu(X)=-g^2\int \frac{d^3p}{(2\pi)^3}\frac{p^\nu}{|\mathbf{p}|}\frac{\partial f(\mathbf{p})}{\partial p^\rho}W^\rho(\mathbf{v},X)\,,\label{WfieldYM}
\end{equation}
with $f(\mathbf{p})$ as in Eq.\,(\ref{f}). These simulations are
reliable as long as the potential's amplitude is not too large,
i.e., $A_a^\mu\ll p_{\text{hard}}/g$, where $p_{\text{hard}}$ is
the characteristic momentum of the hard partons. Note that
Eqs.\,(\ref{Wfield}) and (\ref{WfieldYM}) only describe small
fluctuations around the stationary homogeneous state, and
therefore only a small fraction of the hard particles is actually
influenced by the growing chromo-dynamic field. Hence, one may
consider the hard particles an effective stationary (anisotropic)
background.

The simulations \cite{Dumitru:2005gp,Dumitru:2005hj,Dumitru:2006pz} use the classical, not linearized version of Eqs.\,(\ref{HTLvlasov}). This means that the QGP is treated as a completely classical system, in which partons, that carry classical color charges, interact with the self-consistently generated classical chromodynamic field. In Chapter \ref{chap:wong}, we will discuss this kind of simulation in great detail.

The simulations \cite{Rebhan:2004ur,Dumitru:2005gp} have been performed in 1+1 dimensions, whereas the calculations \cite{Arnold:2005vb,Rebhan:2005re} are full 1+3 dimensional simulations. In most cases the SU(2) gauge group was studied but some SU(3) results, which are qualitatively similar to the SU(2) ones, are given in \cite{Rebhan:2005re}.

The initial conditions used in the simulations \cite{Rebhan:2004ur,Dumitru:2005gp,Arnold:2005vb,Rebhan:2005re,Arnold:2005ef,Dumitru:2005hj,Dumitru:2006pz} are quite similar while the techniques of discretization are different. The initial field amplitudes have a \textsc{Gaussian} white noise distribution, whereas the momentum distribution of the hard particles is strongly anisotropic. For example in the classical simulation \cite{Dumitru:2005gp} the initial parton momentum distribution is chosen as
\begin{equation}
    f(\mathbf{p})\sim \delta(p_x) e^{-\sqrt{p_y^2+p_z^2}/p_{\text{hard}}}\,,\label{dumitrudistr}
\end{equation}
with $p_{\text{hard}}=10$ GeV. The results turn out to be
insensitive to the specific form of the momentum distribution. If
the parton distribution function is written in the form
(\ref{anisodist}), the results are shown
\cite{Romatschke:2003ms,Rebhan:2004ur} to only depend on the
strength of the anisotropy parameterized by $\xi$ and the
\textsc{Debye} mass $m_D$ (Eq.\,(\ref{debyemass})) of the
corresponding isotropic system.

 \begin{figure}[H]
   \begin{center}
      \includegraphics[width=10cm]{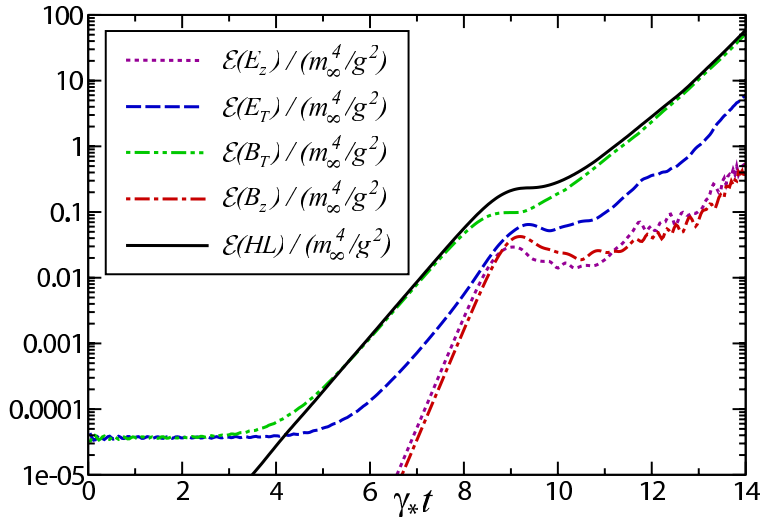}
      \caption{Time evolution of the (scaled) energy density (split into
various electric and magnetic components) which is carried by the
chromo-dynamic field. The simulation is 1+1 dimensional and the
gauge group is SU(2). The parton momentum distribution is squeezed
along the $z$ axis. The solid line corresponds to the total energy
transferred from the particles. $\gamma_*$ is the growth rate of
the fastest growing unstable mode. The figure is taken from
\cite{Rebhan:2004ur}.}
       \label{fig:HLsim1}
   \end{center}
\end{figure}

\begin{figure}[htb]
   \begin{center}
      \includegraphics[width=10cm]{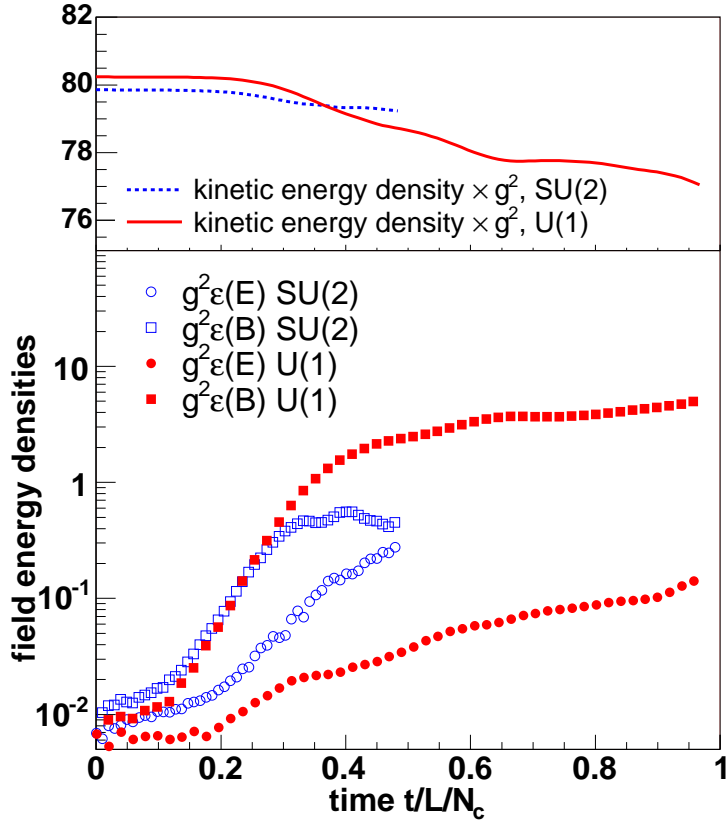}
      \caption{Time evolution of the kinetic energy of particles (upper panel)
and of the energy of electric and magnetic fields (lower panel)
in ${\rm GeV}/{\rm fm}^3$ for the U(1) and SU(2) gauge groups. The
figure is taken from \cite{Dumitru:2005gp}.}
       \label{fig:classim1}
   \end{center}
\end{figure}

The results from the HL-simulation in 1+1 dimensions are shown in
Fig. \ref{fig:HLsim1}, taken from \cite{Rebhan:2004ur}. One
observes an exponential growth of the energy density stored in the
fields, which is dominated by chromo-magnetic fields which are
transverse to the direction of the momentum deficit ($B_T$). Fig.
\ref{fig:classim1}, taken from \cite{Dumitru:2005gp}, shows
results of the classical simulation on a 1+1 dimensional lattice
of physical size $L=40$ fm. As in Fig. \ref{fig:HLsim1} the amount
of field energy grows exponentially and again the magnetic
contribution dominates.
The Abelian (U(1)) and non-Abelian (SU(2)) results are remarkably
similar to each other, which means that the Abelianization,
mentioned in Section \ref{iso-abel}, appears to be very efficient
in 1+1 dimensions.

\begin{figure}[htb]
   \begin{center}
      \includegraphics[width=10cm]{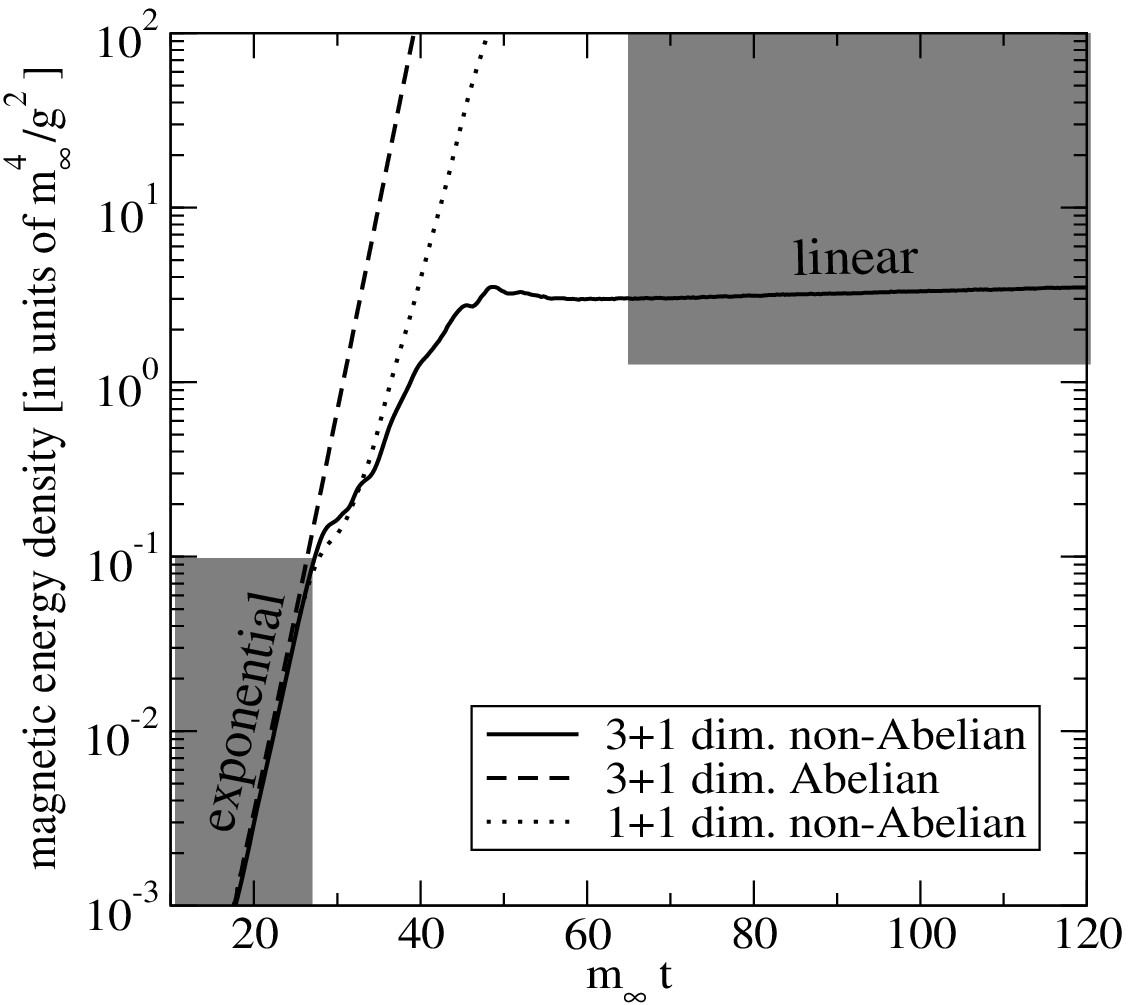}
      \caption{Time evolution of the (scaled) chromo-magnetic energy
density in the 1+3 dimensional simulation. The Abelian result and
that of 1+1 dimensions are also shown. The figure is taken from
\cite{Arnold:2005vb}.}
       \label{fig:HLsim2}
   \end{center}
\end{figure}

\begin{figure}[htb]
\vspace{0.5cm}
   \begin{center}
      \includegraphics[width=10cm]{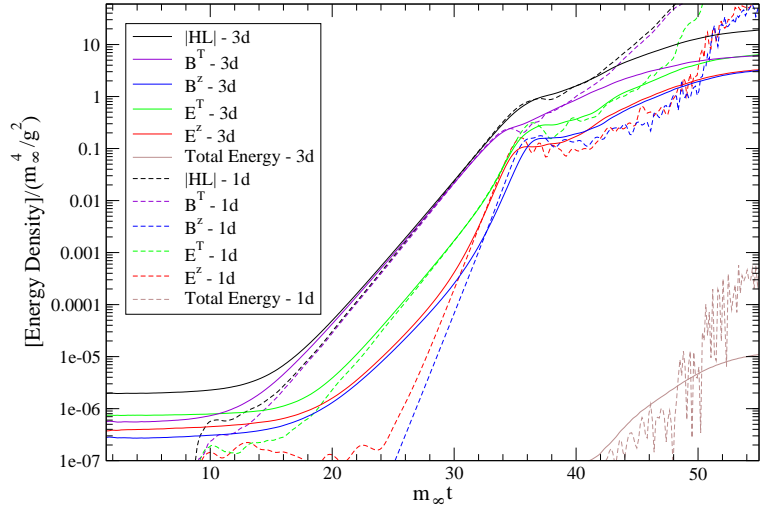}
      \caption{Time evolution of the (scaled) energy density (split into
various electric and magnetic components) of the chromo-dynamic
field in the 1+1 and 1+3 simulations. `HL' denotes the total
energy contributed by hard particles. The figure is taken from
\cite{Rebhan:2005re}.}
       \label{fig:HLsim3}
   \end{center}
\end{figure}

The results of the 1+3 dimensional simulations
\cite{Arnold:2005vb,Rebhan:2005re} are qualitatively different
from the 1+1 dimensional simulations. As seen in Figs.
\ref{fig:HLsim2} and \ref{fig:HLsim3}, taken from
\cite{Arnold:2005vb} and \cite{Rebhan:2005re} respectively, the
growth of the field energy density is exponential only for some
time; then it becomes linear.
It appears that the behavior changes when the field amplitude
becomes of the order $k/g$, where $k$ is the characteristic field
wave vector. Then, the non-Abelian effects start to be important.
The regime of linear growth of the magnetic energy, shown in Figs.
\ref{fig:HLsim2} and \ref{fig:HLsim3}, was studied in more detail
in \cite{Arnold:2005ef}. There it was found that when the
exponential growth of the magnetic energy ends, the
long-wavelength modes associated with the instability stop
growing, but they \emph{cascade} energy towards the ultraviolet in
the form of plasmon excitations. This way a quasi-stationary state
with a power-law distribution $k^{-2}$ of the plasmon mode
population is created. This phenomenon was argued to be very
similar to the \textsc{Kolmogorov} wave turbulence, where
long-wavelength modes transfer their energy to shorter ones
without dissipation. In the classical simulation
\cite{Dumitru:2006pz}, when the field strength is large enough,
the energy drained from the particles by the instability does not
build up exponentially in magnetic fields but instead returns
isotropically to the ultraviolet, not via the quasi-stationary
process, as argued in \cite{Arnold:2005ef}, but via a rapid
\emph{avalanche}.

\begin{figure}[htb]
   \begin{center}
      \includegraphics[width=10cm]{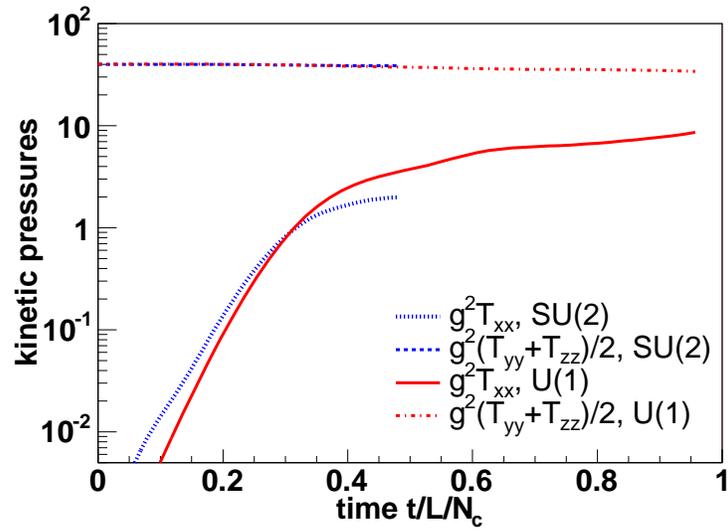}
      \caption{Temporal evolution of the energy-momentum tensor
components $T^{xx}$ and $(T^{yy}+T^{zz})/2$. The Abelian and
non-Abelian results are shown. The figure is taken from
\cite{Dumitru:2005gp}.}
       \label{fig:classimiso}
   \end{center}
\end{figure}

The 1+1 dimensional classical simulation \cite{Dumitru:2005gp}
nicely shows the effect of isotropization \index{Isotropization}
due to the action of the \textsc{Lorentz} force. In Fig.
\ref{fig:classimiso}, taken from \cite{Dumitru:2005gp}, there are
shown the diagonal components of the energy-momentum tensor
\begin{equation}
    T^{\mu\nu}= \int \frac{d^3p}{(2\pi)^3}\frac{p^\mu p^\nu}{E_p} f(\mathbf{p})\,.
\end{equation}
The initial momentum distribution is given by
Eq.\,(\ref{dumitrudistr}), such that $T^{xx}=0$ at $t=0$. Fig.
\ref{fig:classimiso} shows that $T^{xx}$ grows exponentially, but
full isotropy, which requires $T^{xx}=(T^{yy}+T^{zz})/2$, is not
achieved. So far, all mentioned numerical studies have been
dealing with quark-gluon systems of constant volume. In
\cite{Romatschke:2006wg,Rebhan:2008uj} a system that expands
boost-invariantly in one direction is treated and it is shown
numerically as well as analytically that the expansion slows down
the growth of the unstable modes, even if the system is highly
anisotropic. The field amplitude grows like an exponential with
the exponent proportional to $\sqrt{t}$. Further quantitative
analysis of the effect of expansion on the instability growth is
necessary, since it may turn out that they are irrelevant for
heavy-ion collisions in the case that they grow too slowly to
compete with the expansion.

Unstable parton systems under conditions close to those realized
in heavy-ion collisions have recently been studied in
\cite{Romatschke:2005pm,Romatschke:2005ag,Romatschke:2006nk}. The
system was described in terms of the Color Glass Condensate
\index{Color Glass Condensate} approach \cite{Iancu:2003xm} where
small $x$ partons of large occupation numbers, which dominate the
wave functions of incoming nuclei, are treated as classical
\textsc{Yang-Mills} fields. In these simulations the hard modes of
the classical fields play the role of particles. The
instabilities, identified as the \textsc{Weibel} modes, appear to
be generated when the system of \textsc{Yang-Mills} fields expands
into vacuum.

      \chapter{A model approach to the inclusion of collisions}
\epigraphwidth 250pt \epigraph{Sehr leicht zerstreut der Zufall
was er sammelt.}{\emph{Torquato Tasso. Erster Aufzug. Erster Auftritt. Leonore.}\\
Johann Wolfgang von Goethe (1749-1832)} \label{collisionsmodel}
\index{Collisions} As pointed out already in Section
\ref{instabilities}, collisions, being responsible for the
dissipation, are needed to drive the system towards equilibrium,
even if instabilities speed up isotropization. All calculations so
far have been carried out at leading order in perturbation theory,
such that collisions among the hard particles that enter at higher
orders in $g$ could be neglected when concentrating on collective
modes with soft momenta of the order $gT$. However, in heavy-ion
collision experiments at the RHIC and LHC the couplings expected
are of the order $\alpha_s \sim 0.2-0.3$ and higher order terms
will be important. Hence, collisions can not simply be neglected
and their effect on the system's evolution, particularly on the
collective modes, has to be investigated.

In this chapter we give a first quantitative estimate of how the
collisions affect the dispersion relations of the collective modes
and the growth of instabilities in particular. To achieve this, we
introduce a model for the inclusion of collisions based on the
\textsc{Vlasov} equations for QCD combined with a
\textsc{Bhatnagar-Gross-Krook} (BGK)-type collision term
\cite{Bhatnagar:1954}.\footnote{This chapter is based on the work
published in \cite{Schenke:2006xu}.} We will deal with anisotropic
systems with momentum distributions of the kind (\ref{anisodist}),
introduced and discussed in Chapter \ref{anisotropy}. We
concentrate on the case in which the direction of anisotropy is
parallel to the wave vector of the regarded collective mode,
because in this case the growth rate of the magnetic instability
is maximal (see Sections \ref {unstable modes} and
\ref{instabilities}) and analytic expressions for the structure
functions can be found.

\section{Structure functions and dispersion relations}
In the effectively Abelian limit, discussed in Section
\ref{solving}, the equation of motion holds for each color channel
separately. We now include a BGK-type collision term
\cite{Bhatnagar:1954} \index{Collisions!BGK collision term}on the
right hand side of Eq.\,(\ref{HTLvlasov2}), which becomes
\begin{equation}
V\cdot \partial_X \delta f^{i}_a(p,X) + g \theta_{i}
V_{\mu}F^{\mu\nu}_a(X)\partial_{\nu}^{(p)}f^{i}(\mathbf{p})=\mathcal{C}^{i}_a(p,X)\label{trans2}\,\text{,}
\end{equation}
with
\begin{equation}
    \mathcal{C}^{i}_a(p,X)=-\nu\left[f^{i}_a(p,X)-\frac{N^{i}_a(X)}{N^{i}_{\text{eq}}}f^{i}_{\text{eq}}(|\mathbf{p}|)\right]\,\text{,}\label{collision}
\end{equation}
where $f^{i}_a(p,X)=f^{i}(\mathbf{p}) + \delta f^{i}_a(p,X)$. This
term describes how collisions equilibrate the system within a time
proportional to $\nu^{-1}$. We will assume that the collision rate
$\nu$ is independent of momentum and particle species, however,
these assumptions are easily relaxed. Note that these collisions
are not color-rotating. The particle numbers are given by
\begin{align}
    N^{i}_a(X)=\int_{\mathbf{p}} f^{i}_a(p,X) \text{ , ~} N^{i}_{\text{eq}}=\int_{\mathbf{p}} f^{i}_{\text{eq}}(|\mathbf{p}|)=\int_{\mathbf{p}} f^{i}(\mathbf{p})\text{\,.}\label{numbers}
\end{align}
The difference between the BGK collisional kernel
(\ref{collision}) and the conventional relaxation-time
approximation (RTA) is the multiplication of the second term in
brackets by the ratio of the density over the equilibrium density.
RTA simply takes the difference of the distribution function and
the equilibrium distribution function implicitly setting this
ratio to one. The advantage of the BGK kernel over an RTA kernel
lies in the fact that the number of particles is instantaneously
conserved by the BGK collisional-kernel \cite{Manuel:2004gk},
i.e., that
\begin{equation}
    \int_{\mathbf{p}} \mathcal{C}^{i}_a(p,X)=0\text{\,.}
\end{equation}
This simply states that the collisions can only occur if a
particle is present and that only the momentum of the particles
will change as a result, not the particle number.  This condition
is violated by RTA.

The inclusion of a BGK collisional kernel allows for simulation of
the effect of binary collisions with substantial momentum transfer
and is merely an approximation for collisions between the hard
charged particles in a hot quark-gluon plasma. However, it is a
reasonable way to yield a first quantitative answer to the
question of how collisions among hard particles affect the
collective modes of QCD. The effects of such a collision term on
the dispersion relations in the ultra-relativistic case for an
isotropic system were investigated in \cite{Carrington:2003je}.
For now the collision rate $\nu$ is taken to be a free parameter
and we postpone the estimation of its magnitude to the discussions
in Section \ref{discussions}.

Using (\ref{numbers}) we can write
\begin{align}
V\cdot \partial_X \delta f^{i}_a(p,X) + &g \theta_{i}
V_{\mu}F^{\mu\nu}_a(X)\partial_{\nu}^{(p)}f^{i}(\mathbf{p})=\notag\\
&-\nu\left[f^{i}(\mathbf{p})+\delta
f^{i}_a(p,X)-\left(1+\frac{\int_{\mathbf{p}} \delta
f^{i}_a(p,X)}{N^{i}_{\text{eq}}}\right)
    f^{i}_{\text{eq}}(|\mathbf{p}|)\right]\text{\,.}\label{trans3}
\end{align}
Solving for $\delta f^{i}_a(p,X)$ and
\textsc{Fourier}-transforming as in (\ref{HTLvlasov3}) and
(\ref{HTLvlasov4}) leads to the result for the linearized induced
current \index{Induced current} by each particle species $i$. From
Eq.\,(\ref{trans3}) we immediately get
\begin{align}
    (-i\omega+i\mathbf{v}\cdot\mathbf{k}+\nu)\delta f^{i}(p,K) = &-g\theta_iV_{\mu}F^{\mu\nu}(K)\partial_{\nu}^{(p)}f^{i}(\mathbf{p})+\notag\\
    &+\nu(f^{i}_{\text{eq}}(\mathbf{p})-f^{i}(\mathbf{p}))+\nu \frac{f^{i}_{\text{eq}}(\mathbf{p})}{N_{\text{eq}}} \int_{\mathbf{p}^{\prime}} \delta f^{i}(p^{\prime},K)\,\text{,}
\end{align}
where $\delta f^{i}(p,K)$ and $F^{\mu\nu}(K)$ are the
\textsc{Fourier}-transforms of $\delta f^{i}(p,X)$ and
$F^{\mu\nu}(X)$, respectively. This yields
\begin{equation}
    \delta f^{i}(p,K)=\frac{-ig\theta_iV_{\mu}F^{\mu\nu}(K)\partial_{\nu}^{(p)}f^{i}(\mathbf{p})+i\nu(f^{i}_{\text{eq}}(\mathbf{p})-f^{i}(\mathbf{p}))+i\nu f^{i}_{\text{eq}}(\mathbf{p})\left( \int_{\mathbf{p}^{\prime}}\delta f^{i}(p^{\prime},K) \right)/N_{\text{eq}}}{\omega-\mathbf{v}\cdot\mathbf{k}+i\nu}\text{\,.}
\end{equation}
Defining
\begin{equation}
    \delta f^{i}_0(p,K)=\left(-ig\theta_iV_{\mu}F^{\mu\nu}(K)\partial_{\nu}^{(p)}f^{i}(\mathbf{p})+i\nu(f^{i}_{\text{eq}}(\mathbf{p})-f^{i}(\mathbf{p}))\right)D^{-1}(K,\mathbf{v},\nu)\,\text{,}
\end{equation}
with $D(K,\mathbf{v},\nu)=\omega-\mathbf{k}\cdot\mathbf{v}+i\nu$
we can write
\begin{align}
    \delta f^{i}(p,K)=\delta f^{i}_0(p,K)&+i\nu D^{-1}(K,\mathbf{v},\nu)\frac{f^{i}_{\text{eq}}(\mathbf{p})}{N_{\text{eq}}}\int_{\mathbf{p}^{\prime}}\delta f^{i}_0(p^{\prime},K)\notag\\
    &+i\nu D^{-1}(K,\mathbf{v},\nu)\frac{f^{i}_{\text{eq}}(\mathbf{p})}{N_{\text{eq}}} \frac{i\nu}{N_{\text{eq}}} \int_{\mathbf{p}^{\prime}} f^{i}_{\text{eq}}(\mathbf{p}^{\prime})
    D^{-1}(K,\mathbf{v}^{\prime},\nu)
    \int_{\mathbf{p}^{\prime\prime}}\delta f^{i}_0(p^{\prime\prime},K)\notag\\
    &+\ldots
\end{align}
Using the shorthand notation
\begin{equation}
\eta(K)=\int_{\mathbf{p}}\delta f^{i}_0(p,K)\,,
\end{equation}
and
\begin{equation}
\lambda(K,\nu)=\frac{i\nu}{N_{\text{eq}}}\int_{\mathbf{p}}
    f^{i}_{\text{eq}}(\mathbf{p})D^{-1}(K,\mathbf{v},\nu)\,,
\end{equation}
we get
\begin{align}
    \delta f^{i}(p,K)&=\delta f^{i}_0(p,K)+i\nu D^{-1}(K,\mathbf{v},\nu)\frac{f^{i}_{\text{eq}}(\mathbf{p})}{N_{\text{eq}}} \eta(K) \left(1+\lambda+\lambda^2+\ldots\right)\notag\\
&=\delta f^{i}_0(p,K)+i\nu
D^{-1}(K,\mathbf{v},\nu)\frac{f^{i}_{\text{eq}}(\mathbf{p})}{N_{\text{eq}}}
\eta(K)\frac{1}{1-\lambda}\,\text{.}
\end{align}
Using
\begin{equation}
    J_{\text{ind}\,a}^{\mu\, i}(K)=g \int_{\mathbf{p}} V^{\mu} \delta f^{i}_a(p,K)\,\text{,}
\end{equation}
we find the final result for the current:
\begin{align}
    J^{\mu\,i}_{\text{ind}\,a}(K)&=g^2 \int_{\mathbf{p}} V^{\mu}
    \partial^{\beta}_{(p)}
    f^{i}(\mathbf{p})\mathcal{M}_{\gamma\beta}(K,V)D^{-1}(K,\mathbf{v},\nu)A^{\gamma}_a(K) +g\nu \mathcal{S}^{i}(K,\nu) \notag\\
    &~~~~+g \frac{i \nu}{N^{i}_{\text{eq}}}\int_{\mathbf{p}}
    V^{\mu}f^{i}_{\text{eq}}(|\mathbf{p}|)D^{-1}(K,\mathbf{v},\nu)\notag\\
    &~~~~\times g \left[\int_{\mathbf{p}^{\prime}}\partial^{\beta}_{(p^{\prime})}f^{i}(\mathbf{p}^{\prime})
    \mathcal{M}_{\gamma\beta}(K,V^{\prime})D^{-1}(K,\mathbf{v}^{\prime},\nu)A^{\gamma}_a(K)+ g \nu \mathcal{S}^{i}
    (K,\nu)\right]\mathcal{W}_{i}^{-1}(K,\nu)\,\text{,}\label{current}
\end{align}
with
\begin{equation}
\mathcal{M}_{\gamma\beta}(K,V):=g_{\gamma\beta}(\omega-\mathbf{k}\cdot\mathbf{v})-V_{\gamma}K_{\beta}\,\text{,}
\end{equation}
\begin{equation}
D(K,\mathbf{v},\nu):=\omega+i\nu-\mathbf{k}\cdot\mathbf{v}\,\text{,}
\end{equation}
\begin{equation}
\mathcal{S}^{i}(K,\nu):=\theta_{i} \int_{\mathbf{p}}
V^{\mu}[f^{i}(\mathbf{p})-f^{i}_{\text{eq}}(|\mathbf{p}|)]D^{-1}(K,\mathbf{v},\nu)\,\text{,}
\end{equation}
and
\begin{equation}
\mathcal{W}_{i}(K,\nu):=1-\frac{i
\nu}{N^{i}_{\text{eq}}}\int_{\mathbf{p}}f^{i}_{\text{eq}}(|\mathbf{p}|)D^{-1}(K,\mathbf{v},\nu)\text{\,.}
\end{equation}
The total induced current is given by
$J^{\mu}_{\text{ind}\,a}(K)=2 N_c J^{g\,\mu}_{\text{ind}\,a}(K) +
N_f \left[J^{q\,\mu}_{\text{ind}\,a}(K)+
J^{\bar{q}\,\mu}_{\text{ind}\,a}(K)\right]$. It can be simplified
due to the fact that the integrals over $f^{i}_{\text{eq}}$ in
Eq.~(\ref{current}) are independent of the particle species:
\begin{align}
    \frac{1}{N^{i}_{\text{eq}}}\int_{\mathbf{p}}f^{i}_{\text{eq}}(|\mathbf{p}|)D^{-1}(K,\mathbf{v},\nu)&=
    \int \frac{d\Omega}{4\pi}D^{-1}(K,\mathbf{v},\nu)\,\text{,}\notag\\
    \frac{1}{N^{i}_{\text{eq}}}\int_{\mathbf{p}}
    V^{\mu}f^{i}_{\text{eq}}(|\mathbf{p}|)D^{-1}(K,\mathbf{v},\nu)&=\int
    \frac{d\Omega}{4\pi}V^{\mu}D^{-1}(K,\mathbf{v},\nu)\,\text{,}\label{simpl}
\end{align}
where $d\Omega=\sin\theta d\theta d\varphi$. Assuming equal
distributions for quarks and antiquarks, the full linearized
induced current reads \index{Induced current}
\begin{align}
    J^{\mu}_{\text{ind}\,a}(K)&=g^2 \int_{\mathbf{p}}V^{\mu}
    \partial^{\beta}_{(p)}
    f(\mathbf{p})\mathcal{M}_{\gamma\beta}(K,V)D^{-1}(K,\mathbf{v},\nu)A^{\gamma}_a
    + 2 N_c g \nu \mathcal{S}^{g}(K,\nu)\notag\\
    &~~+ g^2 (i \nu) \int
    \frac{d\Omega}{4\pi}V^{\mu}D^{-1}(K,\mathbf{v},\nu)\notag\\
    &~~~~~~~~~~~~~~~~\times \int_{\mathbf{p}^{\prime}}\partial^{\beta}_{(p^{\prime})}f(\mathbf{p}^{\prime})
    \mathcal{M}_{\gamma\beta}(K,V^{\prime})D^{-1}(K,\mathbf{v}^{\prime},\nu)\mathcal{W}^{-1}(K,\nu)
    A^{\gamma}_a \notag\\
    &~~ + 2 N_c g^2 (i \nu^2)  \int \frac{d\Omega}{4\pi}V^{\mu}D^{-1}(K,\mathbf{v},\nu)
    \mathcal{S}^{g}(K,\nu) \mathcal{W}^{-1}(K,\nu)\,\text{,}\label{fullcurrent}
\end{align}
where $\mathcal{W}(K,\nu)=1-i \nu \int
\frac{d\Omega}{4\pi}D^{-1}(K,\mathbf{v},\nu)$ and $f(\mathbf{p})$
as in Eq.~(\ref{f}). The self energy is obtained from
Eq.~(\ref{fullcurrent}) as in Section \ref{solving} via
\begin{equation}
    \Pi^{\mu\nu}_{ab}(K)=\frac{\delta J^{\mu}_{\text{ind}\,a}(K)}{\delta
    A_{\nu}^b(K)}\,\text{,}
\end{equation}
resulting in
\begin{align}
    \Pi^{\mu\nu}_{ab}(K)&=\delta_{ab} g^2 \int_{\mathbf{p}} V^{\mu}
    \partial_{\beta}^{(p)}
    f(\mathbf{p})\mathcal{M}^{\nu\beta}(K,V)D^{-1}(K,\mathbf{v},\nu)\notag\\
    &~~~+\delta_{ab} g^2 (i \nu)\int \frac{d\Omega}{4\pi}V^{\mu}D^{-1}(K,\mathbf{v},\nu)\notag\\
    &~~~~~~~~~~~~~~~~~~~~~\times\int_{\mathbf{p}^{\prime}}
    \partial_{\beta}^{(p^{\prime})}f(\mathbf{p}^{\prime})
    \mathcal{M}^{\nu\beta}(K,V^{\prime})D^{-1}(K,\mathbf{v}^{\prime},\nu)\mathcal{W}^{-1}(K,\nu)
    \,\text{,}\label{selfenergy}
\end{align}
which is diagonal in color and can be shown to be transverse,
i.e., $K_{\mu}\Pi^{\mu\nu}=K_{\nu}\Pi^{\mu\nu}=0$. We will from
now on omit the color indices of $\Pi^{\mu\nu}$. The terms in the
induced current involving $\mathcal{S}_{\nu}$ drive the
distribution into an isotropic equilibrium shape. They represent a
parity conserving current and thus create a zero average
electromagnetic field and therefore do not contribute to
instability growth.

So far in this section, the distribution function $f(\mathbf{p})$
in Eq.~(\ref{selfenergy}) has not yet been specified. We now
assume that $f(\mathbf{p})$ is obtained from an isotropic
distribution function by the rescaling of one direction in
momentum space using (\ref{anisodist}):
\begin{equation}
    f(\mathbf{p})=\sqrt{1+\xi}\,f_{\text{iso}}\left(\mathbf{p}^2+\xi(\mathbf{p}\cdot\mathbf{\hat{n}})^2\right)\,\text{,}
    \label{anisodist2}
\end{equation}
for an arbitrary isotropic distribution function
$f_{\text{iso}}(\mid\!\!\mathbf{p}\!\!\mid)$. As in Section
\ref{tensordecomp} we are able to perform the radial part of the
integrations involving $f(\mathbf{p})$ in Eq.\,(\ref{selfenergy})
by changing variables to
$\tilde{p}^2=p^2(1+\xi(\mathbf{v}\cdot\mathbf{\hat{n}})^2)$. The
result is
\begin{align}\label{piij}
    \Pi^{ij}(K)&=m_D^2\sqrt{1+\xi}\int\frac{d\Omega}{4\pi}v^i\frac{v^l+\xi(\mathbf{v}\cdot\mathbf{\hat{n}})n^l}{(1+\xi(\mathbf{v}\cdot\mathbf{\hat{n}})^2)^2}
    \left[\delta^{jl}(\omega-\mathbf{k}\cdot\mathbf{v})+v^jk^l\right]D^{-1}(K,\mathbf{v},\nu)\notag\\
    &~~~~+(i\nu) m_D^2 \sqrt{1+\xi} \int\frac{d\Omega^{\prime}}{4\pi}
    (v^{\prime})^i D^{-1}(K,\mathbf{v}^{\prime},\nu)\notag\\
    &~~~~\times\int\frac{d\Omega}{4\pi}\frac{v^l+\xi(\mathbf{v}\cdot\mathbf{\hat{n}})n^l}{(1+\xi(\mathbf{v}\cdot\mathbf{\hat{n}})^2)^2}
    \left[\delta^{jl}(\omega-\mathbf{k}\cdot\mathbf{v})+v^jk^l\right]D^{-1}(K,\mathbf{v},\nu)\mathcal{W}^{-1}(K,\nu)\,\text{,}
\end{align}
where $m_D^2$ is given by Eq.\,(\ref{debyemass}):
\begin{equation}
    m_D^2=-\frac{g^2}{2\pi^2}\int_0^{\infty}d\tilde{p}\,\tilde{p}^2\frac{d f_{\text{iso}}(\tilde{p}^2)}{d\tilde{p}}\text{\,.}
\end{equation}
We apply the same tensor decomposition as in the collision free
case (see Section \ref{tensordecomp}). Since the growth rate of
the filamentation instability is the largest when the wave vector
is parallel to the direction of the anisotropy, i.e.,
$\mathbf{k}\parallel\mathbf{\hat{n}}$, we concentrate on this
particular case. Then
\begin{equation}
    \mathbf{k}\cdot\mathbf{v}=k\,\mathbf{\hat{n}}\cdot\mathbf{v}=k
    \cos\theta\,\text{,}
\end{equation}
and $\gamma$ and
$\tilde{n}^2=1-(\mathbf{k}\cdot{\mathbf{\hat{n}}})^2$ in
Eq.\,(\ref{deltag}) vanish identically. To determine the poles of
the propagator (\ref{propagator}), and hence the dispersion
relations, we are now left with two separate equations for the
$\alpha-$ and $\beta-$mode, respectively:
\begin{align}
    k^2-\omega^2+\alpha &= 0\,\text{,}\notag\\
    \beta-\omega^2 &= 0\text{\,.}\label{equationstosolve}
\end{align}
Applying the projections (\ref{contractions}) to the self energy
(\ref{piij}) we determine $\alpha$ and $\beta$. The appearing
integrals can be performed analytically and the final results
simplify to:
        \begin{align}
            \alpha(\omega,k,\xi,\nu)=&\frac{m_D^2}{4}\frac{\sqrt{1+\xi}}{k (1+\xi
            z^2)^2}\times
            \notag\\&~~~\times\bigg\{(k(z^2-1)-iz\nu)(1+\xi z^2)-(z^2-1)(k z(1+\xi)-i\nu)\ln\left[\frac{z+1}{z-1}\right]\notag\\
             &~~~~~~~~-\frac{i}{\sqrt{\xi}}\left[z\nu(1+(3+z^2(1-\xi))\xi) \notag\right.\\
             &~~~~~~~~~~~~~~~~~\left.+ik(1-\xi+z^2(1+\xi(6+z^2(\xi-1)+\xi)))\right] \arctan\sqrt{\xi}\bigg\}\,\text{,}
            \label{alphaeq}\\
            \beta(\omega,k,\xi,\nu)=&m_D^2\sqrt{1+\xi}k(kz-i\nu)^2\times\notag\\
            &~~~\times\bigg\{-2\sqrt{\xi}(1+z^2\xi)+(1+\xi)\Big(2z\sqrt{\xi}\ln\left[\frac{z+1}{z-1}\right]
            +2(z^2\xi-1)\arctan\sqrt{\xi}\Big)\bigg\}\notag\\
            &~~~\times\left(2\sqrt{\xi}(1+z^2\xi)^2k^2\left(2k-i\nu\ln\left[\frac{z+1}{z-1}\right]\right)\right)^{-1}\,\text{,}
             \label{betaeq}
        \end{align}
where we abbreviate $z=(\omega+i\nu)/k$.
\section{Stable modes}
\label{stable}
\begin{figure}[htb]
    \begin{center}
      \begin{minipage}{\textwidth}
            \includegraphics[width=15cm]{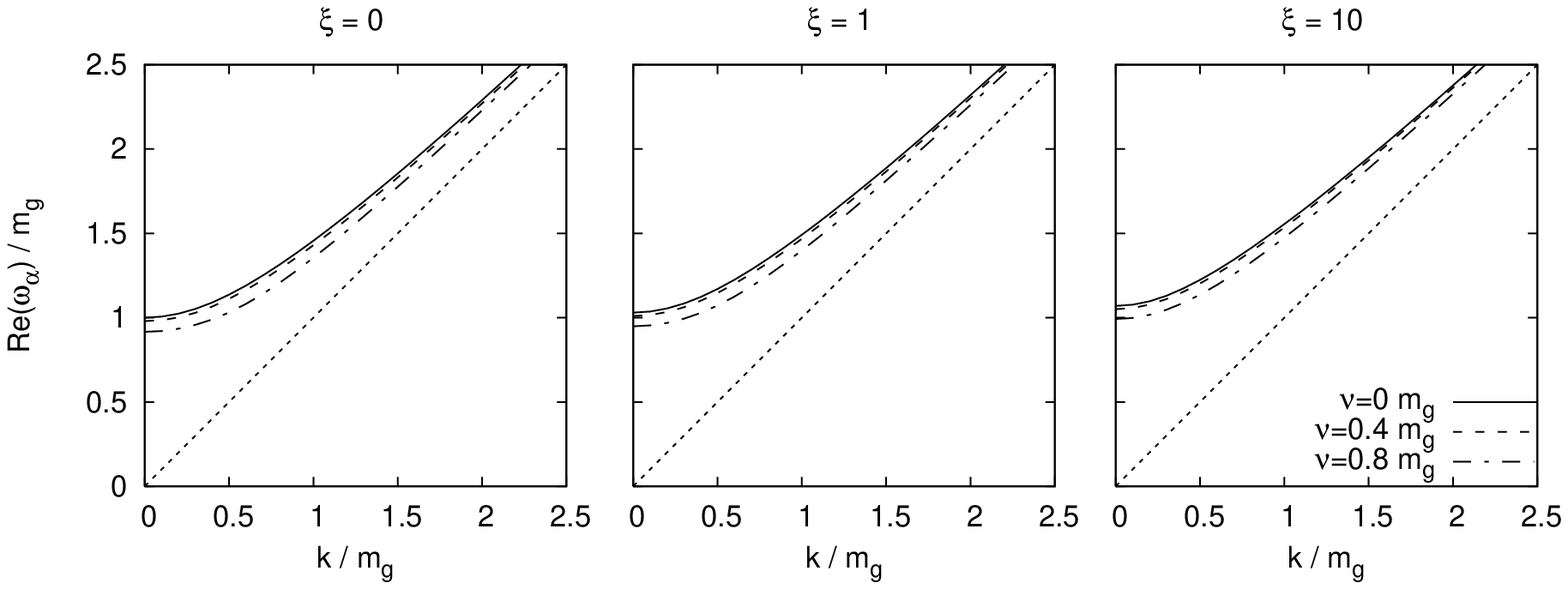}
            \caption{Real part of the dispersion relation for the stable $\alpha$-mode
                     for an anisotropy parameter of $\xi=\{0,1,10\}$ and different collision rates in units of $m_g=m_D/\sqrt{3}$.}
                           \vspace*{5mm}
            \label{fig:stablealphaxire}
      \end{minipage}
      \begin{minipage}{\textwidth}
            \includegraphics[width=15cm]{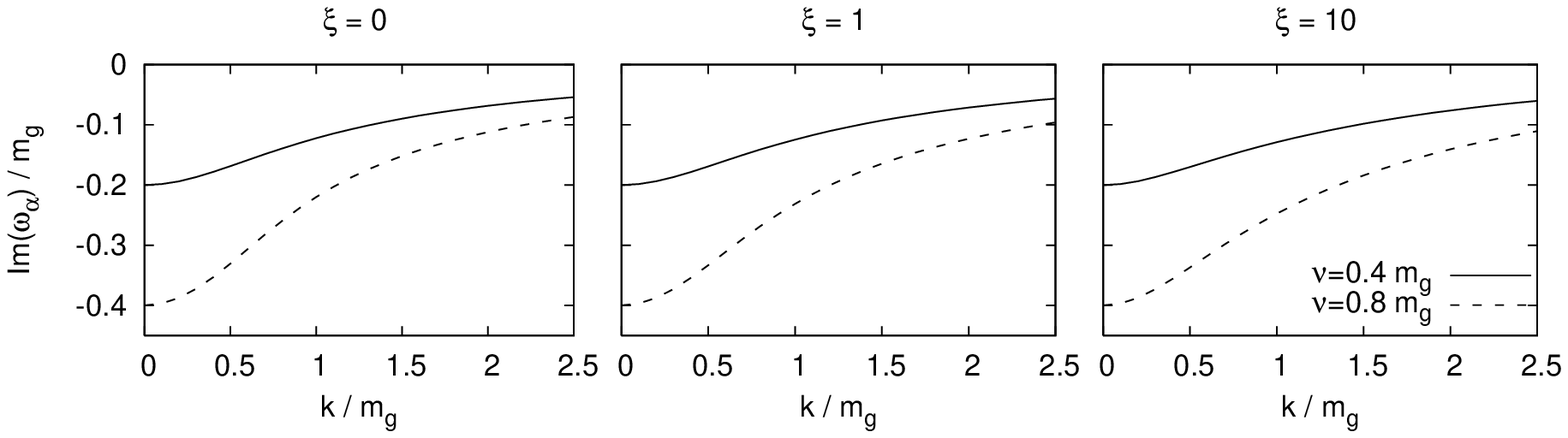}
            \caption{Imaginary part of the dispersion relation for the stable $\alpha$-mode
                     for an anisotropy parameter of $\xi=\{0,1,10\}$ and different collision rates in units of $m_g=m_D/\sqrt{3}$.}
            \label{fig:stablealphaxiim}
      \end{minipage}
    \end{center}
  \end{figure}

\begin{figure}[H]
    \begin{center}
      \begin{minipage}{\textwidth}
        \includegraphics[width=15cm]{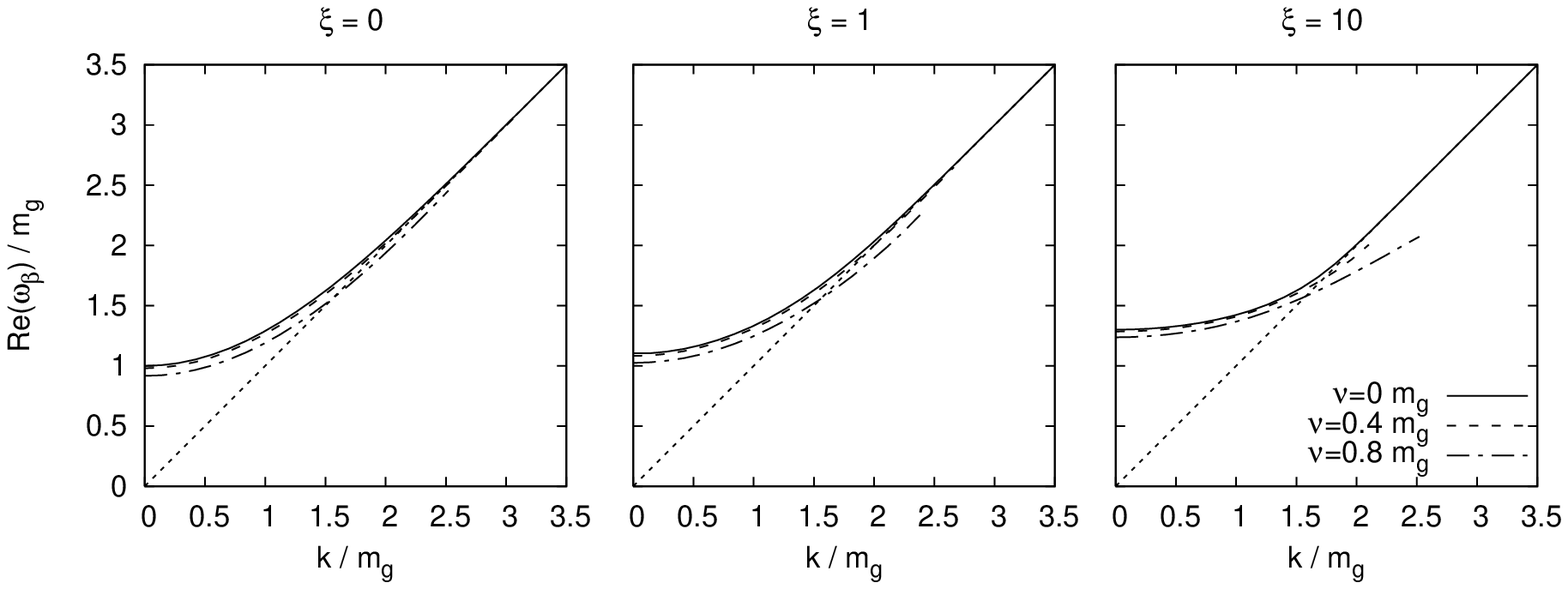}
        \caption{Real part of the dispersion relation for the stable $\beta$-mode for an anisotropy parameter of $\xi=\{0,1,10\}$ and
        different collision rates in units of $m_g=m_D/\sqrt{3}$.
        Note that for finite $\nu$ there is a maximal $k$ beyond which there is no solution.
        It vanishes when $\omega$ crosses the cut in the complex plane,
        which extends from $-k$ to $+k$ at $-i\nu$ (cf. Fig \ref{fig:stablebetaxiim}).}
        \vspace*{5mm}
        \label{fig:stablebetaxire}
      \end{minipage}
      \begin{minipage}{\textwidth}
        \includegraphics[width=15cm]{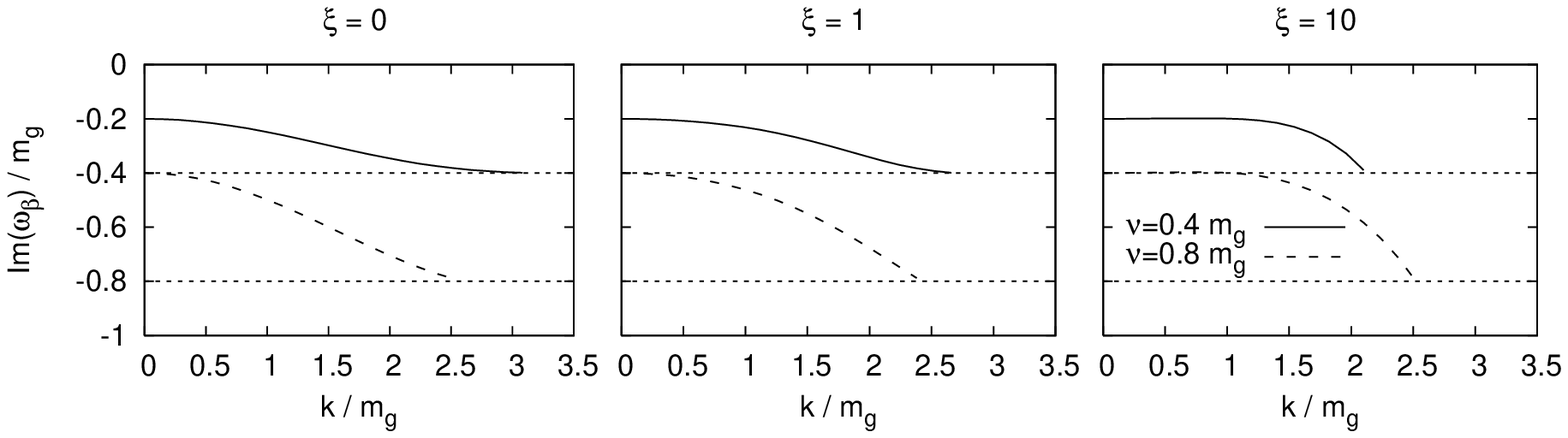}
        \caption{Imaginary part of the dispersion relation for the stable $\beta$-mode for an anisotropy parameter of $\xi=\{0,1,10\}$
        and different collision rates in units of $m_g=m_D/\sqrt{3}$.
        The solution vanishes, when $\omega$ crosses the cut in the complex plane, which extends from $-k$ to $+k$ at $-i\nu$ (indicated by the straight dotted line).}
        \label{fig:stablebetaxiim}
      \end{minipage}
    \end{center}
  \end{figure}

\begin{figure}[H]
    \begin{center}
      \begin{minipage}{\textwidth}
        \includegraphics[width=15cm]{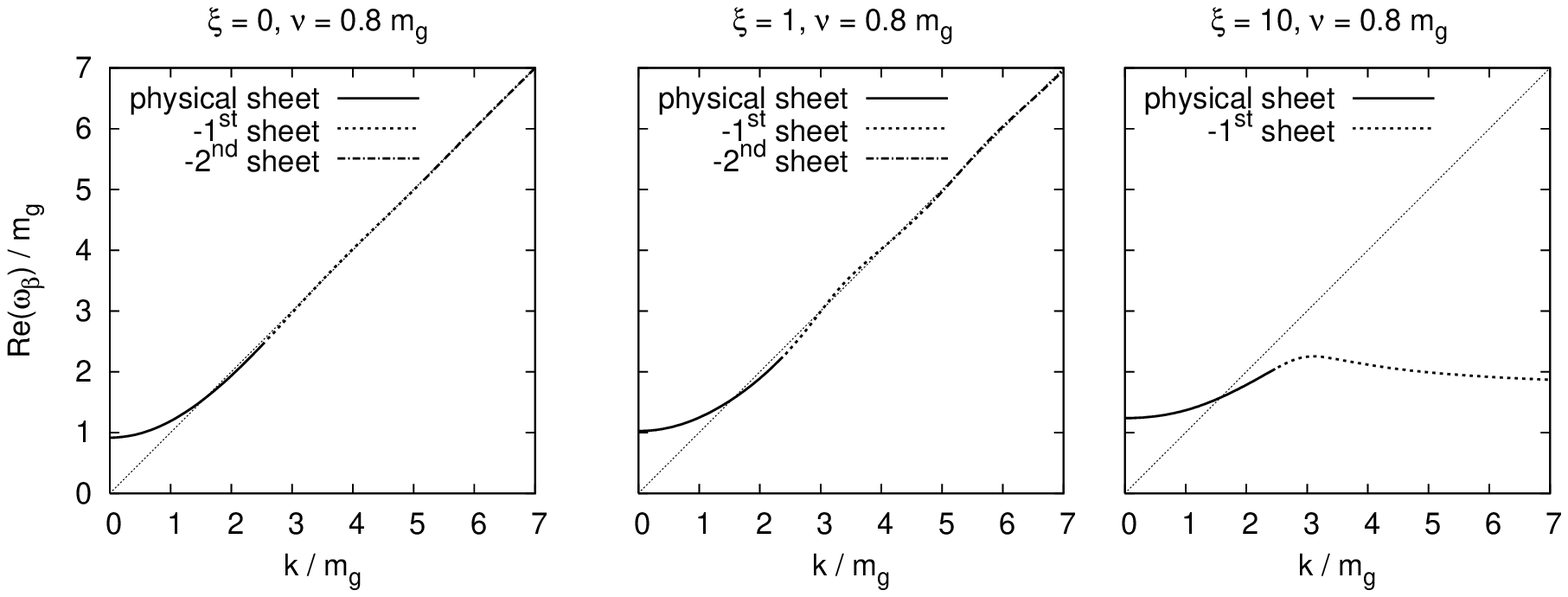}
        \caption{Real part of the dispersion relation for the stable $\beta$-mode for an anisotropy parameter of $\xi=\{0,1,10\}$ and
        $\nu=0.8\,m_g$. It is shown how the solution continues on lower \textsc{\textsc{Riemann}}
        sheets.}
        \vspace*{5mm}
        \label{fig:stablebetahigherxire}
      \end{minipage}
      \begin{minipage}{\textwidth}
        \includegraphics[width=15cm]{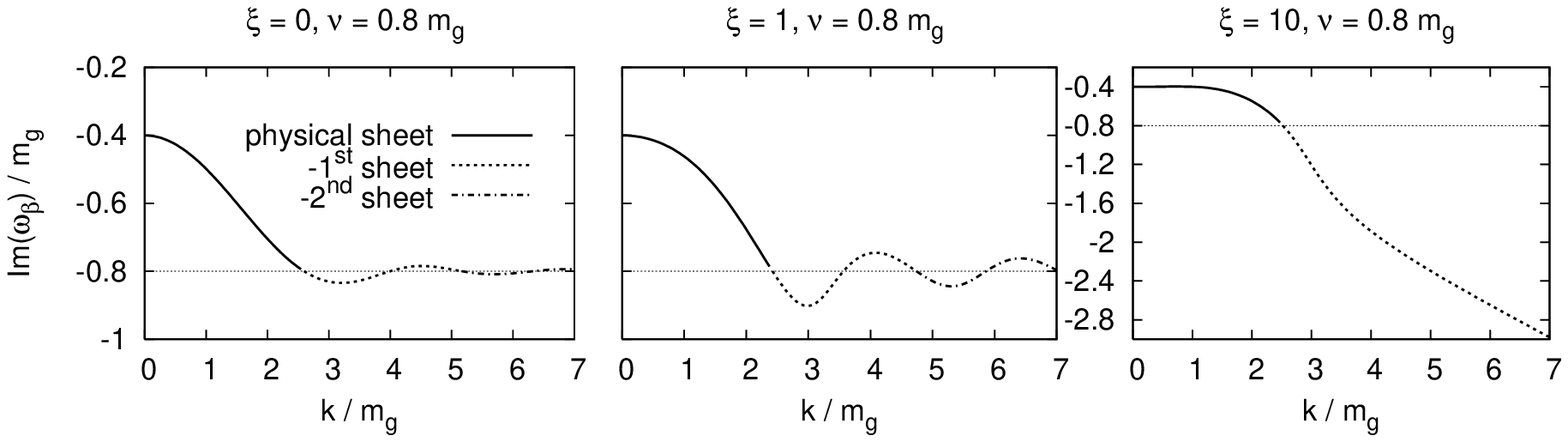}
        \caption{Imaginary part of the dispersion relation for the stable $\beta$-mode for an anisotropy parameter of $\xi=\{0,1,10\}$ and
        $\nu=0.8\,m_g$. It is shown how the solution continues on lower \textsc{Riemann} sheets. Note the different scale for the third plot.}
        \label{fig:stablebetahigherxiim}
      \end{minipage}
    \end{center}
  \end{figure}

    The dispersion relations for all modes are given by the solutions $\omega(k)$ of Eqs.~(\ref{equationstosolve}).
    These solutions are found numerically for
    different values of the collision rate $\nu$.
    The results for the stable transverse ($\alpha$-) mode are
    shown in Figs.~\ref{fig:stablealphaxire} and
    \ref{fig:stablealphaxiim} for two
    anisotropic distributions, $\xi=1$ and
    $\xi=10$, together with the result for the isotropic case ($\xi=0$).
    Note that the finite collision rate
    causes $\omega_{\alpha}$ to become complex with a negative imaginary part corresponding to damping
    of these types of modes.
  The effect of collisions on the stable longitudinal ($\beta$-) modes is more significant.
  The results are presented in Figs.~\ref{fig:stablebetaxire} and
  \ref{fig:stablebetaxiim}. We find that for finite $\nu$ the
  dispersion becomes spacelike (${\rm Re}(\omega)<k$) at large $k$ in contrast to
  the collisionless case, in which ${\rm Re}(\omega)>k$ always holds. This
  behavior is responsible for the fact that the solution vanishes from the physical \textsc{Riemann} sheet
  above some finite $k$. This occurs precisely when the solution for $\omega_{\beta}$ reaches the
  cut between $-k$ and $k$ at $-i\nu$.
  It is however possible to
  find solutions on the unphysical \textsc{Riemann} sheets by replacing
  the logarithm in the structure function with its usual
  analytic continuation \cite{Romatschke:2004jh}:
  \begin{align}
    \ln\left(\frac{z+1}{z-1}\right)=\ln\left(\left|\frac{z+1}{z-1}\right|\right)+i\left[\arg\left(\frac{z+1}{z-1}\right)+2\pi
    N\right]\,\text{,}
  \end{align}
  where $N$ specifies the sheet number. The continuation of the
  solution to the lower \textsc{Riemann} sheets is shown in
  Figs.~\ref{fig:stablebetahigherxire} and
  \ref{fig:stablebetahigherxiim} for a collision rate of
  $\nu=0.8\,m_g \approx 0.46\,m_D$ and different anisotropy
  parameters $\xi$. For smaller anisotropies the solution is found
  to converge to the light cone in an oscillating manner, while
  the imaginary part of $\omega$ oscillates around $\pm\nu$. Between $\xi=2$ and
  $\xi=3$ (for $\nu \approx 0.46\,m_D$) this behavior changes qualitatively to the one shown in
  the case $\xi=10$. With increasing $k$ the real part of $\omega$ moves
  away from the light cone, while the imaginary part drops to large negative
  values with the solution remaining on the $N=-1$ \textsc{\textsc{Riemann}} sheet, such that these modes are strongly damped.

\section{Unstable modes} \label{unstable}
\begin{figure}[htb]
      \begin{center}
        \includegraphics[height=6cm]{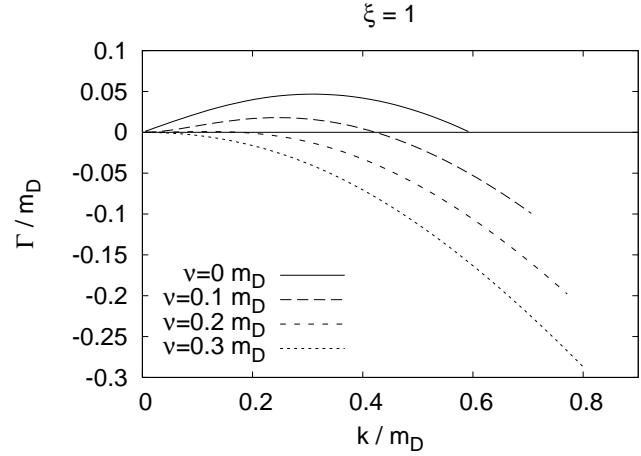}
        \caption{Dependence of the growth rate $\Gamma$ of the unstable transverse ($\alpha-$) mode on the collision rate $\nu$, for an anisotropy parameter $\xi=1$.}
        \label{fig:instablexi1}
      \end{center}
\end{figure}

\begin{figure}[htb]
      \begin{center}
        \includegraphics[height=6cm]{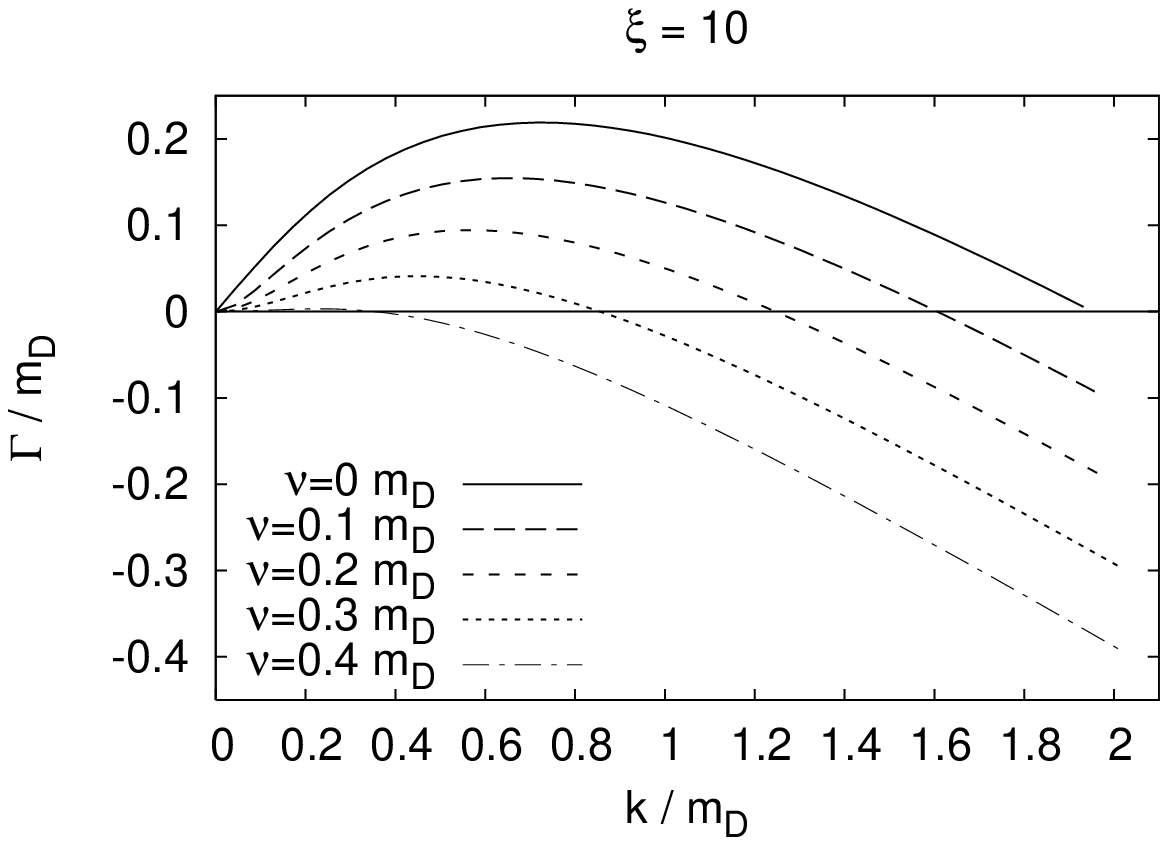}
        \caption{Dependence of the growth rate $\Gamma$ of the unstable transverse ($\alpha-$) mode on the collision rate $\nu$, for an anisotropy parameter $\xi=10$.}
        \label{fig:instablexi10}
      \end{center}
\end{figure}

\index{Collisions!Effect on instabilities}
    We now investigate how the inclusion of collisions affects the
    growth rates of the instabilities, found for the collisionless
    case in Section \ref{unstable modes} and discussed in Section \ref{instabilities}.
    Qualitatively one expects a decrease of the growth rates because the particles,
    which move perpendicular to $\mathbf{k}$ (the particles moving in the $z$-direction in Fig. \ref{fig:mechanism}), can scatter with
    other particles and will no longer be trapped in this direction. Other particles
    can gain a momentum close to perpendicular to
    $\mathbf{k}$ and form a new contribution to the instability.
    However, since the collision term tends to randomize the
    momentum distribution, the growth of $\delta f$ and the magnetic field
    is prevented.
    In order to describe this effect quantitatively, we
    solve Eq.~(\ref{equationstosolve}) for purely imaginary
    $\omega$ and vary the collision rate $\nu$. The solution
    $\omega(k)=i\Gamma(k)$ gives the growth rate $\Gamma(k)$.
    In the case that $\mathbf{\hat{n}}\parallel\mathbf{k}$
    solutions like that only exist for the transverse ($\alpha$-) mode.
    The one with $\Gamma>0$ corresponds to the filamentation instability.
    Results for different values of the collision rate $\nu$ are
    shown in Fig.~\ref{fig:instablexi1} for $\xi=1$ and in
    Fig.~\ref{fig:instablexi10} for $\xi=10$.
    The qualitatively expected effect is nicely reproduced. The
    growth rate decreases with an increasing collision rate as
    does the maximal wave number for an unstable mode.
    One can see that in
    the case where $\xi=1$ already for $\nu$ being around $20\%$ of the \textsc{Debye} mass,
    growth has completely turned into damping and no instability
    can evolve. For $\xi=10$ the collision rate $\nu$ has to be
    slightly larger than $30\%$ of the \textsc{Debye} mass in order to
    prevent growth of a collective mode.

\begin{figure}[t]
  \hfill
  \begin{minipage}[t]{.45\textwidth}
      \begin{center}
        \includegraphics[height=5cm]{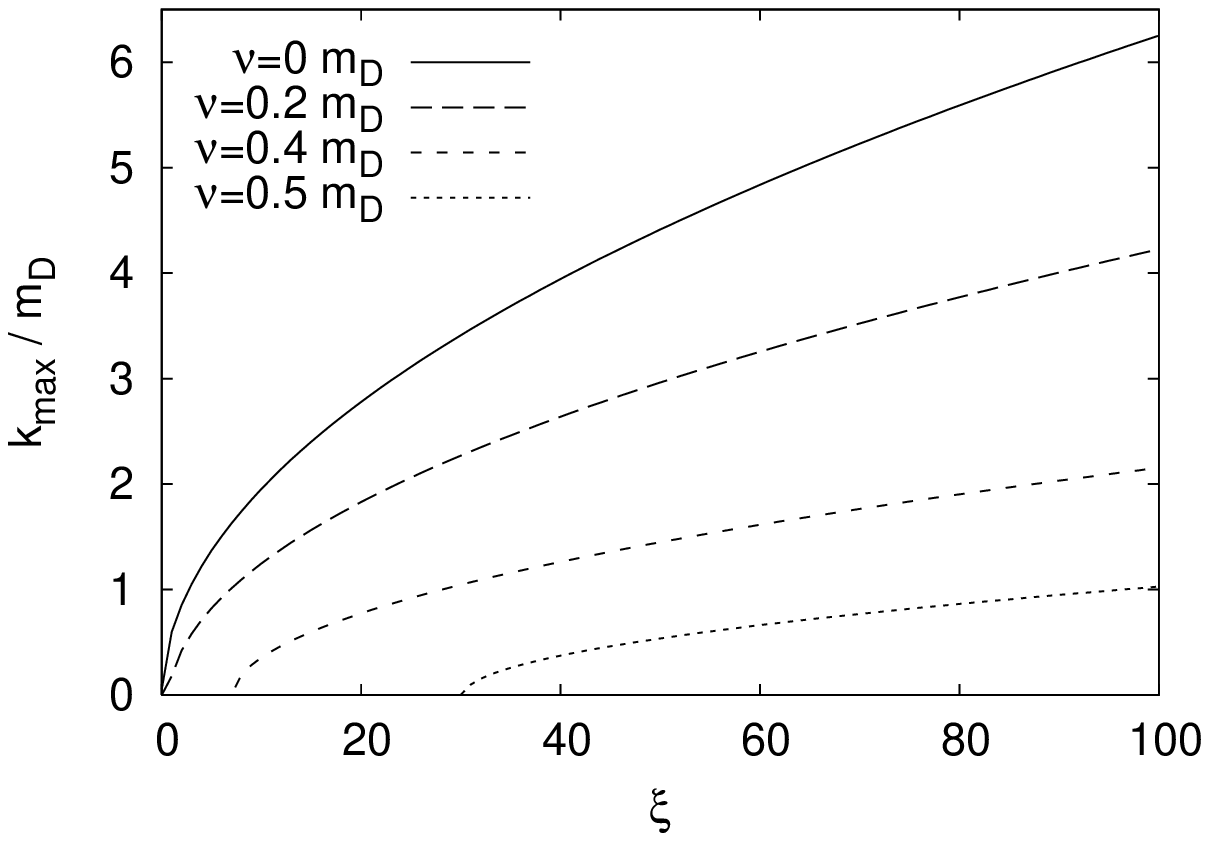}
        \caption{$k_{\text{max}}$ of the maximally unstable mode as a function of the anisotropy parameter $\xi$ for different values of $\nu$.
        For given $\nu$, the momentum anisotropy parameter must be above the critical value for instabilities to occur.}
        \label{fig:kmaxofxi}
      \end{center}
  \end{minipage}
  \hfill
  \begin{minipage}[t]{.45\textwidth}
      \begin{center}
        \includegraphics[height=5cm]{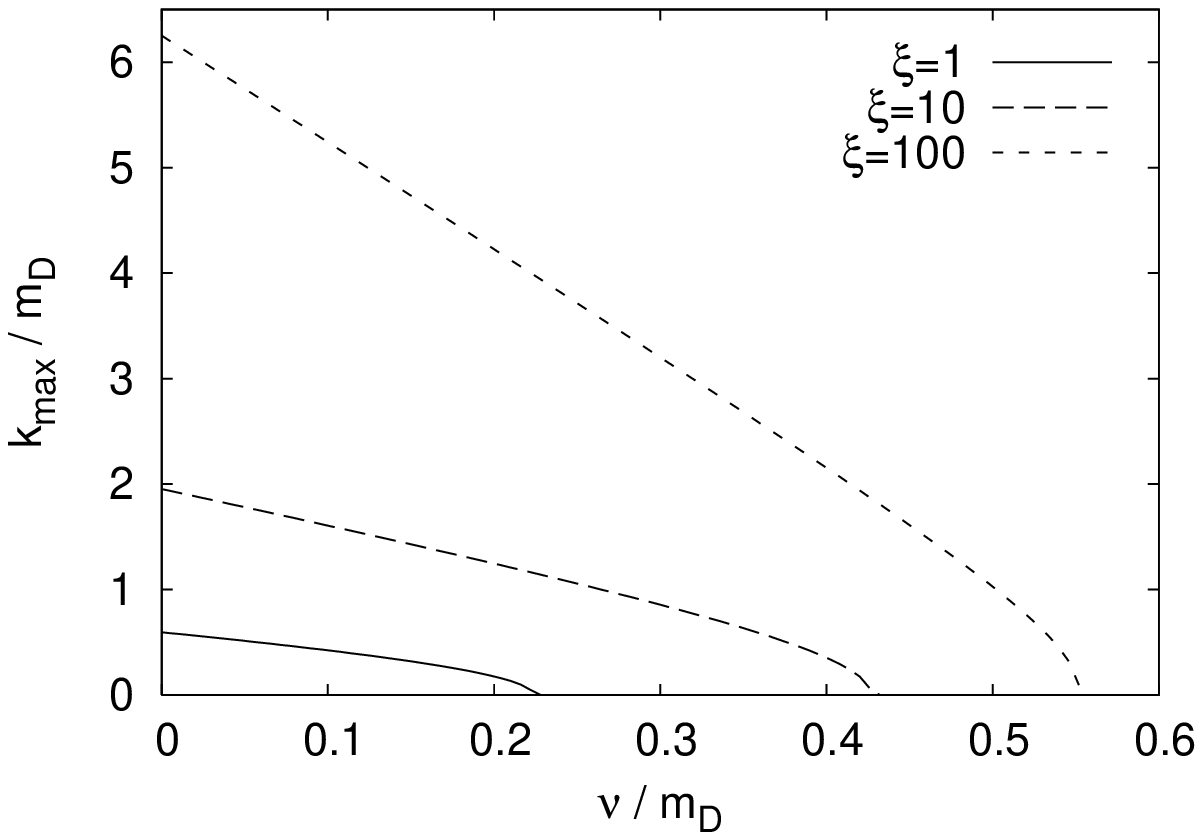}
        \caption{$k_{\text{max}}$ of the maximally unstable mode as a function of the collision rate $\nu$ for given anisotropy parameter $\xi$.
        There exist $\xi$-dependent critical collision rates $\nu_{\text{max}}(\xi)$ above which instabilities can not exist.}
        \label{fig:kmaxofnu}
      \end{center}
  \end{minipage}
  \hfill
\end{figure}

In order to find the wave number $k_{\text{max}}(\xi,\nu)$ at
which the unstable mode spectrum terminates, we take the limit
$\omega \rightarrow 0$ to obtain
\begin{align}
    m_{\alpha}^2=\lim_{\omega \rightarrow 0} \alpha =&
    -\frac{m_D^2}{8} i k
    \frac{\sqrt{1+\xi}}{\sqrt{\xi}(k^2-\nu^2\xi)^2}\times\notag\\
    &\left.~~~~\times\bigg\{-2 i k
    \sqrt{\xi}(k^2-\nu^2\xi)-2\nu
    (k^2+\nu^2)\xi^{3/2}\ln\left(1-\frac{2k}{k-i\nu}\right)\right.\notag\\
    & ~~~~~~~~~~~\left. -2 i
    k\left[k^2(\xi-1)+\nu^2\xi(\xi+3)\right]\arctan(\sqrt{\xi})\right.\bigg\}\text{\,.}\label{ma2}
\end{align}
One of the solutions to the equation
\begin{equation}
    k^2+m_{\alpha}^2=0\,\text{,}\label{kmaxeq}
\end{equation}
which is just the limit $\omega\rightarrow 0$ of the first of the
Eqs.~(\ref{equationstosolve}), is $k_{\text{max}}$. Results for
different $\xi$ and $\nu$ are shown in Figs.~\ref{fig:kmaxofxi}
and \ref{fig:kmaxofnu}. We find that for a given anisotropy
parameter $\xi$, there exists a critical collision rate, above
which instabilities can not occur. This is also true for the limit
$\xi \rightarrow \infty$, as we will show in the following. Taking
$\xi\rightarrow \infty$, Eq.~(\ref{ma2}) becomes
\begin{equation}
    m_{\alpha}^2(\xi\rightarrow
    \infty)=-\frac{\pi}{8}m_D^2\frac{k^2}{\nu^2}\,\text{,}
\end{equation}
which together with Eq.~(\ref{kmaxeq}) gives
\begin{align}
    k^2(\nu^2-\frac{\pi}{8}m_D^2)&=0\text{\,.}
\end{align}
Apart from the result $k=0$, this is solved by
\begin{equation}
    \nu_{\text{max}}(\xi\rightarrow\infty)=\sqrt{\frac{\pi}{8}}m_D\approx 0.6267\, m_D =\Gamma_{\text{max}}(\xi\rightarrow\infty)\,\text{,}
    \label{numaxeq}
\end{equation}
the critical collision rate, above which even in the extremely
anisotropic limit $\xi=\infty$ no instability can occur.

\begin{figure}[t]
      \begin{center}
        \includegraphics[height=6cm]{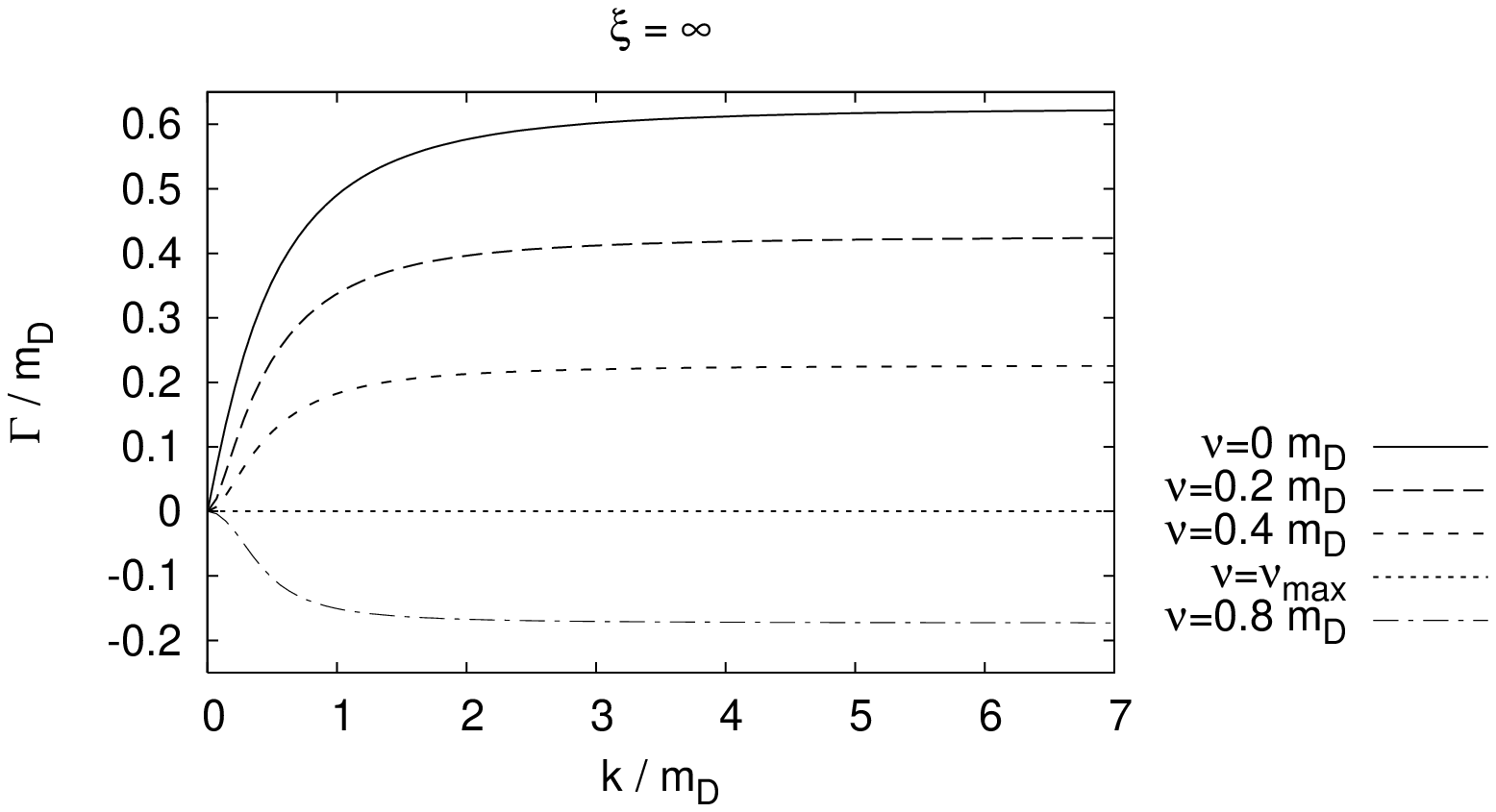}
        \caption{Dependence of the growth rate $\Gamma$ of the unstable transverse ($\alpha-$) mode on the collision rate $\nu$, for the extremely anisotropic limit
        $\xi=\infty$.}
        \label{fig:unstablealphaxiinf}
      \end{center}
  \end{figure}

\begin{figure}[t]
      \begin{center}
        \includegraphics[height=6cm]{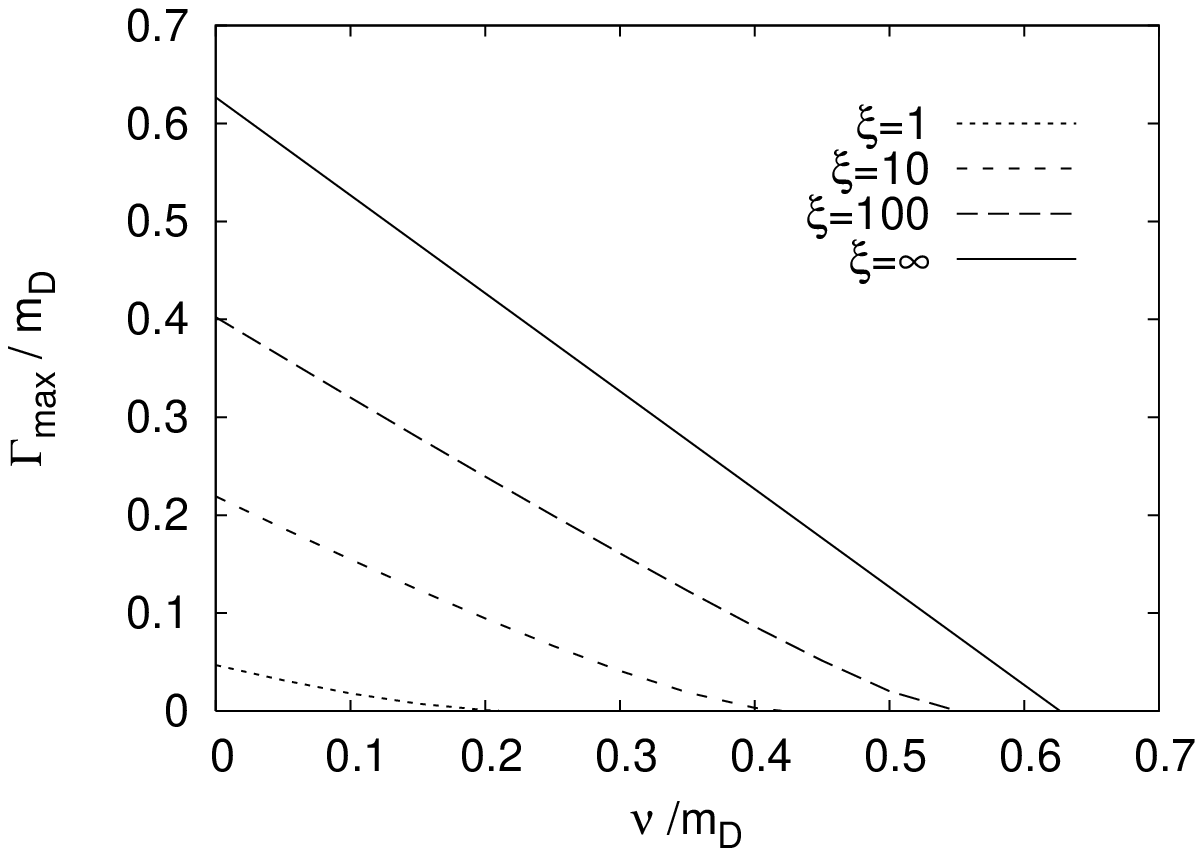}
        \caption{The maximal growth rate of the instability as a function of the collision rate $\nu$.}
        \label{fig:gammamax}
      \end{center}
  \end{figure}

This is also visible in the plot of the growth rate
$\Gamma_{\xi\rightarrow\infty}$, shown in
Fig.~\ref{fig:unstablealphaxiinf}. For
$\nu=\nu_{\text{max}}(\xi\rightarrow\infty)$ the growth rate
becomes zero for all $k$, and for larger $\nu$ only damping
occurs. In this case ($\xi=\infty$) the value of the maximal
collision rate equals that of the maximal growth rate in the
collisionless limit. This simply means that the instability
vanishes completely at the point where the collisions damp at the
same rate at which the instability grows. Note however that this
relation is more complicated in general as shown in
Fig.~\ref{fig:gammamax}, where the dependence of the maximal
growth rate on the collision rate is plotted. In order to make the
instability vanish completely for any $\xi<\infty$, a collision
rate larger than the maximal growth rate of the instability in the
collisionless limit is needed.


\section{Discussions}
\label{discussions} So far the collisional frequency $\nu$ has been
taken to be arbitrary. Because the inclusion of a BGK collisional
term is a phenomenological model for the equilibration of a system
and cannot be derived from first principles this makes it
difficult to fix the magnitude of $\nu$. However, since as treated
here the underlying framework of \textsc{Boltzmann-Vlasov} is
implicitly perturbative, we can attempt to fix $\nu$
perturbatively. However, even this is non-trivial since within
non-Abelian theories there are at least two possible collisional
frequencies to be considered
\cite{Selikhov:1993ns,Bodeker:1998hm,Arnold:1998cy,Bodeker:1999ey,Arnold:2002zm}:
(1) the frequency for hard-hard scatterings which is
parametrically $\nu_{\rm hard} \sim \alpha_s^2 \log \alpha_s^{-1}$
and (2) the frequency for hard-soft scattering which is
parametrically $\nu_{\rm soft} \sim \alpha_s \log \alpha_s^{-1}$.

The hard-hard scatterings correspond to interactions which change
the momentum of a hard particle by ${\cal O}(p_{\rm hard})$ and
therefore represent truly momentum-space isotropizing
interactions.  On the other hand the hard-soft scatterings
correspond to changes in momentum which are order ${\cal O}(g
p_{\rm hard})$. These small angle scatterings occur rather
frequently and it turns out that after traversing one hard
scattering mean free path, $\lambda_{\rm hard} \sim \nu_{\rm
hard}^{-1}$, the typical deflection of the particle is also ${\cal
O}(1)$.\footnote{This, in the end, is the source of the logarithm
in $\nu_{\rm hard}$ above.}  The physics of small-angle scattering
by the soft-background is precisely what is captured by the
hard-loop treatment.  However, the hard-loop framework does not
explicitly take into account that $\nu_{\rm soft}$ is also the
frequency at which there are color-rotating interactions of the
hard particles themselves. One would expect that color-rotation of
the hard particles to have a larger effect on the growth of
instabilities than the momentum-space isotropization via hard-hard
scattering.  That being said, the form of the BGK scattering
kernel does not mix color channels and in that sense cannot be
used to describe the physics of color-rotation of the hard
particles.  For this reason one is lead to the conclusion that
when using the BGK kernel the appropriate damping rate is $\nu
\sim \nu_{\rm hard} \sim \alpha_s^2 \log \alpha_s^{- 1}$.  This
conclusion for the parametric dependence of $\nu$ is also
supported by looking at the leading order result for the shear
viscosity \cite{Thoma:1993vs}.

Even with this conclusion it is hard to say anything quantitative
about $\nu$ since the overall coefficient and the coefficient in
the logarithm are not specified by such a parametric relation.
One could hope that previous calculations contained in
\cite{Thoma:1993vs} of parton interaction rates could be of some
use.  Unfortunately, for a purely gluonic plasma it was found $\nu
= 5.2 \, \alpha_s^2 T \, \log(0.25 \alpha_s^{-1})$, which clearly
cannot be trusted for the values of $\alpha_s$ which are relevant
for heavy-ion collisions ($\alpha_s \sim 0.2-0.4$) since $\nu$
becomes negative for large $\alpha_s$. We note here that the
negativity of this result at large couplings most likely stems
from the strict perturbative expansion of the integrals involved
failing when the hard and soft scales become comparable in
magnitude. Similar erroneous negative values also occur in the
perturbative expressions for heavy-quark collisional energy loss
\cite{Braaten:1991we} when extrapolated to large coupling. A
corrected calculational method which yields positive-definite
results for the heavy-quark energy loss was detailed in
Refs.~\cite{Romatschke:2003vc, Romatschke:2004au}.

Ideally, one would revisit the calculation of the interaction rate
and improve upon the techniques used where necessary.  Short of
such a calculation one cannot say with certainty what the
numerical value of $\nu$ should be and the best we can do is to
play ``games''.  For instance, one could insert a one into the
logarithm appearing in $\nu$ similar to what other authors have
done \cite{Kapusta:1991qp} to obtain $\nu \sim 5.2 \, \alpha_s^2 T
\log(1 + 0.25 \alpha_s^{-1}) \sim 0.1-0.2\,m_D$ for $\alpha_s =
0.2-0.4$.  Using this admittedly specious expression for
large-coupling one can expand and obtain $\lim_{\alpha_s
\rightarrow \infty} \nu/m_D \sim 0.37\, \alpha_s^{1/2}$.  Note
that in the case $N_f=2$ both $\nu$ and $m_D$ increase; however,
the ratio of these two scales is still in the range quoted for
$N_f=0$.  The range, $\nu = 0.1-0.2\,m_D$, places us well below
the threshold needed to turn off instabilities in the case of
extremely anisotropic distribution functions but does imply (see
Fig.~\ref{fig:gammamax}) that for moderately small anisotropies,
$\xi \sim 1$, and large coupling that it is possible for
collisional damping to eliminate the unstable modes from the
spectrum completely. Of course, in the limit of asymptotically
small couplings the ratio $\nu/m_D$ approaches zero and the
collisionless results hold to very good approximation.  In the
opposite limit of strong coupling the estimates here are at best
guesswork and it is indeed possible that the ratio $\nu/m_D$ is
larger than the range we have quoted.  For example, the recent
work of \textsc{Peshier} \cite{Peshier:2004bv,Peshier:2005pp}
implies that $\nu/m_D$ could be as large as 0.5; however, this
number results from a fit of a model assumption to lattice data
and is not directly comparable to the collisional widths
considered here since in their description the gluon width was
assumed to be parametrically given by $\nu_{soft}$.

Additionally, we have to mention the caveat that all the estimates
above rely on full equilibrium thermal field theory calculations.
For the very initial state of the matter created in an
ultra-relativistic heavy-ion collision the system is clearly not
in equilibrium and it is not clear how this estimate will change
as a result.  However, it is of crucial importance to attempt to
estimate the scattering rate in a non-equilibrium setting.

      \chapter{Wong-Yang-Mills simulation including collisions}
\label{chap:wong}
\epigraphwidth 200pt \epigraph{Willst du dich am Ganzen
erquicken,\\ so mu{\ss}t du das Ganze im Kleinsten erblicken.}
{\emph{Gott, Gem\"ut und Welt.}\\
Johann Wolfgang von Goethe (1749-1832)}
\newcommand{\Tr}{\mbox{${\rm Tr}$}}
In this chapter we introduce the real time lattice calculations
that we use to simulate coupled particle-field evolution in
classical $SU(N)$ gauge theories. Because occupation numbers are
large in the infrared, the classical field approximation becomes
reasonable. The physics is non-perturbative but in an essentially
classical way, which gives us hope that analysis of the analogous
classical field theory at finite temperature can help understand
at leading order the quantum system of interest. We derive the
\textsc{Wong-Yang-Mills} equations and prove their equivalence to
the unlinearized \textsc{Vlasov} equation for a microscopic
distribution function. Then we describe the update algorithm used
to solve the coupled system of \textsc{Wong-Yang-Mills} equations
with emphasis on the numerical method of current smearing. Finally
we introduce a new component to the simulation, elastic binary
particle collisions with large momentum exchange, i.e., momentum
exchange larger than that mediated by the fields. For an isotropic
plasma we show that in this scheme particle momentum diffusion is
independent of the lattice spacing $a$, if we adjust initial
energy densities of fields and particles appropriately. The
numerical implementation is based on a code developed by
\textsc{Krasnitz} and \textsc{Nara} and previously used in
\cite{Krasnitz:1998ns,Krasnitz:1999wc,Krasnitz:2001qu,Krasnitz:2002mn,Krasnitz:2003jw,Krasnitz:2002ng,Dumitru:2005gp,Dumitru:2006pz}.

\section{The Wong equations}\label{derivewong}
The classical transport theory for non-Abelian plasmas has been
established by \textsc{Heinz} and \textsc{Elze}
\cite{Heinz:1983nx,Heinz:1984yq,Heinz:1985qe,Elze:1989un}. We have
given a particular derivation of the \textsc{Vlasov} equations in
Chapter \ref{kintheory} based on the approach by \textsc{Blaizot}
and \textsc{Iancu} \cite{Blaizot:1993zk,Blaizot:1993be}. In this
section we will derive the \textsc{Wong} equations
\cite{Wong:1970fu} (see also \cite{Litim:1999id,Litim:1999ns}) and show their equivalence to the previously
derived \textsc{Vlasov} equation in its non-linearized form in
Section \ref{wongtovlasov}. For the numerical simulation with test
particles the \textsc{Wong} equations, together with the
\textsc{Yang-Mills} equation for the soft gluon fields, are the
natural choice because they describe the evolution of a
microscopic system of individual particles. Let us start with the
classical equations of motion in electrodynamics. The
\textsc{Wong} equations are then just the extension to
chromodynamics. Ignoring the dynamical effects of the spin of the
particles, as they are typically small, in electrodynamics we have
for particle $i$:\index{Wong equations}
\begin{equation}\label{easyeqm}
    \dot{\mathbf{x}}_i(t)=\mathbf{v}_i(t)\,,
\end{equation}
i.e., the change of position of particle $i$ is given by its
velocity $\mathbf{v}_i(t)$. The change of the particle's momentum
is given by the \textsc{Lorentz} force
\begin{equation}
    \dot{\mathbf{p}}_i(t)=\mathbf{F}_{\text{Lorentz}}=e\left(\mathbf{E}(t)+\mathbf{v}_i(t)\times\mathbf{B}(t)\right)\,,
\end{equation}
where $e$ is the electric charge of the particle, and $\mathbf{E}$
and $\mathbf{B}$ are the electric and magnetic field,
respectively. If the particle has a non-Abelian $SU(N)$ color
charge $q^a$, with $a=1,2,\ldots, N^2-1$, then the term
corresponding to the \textsc{Lorentz} force term is given by
\begin{equation}\label{sunlorentz}
    \dot{\mathbf{p}}_i(t)=g q^a_i(t)\left(\mathbf{E}^a(t)+\mathbf{v}_i(t)\times\mathbf{B}^a(t)\right)\,.
\end{equation}
As already indicated in Eq. (\ref{sunlorentz}), $q^a_i$ is now a time dependent quantity because the color vector $\vec{q}_i$\footnote{Note that we denote color vectors by $\vec{\cdot}$, while spatial three-vectors are always represented by bold characters to avoid confusion between the two.} may be rotated in color space by the color-electric and color-magnetic fields $\vec{\mathbf{E}}$ and $\vec{\mathbf{B}}$. Hence we need to find another equation of motion describing the time evolution of $\vec{q}_i(t)$. Using that the charge and current densities generated by a particle with color charge $\vec{q}_i$ are given by
\begin{align}
    \vec{\rho}(t) &= g \vec{q}_i(t) \delta(\mathbf{x}-\mathbf{x}_i(t))\,,\label{j1y}\\
    \vec{\mathbf{J}}(t) &= g \vec{q}_i(t) \mathbf{v}_i(t) \delta(\mathbf{x}-\mathbf{x}_i(t))\label{j2y}\,,
\end{align}
we can write
\begin{align}
    \dot{\vec{\rho}}&=\frac{d}{dt} \left(g \vec{q}_i(t) \delta(\mathbf{x}-\mathbf{x}_i(t))\right)\notag\\
                    &=g\dot{\vec{q}}_i(t)\delta(\mathbf{x}-\mathbf{x}_i(t))+g\vec{q}_i(t)\dot{\mathbf{x}}_i(t)\frac{\partial\delta(\mathbf{x}-\mathbf{x}_i(t))}{\partial \mathbf{x}_i(t)}\notag\\
                    &=g\dot{\vec{q}}_i(t)\delta(\mathbf{x}-\mathbf{x}_i(t))+g\vec{q}_i(t)\mathbf{v}_i(t)\left(-\frac{\partial\delta(\mathbf{x}-\mathbf{x}_i(t))}{\partial \mathbf{x}}\right)\notag\\
                    &=g\dot{\vec{q}}_i(t)\delta(\mathbf{x}-\mathbf{x}_i(t))-{\boldsymbol \nabla}\cdot\vec{\mathbf{J}}\,,\label{rhodot}
\end{align}
where we used Eqs. (\ref{easyeqm}) and (\ref{j2y}).
Equation (\ref{rhodot}) can be written as
\begin{equation}
    g\dot{\vec{q}}_i(t)\delta(\mathbf{x}-\mathbf{x}_i(t))=\dot{\vec{\rho}}+{\boldsymbol \nabla}\cdot\vec{\mathbf{J}}=\partial_\mu \vec{J}^\mu\,,
\end{equation}
with the four-current $\vec{J}=(\vec{\rho},\vec{\mathbf{J}})$. For a non-Abelian theory \textsc{Gauss}' law reads
\begin{equation}
    \vec{D}_\mu \vec{J}^\mu=0\,,
\end{equation}
with the covariant derivative $\vec{D}_\mu=\partial_\mu+ig[\vec{A}_\mu,\cdot]$.
Hence
\begin{equation}
    g\dot{\vec{q}}_i(t)\delta(\mathbf{x}-\mathbf{x}_i(t))=\partial_\mu \vec{J}^\mu=\vec{D}_\mu \vec{J}^\mu-ig[\vec{A}_\mu,\vec{J}^\mu]\,.
\end{equation}
Again using the definition (\ref{j2y}), we can write
\begin{align}
    g\dot{\vec{q}}_i(t)\delta(\mathbf{x}-\mathbf{x}_i(t))&=-ig[\vec{A}_\mu(t),g v_i^\mu(t)\vec{q}_i\delta(\mathbf{x}-\mathbf{x}_i(t))]\notag\\
    \dot{\vec{q}}_i(t)&=-igv_i^\mu(t)[\vec{A}_\mu,\vec{q}_i(t)]\,.\label{wongi3}
\end{align}
Equation (\ref{wongi3}) completes \textsc{Wong}'s equations. It describes the time evolution of the color charge $\vec{q}_i(t)$, as it is rotated by the fields. From now on we will drop the explicit indication of color vectors $\vec{\cdot}$.
The \textsc{Wong} equations are coupled self consistently to the \textsc{Yang-Mills} equation for the soft gluon fields
\begin{equation}\label{currentwym}
    D_\mu F^{\mu\nu}=J^\nu=g\sum_i q_i v_i^\nu\delta(\mathbf{x}-\mathbf{x}_i(t))\,,
\end{equation}
with the current that is generated by all particles.
Let us once more write down the \textsc{Wong} equations, this time as one set:
\begin{align}
    \dot{\mathbf{x}}_i(t)&=\mathbf{v}_i(t)\,, \label{wong1}\\
    \dot{\mathbf{p}}_i(t)&=g q^a_i(t)\left(\mathbf{E}^a(t)+\mathbf{v}_i(t)\times\mathbf{B}^a(t)\right)\,,\label{wong2}\\
    \dot{q}_i(t)&=-igv_i^\mu(t)[A_\mu(t),q_i(t)]\,.\label{wong3}
\end{align}

\section{The relation between the Wong and the Vlasov equations}
\label{wongtovlasov} In this section we show that the
\textsc{Wong} equations are equivalent to the non-linearized
\textsc{Vlasov} equation for a microscopic system of individual
particles. To do so let us first recall the \textsc{Vlasov}
equation (\ref{gluovlasov}), derived in Section
\ref{derivevlasov}, and rewrite it in a form that is also often
used in the literature (e.g. \cite{Heinz:1983nx} or
\cite{Kelly:1994dh}), using the color charge $q$ as a variable.
Using the covariant derivative
$D_\mu=\partial_\mu+ig[A_\mu,\cdot]$\footnote{Note that the signs
in front of the force term and the color rotation term in the
\textsc{Vlasov} equation will depend on the choice of the sign in
this definition - this leads to different sign conventions in e.g.
\cite{Heinz:1983nx} and \cite{Kelly:1994dh} and this work.}, we
can expand Eq.(\ref{gluovlasov}) to read
\begin{equation}
    V^\mu \partial_\mu n^g(P,X,q) + ig V^\mu [A_\mu(X),n^g(P,X,q)]+gV^\mu F_{\mu\nu}\partial^\nu_{(p)}n^g(P,X,q)=0\,.
\end{equation}
Since the distribution function $n^g$ is linear in the color charge, we can write $n^g=f=f^a q^a \equiv \sum_a f^aq^a$, with $f^a$ the components of $f$ in a certain color 'direction'. From now on $f$ shall be the complete gluon distribution function. Note that for the components of the color vector $q$ it holds that $q^a\sim T^a$ up to normalization. With that and after multiplication with $|\mathbf{p}|$ we have
\begin{align}
    &P^\mu \partial_\mu f(P,X,q) + ig P^\mu A_\mu^b(X) f^a(P,X,q) [q^b,q^a] + gP^\mu q^a F_{\mu\nu}^a \partial^\nu_{(p)}f(P,X,q)=0\notag\\
    \Leftrightarrow ~ & P^\mu \left[\partial_\mu + g q^a F_{\mu\nu}^a \partial^\nu_{(p)} + g f^{abc} A_\mu^b(X) q^c \partial_{q^a} \right]f(P,X,q)=0\label{qvlasov}\,,
\end{align}
where we used that $[q^b,q^a]=if^{bac}q^c=-if^{abc}q^c$ and
$\partial_{q^a}f=f^a$. This form of the \textsc{Vlasov} equation
\index{Vlasov equation} can now be easily compared to the
\textsc{Wong} equations derived in Section \ref{derivewong}. We
shall do this by using the \textsc{Wong} equations together with
the collisionless equation of motion for the single particle
distribution function $f(t,\mathbf{x},\mathbf{p},q)$ :
\begin{equation}
    \frac{d}{dt}f(t,\mathbf{x},\mathbf{p},q)=0\,,\label{eqmf}
\end{equation}
and by replacing the continuous $f$ by a large number of test particles \cite{Bertsch:1988ik}
\begin{equation}
    f(\mathbf{x},\mathbf{p},q)=\frac{1}{N_{\text{test}}}\sum_i \delta^3(\mathbf{x}-\mathbf{x}_i(t))(2\pi)^3\delta^3(\mathbf{p}-\mathbf{p}_i(t))\delta^{N^2-1}(q-q_i(t))\,.
\end{equation}
With this distribution function Eq.(\ref{eqmf}) becomes
\begin{align}
    \frac{d}{dt}f(\mathbf{x},\mathbf{p},q)&=\frac{1}{N_{\text{test}}}\sum_i\frac{d\mathbf{x}}{dt}\frac{d}{d\mathbf{x}}\delta^3(\mathbf{x}-\mathbf{x}_i(t))(2\pi)^3\delta^3(\mathbf{p}-\mathbf{p}_i(t))\delta^{N^2-1}(q-q_i(t))\notag\\
    &~~~~~~~~~~~~~~~~~+\frac{d\mathbf{p}}{dt} \delta^3(\mathbf{x}-\mathbf{x}_i(t))(2\pi)^3\frac{d}{d\mathbf{p}}\delta^3(\mathbf{p}-\mathbf{p}_i(t))\delta^{N^2-1}(q-q_i(t))\notag\\
    &~~~~~~~~~~~~~~~~~+\frac{d q}{dt}\delta^3(\mathbf{x}-\mathbf{x}_i(t))(2\pi)^3\delta^3(\mathbf{p}-\mathbf{p}_i(t))\frac{d}{dq}\delta^{N^2-1}(q-q_i(t))=0\notag\\
    &\Leftrightarrow-\frac{1}{N_{\text{test}}}\sum_i\frac{d\mathbf{x}_i(t)}{dt}\frac{d}{d\mathbf{x}_i}\delta^3(\mathbf{x}-\mathbf{x}_i(t))(2\pi)^3\delta^3(\mathbf{p}-\mathbf{p}_i(t))\delta^{N^2-1}(q-q_i(t))\notag\\
    &~~~~~~~~~~~~~~~~~~~\,+\frac{d\mathbf{p}_i(t)}{dt} \delta^3(\mathbf{x}-\mathbf{x}_i(t))(2\pi)^3\frac{d}{d\mathbf{p}_i}\delta^3(\mathbf{p}-\mathbf{p}_i(t))\delta^{N^2-1}(q-q_i(t))\notag\\
    &~~~~~~~~~~~~~~~~~~~\,+\frac{d q_i(t)}{dt}\delta^3(\mathbf{x}-\mathbf{x}_i(t))(2\pi)^3\delta^3(\mathbf{p}-\mathbf{p}_i(t))\frac{d}{dq_i}\delta^{N^2-1}(q-q_i(t))=0\notag\,.
\end{align}
Now we use the \textsc{Wong} equations (\ref{wong1})-(\ref{wong3}) to write
\begin{align}\label{wongvlasov}
    &\frac{1}{N_{\text{test}}}\sum_i\left(-\mathbf{v}_i(t)\right)\frac{d}{d\mathbf{x}_i}\delta^3(\mathbf{x}-\mathbf{x}_i(t))(2\pi)^3\delta^3(\mathbf{p}-\mathbf{p}_i(t))\delta^{N^2-1}(q-q_i(t))\notag\\
    &\,+\left(-gq_i^a(t)(\mathbf{E}^a(t)+\mathbf{v}_i(t)\times\mathbf{B}^a(t))\right) \delta^3(\mathbf{x}-\mathbf{x}_i(t))(2\pi)^3\frac{d}{d\mathbf{p}_i}\delta^3(\mathbf{p}-\mathbf{p}_i(t))\delta^{N^2-1}(q-q_i(t))\notag\\
    &\,+\left(ig\mathbf{v}_i^\mu(t)[A_\mu(t),q_i(t)]\right)\delta^3(\mathbf{x}-\mathbf{x}_i(t))(2\pi)^3\delta^3(\mathbf{p}-\mathbf{p}_i(t))\frac{d}{dq_i}\delta^{N^2-1}(q-q_i(t))=0\,.
\end{align}
For every $i$ we have
\begin{align}
    &V_i^\mu \partial_\mu\delta^3(\mathbf{x}-\mathbf{x}_i(t))(2\pi)^3\delta^3(\mathbf{p}-\mathbf{p}_i(t))\delta^{N^2-1}(q-q_i(t))\notag\\
    &\,+g q_i^aV_i^\mu F_{\mu\nu}^a\partial_{(p)}^\nu \delta^3(\mathbf{x}-\mathbf{x}_i(t))(2\pi)^3\delta^3(\mathbf{p}-\mathbf{p}_i(t))\delta^{N^2-1}(q-q_i(t))\notag\\
    &\,+gV_i^\mu f_{abc}A_\mu^b q_i^c\delta^3(\mathbf{x}-\mathbf{x}_i(t))(2\pi)^3\delta^3(\mathbf{p}-\mathbf{p}_i(t))\delta^{N^2-1}(q-q_i(t))=0\,,
\end{align}
where we used that $\mathbf{v}\cdot{\boldsymbol\nabla_x}=V^\mu\partial_\mu$ because $\partial_t f(\mathbf{x},\mathbf{p},q)=0$ for the first part,
$-gq_i^a(t)(\mathbf{E}^a(t)+\mathbf{v}_i(t)\times\mathbf{B}^a(t)){\boldsymbol \nabla_p}=g q_i^a(F_{0k}^a+v_i^j F_{jk}^a)\partial^k_{(p)}=g q_i^a V^\mu F_{\mu\nu}^a\partial^\nu_{(p)}$, because $\partial^0_{(p)} f(\mathbf{x},\mathbf{p},q)=0$ in the second part, and $igV_i^\mu[A_\mu,q_i]=-gV_i^\mu f_{bac}A_\mu^b q_i^c=gv_i^\mu f_{abc}A_\mu^bq_i^c$ in the last component.
Note that the derivatives are written with respect to the variables $\mathbf{x}$, $\mathbf{p}$ and $q$ now instead of $\mathbf{x}_i$, $\mathbf{p}_i$ and $q_i$, which is possible due to the $\delta$-functions and just introduces an overall minus sign, which does not matter.
The $\delta$-functions also allow the replacement of $V^\mu_i$ with $V^\mu$ and $q_i$ with $q$ in every summand and we can use this to rewrite Eq.(\ref{wongvlasov}) to read
\begin{align}
    P^\mu \left[\partial_\mu + g q^a F_{\mu\nu}^a \partial^\nu_{(p)} + g f^{abc} A_\mu^b(X) q^c \partial_{q^a} \right]f(\mathbf{x},\mathbf{p},q)=0\,,
\end{align}
which is just the \textsc{Vlasov} equation (\ref{qvlasov}) for the distribution function $f(\mathbf{x},\mathbf{p},q)$.

\section{Numerical methods for solving the Wong-Yang-Mills system}
In this section we introduce numerical methods for solving the classical field equations coupled to particles and discuss the implementation of the most important parts of the calculation.

\subsection{Lattice simulation}
%{\bf The covariant derivative is defined to be $D_\mu=\partial_\mu-ig[A_\mu,\cdot]$, which has some consequences for the signs within the \textsc{Vlasov} and \textsc{Wong} equations.} The \textsc{Vlasov} equation becomes
%\begin{align}
%   P^\mu \left[\partial_\mu - g q^a F_{\mu\nu}^a \partial^\nu_{(p)} - g f^{abc} A_\mu^b(X) q^c \partial_{q^a} \right]f(P,X,q)=0\label{qvlasovminus}\,,
%\end{align}
%while the \textsc{Wong} equations for the $i$-th test particle now read
%\begin{align}
%    \dot{\mathbf{x}}_i(t)&=\mathbf{v}_i(t)\,, \label{wongs1}\\
%    \dot{\mathbf{p}}_i(t)&=g q^a_i(t)\left(\mathbf{E}^a(t)+\mathbf{v}_i(t)\times\mathbf{B}^a(t)\right)\,,\label{wongs2}\\
%    \dot{q}_i(t)&=igv_i^\mu(t)[A_\mu(t),q_i(t)]\,.\label{wongs3}
%\end{align}
%Of course this has no effect on any physical observable, since the sign dependence will drop out for those.
%For completeness let us once more give the current (\ref{currentwym}) generated by the particles, this time taking into account that we are using $N_{\text{test}}$ test particles per physical particle:
%\begin{equation}
%     J^{a\,\nu}=\frac{g}{N_{\text{test}}}\sum_i q_i^a v_i^\nu\delta(\mathbf{x}-\mathbf{x}_i(t))\,.
%\end{equation}
The time evolution of the \textsc{Yang-Mills} field is determined by the standard Hamiltonian method in $A^0=0$ gauge \cite{Ambjorn:1990pu,Hu:1996sf,Moore:1997sn}. The temporal gauge is particularly useful because it allows for a simple identification of the canonical momentum as the electric field
\begin{equation}
    \mathbf{E}^a=-\dot{\mathbf{A}}^a\,.
\end{equation}
In addition, time-like link variabes $U$, defined below, become simple identity matrices.

The lattice Hamiltonian\index{Kogut-Susskind Hamiltonian}, which we derive in Appendix \ref{kogut}, is given by
\begin{equation}\label{KSH}
    H_L=\frac{1}{2}\sum_i \mathbf{E}_{L\,i}^{a\,2}+\frac{1}{2}\sum_\Box\left(N_c-\text{Re}\text{Tr}U_\Box\right)+\frac{1}{N_{\text{test}\,L}}\sum_j|\mathbf{p}_{L\,j}|\,,
\end{equation}
here including the particle contribution $1/N_{\text{test}\,L}\sum_j|\mathbf{p}_{L\,j}|$.
The plaquette is defined by
\begin{equation}
    U_\Box=U_{x,y}(i)=U_x(i)U_y(i+\hat{x})U_x^\dag(i+\hat{y})U_y^\dag(i)\,,
\end{equation}
with the link variable
\begin{equation}\label{linkdef}
    U_\mu(i)=e^{iagA_\mu(i)}\,.
\end{equation}
The shifts $\hat{x}$ and $\hat{y}$ are one lattice spacing in length and directed into the $x$- or $y$- direction, respectively.
Please refer to Appendix \ref{kogut} for more details. We only point out here that we set $\tau_a=\sigma_a$, the \textsc{Pauli} matrices, without the usual factor of $1/2$, i.e., the commutation relation reads $[\tau^a,\tau^b]=2\delta^{ab}$. Another factor of $1/2$ is absorbed into the $A$-field, which has to be taken into account when calculating the physical fields $\mathbf{E}$ and $\mathbf{B}$ from it.

Eq. (\ref{KSH}) is given in lattice units\index{Lattice units}, which are chosen such that all lattice variables are dimensionless:
\begin{align}\label{latticevariables}
    \mathbf{E}^a_L=\frac{ga^2}{2}\mathbf{E}^a\,, && \mathbf{B}^a_L=\frac{ga^2}{2}\mathbf{B}^a\,, && \mathbf{p}_L=\frac{a}{4}\mathbf{p}\,, && Q_L^a=\frac{1}{2}q^a\,, && N_{\text{test}\,L}=\frac{1}{g^2}N_{\text{test}}\,,
\end{align}
with the lattice spacing $a$. $H_L$ is hence related to the
physical Hamiltonian by $H=4/(g^2a)H_L$ (also see Appendix
\ref{kogut}). To convert lattice variables to physical units we
will fix the lattice length $L$ in fm, which will then determine
the physical scale for $a$. All other dimensionful quantities can
then be determined by using Eqs. (\ref{latticevariables}). The
Hamiltonian (\ref{KSH}) determines the energy density of the
system and allows to determine the equations of motion for the
fields, e.g.,
\begin{equation}
    \frac{d}{dt}\mathbf{E}_L=\left\{\mathbf{E}_L,H_L\right\}\,,
\end{equation}
with the anti-commutator $\{\cdot,\cdot\}$.

Our lattice has periodic boundary conditions.

\subsection{Coordinate and current update}
In the simulation we keep track of every particle's position $\mathbf{x}_i(t)$, momentum $\mathbf{p}_i(t)$ and charge $q_i(t)$. Knowing the $i$-th particle's momentum, its coordinate is updated according to Eq. (\ref{wong1}) for a time step of length $\Delta t$:
\begin{equation}\label{coordupdate}
    \mathbf{x}_i(t+\Delta t)=\mathbf{x}_i(t)+\mathbf{v}_i(t+\Delta t /2) \Delta t\,,
\end{equation}
where $\mathbf{v}_i=\mathbf{p}_i/E_i$, and $E_i=|\mathbf{p}_i|$. Note that the velocity is defined at time $t+\Delta t/2$, i.e., in the middle between the initial and final time ($t$ and $t+\Delta t$). Correspondingly, the momentum update will provide $\mathbf{p}_i$ at time $t+\Delta t/2$.

\begin{figure}[hbt]
  \begin{center}
    \includegraphics[width=8cm]{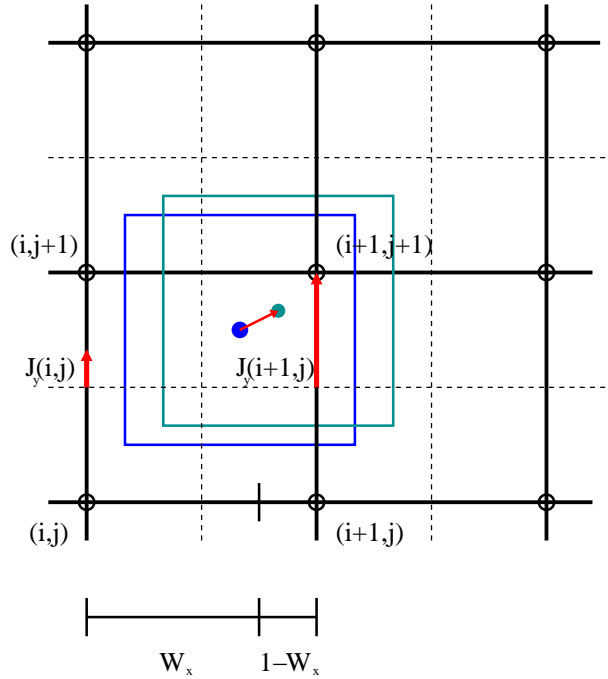}
    \caption{The \textsc{Eastwood} method for calculating the current with charge conservation. It is shown how the smeared particle
    charge moves across the reciprocal (dashed) lattice and in the process produces a current proportional to $W_x$ or $1-W_x$ respectively.
    The created currents in the $x$-direction are not shown for clarity.}
    \label{fig:current}
  \end{center}
\end{figure}

The update of the current is more involved. We employ smoothed
currents as it is common for particle-in-cell (PIC)
\index{Particle-In-Cell} simulations in Abelian plasma physics
\cite{Hockney:1981,Birdsall:1985} in order to avoid numerical
noise. Let us explain the method for a two-dimensional Abelian
system first. Let us consider an Abelian charge $q$ moving from
$(x^t,y^t)$ to $(x^{t+\Delta t},y^{t+\Delta t})$ and define:
\begin{align}
    x_0\equiv x^t\,,  &~~~~x_1\equiv x^{t+\Delta t}=x^t+v_x^{t+\Delta t/2} \Delta t\,,\\
    y_0\equiv y^t\,,  &~~~~y_1\equiv y^{t+\Delta t}=y^t+v_y^{t+\Delta t/2} \Delta t\,,
\end{align}
and
\begin{align}
    i_{x0}\equiv \text{floor}(x_0)\,, & ~~~~i_{x1}\equiv \text{floor}(x_1)\,,\\
    i_{y0}\equiv \text{floor}(y_0)\,, & ~~~~i_{y1}\equiv \text{floor}(y_1)\,,
\end{align}
where the $i$'s are the largest integer values not greater than
$x$ or $y$, respectively. They indicate the cell, the particle is
located in. We assume that a particle does not move further than
one grid spacing $\Delta x$ or $\Delta y$ (or $a$) in one time
step $\Delta t$, i.e., $v_x \Delta t <\Delta x$ etc. Let us first
consider the case when a particle remains in the same cell during
one time step. The charge density at a site $(i,j)$ is obtained by
smearing
\begin{equation}
    \rho(i,j)=\sum_{n=1}^N q_nS_i(x_n)S_j(y_n)\,,
\end{equation}
where the sum runs over all $N$ particles within the cell and $S$ is a form factor. We use the first-order shape factor defined by
\begin{align}
    S_i(x):=\left\{\begin{array}{ll}1-|x-i_x|\, ~~~ &\text{for }     |x-i_x|\leq 1  \\
                                    0\, &\text{for } |x-i_x|> 1  \end{array}\right.
\end{align}
For a single particle with charge $q$ and coordinates $(x,y)$ and $\tilde{x}=x-i_x, \tilde{y}=y-i_y$, we have
\begin{align}
    \rho(i_{x},i_{y})&=q(1-\tilde{x})(1-\tilde{y})\,,\notag\\
    \rho(i_x+1,i_y)&=q\tilde{x}(1-\tilde{y})\,,\notag\\
    \rho(i_x,i_y+1)&=q(1-\tilde{x})\tilde{y}\,,\notag\\
    \rho(i_x+1,i_y+1)&=q\tilde{x}\tilde{y}\,.
\end{align}
A charge flux can be computed from the start point $(x_0,y_0)$ and
end point $(x_1,y_1)$ of the particle movement. Using the
procedure described by \textsc{Eastwood} \cite{Eastwood:1991Co},we
can do this ensuring charge conservation of the assigned current
densities. Let us define
\begin{align}
J_x(i_{x0}+\frac{1}{2},i_{y0})&\equiv J_x(i_{x0},i_{y0})=F_x(1-W_y)\,,\label{jx1}\\
J_x(i_{x0}+\frac{1}{2},i_{y0}+1)&\equiv J_x(i_{x0},i_{y0}+1)=F_x W_y\,,\\
J_y(i_{x0},i_{y0}+\frac{1}{2})&\equiv J_y(i_{x0},i_{y0})=F_y(1-W_x)\,,\\
J_y(i_{x0}+1,i_{y0}+\frac{1}{2})&\equiv J_x(i_{x0}+1,i_{y0})=F_y W_x\,,\label{jy2}
\end{align}
where $F_x$ and $F_y$ represent the charge flux given by
\begin{align}
    F_x=q\frac{x_1-x_0}{\Delta t}\,, ~~~~ F_y=q\frac{y_1-y_0}{\Delta t} ~~~ \text{(Abelian charge here for simplicity)}\,,
\end{align}
and $W$ represents the first-order shape-factor corresponding to the linear weighting function defined at the midpoint
between the start point and the end point:
\begin{align}
    W_x=\frac{x_1+x_0}{2}-i_{x0}\,, ~~~~ W_y=\frac{y_1+y_0}{2}-i_{y0}\,.\label{thews}
\end{align}
Note that the currents are defined not at the lattice sites
themselves but halfway between two grid points as indicated in
Eqs. (\ref{jx1})-(\ref{jy2}). We shall however omit the $1/2$ in
all expressions for clarity as already done in Eqs.
(\ref{jx1})-(\ref{jy2}). The meaning of the definitions
(\ref{jx1}) through (\ref{thews}) is best explained in a graphical
way as done in Fig. \ref{fig:current}. One can see that the larger
the part of the smeared charge that moves across a cell boundary
of the reciprocal lattice that is closest to the point where the
current is defined, the larger the contribution to this current.

One can easily see that the lattice continuity equation
\begin{equation}
    J_x(i,j)-J_x(i-1,j)+J_y(i,j)-J_y(i,j-1)=\frac{\rho^t(i,j)-\rho^{t+\Delta t}(i,j)}{\Delta t}\,,
\end{equation}
is in fact fulfilled for this procedure. Let us check it for the site $(i_x+1,i_y+1)$, considering a single particle moving within cell $(i_x,i_y)$ from $(x_0,y_0)$ to $(x_1,y_1)$ during the time $\Delta t$. The total current into and out of cell $(i_x+1,i_y+1)$ is
\begin{equation}
    J_x(i_x+1,i_y+1)-J_x(i_x,i_y+1)+J_y(i_x+1,i_y+1)-J_y(i_x+1,i_y)\,,
\end{equation}
however, our single particle contributes only to $J_x(i_x,i_y+1)$ and $J_y(i_x+1,i_y)$, such that we have
\begin{align}
   -J_x(i_x,i_y+1)-J_y(i_x+1,i_y) &= -F_xW_y-F_yW_x\notag\\
                                  &= -\frac{q}{\Delta t} (x_1-x_0)\left(\frac{y_1+y_0}{2}-i_y\right)\notag\\
                                  & ~~~~~~~ -\frac{q}{\Delta t} (y_1-y_0)\left(\frac{x_1+x_0}{2}-i_x\right)\notag\\
                                  &= -\frac{q}{\Delta t} (\tilde{x}_1-\tilde{x}_0)\left(\frac{\tilde{y}_1+\tilde{y}_0}{2}\right)
                                     -\frac{q}{\Delta t} (\tilde{y}_1-\tilde{y}_0)\left(\frac{\tilde{x}_1+\tilde{x}_0}{2}\right)\notag\\
                                  &= \frac{q}{\Delta t} \left(\tilde{x}_1\tilde{y}_1-\tilde{x}_2\tilde{y}_2\right)\notag\\
                                  &= \frac{\rho^t(i_{x}+1,i_y+1)-\rho^{t+\Delta t}(i_{x}+1,i_y+1)}{\Delta t}\,.
\end{align}

\begin{figure}[htb]
  \begin{center}
    \includegraphics[height=4cm]{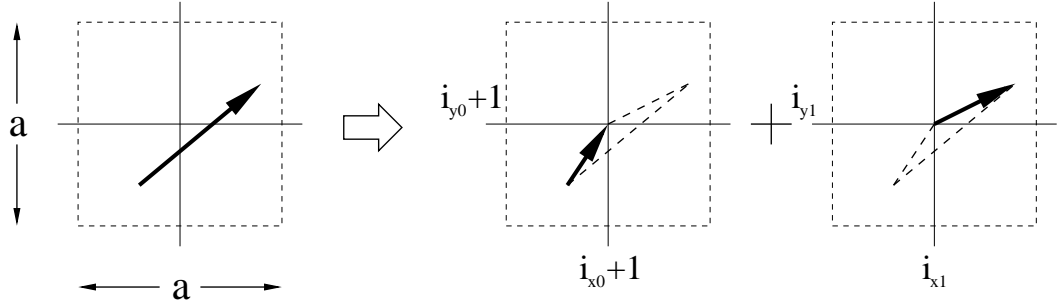}
    \caption{Particle trajectory of the two-dimensional zigzag scheme for $i_{x0}\neq i_{x1}$ and $i_{y0}\neq i_{y1}$.
    One particle moves from $(x_0,y_0)$ to $(i_{x0}+1,i_{y0}+1)=(i_{x1},i_{y1})$ and another particle moves from
    $(i_{x1},i_{y1})$ to $(x_1,y_1)$ during the time from $t$ to $t+\Delta t$. The solid arrows represent particle trajectories. The thin solid lines represent cell meshes.}
    \label{fig:CPC1}
  \end{center}
\end{figure}

\begin{figure}[htb]
  \begin{center}
    \includegraphics[height=4cm]{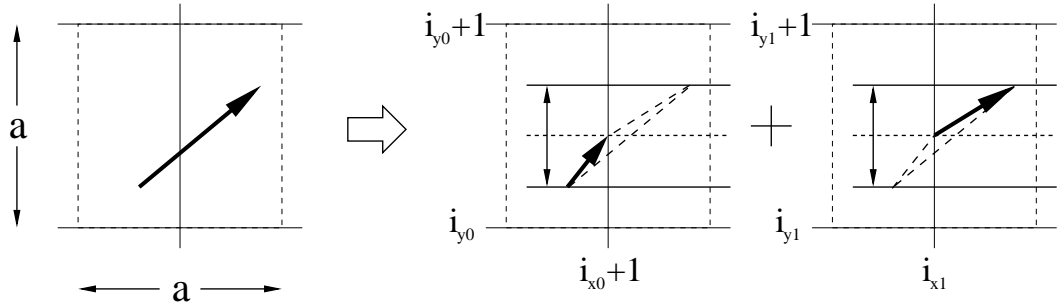}
    \caption{Particle trajectory of the two dimensional zigzag scheme for $i_{x0}\neq i_{x1}$ and $i_{y0}=i_{y1}$.
    One particle moves from $(x_0,y_0)$ to $(i_{x0}+1,(y_1+y_0)/2)=(i_{x1},(y_1+y_0)/2)$, and another particle moves from
    $(i_{x1},(y_1+y_0)/2)$ to $(x_1,y_1)$ during the time from $t$ to $t+\Delta t$. Here $i_{x0}+1=i_{x1}$. The solid arrow represents particle trajectories. The thin solid lines represent cell meshes. Same holds for interchanging $x$ and $y$.}
    \label{fig:CPC2}
  \end{center}
\end{figure}

Now let us consider the case that the particle crosses cell
meshes. To implement that case we decompose the particle movement
with a special assignment pattern as shown in Figs. \ref{fig:CPC1}
and \ref{fig:CPC2}, the so called \emph{zigzag} scheme \cite{Umeda:2003}. If the
mesh is crossed in both directions, the particle moves to the
lattice site first and in a second step to the final position in
the new cell (see Fig. \ref{fig:CPC1}). If only one mesh is
crossed, the particle will first move to the boundary it crosses.
The second coordinate (of the intermediate point on the mesh) is
taken to be the middle point between the initial and final
coordinate in the direction in that no crossing happens (see Fig.
\ref{fig:CPC2}). From there the particle moves to the final
position in the second step. If the particle stays in the same
cell within one time step, the above procedure is not necessary -
however, we formally assume that the movement of the particle is
described in two steps as well. That is, in the first step, the
particle moves from $(x_0,y_0)$ to $((x_0+x_1)/2,(y_0+y_1)/2)$,
and in the second step it moves from $((x_0+x_1)/2,(y_0+y_1)/2)$
to $(x_1,y_1)$ during the time from $t$ to $t+\Delta t$.
 ~\\
 ~\\
Combining all cases, we can define
\begin{align}
x_r:=\left\{\begin{array}{ll}\frac{x_0+x_1}{2}\, ~~~ &\text{for }     i_{x0}=i_{x1}  \\
                            \text{max}(i_{x0},i_{x1})\, &\text{for } i_{x0}\neq i_{x1}  \end{array}\right.
\end{align}
and the analogue for the other directions. \\
By using $(x_r,y_r)$ (in 2D), a charge flux $q(v_x,v_y)$ is decomposed into $(F_{x0},F_{y0})$ and $(F_{x1},F_{y1})$
as follows:
\begin{align}
F_{x0}=q\frac{x_r-x_0}{\Delta t}\,, ~~~~ & F_{y0}=q\frac{y_r-y_0}{\Delta t}\,,\notag\\
F_{x1}=q\frac{x_1-x_r}{\Delta t}=q v_x-F_{x0}\,, ~~~~ & F_{y1}=q\frac{y_1-y_r}{\Delta t}=q v_y-F_{y0}\,.
\end{align}
Now we shall calculate the charge flux at each grid point. Therefore, we apply the shape factors ($W$) defined at the midpoints
between $(x_0,y_0)$ and $(x_r,y_r)$, and $(x_r,y_r)$ and $(x_1,y_1)$, respectively, to assign the segments $(F_{x0},F_{y0})$ and $(F_{x1},F_{y1})$ to the adjacent grid points. The particles exist at the midpoints at time $t+\Delta t/2$ and the first order
shape-factors at the midpoints are defined by
\begin{align}
    W_{x0}=\frac{x_0+x_r}{2}-i_{x0}\,, ~~~~ & W_{y0}=\frac{y_0+y_r}{2}-i_{y0}\,, \notag\\
    W_{x1}=\frac{x_r+x_1}{2}-i_{x1}\,, ~~~~ & W_{y1}=\frac{y_r+y_1}{2}-i_{y1}\,,
\end{align}
Then the segments of the charge flux assigned to the 8 grid points are obtained by the following procedure:
\begin{align}
    J_x(i_{x0},i_{y0})=F_{x0}(1-W_{y0})\,, ~~~~ & J_x(i_{x0},i_{y0}+1)=F_{x0}W_{y0}\,, \notag\\
    J_y(i_{x0},i_{y0})=F_{y0}(1-W_{x0})\,, ~~~~ & J_y(i_{x0}+1,i_{y0})=F_{y0}W_{x0}\,, \notag\\
    J_x(i_{x1},i_{y1})=F_{x1}(1-W_{y1})\,, ~~~~ & J_x(i_{x1},i_{y1}+1)=F_{x1}W_{y1}\,, \notag\\
    J_y(i_{x1},i_{y1})=F_{y1}(1-W_{x1})\,, ~~~~ & J_y(i_{x1}+1,i_{y1})=F_{y1}W_{x1}\,.
\end{align}

\begin{figure}[htb]
  \begin{center}
    \includegraphics[height=8cm]{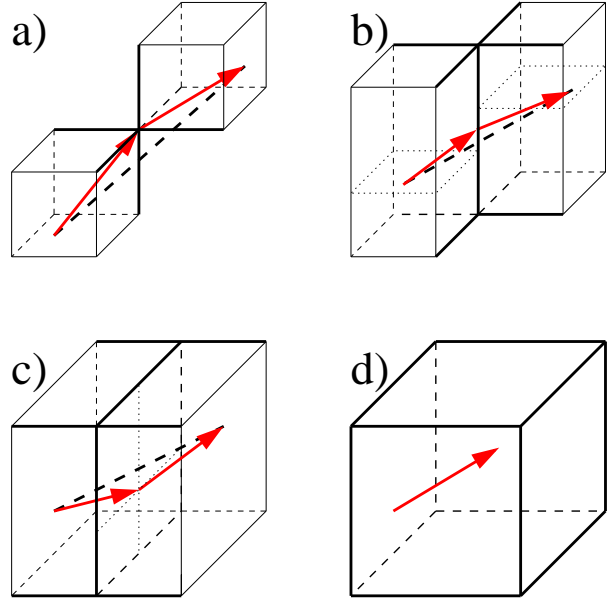}
    \caption{Particle trajectories for the 3D zigzag scheme. The solid arrows represent particle trajectories. The thick solid lines represent cell meshes.
    (a) For $i_{x0}\neq i_{x1}$ and $i_{y0}\neq i_{y1}$ and $i_{z0}\neq i_{z1}$, the particle moves in the first step from $(x_0,y_0,z_0)$ to $(i_{x0}+1,
    i_{y0}+1,i_{z0}+1)=(i_{x1},i_{y1},i_{z1})$, and in the next step from there to $(x_1,y_1,z_1)$ (both steps within the time from $t$ to $t+\Delta t$).
    (b) For $i_{x0}\neq i_{x1}$ and $i_{y0}\neq i_{y1}$ and $i_{z0}=i_{z1}$, the particle moves from $(x_0,y_0,z_0)$ to $(i_{x0}+1,
    i_{y0}+1,\frac{z_0+z_1}{2})$ and then to $(x_1,y_1,z_1)$.
    (c) For $i_{x0}\neq i_{x1}$ and $i_{y0}=i_{y1}$ and $i_{z0}=i_{z1}$, the particle moves
    from $(x_0,y_0,z_0)$ to $(i_{x0}+1,\frac{y_0+y_1}{2},\frac{z_0+z_1}{2})$ and then to $(x_1,y_1,z_1)$.
    (d) For $i_{x0}=i_{x1}$ and $i_{y0}=i_{y1}$ and $i_{z0}=i_{z1}$ the particle moves from $(x_0,y_0,z_0)$ to $(x_1,y_1,z_1)$
    via $(\frac{x_0+x_1}{2},\frac{y_0+y_1}{2},\frac{z_0+z_1}{2})$.}
    \label{fig:CPC3}
  \end{center}
\end{figure}

The procedure described above can be directly generalized to three dimensions. It is worth mentioning that in three dimension we have
four possible cases, i.e., that the particle remains in the same cell within one time step, that it crosses a cell boundary in one direction,
crosses in two directions, or crosses in all three directions. Fig. \ref{fig:CPC3} summarizes all these possibilities and describes how the calculation is done.

Let us now extend the PIC method with current smearing to the case of a non-Abelian plasma, i.e., let us introduce color and call the method CPIC\index{CPIC}, colored-particle-in-cell. Again we consider two dimensions in this discussion for clarity. The currents are now defined by
\begin{eqnarray}
 J_x(i_x,i_y) &=& Q\frac{x_1-x_0}{\Delta t}(1-W_y),
  \qquad J_x(i_x,i_y+1) = Q_{y}\frac{x_1-x_0}{\Delta t}W_y, \\
 J_y(i_x,i_y) &=& Q\frac{y_1-y_0}{\Delta t}(1-W_x), \qquad
  J_y(i_x+1,i_y) = Q_{x}\frac{y_1-y_0}{\Delta t}W_x,
\end{eqnarray}
with the $W$ as above. We have to use the parallel transported charges
\footnote{We omit the term $[A_z,J_z]$ here, because this term
does not appear in 3D and it does not matter in this consideration.}
\begin{equation}
 Q_x  \equiv U_x^\dagger(i_x,i_y)QU_x(i_x,i_y)\,,\qquad
 Q_y  \equiv  U_y^\dagger(i_x,i_y)QU_y(i_x,i_y)\,,
\end{equation}
in order to fulfill the lattice covariant continuity equation
\begin{align}
 \dot{\rho}(i_x,i_y) = &U^\dagger_x(i_x-1,i_y)J_x(i_x-1,i_y)U_x(i_x-1,i_y) - J_x(i_x,i_y) \notag\\
                       &~~ + U^\dagger_y(i_x,i_y-1)J_y(i_x,i_y-1)U_y(i_x,i_y-1) - J_y(i_x,i_y)\,.
\end{align}
The link variable $U_\mu$ was defined in (\ref{linkdef}).
For a single particle moving within cell $(i_x,i_y)$, the continuity equation at the adjacent sites reads
\begin{eqnarray}
\dot{\rho}(i_x,i_y) &=& -J_x(i_x,i_y) - J_y(i_x,i_y)\,, \label{eq:ch1}\\
\dot{\rho}(i_x+1,i_y)&=&
     U^\dagger_x(i_x,i_y)J_x(i_x,i_y)U_x(i_x,i_y) - J_y(i_x+1,i_y)\,,\label{eq:ch2}\\
\dot{\rho}(i_x,i_y+1) &=&
    U^\dagger_y(i_x,i_y)J_y(i_x,i_y)U_y(i_x,i_y) - J_x(i_x,i_y+1)\,. \label{eq:ch3}
\end{eqnarray}
This is easiest to understand if you consider the charge density
$\rho(i_x,i_y)$ as that within one reciprocal lattice cell around
the cite $(i_x,i_y)$ and the currents as those created by a moving
charge in the actual lattice cell $(i_x,i_y)$ that influence that
charge density. See also Fig. \ref{fig:current} to get a better
idea.

Eqs.~(\ref{eq:ch1}), (\ref{eq:ch2}), (\ref{eq:ch3}) are
consistent with the following definitions of the charge densities
\begin{eqnarray}
 \rho(i_x,i_y) &=& Q(1-\tilde{x})(1-\tilde{y})\,,\\
 \rho(i_x,i_y+1) &=& Q_y(1-\tilde{x})\tilde{y}\,,\\
 \rho(i_x+1,i_y) &=& Q_x\tilde{x}(1-\tilde{y})\,.
\end{eqnarray}
However, since a particle's color charge depends on its path, so does
$\rho(i_x+1,i_y+1)$ and we are not able to calculate it from the charge
distribution itself. Rather, we directly employ covariant current
conservation to determine the increment of charge at site $(i_x+1,i_y+1)$
within the time-step. This way, we can satisfy Gauss's law in the
non-Abelian case.

We also have to check that $\Tr(Q^2)$ is conserved by this smearing
method. This is true when the lattice spacing $a$ is small, as the
total charge of a particle is given by
\begin{equation}
 Q_0 = Q(1-x)(1-y) + Q_xx(1-y) + Q_y(1-x)y + [a_pQ_{xy} + (1-a_p)Q_{yx}]xy~,
\end{equation}
where the $a_p$ depend on the path of a particle and
$Q_{xy}=U^\dagger_x(i,j+1)Q_yU_x(i,j+1)$,
$Q_{yx}=U^\dagger_y(i+1,j)Q_xU_y(i+1,j)$.
If we require that $\Tr(Q_0^2)$ be constant, then the
cross terms, for example $\Tr(QQ_x)$, have to vanish.
This is true when $a$ is small, because $\Tr(Q[A,Q])=0$:
\begin{equation}
 \Tr(QQ_x) = \Tr(Q(Q + iga[A_x,Q] + {\cal O}(a^2))) = \Tr(Q^2) +{\cal O}(a^2).
\end{equation}

\subsection{The field updates}
The electric fields $E$ live on the links between the lattice
sites and so do the $U$-fields. The $B$-fields, which can be
calculated from the plaquettes according to
\begin{equation}\label{bfromu}
    U_{\square}=U_{xy}=\exp(ia^2gF_{xy})=\exp(ia^2g(-B_{z}))\approx\mathbf{1}-ia^2gB_z^a{\boldsymbol\sigma}^a\,,
\end{equation}
and analogously for the other two directions, are hence defined within the area enclosed by the plaquette.
The time derivative of
\begin{equation}
    U(i)=e^{iagA(i)}\,,
\end{equation}
depends on whether we use a left or right covariant derivative. We will stick to the prescription, where $\dot{A}(i)$ stands on the left:
\begin{align}
    \dot{U}(i)&=iag\dot{A}(i)U(i)\notag\\
    \Leftrightarrow \dot{U}(i)&=-iagE(i)U(i)\,.
\end{align}
So the update of the $U$-fields is determined by the (color-) electric field.
On the lattice, the electric field is updated using the following expression for $\dot{E}$:
\begin{equation}\label{edotlat1}
    \dot{E}^j=-\frac{i}{2a^3g}\sum_i\left(U_{ij}(x)-U_i^\dag(x-i)U_{ij}(x-i)U_i(x-i)\right)-J^j\,,
\end{equation}
where $J$ is the current calculated in the previous subsection, $i$ and $j$ run over the directions $x$, $y$ and $z$ and the $\tau_a$ are the \textsc{Pauli} matrices $\sigma_a$. Also note that $(x-i)$ means that from position $x$, we move one lattice spacing in the $-i$ direction.
Let us show that in the continuum limit this becomes the usual expression
\begin{equation}
    \dot{E}^j=D_i F^{ij}-J^j=\partial_i F^{ij}+ig[A_i,F^{ij}]-J^j\,.
\end{equation}
Inserting the continuum limits for the plaquettes (see Eq.(\ref{plaquettecont})) and the definitions for the link variables, Eq.(\ref{edotlat1}) becomes
\begin{align}
    \dot{E}^j&=-\frac{i}{2a^3g}\sum_i\left(e^{ia^2gF^{ij}(x)}-e^{-iagA_i(x-i)}e^{ia^2gF^{ij}(x-i)}e^{iagA_i(x-i)}\right)-J^j\notag\\
    &=-\frac{i}{2a^3g}\sum_i\left(e^{ia^2gF^{ij}(x)}-e^{-iagA_i(x-i)+ia^2gF^{ij}(x-i)+\frac{a^3g^2}{2}[A_i,F^{ij}]}e^{iagA_i(x-i)}\right)-J^j\notag\\
    &=-\frac{i}{2a^3g}\sum_i\left(e^{ia^2gF^{ij}(x)}-e^{ia^2gF^{ij}(x-i)+\frac{a^3g^2}{2}[A_i,F^{ij}]-{\frac{a^3g^2}{2}[F^{ij},A_i]+\mathcal{O}(a^4)}}\right)-J^j\notag\\
    &\approx-\frac{i}{2a^3g}\sum_i\left(ia^2gF^{ij}(x)-ia^2gF^{ij}(x-i)-a^3g^2[A_i,F^{ij}]\right)-J^j\notag\\
    &=\frac{1}{2a}\sum_i\left(F^{ij}(x)-F^{ij}(x-i)+iag[A_i,F^{ij}]\right)-J^j\notag\\
    &=\frac{1}{2}\sum_i\left(\frac{F^{ij}(x)-F^{ij}(x-i)}{a}+ig[A_i,F^{ij}]\right)-J^j\notag\\
    &\approx\partial_i F^{ij}+ig[A_i,F^{ij}]-J^j\,.
\end{align}
where we used Eq.(\ref{exprel}) in the first step.
The electric field at time $(t+\Delta t)$ follows directly, using
\begin{equation}\label{Eupd}
    E^j(t+\Delta t)=E^j(t)+\Delta t \dot{E}^j(t+\Delta t/2)\,,
\end{equation}
because the $U$-fields are defined at half time steps. See below.
The $B$-fields do not need an individual update because we can directly obtain them from the updated $U$-fields via Eq.(\ref{bfromu}).
In practice, we use a \emph{leapfrog} algorithm in that the $A$-fields, and with them the $U$-fields, are given at time $(t+\Delta t/2)$.
From this we can directly update $E$ according to Eq.(\ref{Eupd}).
To get the $B$-field at time $(t+\Delta t)$ $U$ is first updated by half a time step $\Delta t/2$, the $B$-field is extracted and the particle momentum updated (to $(t+\Delta t/2)$, see Section \ref{momupdate} below). Then $U$ is evolved by another half time step. The particle coordinates are updated to time $(t+\Delta t)$ using the momentum at time $(t+\Delta t/2)$ and the current at time $(t+\Delta t)$ is calculated.

\subsection{Momentum update}
\label{momupdate} The momentum update is done in a way that
satisfies time reversibility with the so called
\textsc{Buneman-Boris}-method \index{Buneman-Boris-method} (see
\cite{Hockney:1981} and \cite{Birdsall:1985}). The procedure is to
update the momentum in several steps. We write $\mathbf{E}=gq^a \mathbf{E}^a$ and 
$\mathbf{B}=gq^a \mathbf{B}^a$. First calculate
\begin{equation}
    \mathbf{p}_1(t)=\mathbf{p}\left(t-\frac{\Delta t}{2}\right)+\frac{\Delta t}{2} \mathbf{E}(t)\,,
\end{equation}
i.e., take half a time step just using the electric field.
Next define $\mathbf{p}_2(t)$ using
\begin{equation}
    \mathbf{p}_2\left(t\right)=\mathbf{p}\left(t+\frac{\Delta t}{2}\right)-\frac{\Delta t}{2} \mathbf{E}(t)\,,
\end{equation}
and
\begin{align}
    \frac{\mathbf{p}_2(t)-\mathbf{p}_1(t)}{\Delta t}&=\frac{\mathbf{p}_1(t)+\mathbf{p}_2(t)}{2} \times \frac{\mathbf{B}(t)}{e(t)}\notag\\
    \Leftrightarrow \mathbf{p}_2(t)-\mathbf{p}_2(t)\times \frac{\mathbf{B}(t)}{e(t)}\frac{\Delta t}{2}
    &=\mathbf{p}_1(t)+\mathbf{p}_1(t)\times \frac{\mathbf{B}(t)}{e(t)}\frac{\Delta t}{2}\,.\label{timeevp2}
\end{align}
The above expressions make clear that the definition of $\mathbf{p}_2(t)$ is made in order to keep the update time reversible.
Let us solve for $\mathbf{p}_2(t)$ using the shorthand $\tilde{\mathbf{B}}\equiv\frac{\mathbf{B}(t)}{e(t)}\frac{\Delta t}{2}$:
\begin{align}
    \mathbf{p}_2-\mathbf{p}_2\times\tilde{\mathbf{B}}&=\mathbf{p}_1+\mathbf{p}_1\times\tilde{\mathbf{B}}\left.~~~~\right|\times \tilde{\mathbf{B}}\notag\\
    \Leftrightarrow  \mathbf{p}_2\times\tilde{\mathbf{B}}-(\mathbf{p}_2\times\tilde{\mathbf{B}})\times\tilde{\mathbf{B}}
    &=\mathbf{p}_1\times\tilde{\mathbf{B}}+(\mathbf{p}_1\times\tilde{\mathbf{B}})\times\tilde{\mathbf{B}}\notag\\
    \Leftrightarrow \mathbf{p}_2\times\tilde{\mathbf{B}}+ \mathbf{p}_2 \tilde{\mathbf{B}}^2-\tilde{\mathbf{B}}( \mathbf{p}_2\cdot\tilde{\mathbf{B}})
    &=\mathbf{p}_1\times\tilde{\mathbf{B}}- \mathbf{p}_1 \tilde{\mathbf{B}}^2+\tilde{\mathbf{B}}( \mathbf{p}_1\cdot\tilde{\mathbf{B}})\,.
\end{align}
Now replace $\mathbf{p}_2\times\tilde{\mathbf{B}}$ using Eq. (\ref{timeevp2}):
\begin{equation}
    \mathbf{p}_2\times\tilde{\mathbf{B}}=\mathbf{p}_2-\mathbf{p}_1-\mathbf{p}_1\times\tilde{\mathbf{B}}\,,
\end{equation}
Then we have
\begin{align}
    \Leftrightarrow \mathbf{p}_2-\mathbf{p}_1-\mathbf{p}_1\times\tilde{\mathbf{B}}+ \mathbf{p}_2 \tilde{\mathbf{B}}^2-\tilde{\mathbf{B}}( \mathbf{p}_2\cdot\tilde{\mathbf{B}})
    &=\mathbf{p}_1\times\tilde{\mathbf{B}}- \mathbf{p}_1 \tilde{\mathbf{B}}^2+\tilde{\mathbf{B}}( \mathbf{p}_1\cdot\tilde{\mathbf{B}})\,.
\end{align}
Because $\mathbf{p}_2\cdot\tilde{\mathbf{B}}=(\mathbf{p}_1+\mathbf{p}_1\times\tilde{\mathbf{B}}+\mathbf{p}_2\times\tilde{\mathbf{B}})\cdot\tilde{\mathbf{B}}=
         \mathbf{p}_1\cdot\tilde{\mathbf{B}}$, we can write
\begin{align}
    \Leftrightarrow \mathbf{p}_2(1+\tilde{\mathbf{B}}^2)
    &=\mathbf{p}_1+2\mathbf{p}_1\times\tilde{\mathbf{B}}+2\tilde{\mathbf{B}}( \mathbf{p}_1\cdot\tilde{\mathbf{B}})-\mathbf{p}_1 \tilde{\mathbf{B}}^2\,.
\end{align}
Adding $2(\mathbf{p}_1\times\tilde{\mathbf{B}})\times\tilde{\mathbf{B}}$ on both sides yields
\begin{align}
    \mathbf{p}_2(1+\tilde{\mathbf{B}}^2)+2(\mathbf{p}_1\times\tilde{\mathbf{B}})\times\tilde{\mathbf{B}}
    &=\mathbf{p}_1(1-\tilde{\mathbf{B}}^2)+2\mathbf{p}_1\times\tilde{\mathbf{B}}+2(\mathbf{p}_1\times\tilde{\mathbf{B}})\times\tilde{\mathbf{B}}
    +2\tilde{\mathbf{B}}( \mathbf{p}_1\cdot\tilde{\mathbf{B}}) \,.
\end{align}
Now define $\mathbf{p}_3=\mathbf{p}_1+\mathbf{p}_1\times\tilde{\mathbf{B}}$ and write:
\begin{align}
    \Rightarrow &\mathbf{p}_2(1+\tilde{\mathbf{B}}^2)+2\tilde{\mathbf{B}}(\mathbf{p}_1\cdot\tilde{\mathbf{B}})-2\mathbf{p}_1\tilde{\mathbf{B}}^2
    =\mathbf{p}_1(1-\tilde{\mathbf{B}}^2)+2\mathbf{p}_3\times\tilde{\mathbf{B}}
    +2\tilde{\mathbf{B}}( \mathbf{p}_1\cdot\tilde{\mathbf{B}}) \notag\\
    \Leftrightarrow &\mathbf{p}_2(1+\tilde{\mathbf{B}}^2) = \mathbf{p}_1(1+\tilde{\mathbf{B}}^2)+2\mathbf{p}_3\times\tilde{\mathbf{B}}\notag\\
    \Leftrightarrow &\mathbf{p}_2(t) = \mathbf{p}_1(t)+\frac{2}{1+\left(\frac{\mathbf{B}(t)}{e(t)}\frac{\Delta t}{2}\right)^2}
    \mathbf{p}_3(t) \times \frac{\mathbf{B}(t)}{e(t)}\frac{\Delta t}{2}\,,
\end{align}
where we reintroduced the time $t$ explicitly and replaced $\tilde{\mathbf{B}}$ by its definition.
Finally,
\begin{equation}
    \mathbf{p}\left(t+\frac{\Delta t}{2}\right)=\mathbf{p}_2\left(t\right)+\frac{\Delta t}{2} \mathbf{E}(t)\,,
\end{equation}
is the momentum used to determine the velocity in the coordinate update (\ref{coordupdate}).

\section{Inclusion of binary collisions}
\label{wymcollisions} \index{Collisions}
After having discussed the collisionless system and explained the numerical methods for solving the equations of motion we now introduce hard collisions, i.e., particle-particle collisions with hard momentum exchange. For a realistic coupling constant $g$ these interactions become important since the higher orders in $g$ at which they contribute are no longer suppressed, as pointed out already in Chapter \ref{collisionsmodel}. With the collisions included, our system will be similar to that used in parton cascade simulations \cite{PhysRevC.29.2146,Kortemeyer:1995di,Zhang:1998tj,Cheng:2001dz,PhysRevC.40.2611,Molnar:2001ux,Xu:2004gw,Xu:2004mz,Xu:2007aa} with one big advantage. We do not have to cut off (or suppress) the momentum exchange below the \textsc{Debye}-mass ($\sim gT$) which is of the order of the temperature in an equilibrium system ($g \sim 1-2$), but include softer momentum exchanges via particle-field interactions. This way we are also able to study collective phenomena and their contribution to isotropization and thermalization. In particular, we can study systems far from equilibrium for which the scale corresponding to the \textsc{Debye}-mass squared in an isotropic system becomes negative \cite{Romatschke:2003ms}. In this case it can obviously not be used to damp the propagator, i.e., to cut off the momentum exchange in the infrared region, as it is done in the parton cascade simulations. All we need is a separation scale ${k^*}$ between the field and particle degrees of freedom. We will discuss this separation scale in detail in Section \ref{separation}. For now it will serve as a lower bound for the exchanged momenta in the binary elastic particle collisions. All softer momenta will be exchanged via the fields.
\subsection{Cross section and transition rate}
\label{crosssctn} 
\begin{figure}[htb]
  \begin{center}
    \includegraphics[height=2.5cm]{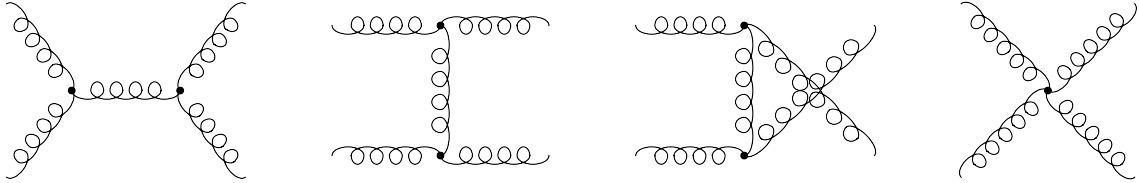}
    \caption{Processes contributing to $gg\rightarrow gg$ scattering.}
    \label{fig:ggline2}
  \end{center}
\end{figure}

\index{Cross section}
We include collisions using the stochastic method introduced and applied in \cite{Danielewicz:1991dh,Lang:1993dh,Xu:2004mz}.
This means that we do not interpret the cross section in a geometrical way as done in \cite{PhysRevC.29.2146,Kortemeyer:1995di,Zhang:1998tj,Cheng:2001dz,PhysRevC.40.2611,Molnar:2001ux}
but determine scattering processes in a stochastic manner by sampling possible transitions in a certain volume per time interval.
This collision algorithm can be extended to include the inelastic
collision processes $gg\leftrightarrow ggg$ as done in \cite{Xu:2004mz,Xu:2007aa}, which will be one of the future goals also for our simulation.
We introduce collisions by adding the collision term ${\cal C}$ to the \textsc{Vlasov} equation (\ref{qvlasov}):
\begin{equation}
    P_1^\mu \left[\partial_\mu + g q^a F_{\mu\nu}^a \partial^\nu_{(p)} + g f^{abc} A_\mu^b(X) q^c \partial_{q^a} \right]f_1(P_1,X,q_1)={\cal C}\,,
\end{equation}
with \index{Collisions!Boltzmann collision term}
\begin{align}
\label{cterm1}
{\cal C} &= \frac{1}{2E_1} \int \frac{d^3p_2}{(2\pi)^3 2E_2} \,
\frac{1}{2} \int \frac{d^3p'_1}{(2\pi)^3 2E'_1}
\frac{d^3p'_2}{(2\pi)^3 2E'_2} f'_1 f'_2 | {\cal M}_{1'2'\to 12} |^2
(2\pi)^4 \delta^{(4)} (p'_1+p'_2-p_1-p_2) \notag \\
& -\frac{1}{2E_1} \int \frac{d^3p_2}{(2\pi)^3 2E_2} \,
\frac{1}{2} \int \frac{d^3p'_1}{(2\pi)^3 2E'_1}
\frac{d^3p'_2}{(2\pi)^3 2E'_2} f_1 f_2 | {\cal M}_{12\to 1'2'} |^2
(2\pi)^4 \delta^{(4)} (p_1+p_2-p'_1-p'_2) \,, \notag \\
\end{align}
with the matrix element ${\cal M}$ including all $gg\rightarrow gg$ events shown in Fig. \ref{fig:ggline2}.
The $1/2$ avoids double counting for identical particles.
 ~\\
When assuming two particles in a
spatial volume element $\Delta^3 x$ with momenta in the range
(${\bf p}_1, {\bf p}_1+\Delta^3 p_1$) and
(${\bf p}_2, {\bf p}_2+\Delta^3 p_2$), the collision rate per unit
phase space for such a particle pair follows from Eq. (\ref{cterm1})
\begin{align}
\label{collrate22}
\frac{\Delta N_{\text{coll}}}{\Delta t \frac{1}{(2\pi)^3} \Delta^3 x
\Delta^3 p_1} &= \frac{1}{2E_1}
\frac{\Delta^3 p_2}{(2\pi)^3 2E_2} f_1 f_2 \nonumber \\
& \times \frac{1}{2} \int \frac{d^3 p^{'}_1}{(2\pi)^3 2E^{'}_1}
\frac{d^3 p^{'}_2}{(2\pi)^3 2E^{'}_2} | {\cal M}_{12\to 1'2'} |^2 (2\pi)^4
\delta^{(4)} (p_1+p_2-p^{'}_1-p^{'}_2).
\end{align}
Expressing distribution functions as
\begin{equation}
\label{distf}
f_i=\frac{\Delta N_i}{\frac{1}{(2\pi)^3} \Delta^3 x \Delta^3 p_i},
\quad i=1,2,
\end{equation}
and employing the usual definition of the cross section
for massless particles \cite{Groot:1980}
\begin{equation}
\label{cs22}
\sigma_{22}= \frac{1}{4s} \int
\frac{d^3 p^{'}_1}{(2\pi)^3 2E^{'}_1} \frac{d^3 p^{'}_2}{(2\pi)^3 2E^{'}_2}
| {\cal M}_{12\to 1'2'} |^2 (2\pi)^4 \delta^{(4)} (p_1+p_2-p^{'}_1-p^{'}_2)
\,,
\end{equation}
\index{Cross section}
one obtains the absolute collision probability in a unit box $\Delta^3 x$
and unit time $\Delta t$
\begin{equation}
\label{p22}
P_{22} = \frac{\Delta N_{\text{coll}}}{\Delta N_1 \Delta N_2} =
\tilde{v}_{\text{rel}} \sigma_{22} \frac{\Delta t}{\Delta^3 x}\,.
\end{equation}
\index{Transition probability} $\tilde{v}_{\text{rel}}=s/2E_1E_2$
denotes the relative velocity\footnote{Please note that this is
not the actual relativistic relative velocity of two particles.
Please see Appendix \ref{appvrel} for a detailed discussion on how
$\tilde{v}_{\text{rel}}$ relates to the true relative
velocity.}\index{Relative velocity}, where $s$ is the invariant
mass of the particle pair. $P_{22}$ can be any number between $0$
and $1$.\footnote{ In practice one has to choose suitable
$\Delta^3 x$ and $\Delta t$ to make $P_{22}$ less than $1$.}
Whether or not a collision occurs is sampled stochastically as
follows: We compare $P_{22}$ with a uniformly distributed random
number between $0$ and $1$. If the random number is less than
$P_{22}$, the collision will occur. Otherwise there will be no
collision between the two particles within the present time step.
Since we employ $N_{\text{test}}$ test particles per particle, we
have to scale the cross section by $\sigma\rightarrow
\sigma/N_{\text{test}}$ , which leads to
\begin{equation}
\label{p22test}
P_{22} = \tilde{v}_{\text{rel}} \frac{\sigma_{22}}{N_{\text{test}}} \frac{\Delta t}{\Delta^3 x}\,.
\end{equation}
To determine this probability, we need to calculate the total cross section $\sigma_{22}$. It follows from the differential cross section, which can be calculated from the graphs in Fig. \ref{fig:ggline2} in perturbative QCD (pQCD) in leading order in the coupling constant $\alpha_s$. To get an idea of how the calculation works, let us write down the matrix element for the first (s-channel) diagram and then skip directly to the final answer.
\begin{figure}[htb]
  \begin{center}
    \includegraphics[height=5cm]{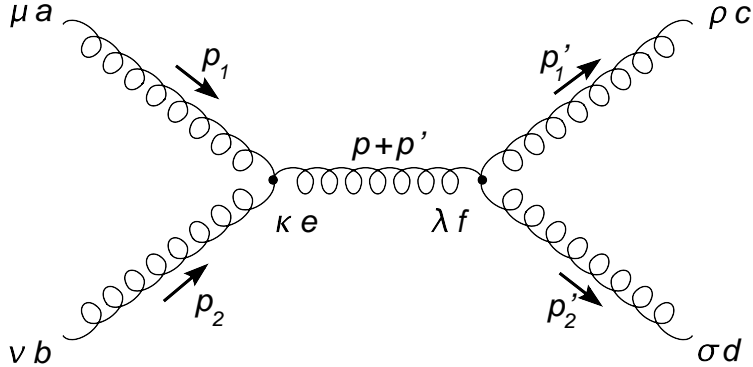}
    \caption{s-channel contribution to $gg\rightarrow gg$ scattering.}
    \label{fig:schannel}
  \end{center}
\end{figure} ~\\
The vector and color indices of the in- and outgoing gluons as
well as that of the intermediate gluon are indicated in Fig.
\ref{fig:schannel}. Using the appropriate \textsc{Feynman} rules,
from this we obtain for the matrix element
\begin{align}
    {\cal M}_{12\rightarrow 1'2'}^{\text{s-channel}}=\epsilon_1^\mu&\epsilon_2^\nu\epsilon_{1'}^{\rho *}\epsilon_{2'}^{\sigma *}\left\{-gf^{aeb}\left[g_{\mu\kappa}(2P_1+P_2)_\nu+g_{\kappa\nu}(-P_1-2P_2)_\mu+g_{\nu\mu}(P_2-P_1)_\kappa\right]\right\}\notag\\
&\times\left\{-gf^{cdf}\left[g_{\rho\sigma}(P_2'-P_1')_\lambda+g_{\sigma\lambda}(-2P_2'-P_1')_\rho+g_{\lambda\rho}(2P_1'+P_2')_\sigma\right]\right\}\notag\\
&\times\left[\frac{-ig^{\kappa\lambda}\delta^{ef}}{(P_1+P_2)^2}\right]\,,
\end{align}
with the gluon polarization vectors $\epsilon$.
After writing down the matrix elements for the three other diagrams, adding them together, taking the complex square, summing over final state polarizations and averaging over initial state polarizations, summing over final state gluon colors and averaging over initial state gluon colors one finds \cite{Combridge:1977dm,Owens:1977sj,Bern:2002tk}
\begin{equation}
    \frac{d\sigma}{dt}=\frac{4\pi\alpha_s^2}{s^2} \frac{N^2}{N^2-1}\left(3-\frac{tu}{s^2}-\frac{su}{t^2}-\frac{st}{u^2}\right)\,,
\end{equation}
with $N$ the number of colors. In particular for SU(2) and SU(3) we have
\begin{align}
    \frac{d\sigma}{dt}&=\frac{16}{3}\frac{\pi\alpha_s^2}{s^2} \left(3-\frac{tu}{s^2}-\frac{su}{t^2}-\frac{st}{u^2}\right)\,&\text{for SU(2),}\label{difcross}\\
    \frac{d\sigma}{dt}&=\frac{9}{2}\frac{\pi\alpha_s^2}{s^2} \left(3-\frac{tu}{s^2}-\frac{su}{t^2}-\frac{st}{u^2}\right)\,&\text{for SU(3).}
\end{align}
The invariant \textsc{Mandelstam} \index{Mandelstam variables} variables are:
\begin{align}
    &s = (P_1+P_2)^2\,,\notag\\
    &t = (P_1-P_1')^2\,,\notag\\
    &u = (P_1-P_2')^2\,.
\end{align}
Some nice relations can be easily derived in the center of mass frame:
\begin{align}
    t=(P_1-P_1')^2=(E_1-E_1')^2-(\mathbf{p}_1-\mathbf{p}_1')^2=\left(\frac{\sqrt{s}}{2}-\frac{\sqrt{s}}{2}\right)^2-\mathbf{q}^2=-\mathbf{q}^2\,,
\end{align}
with the transferred momentum $\mathbf{q}$. Furthermore
\begin{align}\label{tq}
    -t=\mathbf{q}^2=\mathbf{p}_1^2+\mathbf{p}_1'^2-2\mathbf{p}_1\cdot\mathbf{p}_1'=\frac{s}{4}+\frac{s}{4}-\frac{s}{2}\cos\theta
      =\frac{s}{2}(1-\cos\theta)\,.
\end{align}
Because $t$ is an invariant this holds in every frame. Using (\ref{tq}) and $s+t+u=0$ for the massless case, we can express Eq. (\ref{difcross}) by $q^2$ and calculate the total cross section via
\begin{equation}
    \sigma_{22}=\int_{{k^*}^2}^{\frac{s}{2}}\frac{d\sigma}{dq^2}dq^2\,.
\end{equation}
The integral can be done analytically and results in the rather lengthy expression
\begin{align}
     \sigma_{22}=\frac{4\pi\alpha_s^2}{9 {k^*}^2 ({k^*}^2-s)s^4}&\left[-4({k^*}^2)^5+10({k^*}^2)^4s-42({k^*}^2)^3s^2+53({k^*}^2)^2s^3+7{k^*}^2s^4\right.\notag\\
     &~~\left.-12s^5+6{k^*}^2({k^*}^2-s)s^3\log\left(\frac{({k^*}^2)^2}{({k^*}^2-s)^2}\right)\right]\,.
\end{align}

\subsection{Momentum transfer}
\index{Momentum exchange} Now we have all the ingredients to
determine whether or not a collision occurs within one time step.
However, we also have to calculate the transferred momentum
according to the differential cross section $d\sigma/dq^2$. We
perform a \textsc{Lorentz} transformation into the center of mass
frame of the colliding particles and then determine the exchanged
momentum according to the probability distribution
\begin{equation}\label{probdist}
    {\cal P}(q^2)=\frac{1}{\sigma_{22}}\frac{d\sigma}{dq^2}\,,
\end{equation}
which satisfies
\begin{equation}\label{probdistint}
    \int_{{k^*}^2}^{\frac{s}{2}}{\cal P}(q^2)dq^2=1\,.
\end{equation}
To do so, we need to employ the rejection method \cite{Press:1992}.
That is, we need to find an approximation to (\ref{probdist}), ${\cal P}_{\text{app}}(q^2)>{\cal P}(q^2)$, which can be integrated (if possible analytically) and inverted, and is always larger than (\ref{probdist}). With that distribution we can then employ the transformation method \cite{Press:1992} to calculate $q^2$. Finally we need to decide (again stochastically) whether to keep that result or to reject it, with the probability depending on the difference between the exact and the approximate distribution. As the approximate distribution we use the small angle scattering limit of the differential cross section (divided by the total cross section) and shift it to make sure that it always lies above the exact probability distribution (\ref{probdist}).
The transformation method is based on the fundamental transformation law for probabilities and the fact that we can generate random deviates with a uniform probability distribution. For these the probability of generating a number between $x$ and $x+dx$, denoted $u(x)dx$, is given by \begin{align}
u(x)dx=\left\{\begin{array}{ll}dx\, ~~~ &\text{for }     0<x<1 \\
                            0    \,     &\text{otherwise}\end{array}\right.
\end{align}
The distribution $u(x)$ is normalized so that
\begin{equation}
    \int_{-\infty}^{\infty}u(x)dx=1\,.
\end{equation}
Now suppose, we generate a uniform deviate $x$ and take some
function $y(x)$ of it. Then the probability distribution of $q^2$,
denoted ${\cal P}_{\text{app}}(q^2)dq^2$ is determined by the
fundamental transformation law of probabilities, which is simply
\begin{align}
    |{\cal P}_{\text{app}}(q^2)dq^2|=|u(x)dx|\,,
\end{align}
or
\begin{align}
    {\cal P}_{\text{app}}(q^2)=u(x)\left|\frac{dx}{dq^2}\right|\,,
\end{align}
which leads to
\begin{align}
    \int_{{k^*}^2}^{q^2}{\cal P}_{\text{app}}(q'^2)dq'^2=\int_0^xu(x')dx'=x\,,
\end{align}
with $x<x_{\text{max}}$ and
\begin{align}
    x_{\text{max}}=\int_{{k^*}^2}^{\frac{s}{2}}{\cal P}_{\text{app}}(q'^2)dq'^2\,,
\end{align}
which is larger than one because ${\cal P}_{\text{app}}(q^2)>{\cal P}(q^2)$ and Eq. (\ref{probdistint}).
Defining $F({k^*}^2,q^2):= \int_{{k^*}^2}^{q^2}{\cal P}_{\text{app}}(q'^2)dq'^2$, we have
\begin{align}
    F({k^*}^2,q^2)=x \Rightarrow q^2=F^{-1}({k^*}^2,x)\,,
\end{align}
with $F^{-1}$ being the inverse function to $F$. This determines
the transferred momentum $q^2$. However, since we did not use the
exact distribution function ${\cal P}$, we need to perform one
more step. Having determined a $q^2$, we can now easily calculate
${\cal P}_{\text{app}}(q^2)$ and ${\cal P}(q^2)$. Because ${\cal
P}_{\text{app}}(q^2)>{\cal P}(q^2)$, we compare the ratio of the
two with another random number $y$, which is uniformly distributed
on $[0,1]$, reject the $q^2$ if
\begin{equation}
    y>\frac{{\cal P}(q^2)}{{\cal P}_{\text{app}}(q^2)}\,,
\end{equation}
determine a new $q^2$ and repeat the process until a $q^2$ is
accepted. This makes it obvious why we want the approximate
distribution function to be very close to the exact one - if it is
not we will have a lot of rejected $q^2$ and hence longer
computation time.

The exchanged transverse (transverse to the straight line connecting the two particles) momentum is related to $q^2$ and $s$ in the following way
\begin{equation}
    q^2=\frac{s}{2}(1-\cos\theta)\Rightarrow \theta=\arccos\left(1-\frac{2}{s}q^2\right)
\end{equation}
and
\begin{equation}
    q_{\perp}^2=\frac{s}{4}\sin^2\theta=\frac{s}{4}\sin^2\left(\arccos\left(1-\frac{2}{s}q^2\right)\right)\,.
\end{equation}
In the present status of the simulation the particles do not
exchange color in the collisions. As discussed in Section
\ref{discussions}, this may increase the effect of the collisions
on instability growth. The implementation of (classical) color
rotation is one of the future projects.

\section{Momentum space diffusion}
\index{Momentum diffusion}
\index{Jets}
Having introduced hard collisions to the system of coupled
\textsc{Wong-Yang-Mills} equations, we now investigate a first
observable - the momentum space diffusion of hard particles
traversing an isotropic (in momentum space) plasma. We show that
when we match the initial energy density in the field modes to
that of the particles, the obtained results are approximately
independent of the lattice spacing.

\subsection{The separation scale}
\label{separation} \index{Separation scale} We introduced ${k^*}$
as an infrared cutoff on the exchanged momentum in the hard
collisions in Section \ref{wymcollisions}. As previously mentioned
it will serve as a separation scale between soft and hard
exchanged momenta. The scattering processes in the regime of hard
exchanged momentum are described by elastic binary collisions,
while the ones in the regime of soft momentum exchange are
mediated by the fields. A scattering in the soft regime
corresponds to deflection of a particle in the field of the
other(s). In order to avoid double counting of scattering
processes, ${k^*}$ should be on the order of the hardest field
mode that can be represented on the given lattice, ${k^*}\simeq
\pi/a$.

Physically, the separation scale ${k^*}$ should be sufficiently
small so that the soft field modes below ${k^*}$ are highly
occupied~\cite{Ambjorn:1990pu} and hence can be described
classically. On the other hand, ${k^*}$ should be sufficiently
large to ensure that hard modes can be represented by particles
and that collisions are described by~(\ref{cterm1}), which is
valid only for low occupation numbers, since the \text{Bose} term
$(1+f)$ is approximated by 1. Later, when dealing with
anisotropic plasmas, unstable modes arise. These should all be
located below ${k^*}$, i.e., $k_{\text {max}}<{k^*}$. In practice
$g\sim 1$, and we choose ${k^*}$ to be on the order of the
temperature for isotropic systems, and on the order of the hard
transverse momentum scale for anisotropic plasmas. At the same
time ${k^*}$ is determined by the largest available field mode,
which on a cubic lattice is given by $\sqrt{3}\pi/a$. Obviously,
any matching between soft and hard regimes can only be done
approximately, because the lattice on which the field modes are
defined is cubic, while the momentum space cutoff of the particles is
implicitly spherical.

The 'soft' scale in the system, which also sets the
scale for the unstable field growth in anisotropic systems, is described by
\begin{equation}\label{minfty}
  m^2_{\infty} = g^2N_c \int\frac{d^3p}{(2\pi)^3} \frac{f(\mathbf{p})}{|\mathbf{p}|}
             \sim g^2N_c \, \frac{n_g}{p_h}\,,
\end{equation}
analogous to the \textsc{Debye} mass (\ref{debyemass}).
To allow for trustworthy numerical simulations one should have $m_\infty L\gg1$ and
$m_\infty a\ll 1$. The first condition ensures that the relevant
soft modes actually fit on the lattice while the latter ensures that the lattice
can resolve the wavelength $1/m_\infty$ to good
precision.

\subsection{Initial energy density}
\label{inenden} We consider a heat-bath of \textsc{Boltzmann}
distributed particles with a density of $n_g=10/\text{fm}^3$ and
an average particle momentum of $3T=12$~GeV. The reason for not
using a completely thermal but under-saturated distribution is of
technical nature - we have to fulfill the condition $m_\infty a\ll
1$ and at the same time choose low enough densities and number of
lattice sites to keep the computation time within an acceptable
range. The rather extreme ``temperature'' $T$ is also chosen to
satisfy the above conditions on $N_s=32\cdots128$ lattices,
assuming $L=15$~fm.

For a given lattice (resp.\ ${k^*}$) we take the initial
energy density of the thermalized fields to be
\begin{equation}
\varepsilon_\text{fields}=\int \frac{d^3k}{(2\pi)^3} \,k \hat{f}_{\rm Bose}(k)\Theta(k^*-k)\,,
\end{equation}
where 
\begin{equation}
\hat{f}_{\rm Bose}(k)=n_g/(2T^3 \zeta(3))1/(e^{k/T}-1)
\end{equation}
is a Bose
distribution normalized to the assumed particle density $n_g$, and $\zeta$ is the \textsc{Riemann} zeta function.

The initial field amplitudes are sampled from a Gaussian distribution with a width tuned
to the above initial energy density:
\begin{equation}
    \langle A_i^a(x)A_j^b(y)\rangle=\frac{4\mu^2}{g^2}\delta_{ij}\delta^{ab}\delta(\mathbf{x}-\mathbf{y})\,.
\end{equation}
The initial spectrum is then fixed to Coulomb gauge and $A_i\sim 1/k$ (in continuum notation), and also $E_i=0$.
\textsc{Gauss}'s law then implies that the local charge density at time $t = 0$
vanishes. We ensure that any particular initial condition satisfies exact local charge neutrality.
The charge smearing algorithm for SU(2) explicitly exploits (covariant) current conservation and hence
\textsc{Gauss}'s law is satisfied exactly by construction.

\begin{figure}[hbt]
  \begin{center}
    \includegraphics[width=13cm]{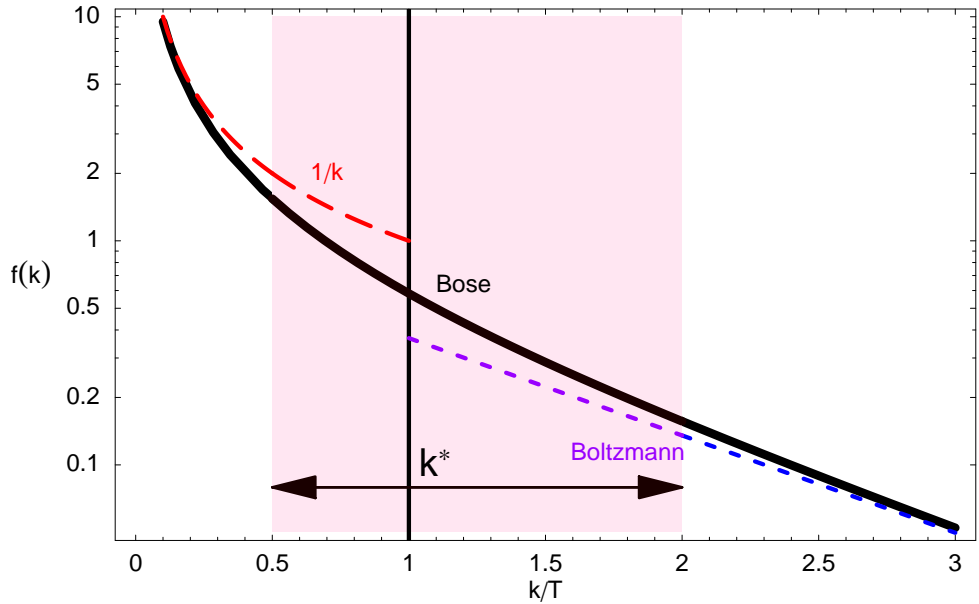}
    \caption{\textsc{Bose} distribution and its low and high-momentum limits, used for the initial fields and particles, respectively. Physically, the separation ${k^*}$ should lie around the temperature $T$. The band between $T/2$ and $2T$ roughly indicates the region whithin which we will vary $k^*$.}
    \label{fig:distributions}
  \end{center}
\end{figure}

The above procedure ensures that there is not a big jump in the
energy density when going from the field to the particle regime.
This way we are able to vary the separation scale ${k^*}$ around
the temperature $T$ by varying the lattice spacing. Fig. \ref{fig:distributions}
shows the distribution of field modes and particles and the separation momentum ${k^*}$ at $T$.

\subsection{Jet momentum diffusion}
\begin{figure}[hbt]
  \begin{center}
    \includegraphics[width=12cm]{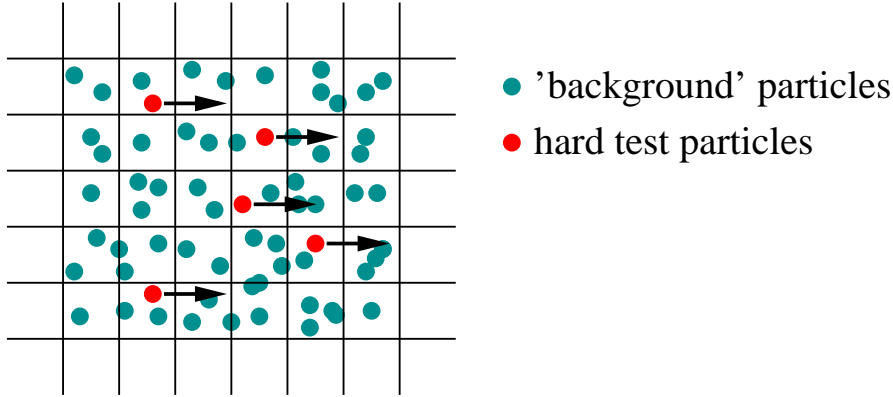}
    \caption{Setup for the measurement of momentum diffusion: High energy jets moving in a bath of particles and fields.}
    \label{fig:jets}
  \end{center}
\end{figure}

\begin{figure}[hbt]
  \begin{center}
    \includegraphics[width=13cm]{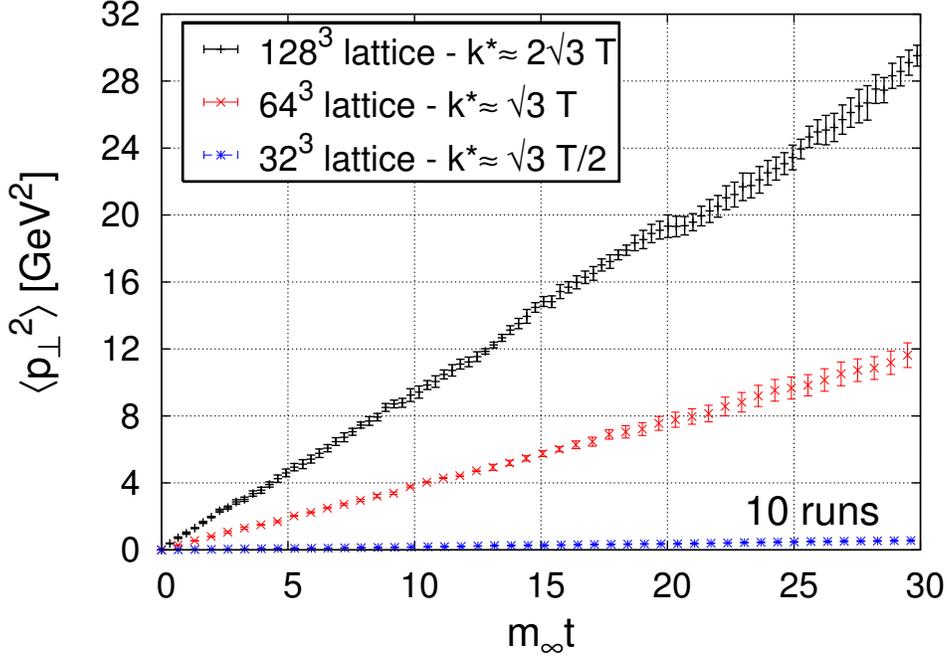}
    \caption{Momentum diffusion caused by
    particle-field interactions only. Additional high-momentum modes
    on larger lattices cause stronger momentum broadening. $T=4$ GeV,
    $g=2$, $n_g=10/\text{fm}^3$, $m_\infty=1.4/$fm.}
    \label{fig:ptnocoll}
  \end{center}
\end{figure}

Having initialized the background particles and fields, we can now add high-momentum test particles with momentum $p/3T\approx 5$ that are oriented in one direction. We initialize with few enough so that they can not influence the background significantly. Fig. \ref{fig:jets} shows the setup with background and jet particles.
We can now measure the momentum broadening $\langle
p_\perp^2\rangle(t)=\langle p_x^2+p_y^2\rangle(t)$ of the
high-energy test particles with initial momentum in the
$z$-direction. First turning off the elastic hard collisions
(${\cal C}=0$) and hence only allowing momentum transfers up to
$q={k^*}$ via the field interactions leads to a strong lattice
(i.e., ${k^*}$) dependence of the results. Fig. \ref{fig:ptnocoll}
shows the transverse momentum squared of the jets versus time,
computed on different lattices. Using half the lattice spacing
corresponds to having twice as many available modes. We use ${k^*}
= \sqrt{3}(T/2, T, 2 T)$ and a temperature of $T=4$ GeV as noted above.
One can clearly see how the availability of harder modes leads to
stronger diffusion on larger lattices.

\begin{figure}[hbt]
  \begin{center}
    \includegraphics[width=13cm]{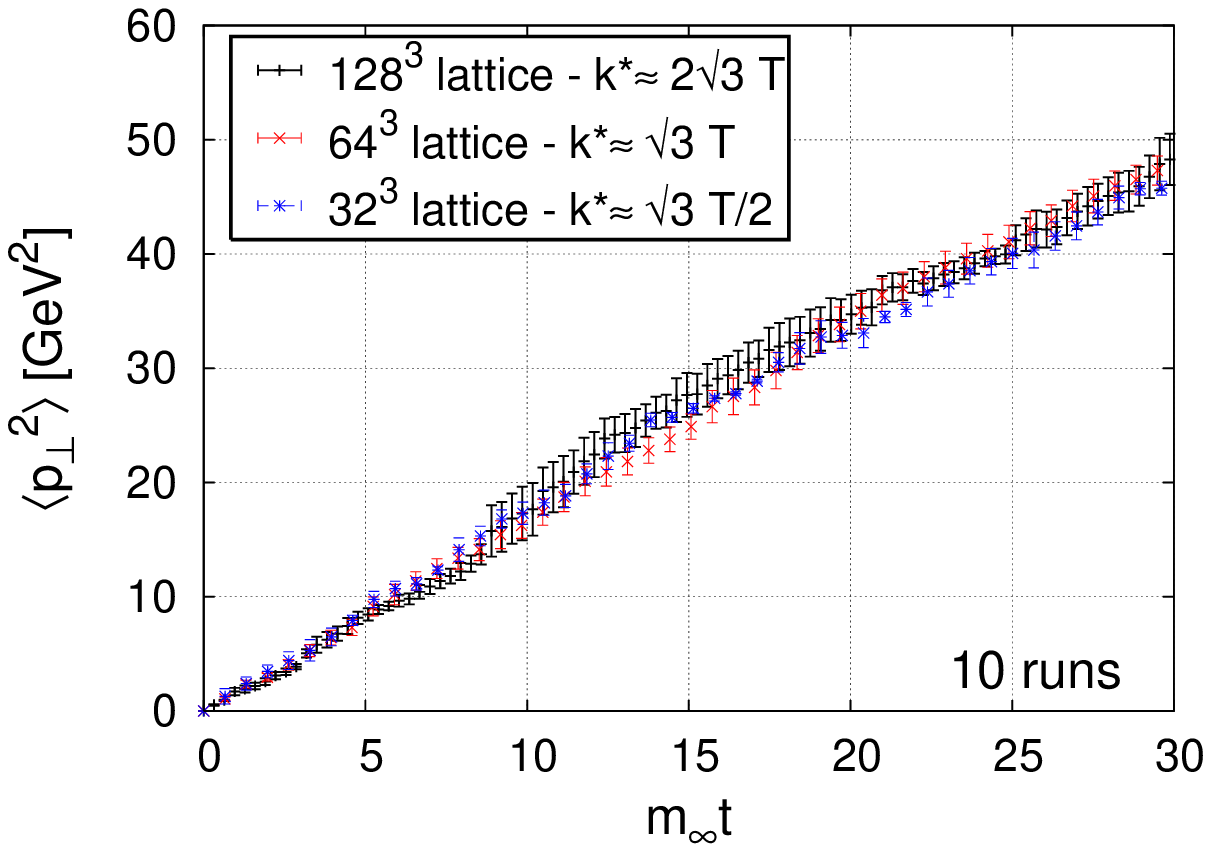}
    \caption{Momentum diffusion caused by
    hard binary collisions and particle-field interactions. $T=4$ GeV,
    $g=2$, $n_g=10/\text{fm}^3$, $m_\infty=1.4/$fm.}
    \label{fig:ptcoll}
  \end{center}
\end{figure}

Now allowing binary collisions with hard momentum exchange in
addition to the deflection by the self-consistently generated
fields and adjusting energy densities as described in Section
\ref{inenden} changes the situation drastically. We find that when
varying the separation momentum $k^*$ by factors of two in both
directions around the central value $T$ (within the band shown in
Fig. \ref{fig:distributions}) the result for $\langle
p_\perp^2\rangle(t)$ is approximately unaffected. For $k^*=T/2$
there are more momentum exchanges described by hard collisions
while for $k^*=2T$ there are more harder field modes present that
take care of those interactions. So within the rather wide range
$T/2<k^*<2T$ we can vary the contribution from field deflections
versus that from binary collisions without affecting the result
for momentum broadening of the jet particles. Fig.
\ref{fig:ptcoll} shows this result using the same different
lattice sizes as in Fig. \ref{fig:ptnocoll}.

Figs.~\ref{fig:ptnocoll} and \ref{fig:ptcoll} show that the relative
contributions to $\langle p_\perp^2\rangle$ from soft and hard
exchanges can depend significantly on $k^*$, even for $p/k^*={\cal
  O}(10)$. It is clear, therefore, that transport coefficients
obtained in the leading logarithmic (LL) approximation from the pure
\textsc{Boltzmann} approach (without soft fields) will be rather sensitive to
the infrared cutoff $k^*$. Fitting the difference of
Fig.~\ref{fig:ptcoll} and Fig.~\ref{fig:ptnocoll} (i.e., the hard
contribution) to the LL formula
\begin{equation}    \label{<kt2>LL}
  \frac{d\langle p_\perp^2\rangle}{dt}  = \frac{C_A}{C_F}
  \frac{g^4}{8\pi}  n_g \log\left(
  C^2\frac{p^2}{k^{* 2}} \right) \,,
\end{equation}
gives $C\simeq 0.43$, $0.41$, $0.31$ for $k^*/T=2\sqrt{3}$,
$\sqrt{3}$, $0.5\sqrt{3}$, respectively.
For the full calculation $C\simeq 0.61 \, k^*/(\sqrt{3}T)$.

A related and frequently used transport coefficient is $\hat
q$~\cite{Baier:1996sk}\index{q@$\hat{q}$}. It is the typical momentum transfer (squared)
per collision divided by the mean-free path $\lambda$, which is nothing but
$\langle p_\perp^2\rangle(t)/t$.
To see this, let us start from the definition of $\hat{q}$
\begin{equation}\label{qhatdef}
    \hat{q}=\frac{1}{\sigma\lambda}\int d^2p_\perp p_\perp^2 \frac{d \sigma}{dp_\perp^2}\,.
\end{equation}
The integral is just $\sigma\langle p_\perp^2\rangle_{\text{1
collision}}$, since $$\frac{1}{\sigma}\frac{d
\sigma}{dp_\perp^2}$$ can be seen as a probability distribution
for a collision with momentum transfer $p_\perp^2$ to occur. To
extract $\hat{q}$ from the accumulated $\langle p_\perp^2\rangle$,
shown in Figs. \ref{fig:ptnocoll} and \ref{fig:ptcoll}, we only
have to divide by the number of collisions that have occurred up
to time $t$, $t/\lambda$. So finally we have
\begin{equation}
    \hat{q}=\frac{\langle p_\perp^2\rangle}{t}\,,
\end{equation}
which is just the slope of the approximately straight lines in Figs. \ref{fig:ptnocoll} and \ref{fig:ptcoll}.

Note that the definition of $\hat{q}$ holds exactly only in the
eikonal limit, where the jet does not change its direction. Away
from this limit, $\hat{q}=\langle p_\perp^2\rangle/t$ is smaller
than the mean momentum transfer squared per collision divided by
the mean-free path, because it considers the actual random walk of
the particle.

\begin{figure}[hbt]
  \begin{center}
    \includegraphics[width=10cm]{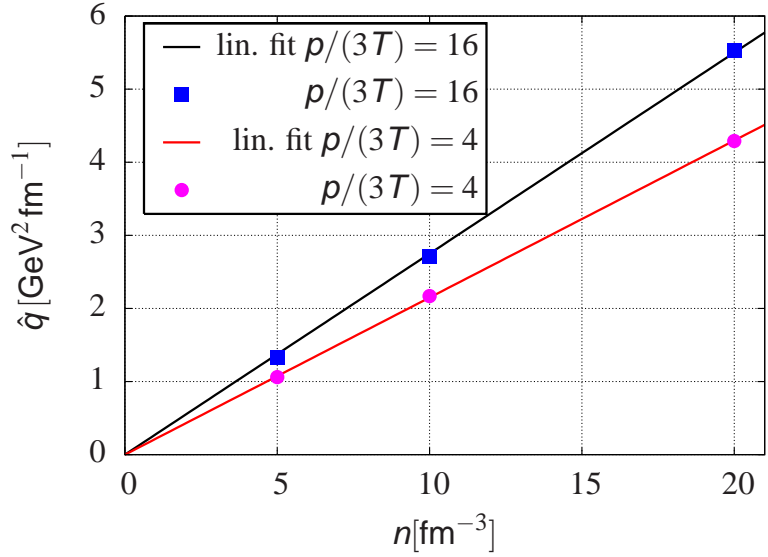}
    \caption{$\hat q$ as a function of the density $n$ at fixed
    $T=4\,$GeV and $p/(3T)\in\{4,16\}$.}
    \label{fig:qhatn}
  \end{center}
\end{figure}

\begin{figure}[hbt]
  \begin{center}
    \includegraphics[width=10cm]{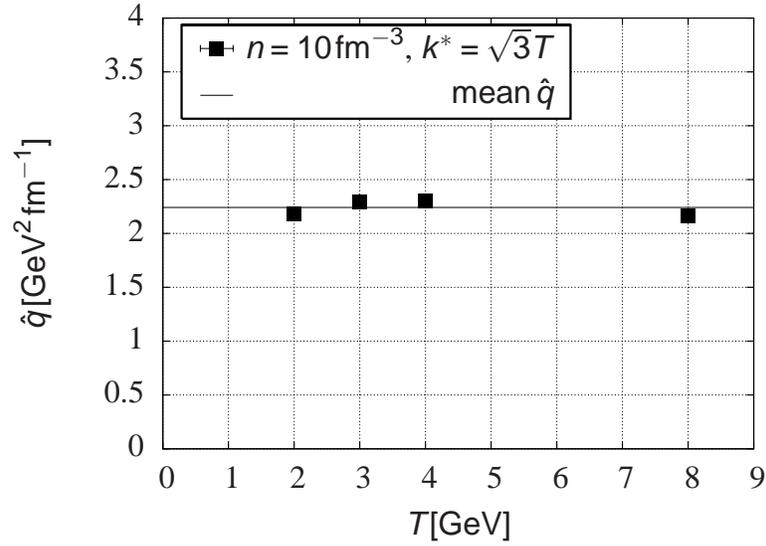}
    \caption{$\hat q$ as a function of $T$ at fixed
    $n_g=10$/fm$^3$ and $p/(3T)=5$.}
    \label{fig:qhatT}
  \end{center}
\end{figure}

From Fig.~\ref{fig:ptcoll}, $\hat q\simeq 2.2$~GeV$^2$/fm for
$n_g=10/$fm$^3$ and $p/(3T)\approx5$. This is the first cut-off
independent result for $\hat{q}$. Its value for $\hat q$ is in the
range extracted from phenomenological analyses of jet-quenching
data from RHIC~\cite{Majumder:2007iu}. As expected from Eq. (\ref{qhatdef}) $\hat{q}$ scales with the
density of the medium $n=1/(\sigma\lambda)$, which is demonstrated in Fig. \ref{fig:qhatn}. 
Further, it depends on the ratio $p/T$, but not on $T$ itself as
demonstrated in Fig. \ref{fig:qhatT}.

We can hence scale to physical densities for a QGP created at RHIC, adjust for
the different color factors in SU(3), and find $\hat{q} \approx 8$
GeV$^2/$fm, at $T=400$ MeV, $E_{\text{jet}}\approx 20$ GeV ($p/3T
= 16$) in a system of quarks and gluons. Here, we have assumed
that the interaction between quarks and gluons is as strong as
that among the gluons. Using the small angle approximation for the
$qg\rightarrow qg$ cross section one finds that it is a factor of
4/9 smaller than that for the $gg\rightarrow gg$ process. This
leads to a modified extrapolated $\hat{q}\approx 5.6$ GeV$^2/$ fm.

      \chapter{Real-time simulation of plasma instabilities}
\epigraphwidth 250pt \epigraph{Es schadet nichts, wenn Starke sich
verst\"arken.} {\emph{Faust II. Vierter Akt. Auf dem Vorgebirg. Faust.}\\
Johann Wolfgang von Goethe (1749-1832)}\label{wyminstabilities}
So far we have considered particle distributions that are
isotropic in momentum space in the simulation. To simulate the
early stage of a quark-gluon plasma, created in a heavy-ion
collision, we need to initialize with an anisotropic momentum
distribution as discussed in Chapter \ref{instabilities}. In this
case, chromo-\textsc{Weibel}-instabilities can occur and in the
following chapter we will show that in fact they do. We will then
concentrate on the effect of collisions on the unstable growth and
discuss isotropization. We close the chapter with the discussion
of the effect of instabilities on jet propagation and potential
observables. \footnote{The main part of this chapter is based on
work published in \cite{Dumitru:2007rp}.}

\index{Instabilities!Weibel instability}
The initial momentum distribution for the hard plasma gluons is now taken to be
\begin{align}
\label{anisof}
f(\mathbf{p})=n_g \left(\frac{2\pi}{p_{\text{h}}}\right)^2 \delta(p_z)
  \exp(-p_\perp/p_{\text{h}})\,,
\end{align}
with $p_\perp=\sqrt{p_x^2+p_y^2}$. This represents a quasi-thermal
distribution in two dimensions with average momentum $=2\,p_{\text{h}}$.
We initialize small-amplitude fields sampled from a
Gaussian distribution and set $k^*\approx p_{\text{h}}$, for the
reasons discussed in Section \ref{separation}.
The band of unstable modes is located below $k^*$.

At the initial time $t=0$ we randomly sample $N_p = N_{\rm test}\, n_g\, a^3$ particles from the
distribution~(\ref{anisof}) at each cell of the
lattice. When $N_p$ is not very large it is useful to ensure
explicitly that the sum of particle momenta in each cell vanishes, for
example by adjusting the momentum of the last particle accordingly.

For the distribution (\ref{anisof}) the mass scale (\ref{minfty}) becomes exactly
\begin{equation}
  m^2_{\infty} = g^2N_c \, \frac{n_g}{p_h}\,.
\end{equation}
As mentioned before in the previous chapter and discussed in detail
using the \textsc{Debye} mass in Chapters \ref{anisotropy} and \ref{collisionsmodel},
this quantity sets the scale for the growth rate of unstable field modes in the linear approximation.

\section{The collisionless limit}
\label{colllesslimit} In the collisionless limit the
chromo-\textsc{Weibel}-instability has been studied in detail in
\cite{Dumitru:2005hj} and \cite{Dumitru:2006pz} within the
\textsc{Wong-Yang-Mills}-simulation. We will now summarize the
most important results and extend the study, showing explicitly
the generation of filaments discussed in Section
\ref{secfilamentation}.

In the Abelian case, it was found that for fields with weak
initial energy density on the order of $0.1 m_\infty^4/g^2$ the
transverse chromo-magnetic and -electric fields grow, the latter
at a slower rate. Modes above $k_z\simeq 12\pi/L$ are stable as
opposed to the result in the linear approximation, where for
extreme anisotropy the spectrum of unstable modes extends to
$k=\infty$ (see Fig. \ref{fig:unstablealphaxiinf}).

\begin{figure}[htb]
\includegraphics[width=3in]{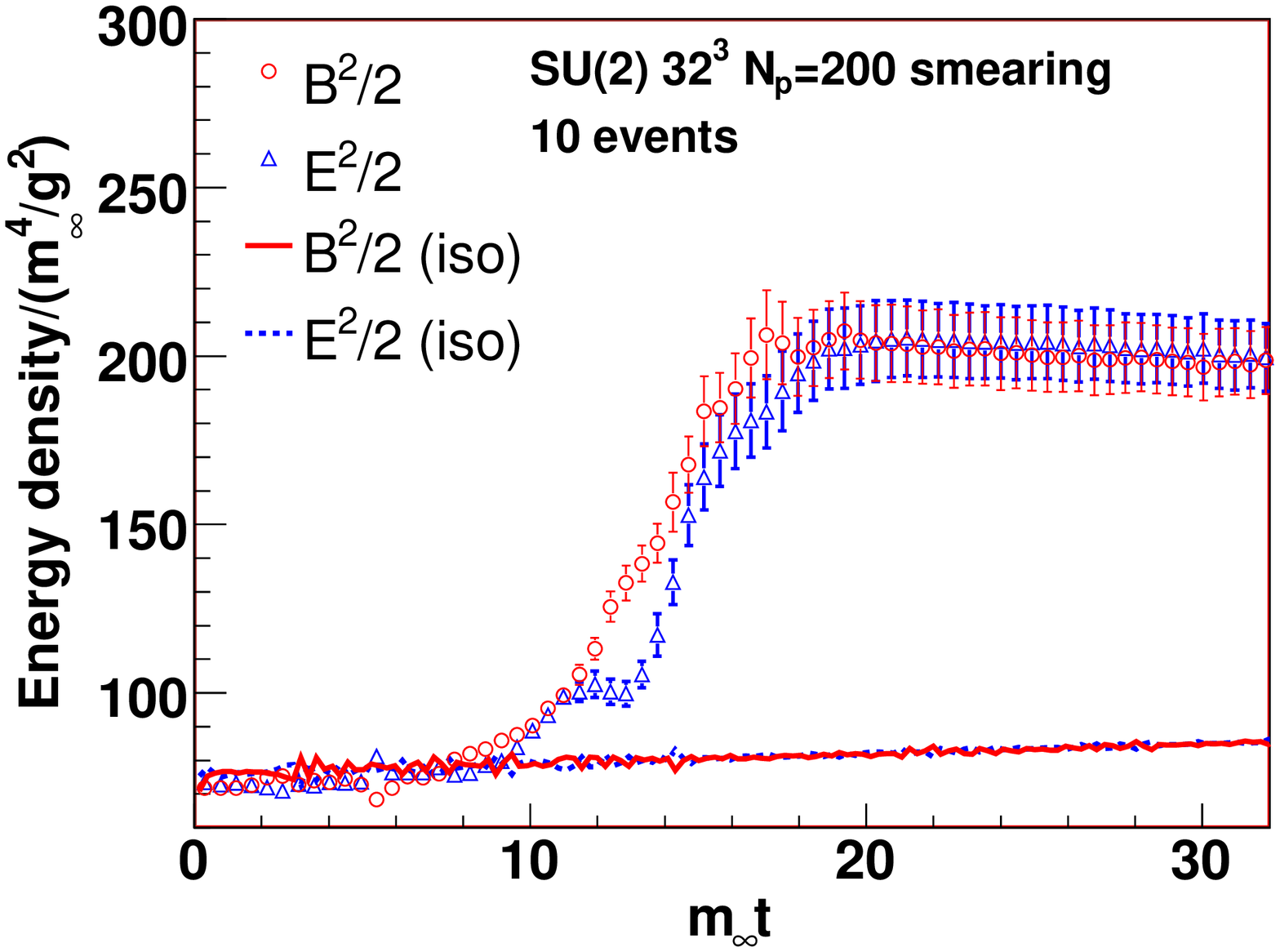}
\includegraphics[width=3in]{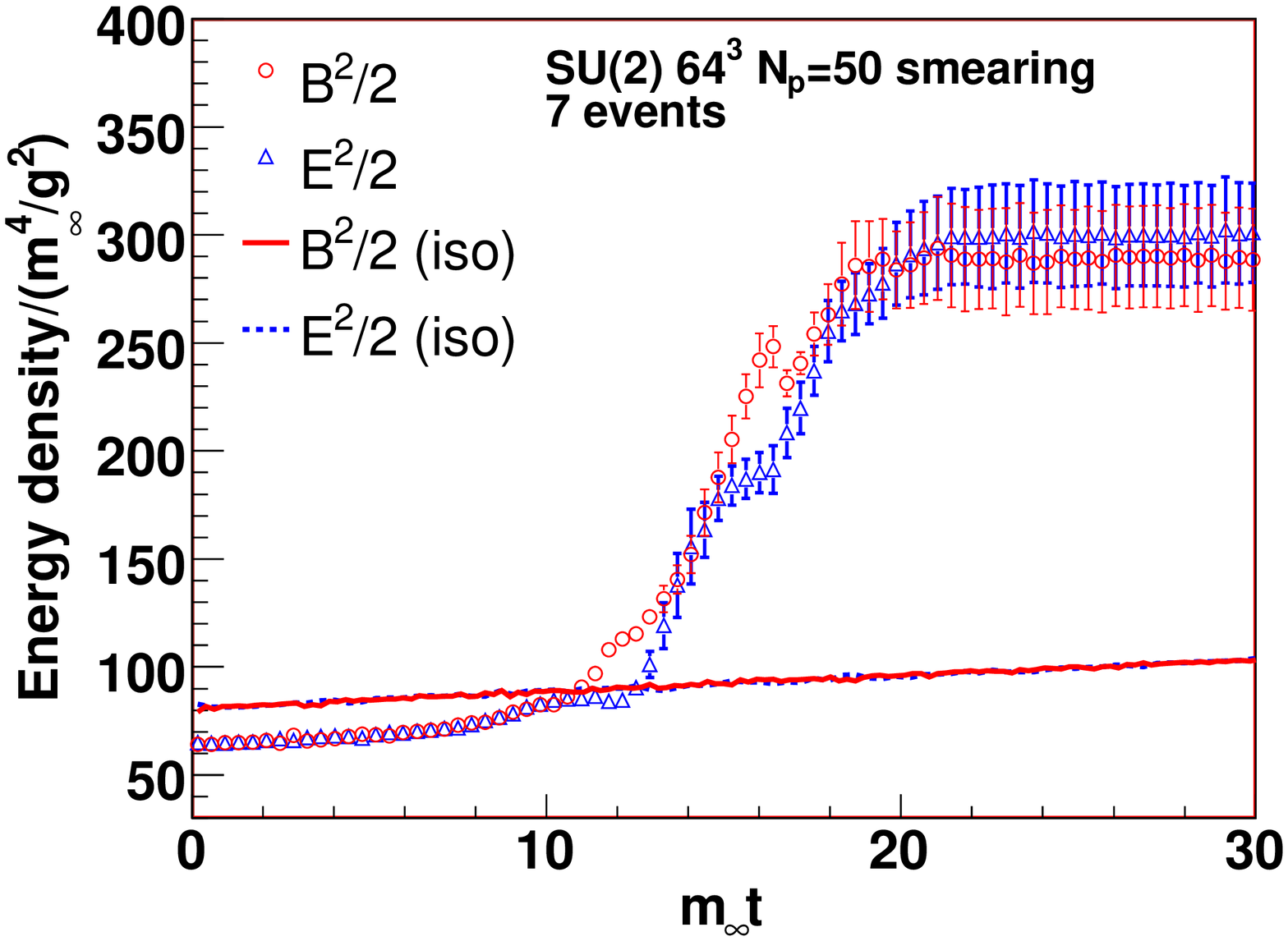}
\caption{Time evolution of the field energy densities for $SU(2)$
gauge group and anisotropic initial particle
momentum distributions. Simulation parameters are $L=5$~fm, $p_{\rm
h}=16$~GeV, $g^2\,n_g=10/$fm$^3$, $m_\infty=0.1$~GeV. From \cite{Dumitru:2006pz}.}
\label{fig:fieldWsu2_weak}
\end{figure}

In the non-Abelian SU(2) case initial field energy densities of
$\sim 10 m_\infty^4/g^2$ and above were studied. In
\cite{Dumitru:2006pz} the initial condition was taken to be
\textsc{Gaussian} random chromo-electric fields, which were
'low-pass' filtered such that only the lower half of available
lattice modes was populated. Fig.~\ref{fig:fieldWsu2_weak} shows
results obtained on various lattices. Runs with an isotropic
particle distribution are shown in each panel as an indication of
our numerical accuracy (error bars are not indicated for the
isotropic runs).  The isotropic runs show nearly constant fields
over the time interval $t\le 30 \,m^{-1}_\infty$. From these
figures it is observed that after a period of rapid growth both
chromo-electric and -magnetic fields settle to an essentially
constant energy density.

Even beyond the hard-loop approximation, the time evolution of non-Abelian fields
stronger than $\sim m_\infty^4/g^2$ differs from that in the
(effectively Abelian) extreme weak-field limit. In particular, a
sustained exponential growth is absent even during the stage where
the backreaction on the particles is weak.
For very strong anisotropies a linear analysis predicts that the exponential field growth (in the weak-field
situation) can perhaps continue until $B^2\sim m_\infty^4/ g^2/
\Delta\theta^2$, with $\Delta \theta^2 = p_z^2 / p_\perp^2 \approx \xi^{-1}$~\cite{Arnold:2005ef}.
For the initial condition~(\ref{anisof}), $\Delta\theta^2=0$ at $t=0$ but grows
to ${\cal O}(10^{-3})$ during the initial transient time with constant
fields ($t\,m_\infty\lesssim10$ in Fig.~\ref{fig:fieldWsu2_weak}) due to
deflection of the particles.  However, such strong fields are not seen
in our PIC simulations. This may be related to the above-mentioned
back-reaction effects which prevent instability of modes near $k_{\rm
 max} \sim m_\infty/\Delta\theta$.

The results shown in Fig.~\ref{fig:fieldWsu2_weak} indicate a
sensitivity to hard field modes at the ultra-violet end of the
Brioullin zone, $k={\cal O}(a^{-1})$, in contrast to the $U(1)$
simulations and to earlier 1d-3v $SU(2)$ simulations~\cite{Dumitru:2005gp,Dumitru:2005hj,Nara:2005fr}.
The energy density contained in the fields at late times increases by a
factor of 1.5 when going from a $16^3$ to $32^3$ to $64^3$ lattice
with the same physical size $L$. Hence, the dynamics of $SU(2)$
instabilities seen here is not dominated entirely by a band of
unstable modes in the infrared but clearly involves a cascade of
energy from those modes to a harder scale
$\Lambda$~\cite{Arnold:2005qs}. However, the simulations shown in
Fig.~\ref{fig:fieldWsu2_weak} indicate that $\Lambda(t)$ grows to
${\cal O}(a^{-1})$ during the period of rapid growth of the field
energy density; otherwise, the final field energy density would not
depend on the lattice spacing.
Spectra of the field modes also show the avalanching to the UV. After the unstable soft modes
have grown, the energy is rapidly transferred to the higher modes \cite{Dumitru:2006pz}.

Although energy conservation will eventually stop the growth of the
fields as the lattice spacing decreases towards the continuum, it does
not solve the following problem. When $\Lambda(t)\sim1/a$, the hard field modes have reached the momentum
scale of the particles, $p_h={\cal O}(a^{-1})$, and so the clean
separation of scales is lost, on which our approach is based. In fact, since the
occupation number (or phase space density) at that scale is of order 1
or less by construction, it is inappropriate to describe modes at that
scale as a classical field. Those perturbative modes should be
converted dynamically into particles at a lower scale $\Lambda_{f\to
p} \ll 1/a$ so that the field energy density and the entire coupled
field-particle evolution is independent of the artificial lattice
spacing. This question will be studied in the future.

\begin{figure}[H]
\begin{center}
\includegraphics[width=7cm]{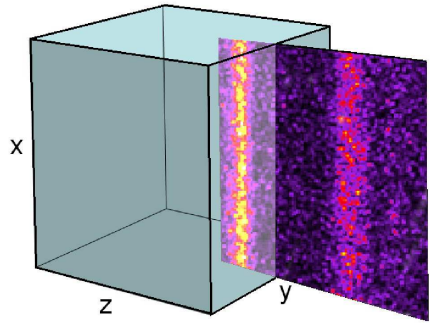}
\caption{The cubic lattice with the pulled out slice we look at. \label{fig:ls}}
\end{center}
\end{figure}

\begin{figure}[H]
\includegraphics[width=14cm]{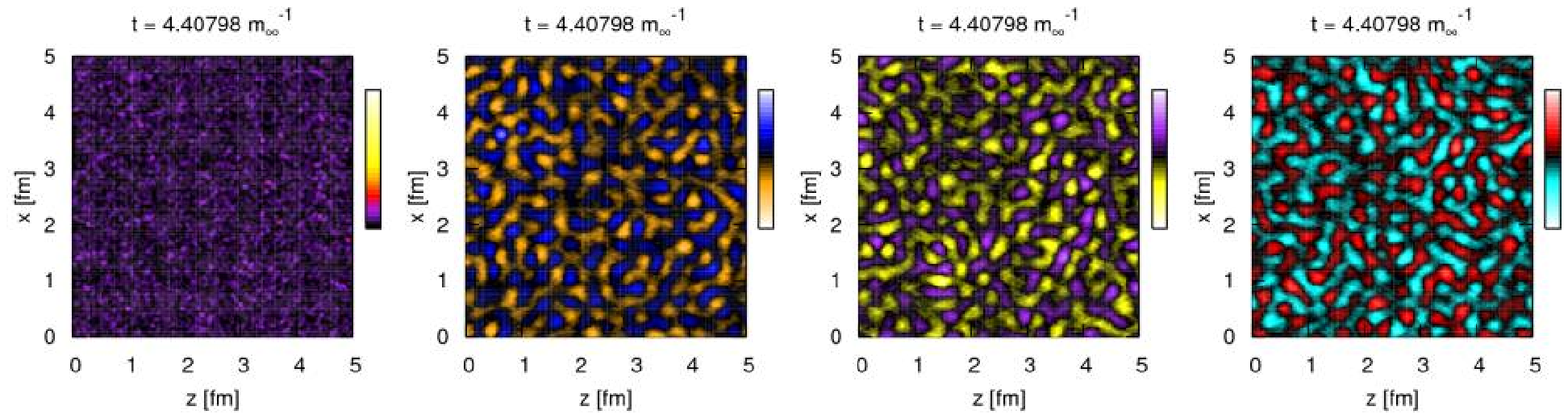}
\includegraphics[width=14cm]{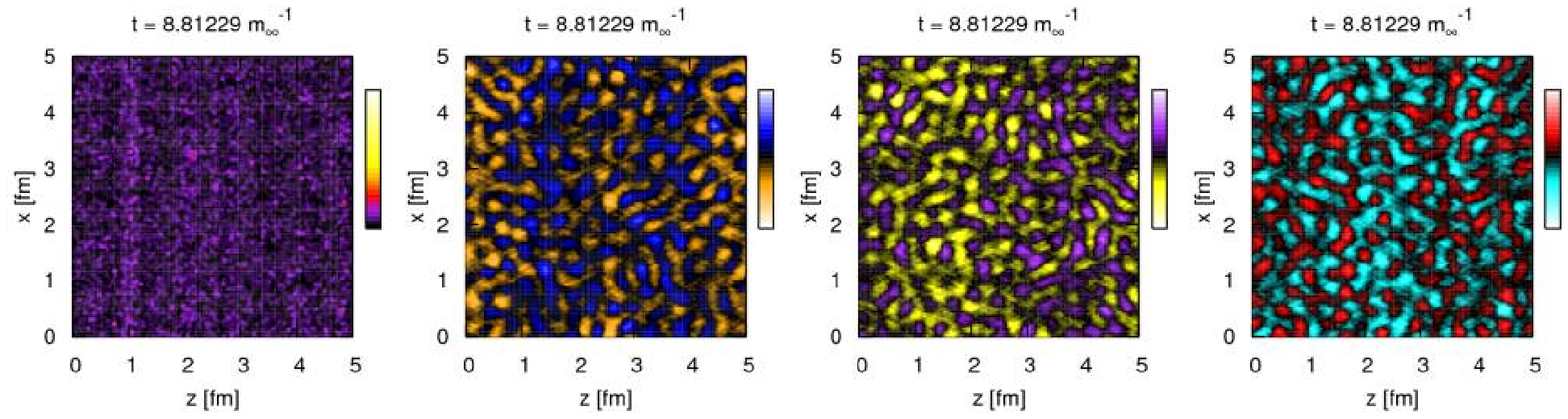}
\includegraphics[width=14cm]{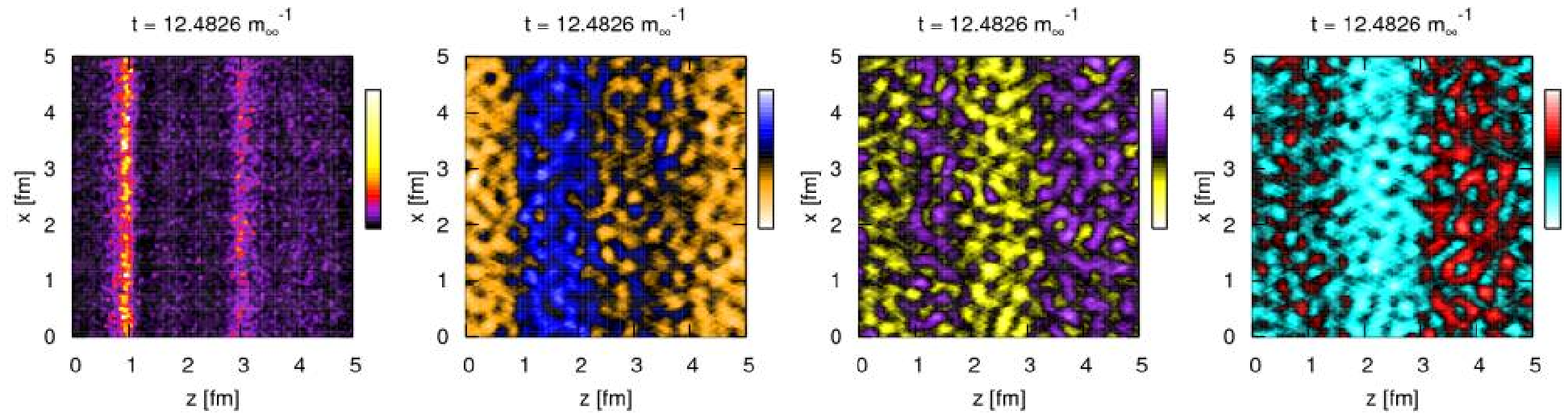}
\includegraphics[width=14cm]{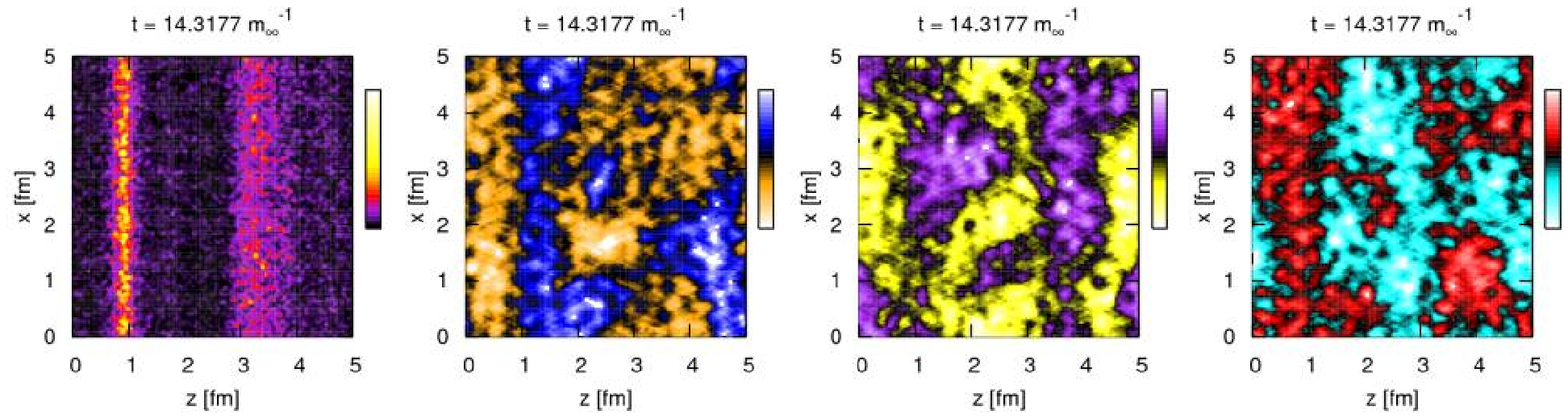}
\includegraphics[width=14cm]{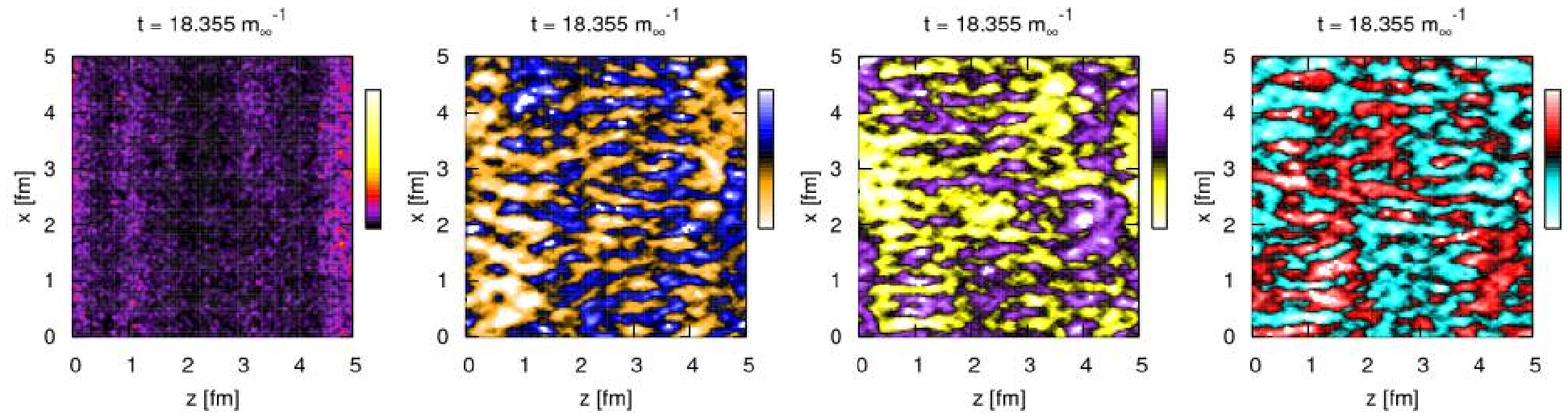}
\caption{Slices in the $x$-$z$-plane at fixed $y=L/2$ of the (from
left to right) current in the $x$-direction, $J_x$, and the three
color components of the chromo-magnetic field in the
$y$-direction, $B_y^1$, $B_y^2$, and $B_y^3$. Time is increasing
from top to bottom. The filaments are nicely visible in the early
stage of the unstable growth in the third row. Please also compare
the indicated times to those in Fig. \ref{fig:fieldWsu2_weak},
which shows the corresponding field energy density vs. time.
Simulation parameters are $L=5$~fm, $p_{\rm h}=16$~GeV,
$g^2\,n_g=10/$fm$^3$, $m_\infty=0.1$~GeV. The scales are in
lattice units and reach from 0 to $5 \cdot 10^{-8}$ for the
current and from $-4 \cdot 10^{-3}$ to $4 \cdot 10^{-3}$ for the
chromo-magnetic fields.} \label{fig:slices}
\end{figure}

We also wish to study the spatial distribution of currents and
fields during the unstable growth. We expect to see the filaments
discussed in Section \ref{secfilamentation} and variation of the
fields (i.e., the wavelength of the unstable modes) on a
lengthscale given by $m_\infty^{-1}$, which with the used
parameters is approximately 0.5 fm. To visualize this directly, we
show a cut through the lattice at fixed $y=L/2$ (see Fig.
\ref{fig:ls}), i.e., the $x$-$z$-plane, at different times in Fig.
\ref{fig:slices}. Shown are the current in the $x$-direction as
well as the three color components of the chromo-magnetic field in
the $y$-direction. One can nicely see that filamentation sets in
after about $t=10 m_\infty^{-1}$, the time at which the
instability starts growing (see Fig. \ref{fig:fieldWsu2_weak}).
Already during the period of growth, which lasts until about
$t\approx 18 m_\infty^{-1}$, the filaments break and the domains
of strong aligned fields become smaller. Finally, at the time when
the saturation sets in, we see directly the populated high
momentum field modes. The observed structure may however be due to
lattice effects, since the saturation depends on the lattice size
as discussed above.

Comparing to Fig. \ref{fig:mechanism} on page
\pageref{fig:mechanism}, which shows the acting forces and
resulting filaments, we find a direct correspondence to the
structure in the simulation (note that since now the direction of
the anisotropy is in the $z$-direction, $x$ and $z$ in Fig.
\ref{fig:mechanism} have to switch places - however, of course the
argument does not change). This nicely shows that the instability
observed in the simulation is indeed due to current filamentation
and hence of the \textsc{Weibel} kind. Note that in order to
compare field amplitudes at different places one would actually
have to parallel transport their values. In other words, since we
are looking at a quantity that is not gauge invariant, the plots
of the $B_y$-components have to be taken with care. In any case
they serve to visualize the qualitative structure of the fields in
a very illuminating fashion.

\section{Instabilities under the influence of collisions}
\label{sec:instcoll} \index{Collisions!Effect on instabilities} We
now study how the inclusion of collisions as described in Section
\ref{wymcollisions} affects instability growth in the
\textsc{Wong-Yang-Mills}-simulation. We expect the growth rate to
be reduced and the number of unstable modes to decrease from the
model calculation in Chapter \ref{collisionsmodel}. The decrease
of the growth rate is now partly because of the randomizing effect
of the collisions, which disturbs the collective behavior, and
because of the faster isotropization, which reduces $\xi\simeq
\langle p_\perp^2\rangle / \langle p_z^2 \rangle$ more rapidly.
With the collision term an explicit dependence on $g$ enters
again. Regardless of the large hard momentum scale used, we set
$g=2$ to simulate the 'worst case' for the instability growth with
the coupling expected at RHIC energies.
\begin{figure}[H]
\begin{center}
\includegraphics[width=10cm]{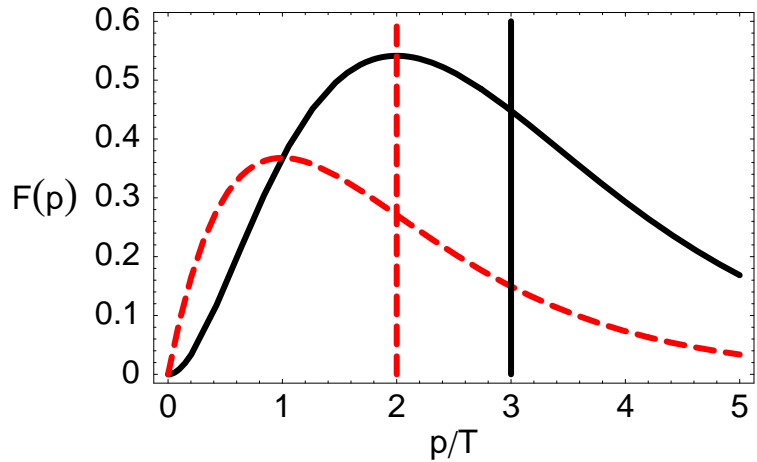}
\caption{Qualitative shape of the quasi thermal particle
distribution functions in three and two dimensions. $F(p)$
includes the phase-space factor: $F(p)=p^2 e^{-p/T}$ (3D) (solid black), $F(p)=p
e^{-p/T}$ (2D) (dashed red), where in the latter case $T=p_{\text{h}}$. The
vertical lines indicate the mean absolute value of the momentum
$\langle p \rangle$ of the distribution.\label{fig:2d3ddist}}
\end{center}
\end{figure}

The problem in these simulations is that we cannot match the
energy densities of the fields and particles as we had done in the
isotropic case. If we did that, the initial field amplitude would
be too large to see any unstable growth at all. However, it may
well be the case in the experimental situation, which is strongly
out of equilibrium, that the distribution functions are far from
their equilibrium shape, such that a matching of the energy
densities based on equilibrium distribution functions would not
make much sense. We initialize weaker fields than the matching
would demand, but choose the separation scale $k^*$ to lie around
the hard momentum scale $p_\text{h}$ of the distribution
(\ref{anisof}). Since we have a two-dimensional distribution the
average momentum is $2 p_{\text{h}}$ as opposed to $3T$ for the
three dimensional \textsc{Boltzmann} distribution. The most
probable momentum is at $p_{\text{h}}$ as opposed to $2T$ (see
Fig. \ref{fig:2d3ddist}). Using the previous arguments for
choosing the separation scale to lie around the temperature, we
should in fact shift this value closer to $p_{\text{h}}/2$,
because the distribution changed its shape. We will do the
calculation for $k^*\approx 0.9\,p_{\text{h}}$ and $k^*\approx
1.7\, p_{\text{h}}$, and note that the more physical choice is $0.9\,p_{\text{h}}$. We
realize different values for $k^*$ by employing different lattice
sizes for the same set of physical parameters.
\begin{figure}[htb]
\begin{center}
\includegraphics[width=12cm]{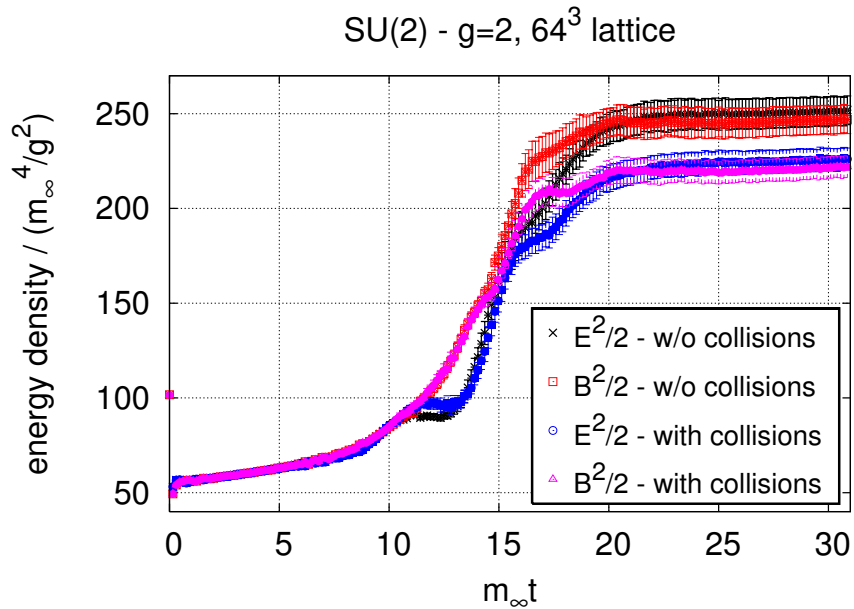}
\caption{The effect of collisions with momentum exchange above
$k^*\approx 1.7\,p_{\text{h}}$ (below that scatterings are mediated by
the fields). Shown are calculations with (lower curves) and
without collisions. \label{fig:collnocoll64}}
\end{center}
\end{figure}
\begin{figure}[htb]
\begin{center}
\includegraphics[width=12cm]{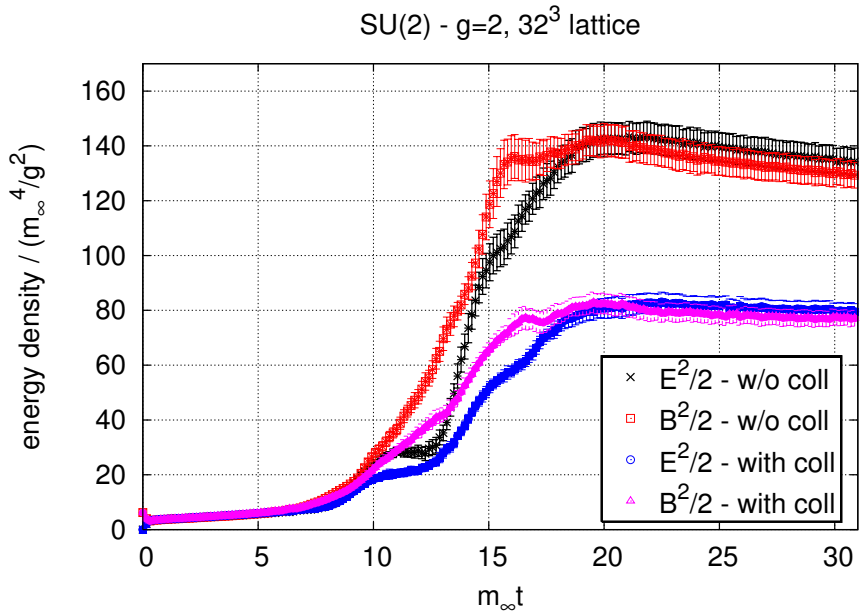}
\caption{The effect of collisions with momentum exchange above
$k^*\approx 0.9\,p_{\text{h}}$ (below that scatterings are mediated
by the fields). Shown are calculations with (lower curves) and
without collisions.\label{fig:collnocoll32}}
\end{center}
\end{figure}

Figs. \ref{fig:collnocoll64} and \ref{fig:collnocoll32} show the
results for a $64^3$ ($k^*\approx 1.7\,p_{\text{h}}$) and a $32^3$
($k^*\approx 0.9\,p_{\text{h}}$), respectively. The reduction of the
saturated value lies between about a factor of 0.6 ($k^*\approx
0.9\,p_{\text{h}}$) and 0.89 ($k^*\approx 1.7\,p_{\text{h}}$). The growth
rate is reduced approximately by a factor 0.85 or 0.95
respectively. This result is obtained from assuming an exponential
behavior between 10 and 20 $m_\infty t$. The comparison to the
results in Chapter \ref{collisionsmodel} is complicated by the
fact that the anisotropy of the system is dynamically changing and
the anisotropy parameter $\xi$ evolves from initially $\infty$
down to about 60 at the time when the instability growth
saturates. From Fig. \ref{fig:gammamax} we can read off that for
the estimated realistic growth rate of $\nu\approx 0.15\, m_D$, a
reduction by $20\%$ is to be expected when the anisotropy stays at
the extreme limit. Hence we find that the two approaches lead to
compatible results: the reduction of instability growth by the
collisions is present, but small enough for the instabilities to
still play an important role in isotropization and equilibration
of the QGP, even if the coupling is not infinitely small and
collisions can not be neglected.

%From the considerations in Chapter \ref{collisionsmodel} we get
%for a dynamically changing $\xi$ an average reduction of about a
%factor of 0.8 - close to the numerical result for the choice
%$k^*\approx p_{\text{h}}/2$. {\bf CHECK AGAIN}

\section{Isotropization}
\label{sec:isotropization}
\begin{figure}[htb]
\begin{center}
\includegraphics[width=12cm]{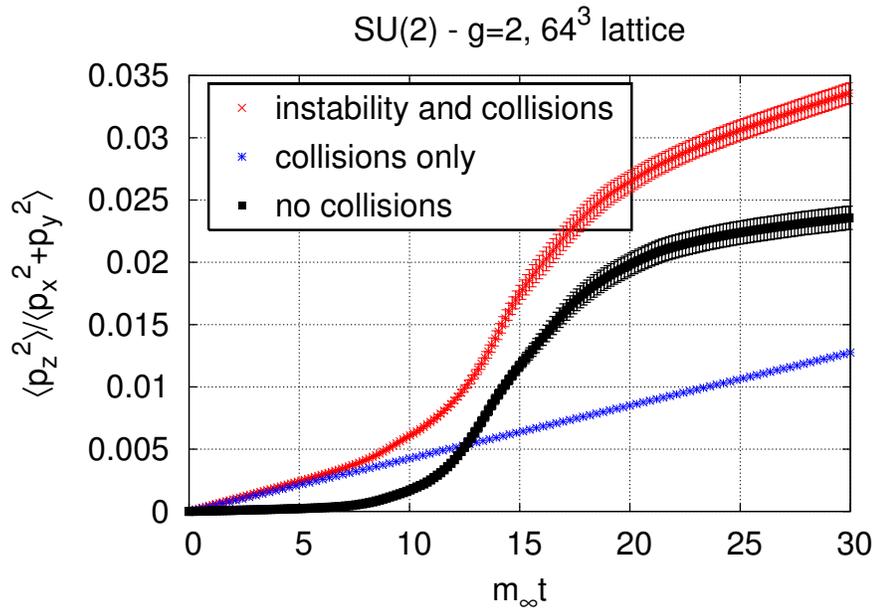}
\caption{Isotropization with time in the case of instability
growth and collisions (red cross), particle-field interactions
only (black square) and collisions only, extrapolated from the
early stage where the particle-field contribution is negligible
(blue star).  \label{fig:pbothg2new}}
\end{center}
\end{figure}

\begin{figure}[htb]
\begin{center}
\includegraphics[width=12cm]{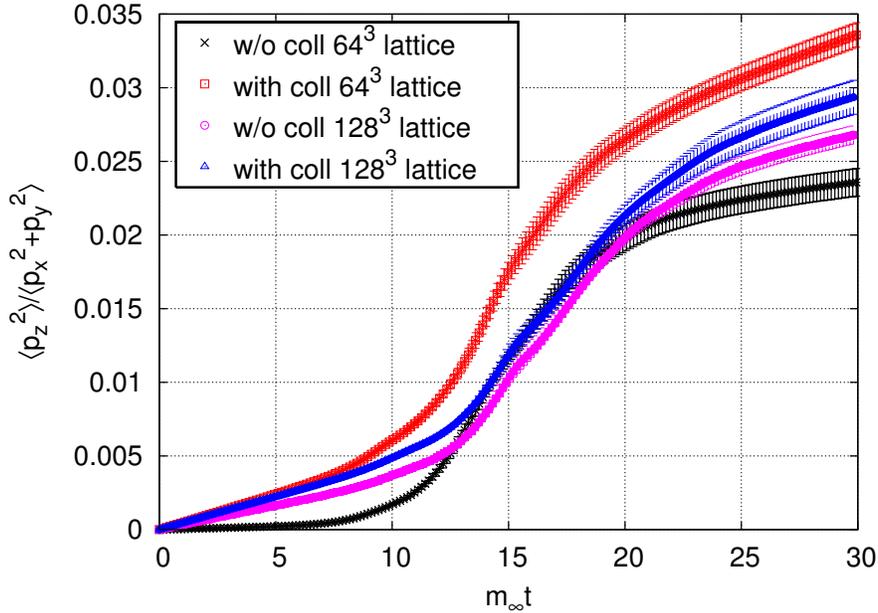}
\caption{Isotropization with time in the case of instability
growth with and without collisions on two different lattices. The
$64^3$-lattice corresponds to $k^*=1.7\,p_{\text{h}}$, the
$128^3$-lattice to $k^*=3.4\, p_{\text{h}}$. In the latter case one
can nicely see that the additional high momentum field modes act
just like the hard binary collisions in the case of the $64^3$
lattice. The few additional large momentum transfer collisions
above $k^*=3.4\, p_{\text{h}}$ do not contribute much to the
isotropization, as seen from the small difference between the pink
(circle) and blue (triangle) curve. \label{fig:p64128g2}}
\end{center}
\end{figure}

\index{Isotropization} It is interesting
to see, at not infinitesimally small coupling, whether the
instability still dominates the isotropization when collisions are
included. For the choice $g=2$, which we chose in spite of the
high temperature to get a contribution of the collisions
comparable to the physical situation in heavy ion collisions, we
find that the unstable growth still dominates the isotropization
when we start with the extremely anisotropic initial distribution
(\ref{anisof}). This is shown in Figs. \ref{fig:pbothg2new} and
\ref{fig:p64128g2}, where we plot the quantity $\langle
p_z^2\rangle/\langle p_x^2 +p_y^2\rangle$ as a measure of the
isotropy of the hard particle distribution. We can relate it to
the previously used $\xi$, defined in Chapter \ref{anisotropy},
via $\xi = \langle p_\perp^2\rangle/\langle p_z^2\rangle-1$. It
starts out at 0 ($\xi=\infty$), because initially the particles do
not have any longitudinal ($z$-direction) momentum at all. First,
we just find a linear rise of this quantity, caused mainly by the
hard collisions with momentum exchange above the separation scale
$k^*$. The weak initial fields do not contribute much, at least
for $k^*\approx p_{\text{h}}$ on the $64^3$-lattice. During the
time of unstable field growth, however, the isotropization happens
much faster than with collisions alone. The isotropizing effect of
the domains of strong chromo-electro-magnetic fields dominates.
These domains are visible in Fig. \ref{fig:slices}, and we will
discuss them in more detail in Sec. \ref{jetunstable}, where we
investigate the effect of the created strong turbulent fields on
jet momentum broadening. The phenomenon, leading to the faster
isotropization of the complete hard particle distribution, is of
course the same.

In Fig. \ref{fig:pbothg2new} the purely collisional contribution
has been extrapolated from the early times when no unstable growth
was present. Fig. \ref{fig:p64128g2} also shows the result
obtained on a $128^3$ lattice, which corresponds to using
$k^*\approx 3.4\,p_{\text{h}}$. The rather few collisions with high
momentum exchange above $3.4\,p_{\text{h}}$ do not change much for the
isotropization when we compare to the collisionless case with only
particle-field interactions. For the more sensible choice
$k^*\approx 1.7\,p_{\text{h}}$ the difference is more significant, as
expected.

In summary, the collective field-particle interaction, particularly the instability, dominates the isotropization process compared to the contribution of binary collisions at coupling $g=2$. When including hard splittings and $gg\leftrightarrow ggg$ processes as in \cite{Xu:2004mz}, this result is likely to be modified and the instability could be less dominating, or not be the most important process at all. This scenario will have to be studied in the future, leading to the most complete simulation of isotropization and thermalization so far.

\newpage

\section{Collective effects on jet propagation}
\label{jetunstable}\index{Jets}
Let us now investigate the effect of the instability on momentum broadening of jets,
including the effect of collisions.

\begin{figure}
    \begin{center}
        \includegraphics[width=12cm]{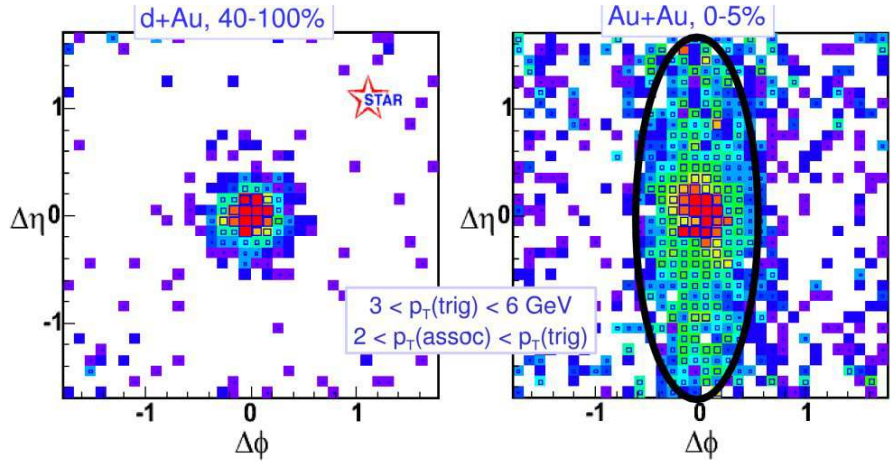}
        \caption{Dihadron correlation functions in azimuth $\Delta \phi$ and
                space-time rapidity $\Delta \eta$, for d+Au and central Au+Au at
                $\sqrt{s}=200$ GeV (figure courtesy J. Putschke (STAR), Proceedings
                of Hard Probes 2006).
                For comparison, on the right plot an ellipse with eccentricity
                $e\simeq \frac{\sqrt{8}}{3}$ which corresponds to a
                ratio $\sqrt{\kappa_z/\kappa_\perp}\sim 3$ is shown \cite{Romatschke:2006bb}.}
    \label{fig:ptcorr}
    \end{center}
\end{figure}

The initial momentum distribution for the hard plasma gluons is again taken to be of the form (\ref{anisof}).
We add additional high momentum particles with $p_x=12\,
p_{\text{h}}$ and $p_x=6\, p_{\text{h}}$, respectively, to investigate
the broadening in the $y$ and $z$ directions via the variances \cite{Moore:2004tg}
\begin{align}
        \kappa_\perp(p_x)&:=\frac{d}{dt}\langle(\Delta p_\perp)^2\rangle\,,\\
        \kappa_z(p_x)&:=\frac{d}{dt}\langle(\Delta p_z)^2\rangle\,.
\end{align}
In our case $\langle \Delta p_\perp^2\rangle=\langle p_y^2 \rangle$ and
$\langle \Delta p_z^2\rangle=\langle p_z^2 \rangle$.
Since for our initial jet profile we have
$\langle(\Delta p_\perp)_0^2\rangle=\langle(\Delta p_z)_0^2\rangle$,
the ratio $\kappa_z/\kappa_\perp$ can be roughly associated with the
ratio of jet correlation widths in azimuth and rapidity \cite{Romatschke:2006bb}:
\begin{equation}
\sqrt{\frac{\kappa_z}{\kappa_\perp}} \approx \frac{\langle\Delta\eta\rangle}{\langle\Delta\phi\rangle}\,.
\end{equation}
Experimental data on dihadron correlation functions for central Au+Au collisions at
$\sqrt{s}=200$~GeV~\cite{Jacobs:2005pk} are consistent with
$\langle\Delta\eta\rangle/\langle\Delta\phi\rangle \approx3$~\cite{Romatschke:2006bb} as is shown in Fig. \ref{fig:ptcorr}. So our result can not quite quantitatively explain the measured ratio.
Photon-jet
correlation measurements should provide additional insight in the future.
\begin{figure}[htb]
  \begin{center}
    \includegraphics[width=12cm]{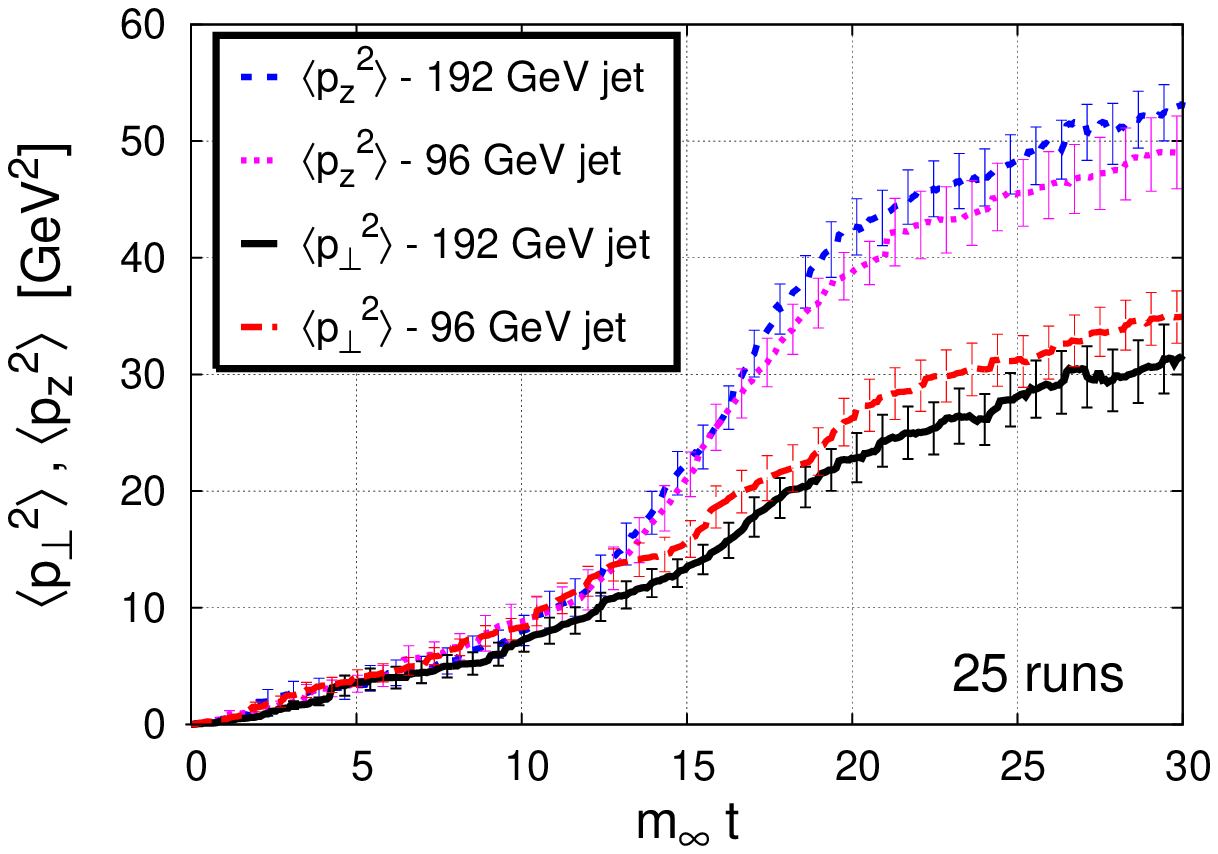}
    \caption{Momentum broadening of a jet in the
      directions transverse to its initial momentum. $p_z$ is directed
      along the beam axis, $p_\perp$ is transverse to the
      beam. Anisotropic plasma, $64^3$ lattice.}
    \label{fig:pxyjet192}
  \end{center}
\end{figure}

Fig.~\ref{fig:pxyjet192} shows the time evolution of $\langle
p_\perp^2 \rangle$ and of $\langle p_z^2 \rangle$. The strong growth
of the soft fields sets in at about $t\simeq 10 \,m_\infty^{-1}$ and
saturates around $t\simeq 20\, m_\infty^{-1}$ due to the finite lattice
spacing, as discussed above, as can be seen in Fig. \ref{fig:fieldWsu2_weak}.
Outside of this time interval the ratio $\kappa_z/\kappa_\perp\approx1$.
During the period of instability, however,
\begin{align}
    \frac{\kappa_z}{\kappa_\perp} \approx 2.3\,,
\end{align}
for both jet energies shown in Fig.~\ref{fig:pxyjet192}.  We find
approximately the same ratio for denser plasmas ($n_g=20/\text{fm}^3$
and $n_g=40/\text{fm}^3$). Reducing the number of lattice sites and
scaling $p_{\text{h}}$ down to 8~GeV in order to still have the separation scale $k^*$
lie around this scale, gives
$\kappa_z/\kappa_\perp\approx2.1$.
However, these latter runs are rather far from the continuum limit and lattice artifacts are
significant.

Initially $\xi\equiv \langle p_\perp^2\rangle /\langle p_z^2\rangle$
is infinite. However, due to collisions and the backreaction of the
fields on the hard particles we find that $\xi\approx 30$ at $t
\approx 20\, m_\infty^{-1}$.  This puts our measured ratio for
$\kappa_z/\kappa_\perp$ in the same range as those calculated
in~\cite{Romatschke:2006bb} for heavy quarks to leading-log accuracy,
where $\kappa_z/\kappa_\perp\simeq 3.6$ was obtained for $\xi=10$ and
velocity $v=0.95$. We emphasize that all of these estimates apply to
the early pre-equilibrium stage before hydrodynamic evolution sets in.

The explanation for the larger broadening along the beam axis is as
follows. In the Abelian case the instability generates predominantly
transverse magnetic fields which deflect the particles in
the $z$-direction~\cite{Majumder:2006wi}.
\begin{figure}[htb]
  \begin{center}
    \includegraphics[width=12cm]{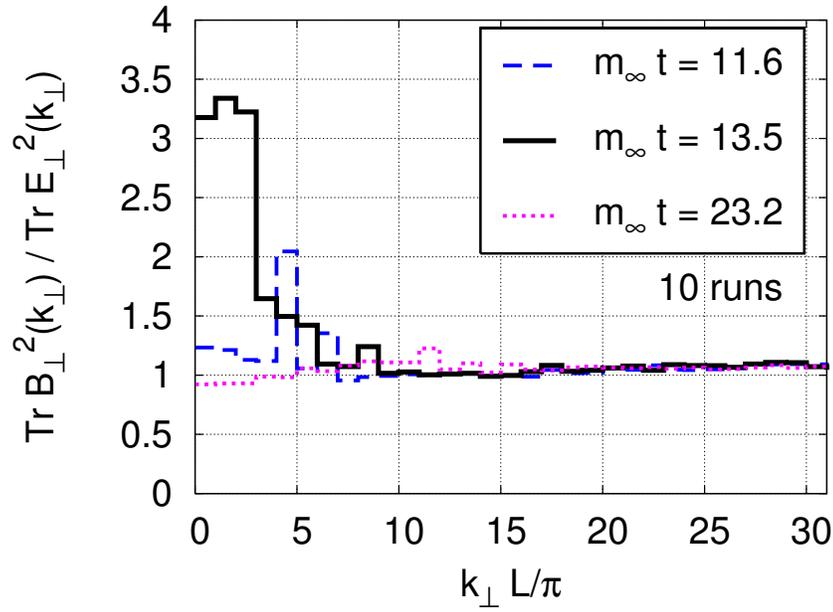}
    \caption{Ratios of fourier transforms of the field energy
      densities (integrated over $k_z$) at various times; the gauge
      potentials were transformed to Coulomb gauge.}
    \label{fig:FTs}
  \end{center}
\end{figure}
In the non-Abelian case, however, on three-dimensional lattices
transverse magnetic fields are much less dominant (see, e.g.\
Fig.~5 in~\cite{Dumitru:2006pz}) although they do form larger
coherent domains in the transverse plane at intermediate times
than $E_\perp$, Fig.~\ref{fig:FTs}. This is also shown directly in
Figs. \ref{fig:slicesdom1} and \ref{fig:slicesdom2}. We see that
the transverse chromo-magnetic fields form the most coherent
domains. The upper right plot in Fig. \ref{fig:slicesdom1} shows
that the transverse chromo-magnetic field is directed in the same
direction over the whole sytem, while the upper plots of Fig.
\ref{fig:slicesdom2} show that for the transverse component of the
chromo-electric field the formation of domains is less pronounced.
The opposite behavior is found for the longitudinal components,
where the effect is less visible, however.

Longitudinal fields and locally non-zero Chern-Simons number $\sim
{\rm Tr}~\mathbf{E} \cdot \mathbf{B}$ emerge as well. However, the
transverse magnetic fields dominate during the time of unstable
growth as can be seen in Fig. \ref{fig:newenxyz}. Furthermore,
Fig.~\ref{fig:ratios} makes it more apparent that also $E_z>B_z$,
aside from $B_\perp>E_\perp$.  Hence, the field configurations are
such that particles are deflected preferentially in the
longitudinal $z$-direction, which helps restore isotropy.
\begin{figure}[htb]
\vspace{-2cm}
 \hfill
  \begin{minipage}[t]{\textwidth}
      \begin{center}
         \begin{minipage}[t]{.45\textwidth}
            \begin{center}
                \includegraphics[width=7cm]{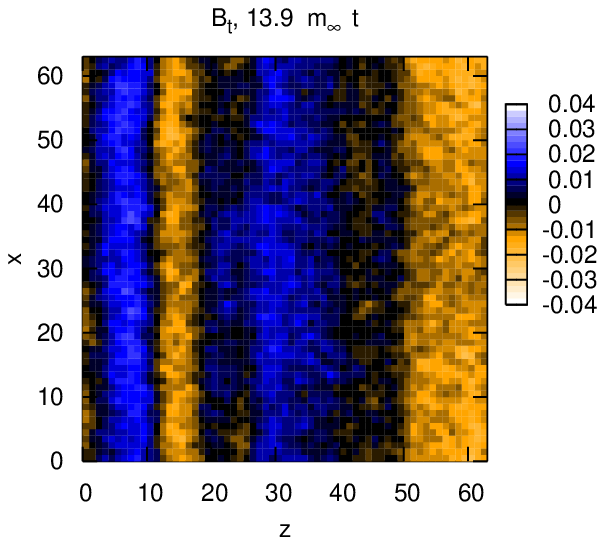}
\vspace{-3cm}
            \end{center}
        \end{minipage}
        \begin{minipage}[t]{.45\textwidth}
            \begin{center}
                \includegraphics[width=7cm]{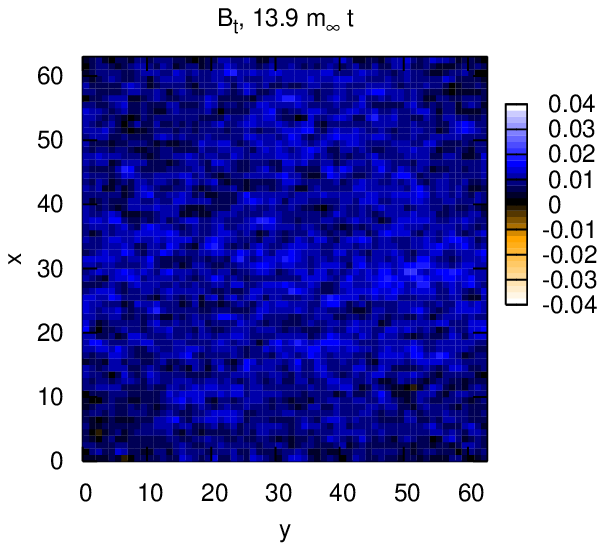}
\vspace{-3cm}
            \end{center}
        \end{minipage}
    \end{center}
  \end{minipage}

  \begin{minipage}[t]{\textwidth}
      \begin{center}
         \begin{minipage}[t]{.45\textwidth}
            \begin{center}
                \includegraphics[width=7cm]{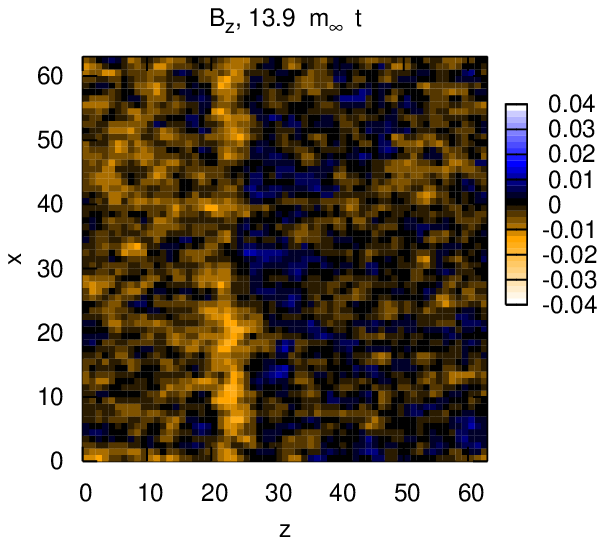}
\vspace{-1.5cm}
            \end{center}
        \end{minipage}
        \begin{minipage}[t]{.45\textwidth}
            \begin{center}
                \includegraphics[width=7cm]{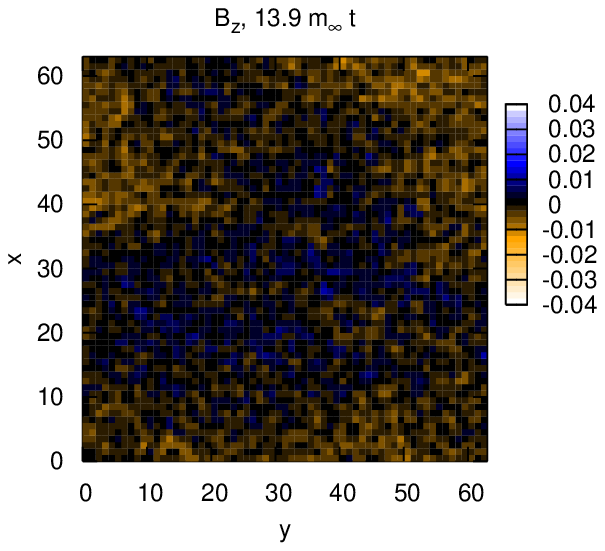}
\vspace{-1.5cm}
            \end{center}
        \end{minipage}
    \end{center}
  \end{minipage}
    \caption{Transverse (upper panel) and longitudinal (lower panel) components of the chromo-magnetic field. The position in the third dimension, which is not shown in the plots, is always fixed to 32. We see strong formation of domains in the transverse component, which spread over the whole range in the transverse direction ($x$-$y$-plane) - see upper right plot. \label{fig:slicesdom1}}
\end{figure}

\begin{figure}[htb]
\vspace{-2cm}
 \hfill
  \begin{minipage}[t]{\textwidth}
      \begin{center}
         \begin{minipage}[t]{.45\textwidth}
            \begin{center}
    \includegraphics[width=7cm]{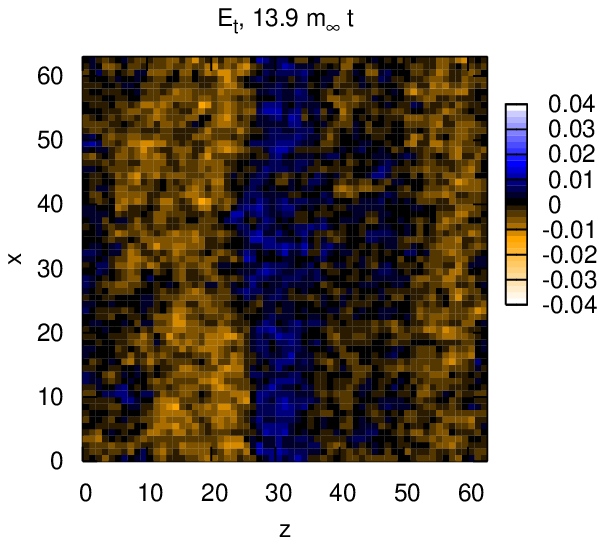}
\vspace{-3cm}
            \end{center}
        \end{minipage}
        \begin{minipage}[t]{.45\textwidth}
            \begin{center}
    \includegraphics[width=7cm]{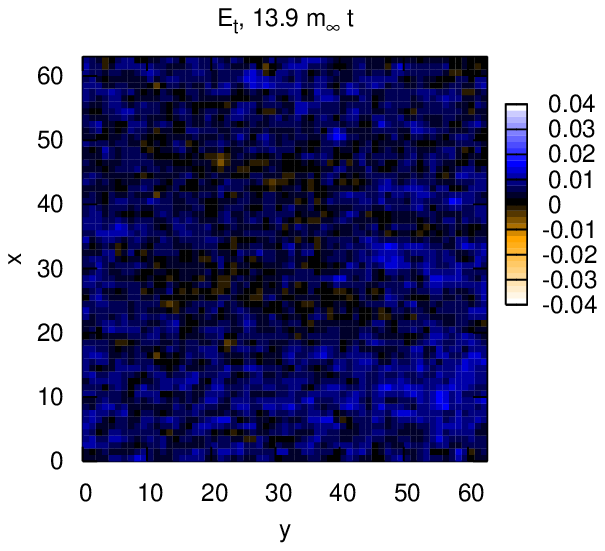}
\vspace{-3cm}
            \end{center}
        \end{minipage}
    \end{center}
  \end{minipage}

  \begin{minipage}[t]{\textwidth}
      \begin{center}
         \begin{minipage}[t]{.45\textwidth}
            \begin{center}
        \includegraphics[width=7cm]{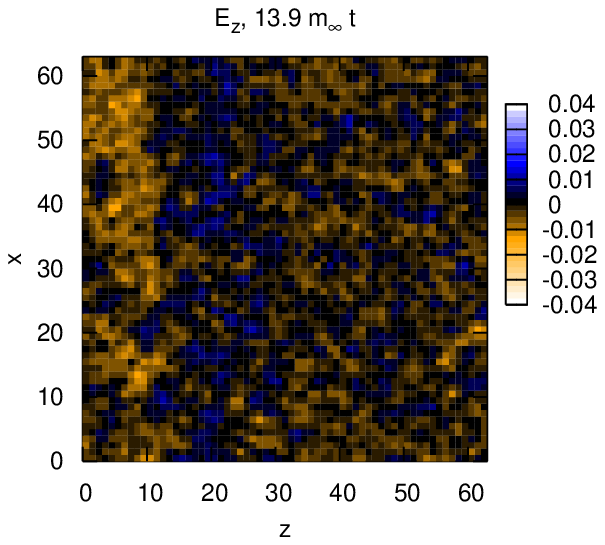}
\vspace{-1.5cm}
            \end{center}
        \end{minipage}
        \begin{minipage}[t]{.45\textwidth}
            \begin{center}
    \includegraphics[width=7cm]{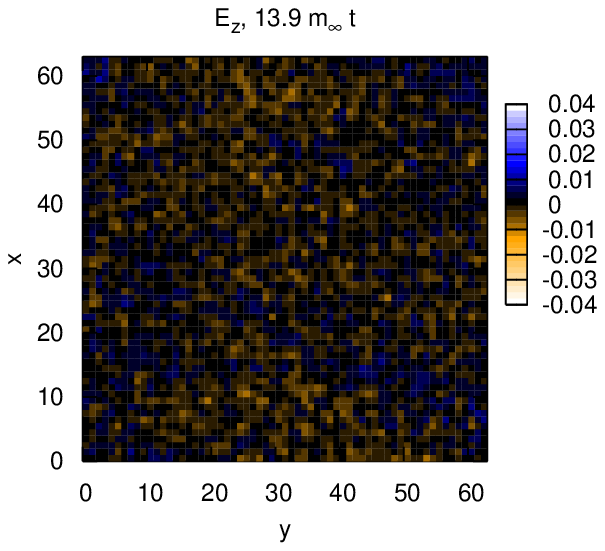}
\vspace{-1.5cm}
            \end{center}
        \end{minipage}
    \end{center}
  \end{minipage}
    \caption{Transverse (upper panel) and longitudinal (lower panel) components of the chromo-electric field. The position in the third dimension, which is not shown in the plots, is always fixed to 32. The formation of domains is less pronounced than for the tranverse chromo-magnetic field. \label{fig:slicesdom2}}
\end{figure}

\begin{figure}[htb]
  \begin{center}
    \includegraphics[width=12cm]{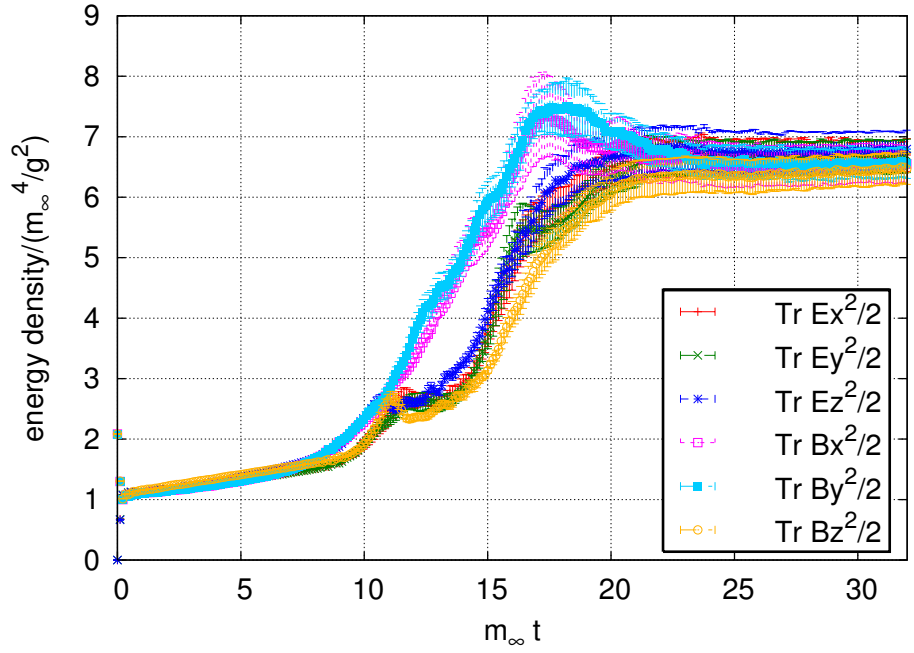}
    \caption{Energy density in the different field components. \label{fig:newenxyz}}
  \end{center}
\end{figure}

\begin{figure}[htb]
  \begin{center}
    \includegraphics[width=12cm]{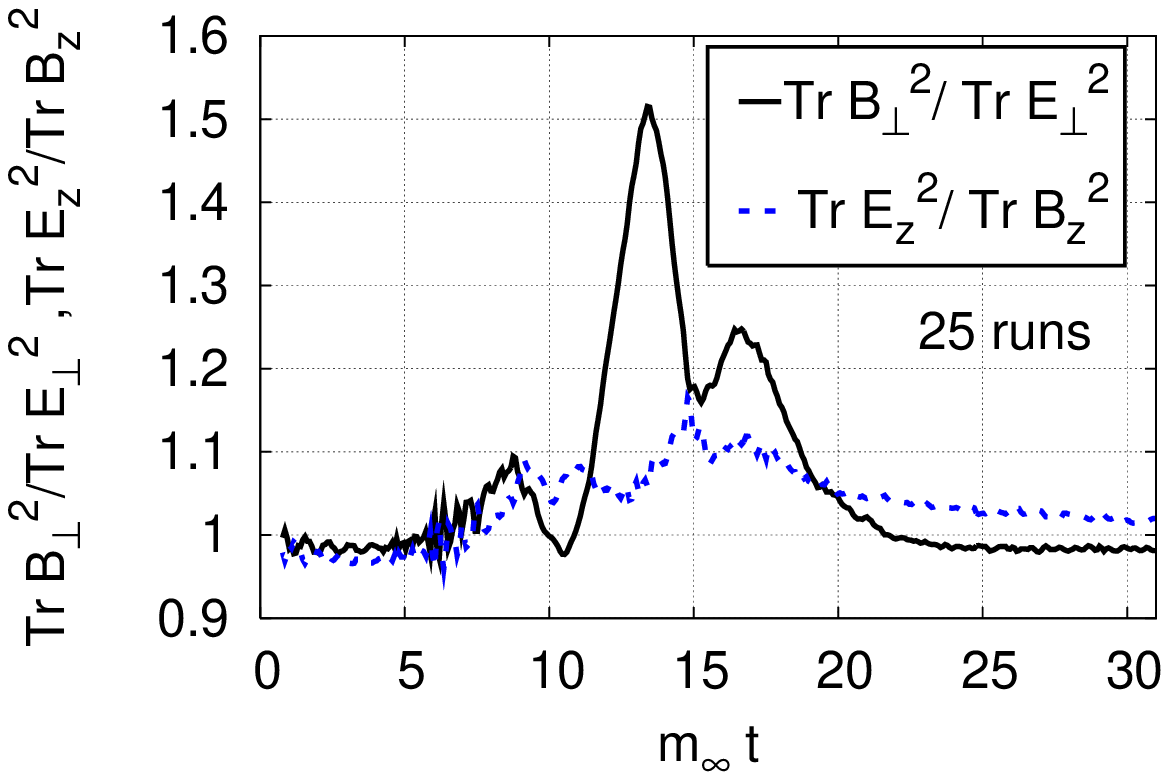}
    \caption{Ratios of field energy densities.}
    \label{fig:ratios}
  \end{center}
\end{figure}

A third contribution to $p_z$ broadening in an expanding plasma, not
considered explicitly here, is due to a longitudinal collective flow
field which ``blows'' the jet fragments to the
side~\cite{Armesto:2004vz}. This mechanism is also available for
collision dominated plasmas with (nearly) isotropic momentum
distribution. However, rather strong flow gradients seem to be
required to reproduce the observed broadening of midrapidity jets (the
flow velocity has to vary substantially {\em within} the narrow jet
cone). In contrast, color fields will naturally deflect particles with
lower momentum by larger angles ($\Delta p\sim E,B$): the jet profile
broadens even if the induced radiation is exactly collinear. It is
therefore important to determine, experimentally, whether the
asymmetric broadening is related to the {\em macroscopic} collective
flow or to an anisotropy of the plasma in the local rest frame.

In summary, we have shown that the collective effects appearing in the anisotropic QGP, as produced in a heavy-ion collision
can lead to jet broadening, which is stronger in the longitudinal direction than in azimuth and thereby offer an explanation
for the experimentally observed near-side ``ridge'', shown in Fig. \ref{fig:ptcorr}.

      \chapter{Fermionic collective modes of an anisotropic quark-gluon plasma}
\label{fermionicmodes} \index{Fermionic collective modes}

After having discussed the gluonic collective modes of the quark-gluon plasma in great detail, we now turn to the fermionic modes and concentrate on the investigation of whether they can be unstable too. We show analytically for two special cases as well as numerically for the general case that there are no fermionic instabilities. This result is similar to the case of the fermionic collective modes in a two-stream system \cite{Mrowczynski:2001az}, where it was also found that fermionic instabilities do not exist.

The absence of unstable fermionic modes is expected on
physical grounds due to the fact that fermion exclusion
precludes the condensation of modes; however, it could be
possible that, through pairing, fermions could circumvent
this as has been predicted \cite{fp1a,fp1b,fp2a,fp2b,fp2c}
and demonstrated \cite{fp3} in superfluid condensation of
cold fermionic atoms. However, this would require a description in
terms of fermionic bound or composite states which are not
included at the level of hard loops so we do not expect to
find any fermionic condensate-like instabilities using this
approximation.
This is verified via an explicit contour
integration of the inverse hard-loop quark propagator for
the two special cases in which we can obtain analytic
expressions for the self-energy. The special cases
considered analytically are (a) the case when the wave
vector of the collective mode is parallel to the anisotropy
direction with arbitrary oblate anisotropy and (b) for all
angles of propagation in the limit of an infinitely oblate
anisotropy.\footnote{This chapter is based on the work published in \cite{Schenke:2006fz}}

\section{Quark self-energy in an anisotropic system}
\label{quarkse}

\newcommand{\beq}{\begin{equation}}
\newcommand{\eeq}{\end{equation}}
\newcommand{\bqa}{\begin{eqnarray}}
\newcommand{\eqa}{\end{eqnarray}}

The integral expression for the retarded quark self-energy for an anisotropic system has been
obtained previously \cite{Mrowczynski:2000ed} and is given by
\bqa \label{q-self} \Sigma(K) = \frac{C_F}{4} g^2 \int_{\bf p} {
f ({\bf p}) \over |{\bf p}|} {P \cdot \gamma \over P\cdot
K} \;,\label{retself}
\eqa
where $C_F \equiv (N_c^2 -1)/2N_c$, and
$
f ({\bf p}) \equiv 2 \left( n({\bf p}) + \bar n ({\bf p})
\right) + 4 n_g({\bf p}) \; .
$
To obtain Eq.\,(\ref{retself}) one computes the leading-order quark self-energy diagram, assuming that (a) the quarks are massless, (b) the external fermion momentum is soft, $k_0\sim k\sim g\,p_{\text{hard}}$, (c) the momentum carried by the internal lines is hard, $p_0\sim p\sim p_{\text{hard}}$, and (d) the distribution function $f$ is symmetric in momentum space, $f(\mathbf{p})=f(-\mathbf{p})$. The hard scale $p_{\text{hard}}$ can be identified with the temperature $T$ in the case of thermal equilibrium but represents an arbitrary scale present in the non-equilibrium system, e.g. the nuclear saturation scale $Q_s$.

To simplify the calculation we follow Ref.~\cite{Romatschke:2003ms} and again choose a
distribution function $f({\bf p})$ given by
\beq
f({\bf p})=f_{\xi}({\bf p}) = N(\xi) \ %
f_{\rm iso}\left(\sqrt{{\bf p}^2+\xi({\bf p}\cdot{\bf \hat n})^2}\right)%
\; .
\label{squashing}
\eeq
Again, note that the appearance of $N(\xi)$ is not relevant for our considerations, but it allows for an easier comparison of systems with different strong anisotropies. Without it, every calculation in this section works analogously up to constant multiplicative factors. Here we use $N(\xi)=\sqrt(1+\xi)$.
Using Eq.~(\ref{squashing}) and performing the change of variables
\begin{equation}
\tilde{p}^2=p^2\left(1+\xi ({\bf v}\cdot{\bf \hat{n}})^2\right) \; ,
\label{variablechange}
\end{equation}
we obtain
\bqa \label{q-self2}
\Sigma(K) = m_q^2 \sqrt{1+\xi}
\int {d\Omega\over4\pi} \left(1+\xi (\hat{\bf p}\cdot\hat{\bf n})^2 \right)^{-1}
{P \cdot \gamma \over P\cdot K} \;,
\eqa
where
\bqa
m_q^2 = {g^2 C_F \over 8 \pi^2} \int_0^\infty dp \,
    p \, f_{\rm iso}(p) \; .\label{mqdefinition}
\eqa
We then decompose the self-energy into four contributions
\bqa \Sigma(K) = \gamma^0 \Sigma_0 + {\boldsymbol\gamma}\cdot{\mathbf
\Sigma}\; .
\eqa

The fermionic collective modes are determined by finding all
four-momenta for which the determinate of the inverse propagator
vanishes
\bqa
{\rm det}\,S^{-1} = 0 \; ,
\eqa
where
\bqa
i S^{-1}(P) &=& \gamma^\mu p_\mu - \Sigma \, , \nonumber \\
&\equiv& \gamma^\mu A_\mu \;.
\eqa
with $A(K)=(k_0 - \Sigma_0,{\bf k} - {\bf \Sigma})$.
Using the fact that ${\rm det}(\gamma^\mu A_\mu) = (A^\mu A_\mu)^2$
and defining $A_s^2 = {\bf A}\cdot{\bf A}$ we obtain
\bqa
A_0 = \pm A_s \, .
\label{fermiondisp}
\eqa

In practice, we can define the $z$-axis to be in the $\hat{\bf n}$
direction and use the azimuthal symmetry to restrict our consideration
to the $x\!-\!z$ plane.  In this case we need only three functions
instead of four
\bqa
\Sigma_0(w,k,\theta_n,\xi) &=& {1\over2} m_q^2 \sqrt{1+\xi} \int_{-1}^1 dx
    { R(w-k\cos\theta_n x,k\sin\theta_n\sqrt{1-x^2})
     \over 1 + \xi x^2 } \; , \nonumber \\
\Sigma_x(w,k,\theta_n,\xi) &=& {1\over2} m_q^2 \sqrt{1+\xi}
\int_{-1}^1 dx
    { \sqrt{1-x^2} S(w-k\cos\theta_n x,k\sin\theta_n\sqrt{1-x^2})
     \over 1 + \xi x^2 } \; , \nonumber \\
\Sigma_z(w,k,\theta_n,\xi) &=& {1\over2} m_q^2 \sqrt{1+\xi}
\int_{-1}^1 dx
    { x R(w-k\cos\theta_n x,k\sin\theta_n\sqrt{1-x^2})
     \over 1 + \xi x^2 } \; ,\label{selfenergies}
\eqa
where
\bqa
R(a,b) &=& \left(\sqrt{a+b+i\epsilon}\sqrt{a-b+i\epsilon}\right)^{-1} \, , \nonumber \\
S(a,b) &=& {1\over b}\left[a R(a,b) - 1\right] \, . 
\eqa
  \begin{figure}[htb]
    \begin{center}
        \includegraphics[height=5cm]{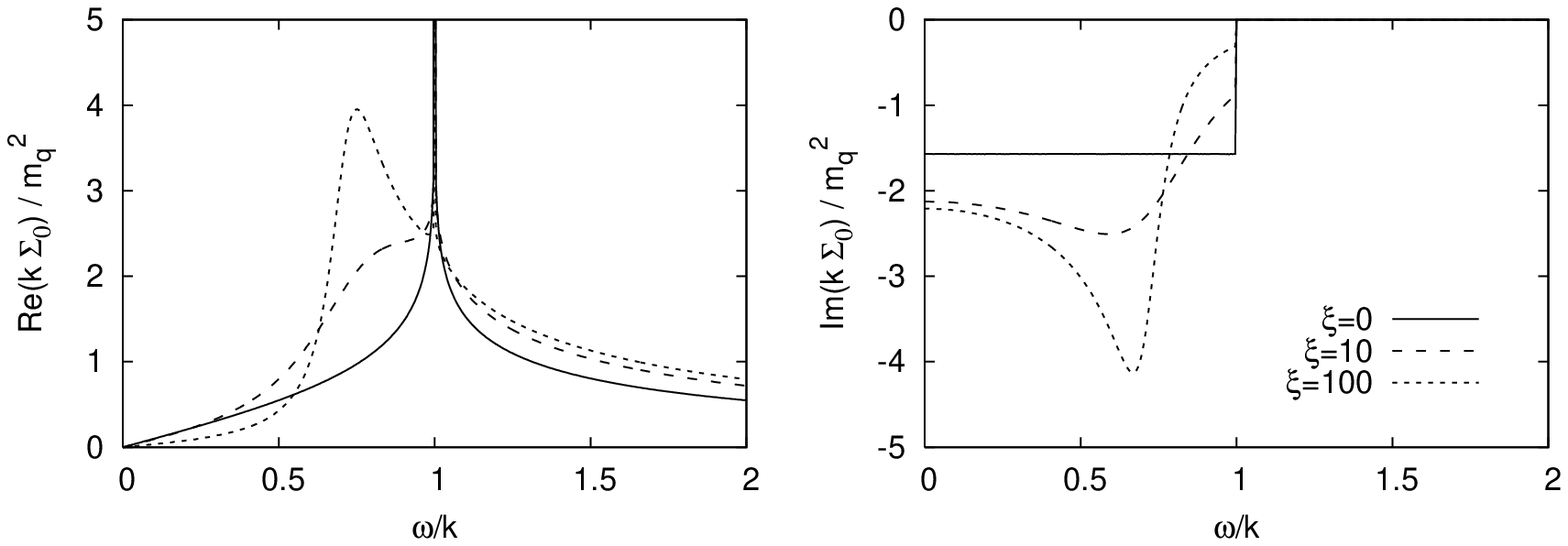}
        \caption{Real and imaginary part of $\Sigma_0$ as a function of $\omega/k$ for $\theta_n=\pi/4$ and $\xi=\{0,10,100\}$.}
        \label{fig:sigma0}
        %\vspace{.3in}
    \end{center}
  \end{figure}
  \begin{figure}[htb]
    \begin{center}
        \includegraphics[height=5cm]{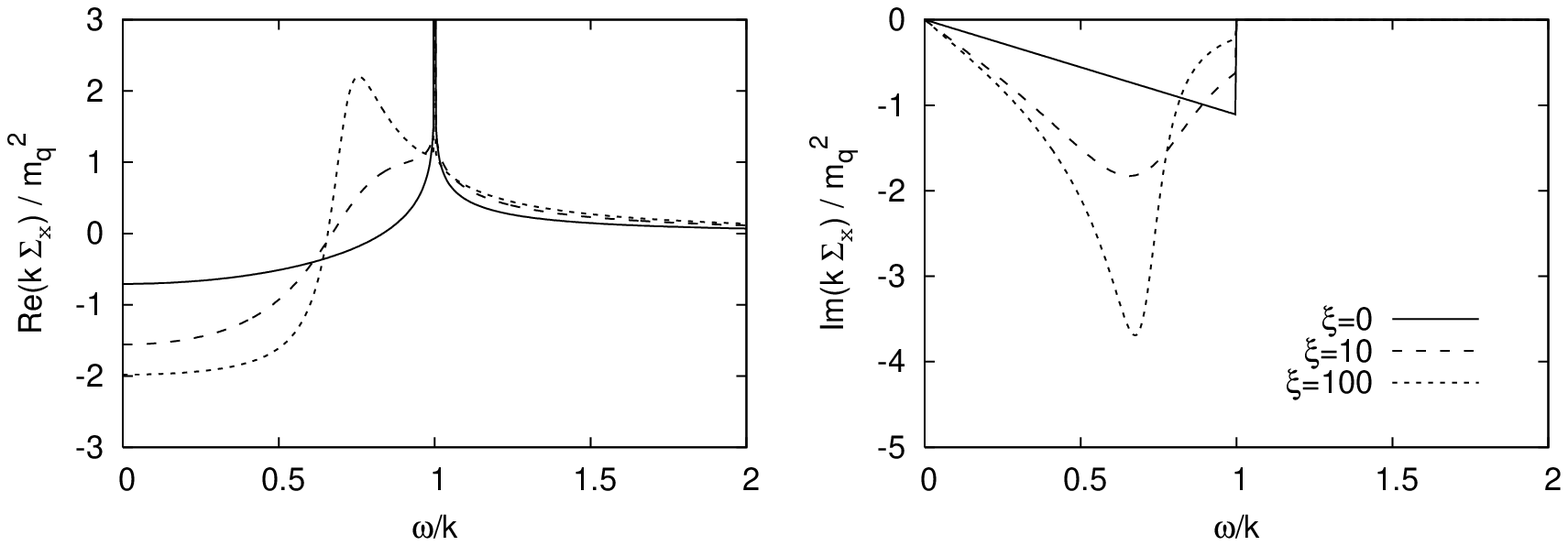}
        \caption{Real and imaginary part of $\Sigma_x$ as a function of $\omega/k$ for $\theta_n=\pi/4$ and $\xi=\{0,10,100\}$.}
        \label{fig:sigma1}
        %\vspace{.3in}
    \end{center}
  \end{figure}
  \begin{figure}[htb]
    \begin{center}
        \includegraphics[height=5cm]{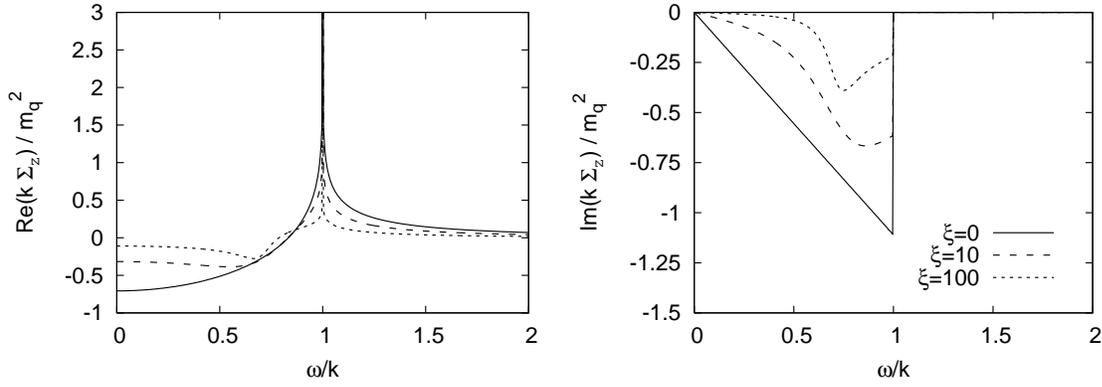}
        \caption{Real and imaginary part of $\Sigma_z$ as a function of $\omega/k$ for $\theta_n=\pi/4$ and $\xi=\{0,10,100\}$.}
        \label{fig:sigma2}
    \end{center}
  \end{figure}
  \newpage
In Figs.~\ref{fig:sigma0} through \ref{fig:sigma2} we plot the
real and imaginary parts of the quark self-energies $\Sigma_0$,
$\Sigma_x$, and $\Sigma_z$ for $\xi=\{0,10,100\}$. From these
Figures we see that the spacelike quark self-energy is strongly
affected by the presence of an anisotropy with a peak appearing at
$\omega/k = \sin\theta_n$ for strong anisotropies. To further
illustrate this in Fig.~\ref{fig:sigmaxi100} we have plotted
$\Sigma_0$ for $\xi=100$ and $\theta_n=\{0,\pi/4,\pi/2\}$. From
this Figure we see that there is a large directional dependence of
the spacelike quark self-energy.  Note that this could have a
measurable impact on quark-gluon plasma photon production during
the early stages of evolution since screening of infrared
divergences in leading order photon production amplitudes requires
as input the hard-loop fermion propagator for spacelike momentum.
We return to this point in Chapter \ref{chap:photons}
and calculate photon emission from an anisotropic quark-gluon plasma.
Assuming the necessary measurements of the rapidity dependence of the thermal
photon spectrum could be performed, photon emission could provide an
excellent measure of the degree of momentum-space anisotropy in
the partonic distribution functions at early stages of a heavy-ion
collision.

  \begin{figure}[t]
    \begin{center}
        \includegraphics[height=5.25cm]{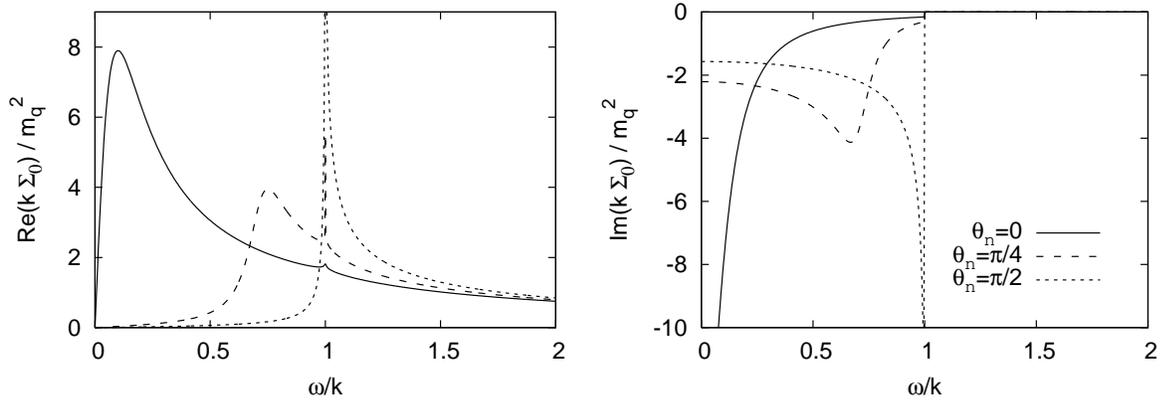}
        \caption{Real and imaginary part of $\Sigma_0$ as a function of $\omega/k$ for $\xi=100$ and $\theta_n=\{0,\pi/4,\pi/2\}$.}
        \label{fig:sigmaxi100}
    \end{center}
  \end{figure}

\begin{figure}[t]
    \begin{center}
        \includegraphics[width=12cm]{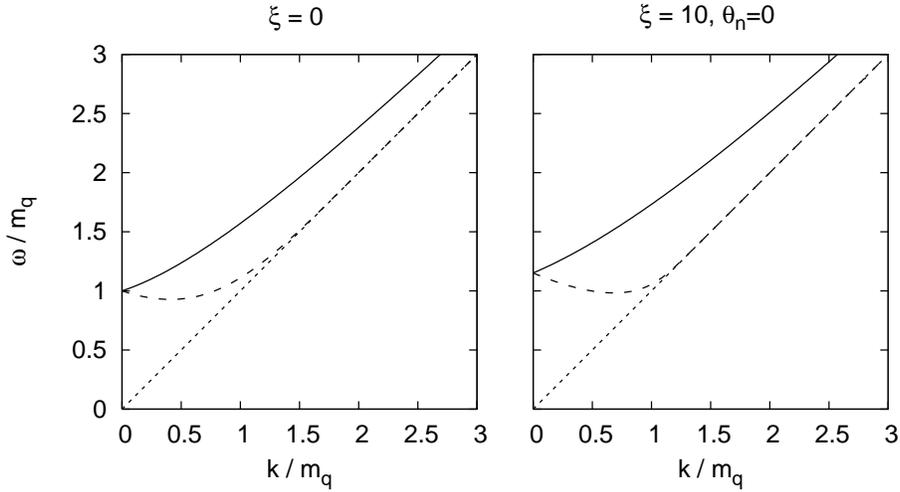}
        \caption{Fermionic dispersion relations for the istropic case and the case $\xi=10$ and $\theta_n=0$.}
        \label{fig:fermdisp}
    \end{center}
  \end{figure}

For general $\xi$ and $\theta_n$ we have to evaluate the integrals
given in Eq.~(\ref{selfenergies}) numerically. To find the
collective modes we then numerically solve the fermionic
dispersion relations \index{Dispersion relations} given by
Eq.~(\ref{fermiondisp}). As in the isotropic case, for real
timelike momenta ($\mid\!\!\omega\!\!\mid>\mid\!\!k\!\!\mid$,
${\rm Im}(\omega/k)=0$) there are two stable quasiparticle modes
which result from choosing either plus or minus in
Eq.~(\ref{fermiondisp}).\footnote{Note that there are four
solutions to the dispersion relations since each solution exists
at both positive and negative $\omega$.} The results for the
isotropic case ($\xi=0$) and the case where $\xi=10$ are shown in
Fig. \ref{fig:fermdisp}

\section{Analytical investigation}
We have looked for modes in the upper- and lower-half planes and numerically we
find none. In the following we explicitly count the
number of modes using complex  contour integration and
demonstrate that there are no unstable collective modes in
two special cases.
\subsection{Special case: $\mathbf{k}\parallel\mathbf{\hat{n}}$}
\label{sec:nyquist} Let us consider the special case where the
momentum of the collective mode is in the direction of the
anisotropy $\mathbf{k}\parallel\mathbf{\hat{n}}$, i.e.,
$\theta_n=0$. In this case the integrals in Eq.
(\ref{selfenergies}) can be evaluated analytically. $\Sigma_x$
becomes zero, while the other components read
\begin{align}
    \Sigma_0(\omega,k,\theta_n=0,\xi)&=\frac{1}{2}m_q^{2}\frac{\sqrt{1+\xi}}{\xi\omega^2+k^2}\left[2\sqrt{\xi}\omega\arctan\sqrt{\xi}+k\ln\left(\frac{\omega+k}{\omega-k}\right)\right]\notag\\
    \Sigma_z(\omega,k,\theta_n=0,\xi)&=\frac{1}{2}m_q^{2}\frac{\sqrt{1+\xi}}{\xi\omega^2+k^2}\left[-2\frac{1}{\sqrt{\xi}}k\arctan\sqrt{\xi}+\omega\ln\left(\frac{\omega+k}{\omega-k}\right)\right]\notag\,\text{.}\\
\end{align}
Eq. (\ref{fermiondisp}) simplifies to
\begin{equation}
    \omega-\Sigma_0=\pm(k-\Sigma_z)
    \label{fermiondispspecial}\,\text{.}
\end{equation}

\subsubsection*{Nyquist analysis}
\begin{figure}[t]
    \begin{center}
        \includegraphics[height=6cm]{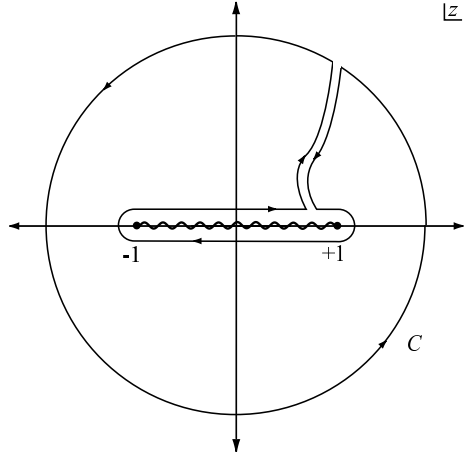}
        \caption{Contour $C$ in the complex $z$-plane used for the Nyquist analysis.}
        \label{fig:nyquistcontour}
    \end{center}
  \end{figure}

\index{Nyquist analysis}
We now show analytically for this special case that unstable modes
do not exist. This is done by a \textsc{Nyquist} analysis of the
following function:
\begin{equation}
    f_{\mp}(\omega,k,\xi)=\omega-\Sigma_0(\omega,k,\xi)\mp\left[k-\Sigma_z(\omega,k,\xi)\right]\label{function}\,\text{.}
\end{equation}
In practice, that means that we evaluate the contour integral
\begin{equation}
    \frac{1}{2\pi i}\oint_C dz\frac{f^{\prime}_{\mp}(z)}{f_{\mp}(z)}=N-P\,\text{,}\label{nyquist}
\end{equation}
which gives the numbers of zeros $N$ minus the number of poles $P$
of $f_{\mp}$ in the region encircled by the closed path $C$. In
Eq. (\ref{nyquist}) and in the following, we write the functions
$f_{\mp}$ in terms of $z=\omega/k$ and for clarity do not always
state the explicit dependence of $f_{\mp}$ on $k$ and $\xi$.
Choosing the path depicted in Fig. \ref{fig:nyquistcontour}, which
excludes the logarithmic cut for real $z$ with $z^2<1$ of the
function (\ref{function}), leads to $P=0$ and the left hand side
of Eq. (\ref{nyquist}) equals the number of modes $N$. Evaluation
of the respective pieces of the contour $C$ for each $f_{-}$ and
$f_{+}$ leads to
\begin{equation}
    N_{\mp}=1+0+0+1=2\,\text{,}\label{number}
\end{equation}
such that for the total number we get is $N=N_{-}+N_{+}=4$, which
corresponds to the stable modes (two for
positive $\omega$ and two for negative $\omega$). The four contributions in
(\ref{number}) are the following:
\begin{enumerate}
    \item The first $1$ results from integration along the large circle at $|z|\gg 1$.
    \item The first zero is the contribution from the path connecting
          the large circle with the contour around $z=\pm 1$.
    \item The second zero stems from the two small half-circles around $z=\pm 1$
    \item The last $1$ is obtained from integration along the straight lines running
          infinitesimally above and below the cut between $z=-1$ and
          $z=1$. See below for details on this integration.
\end{enumerate}

The last contribution can be evaluated using
\begin{equation}
    \int_{-1 + i\epsilon}^{1 + i\epsilon}
    dz\frac{f^{\prime}_{\mp}(z)}{f_{\mp}(z)}=\ln\frac{f_{\mp}(1 + i\epsilon)}{f_{\mp}(-1 + i\epsilon)}+2\pi i n\,\text{,}\label{straightlines}
\end{equation}
for the line above and the corresponding expression for the line
below the cut. $n$ is the number of times the function $f_{\mp}$
crosses the logarithmic cut located on the real axis, running from
zero to minus infinity. This cut is due to the appearance of the
logarithm on the right hand side of Eq. (\ref{straightlines}). In
the sum of the line integrations above and below the cut diverging
contributions from the first part on the right hand side of Eq.
(\ref{straightlines}) cancel and we are left with a contribution
of $2\pi i$ for each function. Furthermore it is necessary to show
that neither $f_{-}$ nor $f_{+}$ crosses the cut. The proof is
given in some detail for $f_{-}$ and is performed analogously for
$f_{+}$. From Eq. (\ref{function}) we find for $f_{-}$:
\begin{align}
    f_{-}(z,k,\xi)&=z-1+\frac{\sqrt{1+\xi}}{2(1+\xi
    x^2)}\frac{1}{k^2}\left[-2\left(z+\frac{1}{\xi}\right)\sqrt{\xi}\arctan\sqrt{\xi}+(z-1)\ln\left(\frac{z+1}{z-1}\right)\right]\,\text{.}
\end{align}
We want to study whether this function crosses the real axis in
the range $\text{Re}[z]\in[-1,1]$ for $\text{Im}[z]\rightarrow0$,
i.e., whether the imaginary part of $f_{-}$ changes sign in that
range. On the straight line infinitesimally above the cut the
imaginary part of $f_{-}$ is given by
\begin{equation}
    \text{Im}\left[
    \lim_{\epsilon\rightarrow0}f_{-}(x+i\epsilon,k,\xi)\right]=-\frac{\pi}{2}\frac{\sqrt{1+\xi}(x-1)}{k^2(1+\xi
    x^2)}\,\text{,}\label{impartf}
\end{equation}
for real $x$. It is only zero for $x=1$, which means that the
function $f_-$ can not cross but merely touch the cut within the
limits of the integration. On the straight line below the cut we
get the same result (\ref{impartf}) with a minus sign. For
$f_{+}$, we find that the imaginary part in the regarded range
only becomes zero for $x=-1$, which means that the logarithmic cut
is not crossed within $[-1,1]$ either. Hence we have proved for
the case $\mathbf{k}\parallel\mathbf{\hat{n}}$ that there are no
more solutions than the four stable modes. In
particular we have shown that unstable fermionic modes can not
exist.

\subsection{Large-$\xi$ limit}
\label{sec:largexi}
 In the extremely anisotropic case where
$\xi\rightarrow \infty$ the self-energies for arbitrary angle
$\theta_n$ can be calculated explicitly. The distribution function
(\ref{squashing}) becomes~\cite{Romatschke:2003yc}
\begin{equation}
    \lim_{\xi\rightarrow\infty}
    f_{\xi}(\textbf{p})=\delta(\mathbf{\hat{p}}\cdot\mathbf{\hat{n}})\int_{-\infty}^{\infty}dx
    f_{\text{iso}}\left(p\sqrt{1+x^2}\right)\,\text{.}\label{largexif}
\end{equation}
With $\mathbf{\hat{n}}$ in the $z$-direction this implies that
$\mathbf{p}$ lies in the $x$-$y$-plane only. As in Section
\ref{quarkse}, due to azimuthal symmetry, we consider the case
where $\mathbf{k}$ lies in the $x$-$z$-plane only. Using
(\ref{largexif}) we obtain from Eqs. (\ref{selfenergies})
\begin{align}
    \Sigma_0(\omega,k,\theta_n)&=\frac{\pi}{2}m_q^2\frac{1}{\sqrt{\omega+k \sin\theta_n}\sqrt{\omega-k
    \sin\theta_n}}\,\text{,}\notag\\
    \Sigma_x(\omega,k,\theta_n)&=\frac{\pi}{2 k \sin \theta_n}m_q^2\left(\frac{\omega}{\sqrt{\omega+k \sin\theta_n}\sqrt{\omega-k
    \sin\theta_n}}-1\right)\,\text{.}\label{lxsigma}
\end{align}
Since $p_z$ is always zero, $\Sigma_z$ vanishes. Eq.
(\ref{fermiondisp}) now becomes
\begin{equation}
    \omega-\Sigma_0=\pm\sqrt{(k_x-\Sigma_x)^2+k_z^2}\,\text{.}\label{lxdisp}
\end{equation}

\subsubsection*{Nyquist analysis}
\index{Nyquist analysis} Again, we only find four stable modes and
will now show analytically that these are the only solutions in
the large $\xi$-limit for arbitrary angle $\theta_n$. The cut
resulting from the complex square roots in (\ref{lxsigma}) can be
chosen to lie between $z = -\sin\theta_n$ and $z = \sin\theta_n$
on the real axis. The \textsc{Nyquist} analysis can then be
performed analogously to that in Section \ref{sec:nyquist} with
the contour in Fig. \ref{fig:nyquistcontour} adjusted such that
the inner path still runs infinitesimally close around the cut.
Using this path in the evaluation of Eq. (\ref{nyquist}) for the
functions
\begin{equation}
    f_{\mp}(\omega,k,\theta_n)=\omega-\Sigma_0\mp
    \sqrt{(k_x-\Sigma_x)^2+k_z^2}\,\text{,}\label{lxfunc}
\end{equation}
we find the number of solutions to Eq. (\ref{lxdisp}) to be
\begin{equation}
    N_{\mp}=1+0+\frac{1}{4}+\frac{1}{4}+\frac{1}{2}=2\,\text{,}\label{lxsol}
\end{equation}
so that again there are $N=N_++N_-=4$ solutions, which are the
known stable modes. The decomposition in (\ref{lxsol}) is done as
follows:
\begin{enumerate}
    \item The first contribution to $N_{\mp}$ comes from integration along the large outer
          circle at $|z|\gg 1$.
    \item The zero stems from the paths connecting the outer and the
          inner circle.
    \item The two contributions of $1/4$ result from integrations along the small
          circles around $-\sin\theta_n$ and $\sin\theta_n$.
    \item The last contribution of $1/2$ comes from integration along the straight lines running infinitesimally
          close above and below the cut. We discuss this part in further detail below.
\end{enumerate}
The last contribution can be obtained using Eq.
(\ref{straightlines}). For the evaluation of the limit
$\epsilon\rightarrow 0$ it is essential to note that the $f_{\mp}$
behave like $\ln \epsilon$ or $1/(\ln \epsilon)$ (depending on
which function is evaluated on which line) and are both negative
as $\epsilon \rightarrow 0$. This results in a contribution of $+i\pi$ for each
function and integration, because in all cases the imaginary part
of both functions can be shown to be positive in the regarded
limit. All other contributions, including the diverging parts $\pm
\ln(-\ln\epsilon)$ cancel in the sum of the results from the upper
and lower line.

Again, we need to show that the functions $f_{\mp}$ do not cross
the logarithmic cut for $z \in [-\sin\theta_n,\sin\theta_n]$,
i.e., that $n=0$ in Eq. (\ref{straightlines}). It is possible to
find an analytic expression for the imaginary part of $f_{\mp}$
using
\begin{equation}
    \text{Im}\sqrt{x+i y}=\frac{1}{\sqrt{2}}\,\text{sgn}(y)\sqrt{\sqrt{x^2+y^2}-x}\,\text{,}
\end{equation}
for the imaginary part of the square root appearing in
(\ref{lxfunc}) with real $x$ and $y$. Then the only solutions to
\begin{equation}
    \text{Im}f_{\mp}(z)=0
\end{equation}
are found analytically to be $\text{Re}(z)=\sin\theta_n$ and
$\text{Re}(z)=-\sin\theta_n$ for $f_-$ and $f_+$ respectively.
This means that the cut is not crossed during the integration
along the straight lines and that the contribution from this piece
is in fact $1/2$.

\section{Summary}
In this chapter we have extended the exploration of the
collective modes of an anisotropic quark-gluon plasma by
studying the quark collective modes. Specifically, we
derived integral expressions for the quark self-energy for
arbitrary anisotropy and evaluated these numerically using
the momentum-space rescaling (\ref{squashing}).
In the direct numerical calculation we found only real
timelike fermionic modes and no unstable modes. Additionally
using complex contour integration we have proven
analytically in the cases (a) when the wave vector of
the collective mode is parallel to the anisotropy direction
with arbitrary oblate anisotropy and (b) for all angles of
propagation in the limit of an infinitely oblate anisotropy
that there are no fermionic unstable modes.

      \chapter{Photon production from an anisotropic quark-gluon plasma}
\epigraphwidth 250pt \epigraph{Wo viel Licht ist, ist starker
Schatten - doch w\"ar's mir willkommen. Wollen sehn, was es gibt.}{\emph{G\"otz von Berlichingen, Erster Aufzug.\\
Jagsthausen. G\"otzens Burg. G\"otz.}\\
Johann Wolfgang von Goethe (1749-1832)} \label{chap:photons}
\index{Photon production} In this chapter we calculate photon
production from a quark-gluon plasma which is anisotropic in
momentum space including the \textsc{Compton} scattering,
$q(\bar{q}) \, g \rightarrow q(\bar{q}) \, \gamma$, and
annihilation, $q \, \bar{q} \rightarrow g \, \gamma$, processes.
We show that for a quark-gluon plasma which has a momentum-space
anisotropy the photon production rate has an angular dependence
which is peaked transverse to the beamline. Convolving this with a
model for the fireball evolution, we calculate the total photon
yield and compare to other contributions like prompt photons and
those from jet fragmentation. We discuss to which extent and under
which circumstances the results can be used to experimentally
determine the degree of momentum-space isotropy of a quark-gluon
plasma produced in relativistic heavy-ion
collisions.\footnote{This chapter is based on the work published
in \cite{Schenke:2006yp}}

\section{Medium photon production rate}
To lowest order photons are produced via annihilation and
\textsc{Compton} scattering processes
\begin{align}
 q + \bar q &\to g + \gamma, \label{annprocess}\\
 q (\bar q) + g &\to q (\bar q) + \gamma \, \label{compprocess}.
\end{align}
We do not include the bremsstrahlung \index{Bremsstrahlung} contribution which in an
equilibrium plasma also contributes at leading order in the
coupling constant due to enhancements from collinear photon
radiation
\cite{Aurenche:1998nw,Aurenche:1999tq,Arnold:2001ba,Arnold:2001ms}.
This contribution has been omitted because it is currently not
known how to resolve the problem that the presence of unstable
modes causes unregulated singularities to appear in matrix
elements which involve soft gluon exchange. Recent work
\cite{Romatschke:2006bb} has suggested that these singularities
could be shielded by next-to-leading order (NLO) corrections to the
gluonic polarization tensor, however, the detailed evaluation of
these NLO corrections has not been performed to date.  Absent
explicit calculation, our naive expectation would be that these
NLO corrections would yield similar angular dependence as the
annihilation and \textsc{Compton} scattering contributions since
they are also peaked at nearly collinear angles.

In a thermal medium the production rate is given by
\cite{Weldon:1983jn,McLerran:1984ay,gk91,Kapusta:1991qp}
\begin{equation}
    E\frac{dR}{d^3q}=-\frac{2}{(2\pi)^3}\text{Im}\Pi_\mu^{\mu}\frac{1}{e^{E/T}-1}\,,\label{productionrateselfenergy}
\end{equation}
with the retarded photon self energy $\Pi_{\mu\nu}$. It is valid
to all orders in the strong interaction $\alpha_s$ and to leading
order in $e^2$, as in its derivation it was assumed that the
produced photons do not interact with the medium after they have
been produced. For an anisotropic system, we adopt the
\textsc{Keldysh} formulation \index{Keldysh formalism}of quantum
field theory, which is appropriate for systems away from
equilibrium \cite{Ke64,Ke65,Chou:1984es}. In this formalism both
propagators and self energies have a $2\times 2$ matrix structure.
The components $(12)$ and $(21)$ of the self-energy matrices are
related to the emission and absorption probability of the particle
species under consideration
\cite{Chou:1984es,Mrowczynski:1992hq,Calzetta:1986cq}. Due to the
almost complete lack of photon absorbing back reactions, the rate
of photon emission can be expressed as \cite{Baier:1997xc} \index{Photon production!Production rate}
\begin{equation}
 E\frac{dR}{d^3 q} = \frac{i}{2(2\pi)^3} {\Pi_{12}}_\mu^\mu (Q) \, ,
 \label{photonrate}
\end{equation}
from the trace of the (12)-element $\Pi_{12}=\Pi^<$ of the
photon-polarization tensor.

\begin{figure}[t]
      \begin{center}
        \includegraphics[width=14cm]{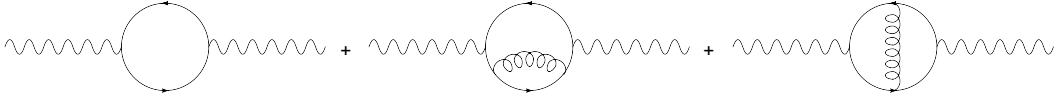}
        \caption{One- and two-loop contributions to the photon self energy in QCD}
        \label{fig:loopexpansion}
      \end{center}
  \end{figure}

If the photon self energy is approximated by carrying out a loop
expansion to some finite order then Eq.
(\ref{productionrateselfenergy}) is equivalent to a description in
terms of relativistic kinetic theory. Generally, when $\mathcal{M}$ is the
amplitude for a reaction of $m$ particles to $n$ particles plus one
photon then the contribution of this reaction to the differential
photon production rate is given by
\begin{equation}
    \mathcal{N}\int
    d\Phi(2\pi)^4\delta(p_{\text{in}}^\mu-p_{\text{out}}^\mu)|\mathcal{M}|^2\,,\label{generalrate}
\end{equation}
with the phase space $d\Phi$ given by a factor of
$d^3p/(2p(2\pi)^3)$ for each particle, a \textsc{Bose-Einstein} or
\textsc{Fermi-Dirac} distribution for each particle in the initial
state and a \textsc{Bose} enhancement or \textsc{Pauli}
suppression factor for each particle in the final state. The
overall degeneracy factor depends on the specific reaction.
Expanding the self energy up to $L$ loops is equivalent to
computing the contribution from all reactions with $m+n\leq L+1$,
with each amplitude calculated to order $g^{L-1}$. In our case
this can be seen when applying cutting rules to the loop diagrams
shown in Fig.\,\ref{fig:loopexpansion}, which yields graphs
corresponding to the following processes: Cutting the one-loop
diagram gives zero for an on-shell photon since the process
$q\bar{q}\to \gamma$ has no phase space. Certain cuts of the
two-loop diagrams give order $g^2$ corrections to this nonexistent
reaction. Other cuts correspond to the reactions $q\bar{q}\to
g\gamma$, $q g\to q\gamma$ and $\bar{q}g\to\bar{q}\gamma$, the
annihilation and \textsc{Compton} scattering processes discussed
above and shown in Figs. \ref{fig:annihilationDiagram} and
\ref{fig:comptonDiagram}.
Following the prescription (\ref{generalrate}), we find the
contributions of these processes to the rate to be
\begin{align}
E \frac{d R_{i}}{d^3q} &=\mathcal{N} \int_{{\bf k}_1}
f_1(\mathbf{k}_1)
        \int_{{\bf k}_2}f_2(\mathbf{k}_2)
        \int_{{\bf k}_3} (1 \pm f_3(\mathbf{k}_3))
        (2\pi)^4\delta^{(4)}(K_1 + K_2 - Q - K_3) \overline{\left|\mathcal{M}_{i}\right|^2}\,,
\label{eq:rate1}
\end{align}
with $\int_{{\bf k}_i}=\int d^3k_i/(2k_i(2\pi)^3)$. The $f_i$ are
the appropriate distribution functions (\textsc{Bose-Einstein} or
\textsc{Fermi-Dirac} in equilibrium) and there is either a
\textsc{Bose} enhancement or \textsc{Pauli} blocking term
depending on the nature of the strongly interacting particle in
the final state, as mentioned above. Putting in the right distribution functions, the
appropriate matrix element, and the correct degeneracy factor
$\mathcal{N}$ yields the annihilation and \textsc{Compton}
scattering contribution.

Let us calculate $\mathcal{N}$: For the annihilation process we
have a quark and an anti-quark in the initial state, which leads
to a color degeneracy of $3\cdot 3=9$, furthermore, we have to
multiply by 4 for the spin degeneracy, and summing over $u$ and
$d$ quarks yields a factor of $$\sum_q
e_q^2=\left(\frac{1}{3}\right)^2+\left(\frac{2}{3}\right)^2=\frac{5}{9}\,.$$
This makes an $\mathcal{N}$ of $9\cdot 4\cdot 5/9=20$. For the
\textsc{Compton} scattering part we have a quark and a gluon in
the initial state, meaning that we have to multiply by $3\cdot
8=24$, we take times 4 for initial spins/helicities and times 2
for both quark and anti quark scattering, and again have a factor
of $5/9$ from the sum over the squares of the quark charges. This
yields an $\mathcal{N}$ of $24\cdot 4\cdot 2\cdot 5/9=320/3$.

\subsection{Cross sections}
\begin{figure}[t]
  \hfill
  \begin{minipage}[t]{.45\textwidth}
      \begin{center}
        \includegraphics[width=4.8cm]{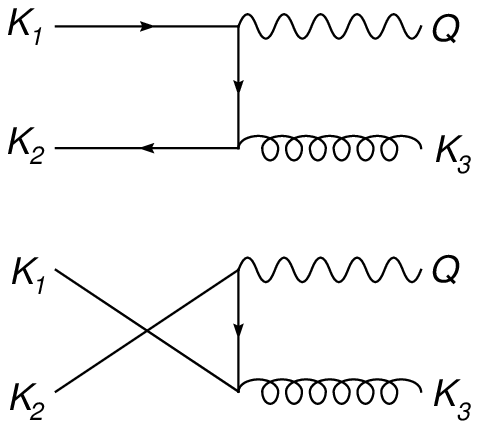}
        \caption{Annihilation diagram with momentum labels.  All
        four-momenta indicated flow to the right so that $K_1 + K_2 = Q +
        K_3$.} \label{fig:annihilationDiagram}
      \end{center}
  \end{minipage}
  \hfill
  \begin{minipage}[t]{.45\textwidth}
      \begin{center}
        \includegraphics[width=4.8cm]{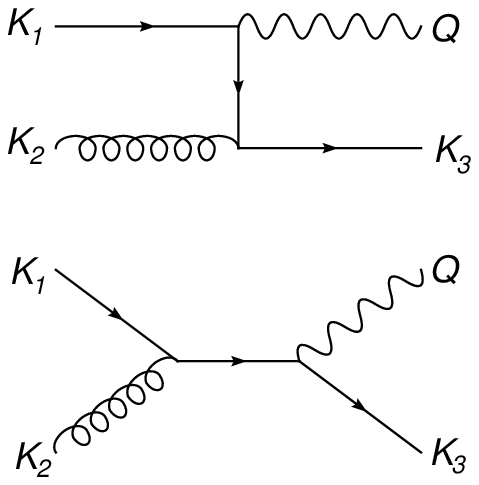}
        \caption{\textsc{Compton} diagrams for hard photon production with momentum
        labels.  All four-momenta indicated flow to the right so that $K_1
        + K_2 = Q + K_3$.  Anti-quark graphs are not shown but are
        included in the calculation.} \label{fig:comptonDiagram}
      \end{center}
  \end{minipage}
  \hfill
\end{figure}

\index{Cross section} We now calculate the cross sections for the
two processes (\ref{annprocess},\ref{compprocess}) in vacuum,
giving the matrix elements needed to compute (\ref{eq:rate1}). We
have to evaluate the \textsc{Feynman} diagrams shown in Figs.
\ref{fig:annihilationDiagram} and \ref{fig:comptonDiagram}

We go through the calculation of the \textsc{Compton} scattering
\index{Compton scattering} process in detail and give
the result of the analogous calculation of the annihilation
diagram.

\subsubsection{QED case}
For simplicity let us first consider the QED process
\begin{equation}
    \gamma\,e^-\to \gamma e^-\,,
\end{equation}
which apart from the color factors that we shall discuss later is
equivalent to the process (\ref{compprocess}). We set the electron
mass to zero as we shall do with the quark masses. Then, using the
\textsc{Feynman} rules for QED (see \cite{hm84}), the amplitudes
of the two diagrams in Fig.\,\ref{fig:comptonDiagram} read: \index{Matrix element}
\begin{align}
    -i\mathcal{M}_1&=\bar{u}^{(s')}(K_3)\left[\varepsilon_\mu(ie\gamma^\mu)\frac{i(\slashed{K}_1-\slashed{Q})}{(K_1-Q)^2}(ie\gamma^\nu)\varepsilon_\nu'^*\right]u^{(s)}(K_1)\,,\\
    -i\mathcal{M}_2&=\bar{u}^{(s')}(K_3)\left[\varepsilon_\nu'^*(ie\gamma^\nu)\frac{i(\slashed{K}_1+\slashed{K}_2)}{(K_1+K_2)^2}(ie\gamma^\mu)\varepsilon_\mu\right]u^{(s)}(K_1)\,,
\end{align}
where $u$ is the positive energy four-spinor solution of the
\textsc{Dirac} equation with spin index $s=\{1,2\}$,
$\bar{u}=u^{\dag}\gamma^0$ and $\varepsilon^\mu$ is the
polarization vector of the free photon, which appears in the
solutions $A^\mu=\varepsilon^\mu(\mathbf{q})e^{-iqx}$ to the
(free) \textsc{Maxwell} equation $\Box^2 A^\mu=0$. $e$ is the
electron charge. The ingoing electron is assigned the
four-momentum $K_1$ and spin $s$, the outgoing electron has
four-momentum $K_3$ and spin $s'$ according to Fig.\,
\ref{fig:comptonDiagram}. The ingoing and outgoing photons have
four-momentum $K_2$ and $Q$ and the polarization vector
$\varepsilon$ and $\varepsilon'$, respectively.

The invariant \textsc{Mandelstam} \index{Mandelstam variables}
variables are:
\begin{align}
    &s = (K_1+K_2)^2=2\,K_1\cdot K_2=2\,Q\cdot K_3\notag\\
    &t = (K_1-Q)^2=-2\,K_1\cdot Q=-2\,K_2\cdot K_3\notag\\
    &u = (K_1-K_3)^2=-2\,K_1\cdot K_3=-2\,K_2\cdot Q\,.
\end{align}
With these definitions the invariant amplitudes can be written as
\begin{align}
    \mathcal{M}_1&=\frac{1}{t}\varepsilon_\nu'^*\varepsilon_\mu\,e^2\,\bar{u}^{(s')}(K_3)\gamma^\mu(\slashed{K}_1-\slashed{Q})\gamma^\nu
    u^{(s)}(K_1)\,,\\
    \mathcal{M}_2&=\frac{1}{s}\varepsilon_\nu'^*\varepsilon_\mu\,e^2\,\bar{u}^{(s')}(K_3)\gamma^\nu(\slashed{K}_1+\slashed{K}_2)\gamma^\mu
    u^{(s)}(K_1)\,.
\end{align}
We want to determine the unpolarized cross section and therefore
have to average $\left|\mathcal{M}_1+\mathcal{M}_2\right|^2$ over
initial electron spins and photon polarizations and sum over the
final ones. Because we are dealing with real photons we can make
the replacement \cite{hm84}
\begin{equation}
    \sum_{\text{transverse}}\varepsilon_\alpha^*\varepsilon_\beta\rightarrow
    -g_{\alpha\beta}\,,
\end{equation}
where the sum is over the physical transverse polarizations, and a
similar one for the outgoing polarizations. This yields for the
squared s-channel amplitude:
\begin{equation}
    \overline{\left|\mathcal{M}_2\right|^2}=\frac{e^4}{4
    s^2}\sum_{s,s'}\left[\bar{u}^{(s')}(K_3)\gamma^\nu(\slashed{K}_1+\slashed{K}_2)\gamma^\mu
    u^{(s)}(K_1)\right]\left[\bar{u}^{(s)}(K_1)\gamma_\mu(\slashed{K}_1+\slashed{K}_2)\gamma_\nu
    u^{(s')}(K_3)\right]\,,\label{averageedmatrix1}
\end{equation}
where the factor $1/4$ comes from averaging over the two initial
electron spins and two initial photon polarizations. In the next
step we calculate the spin sum in (\ref{averageedmatrix1}), using
the spinor completeness relation \cite{hm84,PS}:
\begin{equation}
    \sum_{s=1,2}u^{(s)}(K_1)\bar{u}^{(s)}(K_1)=\slashed{K}_1+m\xrightarrow{m=0}\slashed{K}_1
\end{equation}
We find
\begin{align}
 \overline{\left|\mathcal{M}_2\right|^2}&=\frac{e^4}{4 s^2}
 \text{Tr}
 \left[\slashed{K}_3\gamma^\nu(\slashed{K}_1+\slashed{K}_2)\gamma^\mu\slashed{K}_1\gamma_\mu(\slashed{K}_1+\slashed{K}_2)\gamma_\nu\right]\notag\\
 &=\frac{e^4}{4 s^2} \text{Tr}
 \left[\gamma_\nu\slashed{K}_3\gamma^\nu(\slashed{K}_1+\slashed{K}_2)\gamma^\mu\slashed{K}_1\gamma_\mu(\slashed{K}_1+\slashed{K}_2)\right]\notag\\
 &=\frac{e^4}{4 s^2} \text{Tr}
 \left[(- 2)\slashed{K}_3(\slashed{K}_1+\slashed{K}_2)(-2)\slashed{K}_1(\slashed{K}_1+\slashed{K}_2)\right]\notag\\
 &=\frac{e^4}{s^2} \text{Tr}
 \left[\slashed{K}_3\slashed{K}_2\slashed{K}_1\slashed{K}_2\right]\notag\,,
\end{align}
where we used the cyclic invariance of the trace in the first
step, the relation $\gamma_\nu\gamma^\mu\gamma^\nu=-2\gamma^\mu$
in the second, and the fact that $\slashed{K}\slashed{K}=0$ in the
third. Using
$$\text{Tr}(\slashed{a}\slashed{b}\slashed{c}\slashed{d})=4\left[(a\cdot
b)(c\cdot d)-(a\cdot c)(b\cdot d)+(a\cdot d)(b\cdot c)\right]\,,$$
we can further write
\begin{align}
 \overline{\left|\mathcal{M}_2\right|^2}&=8\frac{e^4}{s^2}(K_2\cdot
 K_3)(K_1\cdot K_2)\notag\\
 &=2 e^4 \left(-\frac{t}{s}\right)\,.
\end{align}
Similarly we obtain
\begin{align}
 \overline{\left|\mathcal{M}_1\right|^2} &=2 e^4
 \left(-\frac{s}{t}\right)\,,\\
 \overline{\mathcal{M}_1\mathcal{M}_2^*}&=0\,,
\end{align}
such that the full spin-averaged squared matrix element is given
by
\begin{equation}
\overline{\left|\mathcal{M}\right|^2}=\overline{\left|\mathcal{M}_1+\mathcal{M}_2\right|^2}=-32\pi^2\alpha^2\left(\frac{t}{s}+\frac{s}{t}\right)\,,\label{compmatrix1}
\end{equation}
using $\alpha=e^2/(4\pi)$. 
%From this result we infer that at high
%energies (large $s$) the cross section peaks as $t\to 0$. We can
%therefore approximate the cross section by its forward peak and
%get
%\begin{equation}
%\frac{d\sigma}{dt}=\frac{1}{16\pi
%s^2}\overline{\left|\mathcal{M}\right|^2}\,.
%\end{equation}

The calculation for the annihilation process is completely
analogous and yields
\begin{equation}
\overline{\left|\mathcal{M}\right|^2}=32\pi^2\alpha^2\left(\frac{u}{t}+\frac{t}{u}\right)\,.\label{annmatrix1}
\end{equation}

\subsubsection{QCD case}
\begin{figure}[t]
      \begin{center}
        \includegraphics[height=4cm]{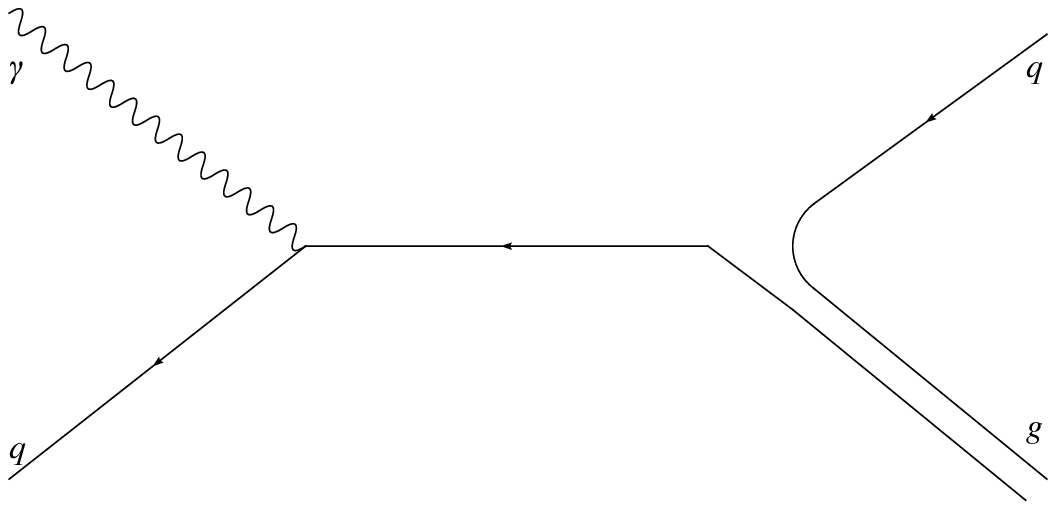}
        \caption{Color lines for $g q\to q \gamma$}
        \label{fig:colorlines}
      \end{center}
  \end{figure}

The result for the spin averaged squared matrix elements in QCD
follows directly from the ones in QED on (1) substitution of
$\alpha\to e_q^2\alpha\alpha_s$ because we now have one photon and
one gluon vertex each with a quark charge $e_q$ (the summation
over both $u$ and $d$ quarks will be done later), and (2) the
multiplication with a color factor, which we shall now derive.
\index{Color factor} One way to determine the color factor is
simply to count all possible color combinations, sum them up and
then divide by the number of possibilities for the color
configuration in the initial state. An easy way to do this is by
counting the color lines in Fig.\,\ref{fig:colorlines}. There are
two, and the same holds for all other diagrams considered here.
Each line can take on three colors such that we have 9
possibilities of combining them. However we have to subtract one
possibility, which corresponds to the SU(3) color singlet state of
the gluon. This singlet does not carry color and cannot mediate
between color charges. Hence we are left with 8 possibilities.
This we have to divide by a factor of two which comes from the
(unfortunate) historical definition of $\alpha_s$ \cite{hm84}. For
the \textsc{Compton} scattering diagram we have a quark and a
gluon in the initial state, which means that for averaging over
color we have to divide by 3 for the colors of the quark and by 8
for all possible color combinations of the gluon. This leads to
the spin and color averaged matrix element
\begin{equation}
\overline{\left|\mathcal{M}\right|^2}=-32\pi^2e_q^2\alpha\alpha_s
\frac{1}{6}\left(\frac{t}{s}+\frac{s}{t}\right)\,.\label{compmatrix}
\end{equation}
The annihilation diagram has two quarks in the initial state,
which means that we have to divide by 9, 3 for each quark, when we
want to average over initial colors. This yields
\begin{equation}
\overline{\left|\mathcal{M}\right|^2}=32\pi^2e_q^2\alpha\alpha_s\frac{4}{9}\left(\frac{u}{t}+\frac{t}{u}\right)\,.\label{annmatrix}
\end{equation}
Alternatively one can perform the summation over initial and final
states by taking the trace over the squared matrix element. Each
diagram includes a gluon vertex yielding a factor of
$-igt^a\gamma^\mu$. This means the color trace is to be taken over
$t^a t^a$ giving
\begin{equation}
    \text{Tr}(t^at^a)=C_2(N)\text{Tr}(\mathbf{1})=\frac{N^2-1}{2N}\cdot
    3=4\,,
\end{equation}
with the quadratic \textsc{Casimir} operator $C_2$ and $N=3$ for
SU(3). What is left to do is dividing by the number of color
combinations in the initial state for averaging and we get the
same results as above.

\section{Infrared divergence and Braaten-Yuan method} The matrix
elements (\ref{compmatrix}) and (\ref{annmatrix}) have poles at
$t$ and/or $u=0$, which causes infrared divergent production rates
(\ref{eq:rate1}). Many-body effects have to be included in the
calculation to account for the screening of these divergences.
This is done by first excluding the contribution from soft
momentum exchange $p<p^*$, with the cutoff on $p^*$, and
separately calculating the soft contribution $p<p^*$ by using
HL-resummed propagators instead of bare ones. This cures the
infrared divergence and in the limit $g\to 0$ the sum of hard and
soft contributions becomes independent of the cutoff $p^*$. This
procedure has first been used for the calculation of the rate of
energy loss of a hot plasma from the \textsc{Primakoff} production
of axions by \textsc{Braaten} and \textsc{Yuan} in
\cite{Braaten:1991dd}. For the calculation of high energy medium
photon production from an isotropic QGP the method has been
applied by \textsc{Baier} et.al \cite{Baier:1991em} and
\textsc{Kapusta} et.al in \cite{Kapusta:1991qp}. We will summarize
their results in Sec. \ref{sec:isotropiclimit} and compare to our
more general result in the limit that the anisotropy,
characterized by the parameter $\xi$, goes to zero.

\subsection{Hard contribution}
\index{Photon production!Hard contribution}
\subsubsection{Annihilation Diagrams}
From (\ref{eq:rate1}), (\ref{annmatrix}) and remembering that
$\mathcal{N}=20$ we find the rate of photon production from the
quark annihilation diagram shown in
Fig.~\ref{fig:annihilationDiagram} to be
\bqa E \frac{d R_{\rm ann}}{d^3q} &=&
    {64 \pi^3 \frac{5}{9} \alpha_s \alpha} \int_{{\bf k}_1} \frac{f_q({\bf k}_1)}{k_1}
        \int_{{\bf k}_2} \frac{f_q({\bf k}_2)}{k_2}
        \int_{{\bf k}_3} \frac{1 + f_g({\bf k}_3)}{k_3} \nonumber \\
    && \hspace{+24mm} \times \,
        \delta^{(4)}(K_1 + K_2 - Q - K_3) \left[ \frac{u}{t} + \frac{t}{u} \right]
        \; ,
\label{eq:annihilation1}
\eqa
where we have assumed that the distribution function for quarks
and anti-quarks is the same, $f_q = f_{{\bar q}}$, which allows us
to simplify Eq.~(\ref{eq:annihilation1}) by combining the two
terms in square brackets after interchanging $K_1$ and $K_2$ in
the second term. This operation casts the second term into the
exact same form as the first so that we simply pick up a factor of
2.

As discussed earlier we will be confronted with an infrared
divergence and hence begin by first changing variables in the
first integration to $P \equiv K_1 - Q$ and introduce the infrared
cutoff $p^*$ on the integration over the exchanged three-momentum
$p$. We also split the zero and vector components of the delta
function as $\delta^{(4)}(P + K_2 - K_3) = \delta(\omega + k_2 -
k_3) \delta^{(3)}({\bf p}+ {\bf k}_2 - {\bf k}_3)$ where $\omega
\equiv \left|{\bf p}+{\bf q}\right| - q$.  We then use the vector
part of the delta function to perform the ${\bf k}_3$ integration.
Doing this and relabelling ${\bf k}_2 \rightarrow {\bf k}$ in
order to simplify the notation we obtain

\bqa E \frac{d R_{\rm ann}}{d^3q} &=&
       {16 \frac{5}{9} \alpha_s \alpha}
        \int_{{\bf p}} \frac{ f_q({\bf p}+{\bf q}) }{ \left|{\bf p}+{\bf q}\right| }
        \int_{{\bf k}} \frac{ f_q({\bf k}) }{ k }
        \frac{ 1 + f_g({\bf p}+{\bf k}) }{ \left|{\bf p}+{\bf k}\right| } \nonumber \\
    && \hspace{24mm} \times \,
        \delta(\omega + k - \left|{\bf p}+{\bf k}\right|) \, \Theta(p-p^*) \left[ \frac{u}{t} \right]
        \; .
\eqa

To proceed further we choose spherical coordinates with the
anisotropy vector $\hat{\bf n}$ defining the $z$-axis.  Exploiting
the azimuthal symmetry about the $z$-axis we also choose ${\bf q}$
to lie in the $x\!-\!z$ plane. It is then possible to reexpress
the remaining delta function as
\begin{align}
\delta(\omega + &k - \left|{\bf p}+{\bf k}\right|)
=\frac{\left|{\bf p}+{\bf k}\right|}{ p k \sin\theta_p
\sin\theta_k \left| \sin(\phi_i - \phi_p) \right|}
    \sum_{i} \delta(\phi_k - \phi_i) \; ,
\label{eq:deltaphi1}
\end{align}
where $\phi_i$ are defined through the transcendental equation
\beq \cos(\phi_i - \phi_p) =
    \frac{\omega^2 - p^2 + 2 k (\omega - p \cos\theta_p \cos\theta_k)}{2 p k \sin\theta_p \sin\theta_k}  \; .
\label{eq:phiksol} \eeq
Eq.~(\ref{eq:deltaphi1}) can be made more explicit using
$\sin^2(\phi_i - \phi_p) = 1 - \cos^2(\phi_i - \phi_p)$ yielding
\beq \delta(\omega + k - \left|{\bf p}+{\bf k}\right|) =
    2 \left|{\bf p}+{\bf k}\right| \chi^{-1/2} \, \Theta(\chi) \, \sum_{i=1}^2 \delta(\phi_k - \phi_i)
     \; ,
\label{eq:deltaphi2} \eeq
with
\begin{align}
\chi \equiv 4 p^2 k^2 \sin^2&\theta_k \sin^2\theta_p - \left[\omega^2 - p^2 + 2k(\omega - p \cos\theta_p \cos\theta_k)\right]^2
\end{align}
and where we have indicated that there are two solutions to
Eq.~(\ref{eq:phiksol}) when $\chi>0$.  Making these substitutions
and expanding out the phase-space integrals explicitly gives
\begin{align}
 E \frac{d R_{\rm ann}}{d^3q} &=
    \frac{\frac{5}{9} \alpha_s \alpha}{2 \pi^6}
      \sum_{i=1}^{2}
      \int_{p^*}^\infty \, dp \, p^2 \,
      \int_{-1}^{1} \, d\cos\theta_p \,
      \int_{0}^{2\pi} \, d\phi_p \;
      \frac{ f_q({\bf p}+{\bf q}) }{ \left|{\bf p}+{\bf q}\right| }
        \nonumber \\
        & \hspace{0.5cm} \times \;
      \int_0^\infty \, dk \, k \,
      \int_{-1}^{1} \, d\cos\theta_k \;
        f_q({\bf k}) \left[ 1 + f_g({\bf p}+{\bf k}) \right]
        \chi^{-1/2} \, \Theta(\chi) \left[ \frac{u}{t} \right] \Biggr|_{\phi_k = \phi_i}
        \; ,
\label{hardannihilationrate} \end{align}
with $t = \omega^2 - p^2$, $u = (k-q)^2 - ({\bf k}-{\bf q})^2$,
and recalling $\omega = \left|{\bf p}+{\bf q}\right| - q$.

\subsubsection{Compton Scattering Diagrams}

Again assuming $f_{\bar q}=f_q$ the rate of photon production from
the Compton scattering diagrams shown in
Fig.~\ref{fig:comptonDiagram} can be expressed as
\bqa E \frac{d R_{\rm com}}{d^3q} &=&
    -{128 \pi^3 \frac{5}{9} \alpha_s \alpha} \int_{{\bf k}_1} \frac{f_q({\bf k}_1)}{k_1}
        \int_{{\bf k}_2} \frac{f_g({\bf k}_2)}{k_2}
        \int_{{\bf k}_3} \frac{1 - f_q({\bf k}_3)}{k_3} \nonumber \\
    && \hspace{28mm} \times
        \delta^{(4)}(K_1 + K_2 - Q - K_3) \left[ \frac{s}{t} + \frac{t}{s} \right]
        \; ,
\label{eq:compton1} \eqa
where we have again assumed massless quarks, and the
\textsc{Mandelstam} variables are defined by $s \equiv (K_1 +
K_2)^2$ and $t \equiv (K_1 - Q)^2$.  Performing a change of
variables to $P \equiv K_1-Q$, redefining $K_2 \rightarrow K$, and
evaluating the delta function as in the annihilation case above
gives

\begin{align} E \frac{d R_{\rm com}}{d^3q} &=
    -\frac{\frac{5}{9} \alpha_s \alpha}{2\pi^6}
      \sum_{i=1}^{2}
      \int_{p^*}^\infty \, dp \, p^2 \,
      \int_{-1}^{1} \, d\cos\theta_p \,
      \int_{0}^{2\pi} \, d\phi_p \;
      \frac{ f_q({\bf p}+{\bf q}) }{ \left|{\bf p}+{\bf q}\right| }
        \nonumber \\
        & \hspace{0.5cm} \times \;
      \int_0^\infty \, dk \, k \,
      \int_{-1}^{1} \, d\cos\theta_k \;
        f_g({\bf k}) \left[ 1 - f_q({\bf p}+{\bf k}) \right]
        \chi^{-1/2} \, \Theta(\chi) \left[ \frac{s}{t} + \frac{t}{s} \right] \Biggr|_{\phi_k = \phi_i}
        \; ,
\label{hardcomptonrate} \end{align}

where here $t = \omega^2 - p^2$, $s = (\omega+q+k)^2 - ({\bf
p}+{\bf q}+{\bf k})^2$, and recalling $\omega = \left|{\bf p}+{\bf
q}\right| - q$.

\subsubsection{Total Hard Contribution}
\label{sec:hardtotal}

The total photon production from processes which have a hard
momentum exchange is given by the sum of
Eqs.~(\ref{hardannihilationrate}) and (\ref{hardcomptonrate}).
\bqa E \frac{d R_{\rm hard}}{d^3q} = E \left( \frac{d R_{\rm
ann}}{d^3q} + \frac{d R_{\rm com}}{d^3q} \right)
\label{hardphotonrate} \eqa
To evaluate the five-dimensional integrals in
Eqs.~(\ref{hardannihilationrate}) and (\ref{hardcomptonrate}) we
use monte-carlo integration.

The total hard contribution (\ref{hardphotonrate}) has a
logarithmic infrared divergence as $p^* \rightarrow 0$. This
logarithmic infrared divergence will be cancelled by a
corresponding ultraviolet divergence in the soft contribution so
that in the limit $g \rightarrow 0$ there will be no dependence of
the total (soft + hard) rate on the separation scale $p^*$.  In the
next section we present the calculation of the soft part which we
will then combine with the hard result to obtain the total photon
production rate.

\subsection{Soft contribution}
\label{sec:softpart}\index{HL approximation}\index{Photon
production!Soft contribution} We now turn to the calculation of
the previously excluded infrared divergent contribution from soft
momentum exchange. The infrared divergence in the photon
production rate is caused by a diverging differential cross
section when the momentum transfer goes to zero. Long-ranged
forces are usually screened by many-body effects at finite
temperature. In the high temperature limit these effects can be
included by using the HL-resummed propagators instead of free
ones. This way we will be able to derive an infrared finite
result. The rate for the soft part is given by Eq.
(\ref{photonrate}):
\begin{equation}
 E\frac{dR_{\rm soft}}{d^3 q} = \frac{i}{2(2\pi)^3} {\Pi^{<}}_\mu^\mu (Q) \, ,
 \label{softphotonrate}
\end{equation}
from the trace of the (12)-element $\Pi^{<}=\Pi_{12}$ of the
photon-polarization tensor.
\begin{figure}[htb]
      \begin{center}
        \includegraphics[height=5cm]{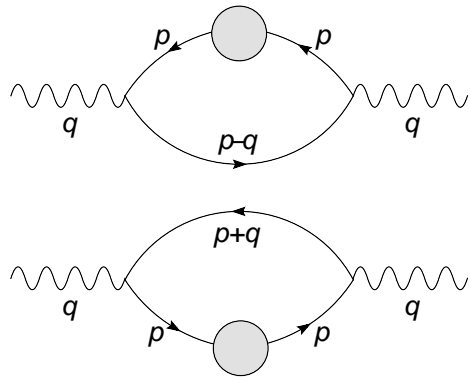}
      \end{center}
  \caption{The photon polarization tensor $\Pi^{<}$ for soft momentum $p\sim gT$. The blob indicates the hard-loop resummed quark propagator.}
  \label{fig:loops}
\end{figure}
After the replacement of free propagators with the HL-resummed
ones, the photon polarization tensor is calculated from the
\textsc{Feynman} graphs in Fig. \ref{fig:loops}, resulting in
\begin{align}\label{iPi}
 i \Pi_{12}&_\mu^\mu(Q) = e^2 \frac{5}{9} N_c \int \frac{d^4 p}{(2\pi)^4}
 \text{Tr}
 \left[
 \gamma^\mu \left. iS_{12}^\star(P) \right|_{HL} \gamma_{\mu} i S_{21}(P-Q) \right.\notag\\
 &~~~~~~~~~~~~~~~~~~~~~~~~~~~~~~~~~~~~~~\left.+
 \gamma^\mu iS_{12}(P+Q) \gamma_\mu \left. iS_{21}^\star(P) \right|_{HL}
 \right]\,,
\end{align}
where $e_q^2=5/9$ comes from the sum over quark charges. $S_{12}$ and $S_{21}$ are the
free fermion propagators for massless quarks, to be read off the matrix propagator
\begin{align}
    S(K)=&\slashed{K}\left[\left(\begin{array}{cc}
        \frac{1}{K^2+i\epsilon} & 0 \\
        0 & \frac{-1}{K^2-i\epsilon}\end{array}
        \right)+2\pi i \delta(K^2)\left( \begin{array}{cc}
        f_{\text{F}}(K) & -\theta(-k_0)+f_{\text{F}}(K)  \\
        -\theta(k_0)+f_{\text{F}}(K) & f_{\text{F}}(K)\end{array}
        \right)\right]\text{,}\label{prop}
\end{align}
with the general fermion distribution function $f_{\text{F}}(K)$,
and propagators with a HL subscript are the full propagators in
the hard-loop approximation. These satisfy a fluctuation
dissipation relation given by
(see Appendix \ref{fdrappendix} on the validity of this expression in our framework)
\begin{equation}\label{fdr}
    S_{12/21}^\star\left.(P)\right|_{HL}=\left.S_{\text{ret}}^\star(P)\right|_{HL}\Sigma_{12/21}(P)\left.S_{\text{adv}}^\star(P)\right|_{HL}\text{\,.}
\end{equation}
The HL resummed retarded propagator reads
\begin{equation}
    \left.S_{\text{ret}}^\star(P)\right|_{HL}=\frac{1}{\slashed{P}-\Sigma(P)}\text{\,,}
\end{equation}
where $\Sigma(P)$ is the retarded self-energy given in
Eq.\,(\ref{q-self}). The advanced propagator follows analogously
with the advanced self-energy. Note that Eq.(\ref{fdr}) is
equivalent to the expression for $S_{12/21}^\star$ in
\cite{Baier:1997xc} (see \cite{Greiner:1998vd}).
To one loop order $\Sigma_{12}$ is given by
\begin{align}
    \Sigma_{12}(P)=2 i g^2 C_\text{F} \int \frac{d^4 k}{(2\pi)^4}
    S_{12}(K)\Delta_{12}(P-K)\text{\,,}
\end{align}
where $\Delta_{12}$ is the (12)-element of the matrix boson
propagator given by
\begin{align}
    \Delta(K)=\left(\begin{array}{cc}
        \frac{1}{K^2-m^2+i\epsilon} & 0 \\
        0 & \frac{-1}{K^2-m^2-i\epsilon}\end{array}
        \right)-2\pi i \delta(K^2-m^2)\left( \begin{array}{cc}
        f_{\text{B}}(K) & \theta(-k_0)+f_{\text{B}}(K)  \\
        \theta(k_0)+f_{\text{B}}(K) & f_{\text{B}}(K)\end{array}
        \right)\text{\,.}\label{propboson}
\end{align}
With the anisotropic distribution function (\ref{squashing})
$\Sigma_{12}$ can be evaluated in the hard-loop approximation to
read
\begin{align}
\Sigma_{12}^{\mu}(P)&=i \frac{g^2 C_{\text{F}}}{(2\pi)^2}\int d\tilde{k}\int_{0}^{2\pi} d\phi
\int_{-1}^{+1} dx \, \frac{\tilde{k}^2}{(1+\xi x^2)^{3/2}} \nonumber \\
& \hspace{2cm} \times \left[\!\left.\frac{k^{\mu}}{k}\right|_{k_0=k}\!\!\delta\left(g_{-}\right)  N(\xi)f_{F}^{\text{iso}}(\tilde{k})\left( N(\xi)f_{\text{B}}^{\text{iso}}(\tilde{k})+1\right)\right.\notag\\
& \hspace{2.5cm} \left.+\left.\frac{k^{\mu}}{k}\right|_{k_0=-k}\!\!\delta\left(g_{+}\right) N(\xi)f_{B}^{\text{iso}}(\tilde{k})\left( N(\xi)f_{\text{F}}^{\text{iso}}(\tilde{k})-1\right)\!\right]\text{\,,}
\nonumber \\
\end{align}
where
\beq
g_\pm = 2\frac{\tilde{k}}{\sqrt{1+\xi
    x^2}}\left[\pm p_0+p\left(\sin\theta_n\sqrt{1-x^2}\cos\phi+\cos\theta_n
    x\right)\right]
\eeq
  \begin{figure}[t]
    \begin{center}
        \includegraphics[height=6cm]{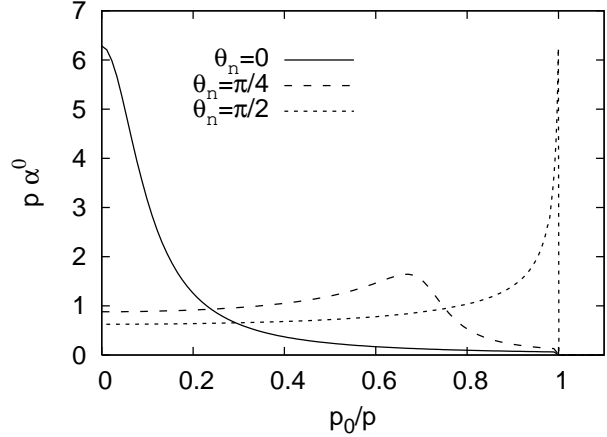}
        \caption{$\theta_n$-dependent part $\alpha^0$ of the self-energy $\Sigma_{12}^0$. $\theta_n\in\{0,\pi/4,\pi/2\}$ and $\xi=100$.}
        \label{fig:sigma12-0-xi100}
    \end{center}
  \end{figure}
and we chose $\mathbf{p}$ to lie in the $x-z$-plane and used the
change of variables (\ref{variablechange}) for $k$. Note that in the hard-loop
limit one can ignore the quark masses and hence they have been explicitly set to
zero above.  The term
$k^{\mu}/k$ does not depend on $k$ and is given by $(\pm 1, \sin
\theta \cos \phi, \sin \theta \sin \phi, \cos \theta)$. Evaluation
of the $\delta$-function leads to
\begin{align}
    -i \Sigma_{12}^{\mu}(P,\theta_n,\xi)=&A\, \alpha^{\mu}(P,\theta_n,\xi)+B\, \beta^{\mu}(P,\theta_n,\xi)\text{\,,}\label{alphabeta}
\end{align}
with
\begin{align}
   \alpha^{\mu}%(P,\theta_n,\xi)
   &=\int d\phi\, \sum_i \left.\frac{k^{\mu}}{k}\right|_{k_0=k}\left|\frac{(1-x_i^2)^{1/2}}{(1-x_i^2)^{1/2}(p
    \cos\theta_n+p_0\xi x_i)-p \sin\theta_n x_i(1+\xi)
    \cos\phi}\right|\theta(1-x_i^2)\,,\notag\\
   \beta^{\mu}%(P,\theta_n,\xi)
   &=\int d\phi\, \sum_i \left.\frac{k^{\mu}}{k}\right|_{k_0=-k}\left|\frac{(1-\tilde{x}_i^2)^{1/2}}{(1-\tilde{x}_i^2)^{1/2}(p
    \cos\theta_n-p_0\xi \tilde{x}_i)-p \sin\theta_n \tilde{x}_i(1+\xi)
    \cos\phi}\right|\theta(1-\tilde{x}_i^2)\,,
\end{align}
where the $x_i$ and $\tilde{x}_i$ are solutions to $\frac{\tilde{k}}{\sqrt{1+\xi
    x^2}}\left[-p_0+p\left(\sin\theta_n\sqrt{1-x^2}\cos\phi+\cos\theta_n
    x\right)\right]=0$ and $\frac{\tilde{k}}{\sqrt{1+\xi
    x^2}}\left[p_0+p\left(\sin\theta_n\sqrt{1-x^2}\cos\phi+\cos\theta_n
    x\right)\right]=0$, respectively, and
\begin{align}
    A&=\frac{g^2 C_{\text{F}}}{8\pi^2}\int dk\,k\,  N(\xi)f_{F}^{\text{iso}}(k)\left( N(\xi)f_{\text{B}}^{\text{iso}}(k)+1\right)\text{\,,}\\
    B&=\frac{g^2 C_{\text{F}}}{8\pi^2}\int dk\,k\,  N(\xi)f_{B}^{\text{iso}}(k)\left( N(\xi)f_{\text{F}}^{\text{iso}}(k)-1\right)\text{\,.}
\end{align}
There can be $N\in\{0,1,2\}$ solutions for both $x_i$ and $\tilde{x}_i$, depending on the parameters $p,p_0,\theta_n$ and $\phi$.
%In particular $\left[(p_0/p-x_i\cos\theta_n)\csc\theta_n\sec\phi\right]\geq0$ has to hold for $x_i$ to be a solution.
Note that $\frac{k^{\mu}}{k}$ is also given in terms of the $x_i$.
It is easily verified that
\begin{equation}
    \alpha^{\mu}(P,\theta_n,\xi)=-\beta^{\mu}(P,\theta_n,\xi)\text{\,,}
\end{equation}
  \begin{figure}[t]
    \begin{center}
        \includegraphics[height=6cm]{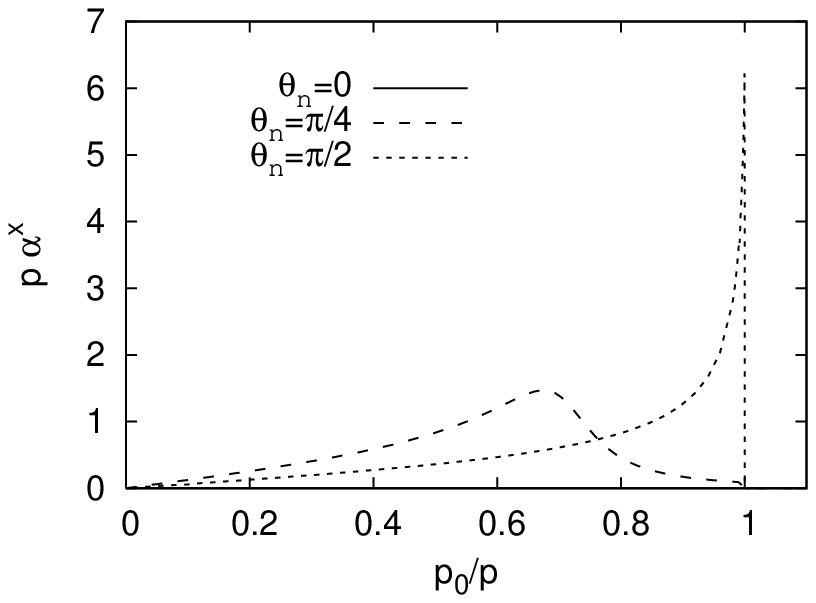}
        \caption{$\theta_n$-dependent part $\alpha^x$ of the self-energy $\Sigma_{12}^x$. $\theta_n\in\{0,\pi/4,\pi/2\}$ and $\xi=100$. For $\theta_n=0$ $\alpha^x=0$.}
        \label{fig:sigma12-x-xi100}
    \end{center}
  \end{figure}
    \begin{figure}[t]
    \begin{center}
        \includegraphics[height=6cm]{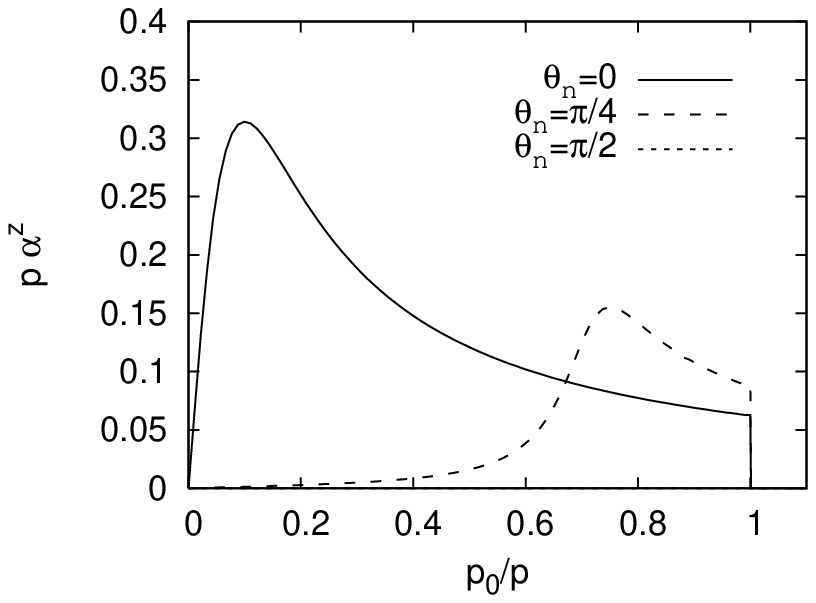}
        \caption{$\theta_n$-dependent part $\alpha^z$ of the self-energy $\Sigma_{12}^z$. $\theta_n\in\{0,\pi/4,\pi/2\}$ and $\xi=100$. For $\theta_n=\pi/2$ $\alpha^z=0$.}
        \label{fig:sigma12-z-xi100}
    \end{center}
  \end{figure}
such that Eq. \,(\ref{alphabeta}) greatly simplifies to read
\begin{equation}
    -i \Sigma_{12}^{\mu}(P,\theta_n,\xi) = (A-B)\, \alpha^{\mu}(P,\theta_n,\xi)\text{\,,}\label{sigma12}
\end{equation}
where
\begin{equation}
    A-B= {g^2 C_F \over 8 \pi^2} N(\xi) \left[\int_0^\infty dk \,
    k \, \left(f^{\rm iso}_B(k)+f^{\rm iso}_F(k)\right) \right]=\frac{1}{4}m_q^2 N(\xi)\; ,
\end{equation}
assuming equal quark and anti-quark distributions. We did not
present the analogous explicit calculation of $\Sigma_{21}$, but
find for it the same result as for $\Sigma_{12}$ with $A$ and $B$
interchanged. We also verified that $\Sigma_{12}$ and
$\Sigma_{21}$ fulfill the general relation
\begin{equation}
    \Sigma_{21}-\Sigma_{12}=2i\,\text{Im}\,\Sigma\text{\,,}
\label{sigrel}
\end{equation}
with the retarded self-energy $\Sigma$ given in Sec. \ref{quarkse}.
Furthermore, since $\Sigma_{21}$ is given by Eq.~(\ref{sigma12}) with $A$ and $B$ interchanged
it follows within the hard-loop approximation that with
the form of the anisotropic distribution function assumed here it always holds that
\begin{equation}
    \Sigma_{12}=-\Sigma_{21}\text{\,,}\label{kms}
\end{equation}
which can be seen as a high-temperature limit of the
\textsc{Kubo-Martin-Schwinger} (KMS) relation \index{KMS
relation}in equilibrium, but also holds for finite $\xi$ and hence
for non-equilibrium systems. Note that in equilibrium the relation
(\ref{kms}) does not contradict the usual KMS-relation, but is an
approximation to it. In equilibrium, the KMS-relation is given by
\begin{equation}
    \Sigma_{12}(P)=-e^{-p_0/T}\Sigma_{21}(P)\text{\,,}\label{kmseq}
\end{equation}
with the temperature $T$. In our case the soft momentum $p$ is of
the order $gT$, such that an expansion of the exponential gives
$1$ as in Eq.\,(\ref{kms}) plus terms that are suppressed by
higher orders of $g$. Eqs.\,(\ref{sigrel}) and (\ref{kms}) show
that in order to calculate the hard-loop photon production rate
from an anisotropic plasma one need only know the retarded
self-energy. We plot the functions $\alpha$ for an anisotropy
parameter of $\xi=100$ and different angles $\theta_n$ in Figs.
\ref{fig:sigma12-0-xi100}, \ref{fig:sigma12-x-xi100} and
\ref{fig:sigma12-z-xi100} to emphasize the strong angular
dependence once more.

Now that we have calculated $ \Sigma_{12}$ and $\Sigma_{21}$ and expressed it in terms of the retarded self energy:
\begin{equation}
    \Sigma_{12}=-\Sigma_{21}=-i\,\text{Im} \Sigma\,,
\end{equation}
we can rewrite Eq.\,(\ref{fdr}) to find
\begin{equation}
    \left.i\,S_{12}^\star\right|_{HL}=\left.-i\,S_{21}^\star\right|_{HL}=\Lambda_{\alpha \beta \gamma}(P)\gamma^\alpha
    \gamma^\beta \gamma^\gamma\,,
\end{equation}
where
\begin{equation}\label{lambdadef}
    \Lambda_{\alpha \beta
    \gamma}(P)=\frac{P_\alpha-\Sigma_\alpha(P)}{(P-\Sigma(P))^2}\,\text{Im}\,\Sigma_\beta(P)\frac{P_\gamma-\Sigma^*_\gamma(P)}{(P-\Sigma^*(P))^2}\,,
\end{equation}
with the complex conjugate of the self energy $\Sigma^*$.
With that Eq.\,(\ref{iPi}) becomes
\begin{align}\label{iPi2}
 i \Pi_{12}&_\mu^\mu(Q) = e^2 \frac{5}{9} N_c \int \frac{d^4 p}{(2\pi)^4}
 \text{Tr} \left[ \gamma^\mu \Lambda_{\alpha\beta\gamma}(P)\gamma^\alpha
    \gamma^\beta \gamma^\gamma \gamma_{\mu} i S_{21}(P-Q) \right.\notag\\
 &~~~~~~~~~~~~~~~~~~~~~~~~~~~~~~~~~~~~\left.- \gamma^\mu iS_{12}(P+Q) \gamma_\mu
 \Lambda_{\alpha\beta\gamma}(P)\gamma^\alpha
    \gamma^\beta \gamma^\gamma
 \right]\,.
\end{align}
In the HL approximation and assuming a ${\bf k}$ dependent
distribution function that satisfies $f({\bf k})=f(-{\bf k})$, we
can write for the free propagator
\begin{align}\label{freeHL}
    i
    S_{21}(P-Q)=S_{21\,\nu}(P-Q)\gamma^\nu&=-2\pi(\slashed{P}-\slashed{Q})\delta((P-Q)^2)f_F(P-Q)\notag\\
               &\simeq -2\pi(\slashed{P}-\slashed{Q})\frac{1}{2q}\delta(p_0-p ({\bf \hat{p}}\cdot {\bf \hat{q}}))f_F({\bf q})\,,\\
    i S_{12}(P+Q)=S_{12\,\nu}(P-Q)\gamma^\nu&=-2\pi(\slashed{P}+\slashed{Q})\delta((P-Q)^2)f_F(P+Q)\notag\\
    \label{freeHL2}
               &\simeq -2\pi(\slashed{P}+\slashed{Q})\frac{1}{2q}\delta(p_0-p ({\bf \hat{p}}\cdot {\bf \hat{q}}))f_F({\bf
               q})\,,
\end{align}
where the $\theta$-functions could be neglected in both cases
because $q_0\gg p_0$.

To evaluate the trace in Eq.\,(\ref{iPi2}), we use its cyclic invariance and the identities
\begin{align}
    \gamma^\mu\gamma^\alpha\gamma^\beta\gamma^\gamma\gamma_\mu&=\gamma_\mu\gamma^\alpha\gamma^\beta\gamma^\gamma\gamma^\mu=-2\gamma^\gamma\gamma^\beta\gamma^\alpha\,,\\
    \text{Tr}\left[\gamma^\gamma \gamma^\beta \gamma^\alpha \gamma^\nu\right]&=4\left(g^{\gamma\beta}g^{\alpha\nu}-g^{\gamma\alpha}g^{\beta\nu}+g^{\gamma\nu}g^{\beta\alpha}\right)\,.
\end{align}
This, together with Eqs.\,(\ref{freeHL}) and (\ref{freeHL2}), leads to the expression
\begin{align}\label{iPi3a}
 i \Pi_{12}&_\mu^\mu(Q) = -e^2 \frac{5}{9} N_c \frac{8 f_F({\bf q})}{q} \int \frac{d^3
 p}{(2\pi)^3}\,Q_\nu \tilde{\Lambda}^{\nu}(P)\,,
\end{align}
where
\begin{equation}\label{tillambda}
    \tilde{\Lambda}^{\nu}(P)=\left(\Lambda^{\nu\alpha}_{~~\alpha}(P)-\Lambda_\alpha^{~\nu\alpha}(P)
    +\Lambda_\alpha^{~\alpha\nu}(P)\right)_{p_0=p ({\bf \hat{p}}\cdot {\bf
    \hat{q}})}\,.
\end{equation}
This can be further simplified when realizing that $Q_\nu \tilde{\Lambda}^{\nu}=\text{Im}\left(Q_\nu \left. S_{\text{ret}}^{\star\,\nu} \right|_{HL}\right)$ \cite{Ipp:2007ng}\footnotemark.
The final expression then becomes
\begin{align}\label{iPi3}
 i \Pi_{12}&_\mu^\mu(Q) = -e^2 \frac{5}{9} N_c \frac{8 f_F({\bf q})}{q} \int^{p^*} \frac{d^3
 p}{(2\pi)^3}\,\text{Im}\left(Q_\nu \left. S_{\text{ret}}^{\star\,\nu} \right|_{HL}(P)\right)\,,
\end{align}
where the ultra-violet cutoff $p^*$ is imposed on the radial part of the integral. 
\footnotetext{I thank A. Ipp for pointing this out to me.}

\subsection{Total contribution}
The total photon rate is found by adding the hard
(\ref{hardphotonrate}) and soft (\ref{softphotonrate}) rates
\bqa E \frac{d R}{d^3q} = E \left( \frac{d R_{\rm hard}}{d^3q} +
\frac{d R_{\rm soft}}{d^3q} \right) \, .
\label{totalphotonrate}\eqa
In the following sections we will discuss its dependence on the
cutoff $p^*$ and compare to previously found analytic results \cite{Kapusta:1991qp,Baier:1991em} in
the isotropic limit.

As in the previous chapters we now choose the anisotropic
distribution functions to be given by isotropic distributions that
are squeezed or stretched along one direction in momentum space,
i.e., $$ f_i({\bf k},\xi,p_{\rm hard}) =
   f_{i,\rm iso}\left(\sqrt{{\bf k}^2+\xi({\bf k}\cdot{\bf \hat n})^2},p_{\rm hard}\right) ,
$$
where $i\in\{ q ,g \}$, $p_{\rm hard}$ is a hard momentum scale
which appears in the distribution functions, ${\bf \hat n}$ is the
direction of the anisotropy, and $\xi>-1$ is a parameter
reflecting the strength and type of the anisotropy~\cite{Romatschke:2003ms}. 

In the following we will
suppress the explicit dependence on $\xi$ and $p_{\rm hard}$. We
specialize to the case $\xi>0$ which is the relevant one for heavy-ion collisions after times
$\tau\,\gtrsim$ 0.2 fm/c. For the arbitrary isotropic distribution
functions, $f_{i,\rm iso}$, which are contracted along the
$z$-axis, we choose \textsc{Bose-Einstein} and
\textsc{Fermi-Dirac} distributions in the case of gluons and
quark/anti-quarks, respectively. 
\newpage
Using these distribution
functions we are able to reproduce the equilibrium results, which
have been obtained previously \cite{Baier:1991em,Kapusta:1991qp}
by taking the limit $\xi \rightarrow 0$.

\subsection{The isotropic limit} \label{sec:isotropiclimit} \index{Isotropic limit}
%%%%%%%%%%%%%%%%%%%%%%%%%%%%%%%%%%%%%%%%%%%%%%%%%%%%%%%%%%%%%%%%%%%%%%%%%%%%%%
\begin{figure}[htb]
\begin{center}
\includegraphics[height=7cm]{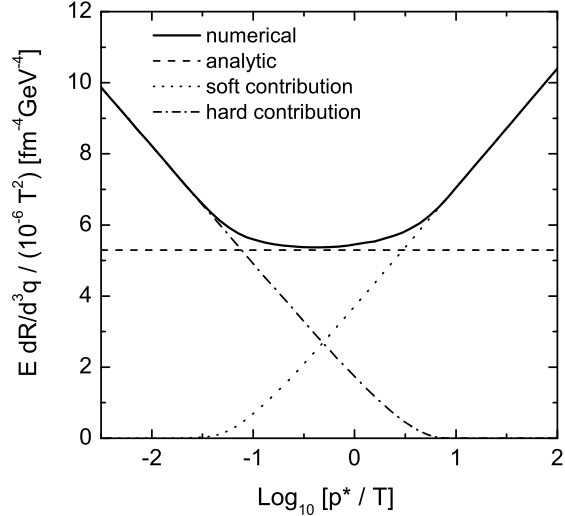}
\end{center}
\caption{Full photon rate for $\xi=0$ and $g=0.1$ as a function of
the cutoff $p^*$ compared to the totally cutoff independent
analytic approximation Eq. (\ref{isofull}). Around the geometrical
mean of the hard and soft momentum scale, $\sqrt{g}T$, the rate is
nearly independent of the cutoff.} \label{fig:pstarplot}
\end{figure}
%%%%%%%%%%%%%%%%%%%%%%%%%%%%%%%%%%%%%%%%%%%%%%%%%%%%%%%%%%%%%%%%%%%%%%%%%%%%%%
%%%%%%%%%%%%%%%%%%%%%%%%%%%%%%%%%%%%%%%%%%%%%%%%%%%%%%%%%%%%%%%%%%%%%%%%%%%%%%
\begin{figure}[htb]
\begin{center}
\includegraphics[height=7cm]{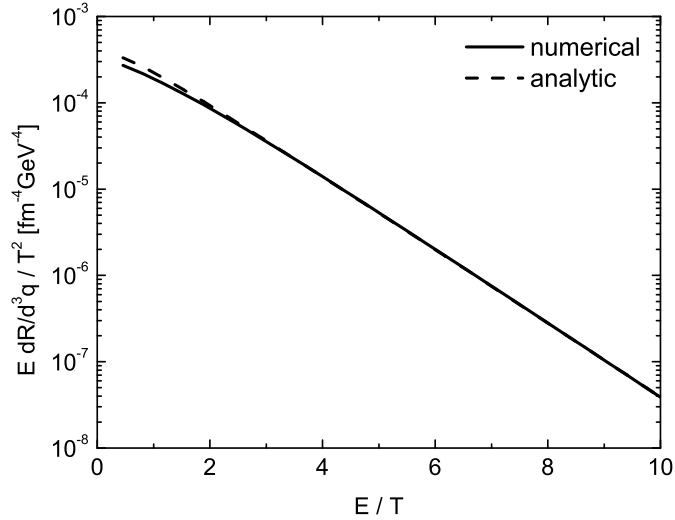}
\end{center}
\caption{Full photon rate for $\xi=0$ and $g=0.1$ as a function of
$E/T$ compared to the analytic expression Eq. (\ref{isofull}). The
cutoff dependence is smaller than the width of the full line.}
\label{fig:qplot}
\end{figure}
%%%%%%%%%%%%%%%%%%%%%%%%%%%%%%%%%%%%%%%%%%%%%%%%%%%%%%%%%%%%%%%%%%%%%%%%%%%%%%
In the
case of isotropic quark/anti-quark and gluon distribution
functions, Eqs. (\ref{eq:annihilation1}) and (\ref{eq:compton1})
can be further simplified by transforming to $s-$ and $t-$
variables as explained in \cite{Staadt:1985uc}. This has been done
in \cite{Baier:1991em} and Eq. (\ref{hardphotonrate}) becomes
\begin{equation}
E \frac{d R_{\rm hard}}{d^3q} \simeq
\frac{\frac{5}{9}\alpha\alpha_s}{2\pi^2} T^2
e^{-E/T}\left[\ln\frac{ET}{p^{*\,2}}+\frac{3}{2}+\frac{\ln
2}{3}-\gamma+\frac{\zeta'(2)}{\zeta(2)}\right]\,,\label{isohard}
\end{equation}
where $T$ is now the temperature of the equilibrated system,
$\gamma$ is \textsc{Euler}'s constant and $\zeta$ is the
\textsc{Riemann} zeta function:
$$ \frac{\zeta'(2)}{\zeta(2)}\simeq-0.57$$
They also calculated the soft part using hard thermal loop quark
self energies in a calculation analogous to ours in Sec.
\ref{sec:softpart}, but assuming equilibrium. The result is
\begin{equation}
E\frac{dR_{\rm soft}}{d^3 q} \simeq
\frac{\frac{5}{9}\alpha\alpha_s}{2\pi^2} T^2
e^{-E/T}\left[\ln\frac{p^{*\,2}}{m_q^2}-1.31\right]\,,\label{isosoft}
\end{equation}
with $m_q^2$ as defined in (\ref{mqdefinition}). When adding
(\ref{isohard}) and (\ref{isosoft}) the $p^{*\,2}$ under the
logarithms cancel exactly and the result is independent of the
cutoff:
\begin{equation}
E\frac{dR}{d^3 q} \simeq \frac{\frac{5}{9}\alpha\alpha_s}{2\pi^2}
T^2 e^{-E/T}\ln\left(\frac{0.23\,E}{\alpha_s\,
T}\right)\,,\label{isofull}
\end{equation}
which holds in the limit of arbitrarily small gauge coupling $g$
and cutoff $p^*$, as well as $E>T$.

We will now compare the limiting case of vanishing anisotropy of
the rate (\ref{totalphotonrate}) to (\ref{isofull}) as a
consistency check. First let us focus on the case of small
coupling, choosing $g=0.1$, and explore the cutoff dependence of
the numerically calculated rate (\ref{totalphotonrate}). It is
shown together with the comparison to the analytic approximation
given by (\ref{isofull}) in Fig. \ref{fig:pstarplot}. We find that
around the intermediate scale $\sqrt{g}T$, which is the
geometrical mean of the soft momentum scale $gT$ and the hard
momentum scale $T$, the rate develops a plateau close to the
analytic result. This is due to a cancellation of the
$p^*$-dependent parts in the soft and the hard contribution in
this regime: The same cancellation that happens in the analytic
result for all values of $p^*$. Outside the plateau the rate rises
logarithmically with $p^*$. A very similar behavior has been found
in the investigation of the energy loss of a heavy fermion in a
QED- or QCD-plasma by \textsc{Romatschke} and \textsc{Strickland}
\cite{Romatschke:2003vc,Romatschke:2004au}. To get a feeling for
the cutoff dependence, we evaluate the rate at the minimum
$p^*_{\text{min}}$, which we determine numerically and at two more
values, half the value at the minimum, $0.5 p^*_{\text{min}}$, and
twice that value, $2 p^*_{\text{min}}$. This choice for the
variation of $p^*$ is completely arbitrary - one could as well
choose a smaller value. However, we decided to be conservative and
use a factor of two. The $E/T$-dependence of the numerical result
compared to the analytic one is shown in Fig. \ref{fig:qplot}. The
cutoff dependence for the case of small coupling, $g=0.1$ is
smaller than the width of the line and hence does not show up as a
band in the plot.

\begin{figure}[htb]
\begin{center}
\includegraphics[height=7cm]{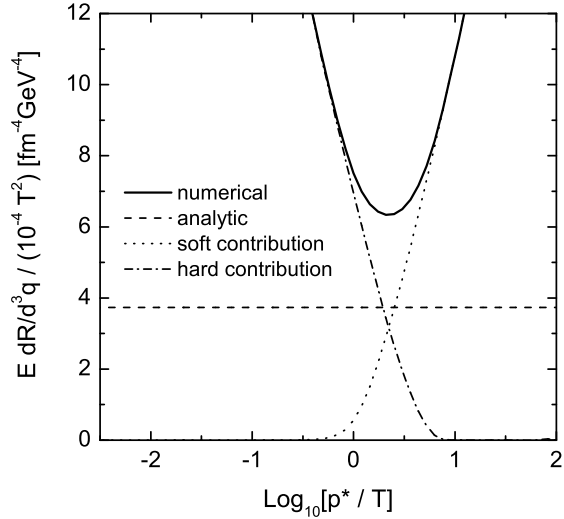}
\end{center}
\caption{Full photon rate for $\xi=0$ and $g=2$ as a function of
the cutoff $p^*$ compared to the totally cutoff independent
analytic approximation Eq. (\ref{isofull}). The plateau is now
merely a minimum, because its width decreases with increasing
coupling. Also, the difference to the approximate analytic result
increases.} \label{fig:pstar2plot}
\end{figure}

\begin{figure}[htb]
\begin{center}
\includegraphics[height=7cm]{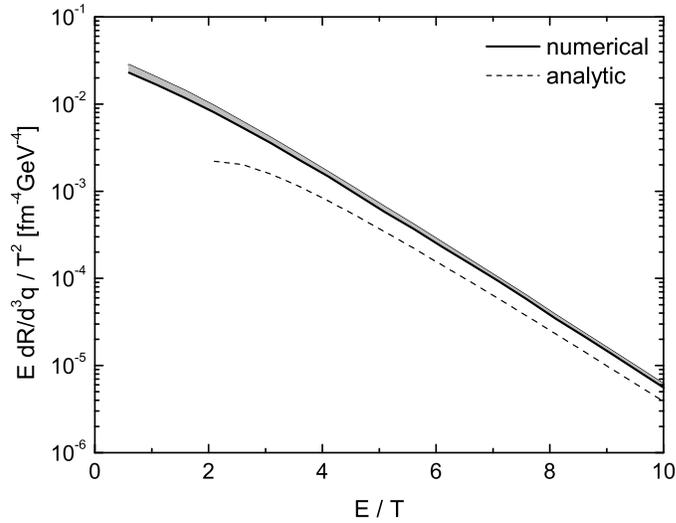}
\end{center}
\caption{Full photon rate for $\xi=0$ and $g=2$ as a function of
$E/T$ compared to the analytic expression Eq. (\ref{isofull}). The
grey band indicates the uncertainty of the result with the
variation of the cutoff $p^*$ by factors of two around the value
at the minimum. } \label{fig:q2plot}
\end{figure}

Next, we move on to a more realistic coupling of $g=2$ and do the
same as before, i.e., explore the cutoff dependence of the rate
(Fig. \ref{fig:pstar2plot}) and plot the rate versus $E/T$,
varying the cutoff by factors of two around the minimal value
(Fig. \ref{fig:q2plot}). Both figures show that the cutoff
dependence increases with increasing coupling $g$, which can be
expected since our theory is only precise for small couplings.
However, the \textsc{Braaten-Yuan}-Method provides a good means to
measure the uncertainty of the final results caused by applying
perturbation theory to systems with large coupling.
%%%%%%%%%%%%%%%%%%%%%%%%%%%%%%%%%%%%%%%%%%%%%%%%%%%%%%%%%%%%%%%%%%%%%%%%%%%%%%
%%%%%%%%%%%%%%%%%%%%%%%%%%%%%%%%%%%%%%%%%%%%%%%%%%%%%%%%%%%%%%%%%%%%%%%%%%%%%%
%%%%%%%%%%%%%%%%%%%%%%%%%%%%%%%%%%%%%%%%%%%%%%%%%%%%%%%%%%%%%%%%%%%%%%%%%%%%%%
%%%%%%%%%%%%%%%%%%%%%%%%%%%%%%%%%%%%%%%%%%%%%%%%%%%%%%%%%%%%%%%%%%%%%%%%%%%%%%

\section{Results for the anisotropic case}
%%%%%%%%%%%%%%%%%%%%%%%%%%%%%%%%%%%%%%%%%%%%%%%%%%%%%%%%%%%%%%%%%%%%%%%%%%%%%%
\begin{figure}[htb]
\begin{center}
\includegraphics[height=8cm]{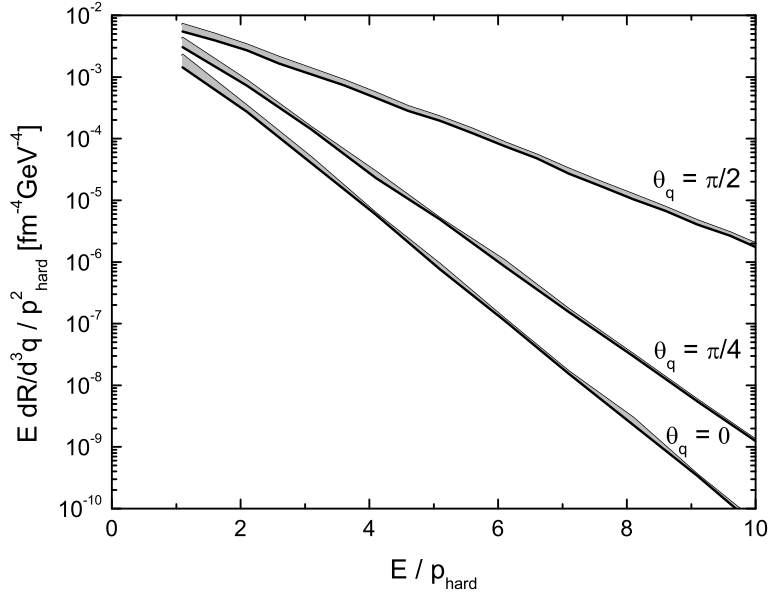}
\end{center}
\vspace{-5mm} \caption{Photon rate for $\xi=10$ and $\alpha_s=0.3$
as a function of energy for three different photon angles,
$\theta_q = \{0,\pi/4,\pi/2\}$, corresponding to rapidities $y =
\{\infty,0.88,0\}$, respectively.} \label{fig:qPlot}
\end{figure}
%%%%%%%%%%%%%%%%%%%%%%%%%%%%%%%%%%%%%%%%%%%%%%%%%%%%%%%%%%%%%%%%%%%%%%%%%%%%%%

%%%%%%%%%%%%%%%%%%%%%%%%%%%%%%%%%%%%%%%%%%%%%%%%%%%%%%%%%%%%%%%%%%%%%%%%%%%%%%
\begin{figure}[htb]
\begin{center}
\includegraphics[height=8cm]{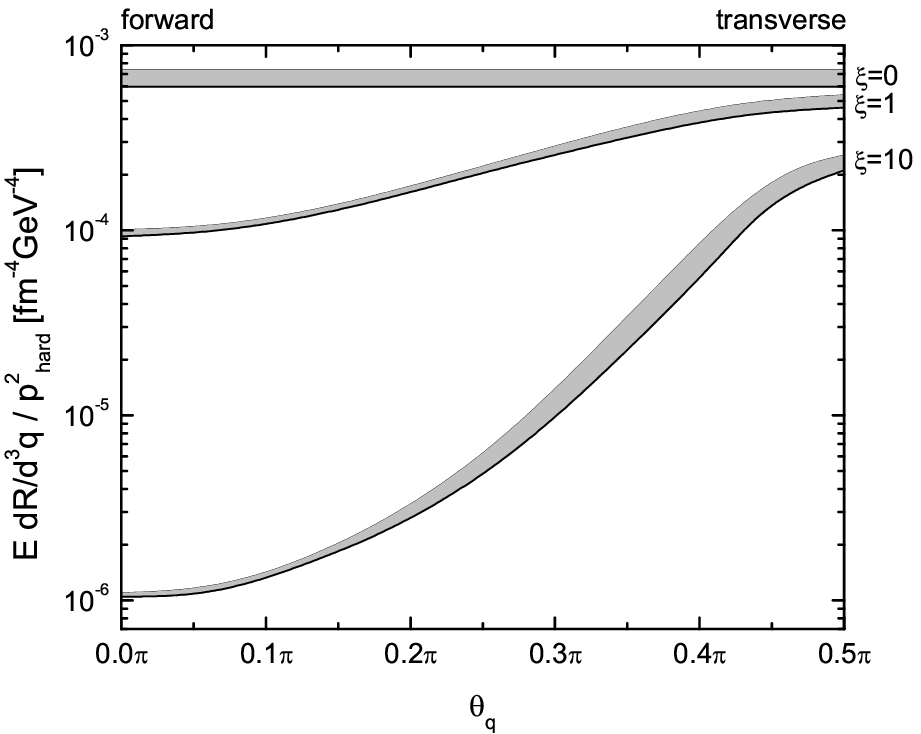}
\end{center}
\vspace{-5mm} \caption{Photon rate for $\xi=\{0,1,10\}$ as a
function of the photon angle, $\theta_q$, for $E/p_{\rm hard} = 5$
and $\alpha_s=0.3$.} \label{fig:thetaPlot}
\end{figure}
%%%%%%%%%%%%%%%%%%%%%%%%%%%%%%%%%%%%%%%%%%%%%%%%%%%%%%%%%%%%%%%%%%%%%%%%%%%%%%

In Fig.~\ref{fig:qPlot} we plot the dependence of the total
anisotropic photon production rate as a function of photon energy,
$E$, for three different photon propagation angles assuming
$\xi=10$ and $\alpha_s=0.3$.  The shaded bands indicate our
estimated theoretical uncertainty which is determined by varying
the hard-soft separation scale $p^*$ by a factor of two around its
central value which is parametrically $p^* \sim \sqrt{g}\,p_{\rm
hard}$.  We have scaled everywhere by the arbitrary hard-momentum
scale $p_{\rm hard}$ which appears in the quark and gluon
distribution functions.  This scale will be, in general, time
dependent with its value set to the nuclear saturation scale,
$Q_s$, at the earliest times and to the plasma temperature at late
times.  For RHIC $Q_s \sim 1.4\!-\!2\;{\rm GeV}$ and expected
initial plasma temperatures are $T_o \sim 300\!-\!400\;{\rm MeV}$.
As can be seen from this Figure there is a clear dependence of the
spectrum on photon angle with the difference increasing as the
energy of the photon increases.

To compare the angular dependence at different $\xi$ in
Fig.~\ref{fig:thetaPlot} we show the dependence of the photon
production rate on angle at a fixed photon energy $E/p_{\rm
hard}=5$ and $\alpha_s=0.3$.  As can be seen from
Fig.~\ref{fig:thetaPlot} the difference between the forward and
transverse production rates increases as $\xi$ increases.  To
summarize the effect we define the photon anisotropy parameter, $
{\cal R}_\gamma \equiv
\left(dR/d^3q|_{\theta_q=0}\right)/\left(dR/d^3q|_{\theta_q=\pi/2}\right)
$. This ratio is one at all energies if the plasma is isotropic.
For anisotropic plasmas it increases as the anisotropy of the
system and/or the energy of the photon increases.  Due to the
limited experimental rapidity acceptance one could define this
ratio at a lower angle, e.g.~$\pi/4$.  However, the ability to
resolve anisotropies increases as the sensitivity to forward
angles increases so the best experiment would be to compare the
most forward photons possible with transverse photon emission.

%%%%%%%%%%%%%%%%%%%%%%%%%%%%%%%%%%%%%%%%%%%%%%%%%%%%%%%%%%%%%%%%%%%%%%%%%%%%%%
\begin{table}[htb]
\begin{center}
\begin{tabular}{|c|c|c|c|}
\hline
${\cal R}_\gamma$ & $\;\xi=0\;$ & $\;\xi=1\;$ & $\;\xi=10\;$ \\
\hline
$\;E/p_{\rm hard}=5\;$ & 1 & 4.7 & $2\cdot10^2$ \\
$\;E/p_{\rm hard}=10\;$ & 1 & 34 & $3\cdot10^4 $ \\
\hline
\end{tabular}
\caption{Photon anisotropy parameter ${\cal R}_\gamma$ for
different photon energies and anisotropy parameters assuming
$\alpha_s=0.3$.} \label{tab:kappa}
\end{center}
\end{table}

\section{Total photon yield}
\label{sec:yields}
%%%%%%%%%%%%%%%%%%%%%%%%%%%%%%%%%%%%%%%%%%%%%%%%%%%%%%%%%%%%%%%%%%%%%%%%%%%%%%
%\begin{widetext}      \begin{center}
\begin{figure*}[t]
\includegraphics[width=15cm]{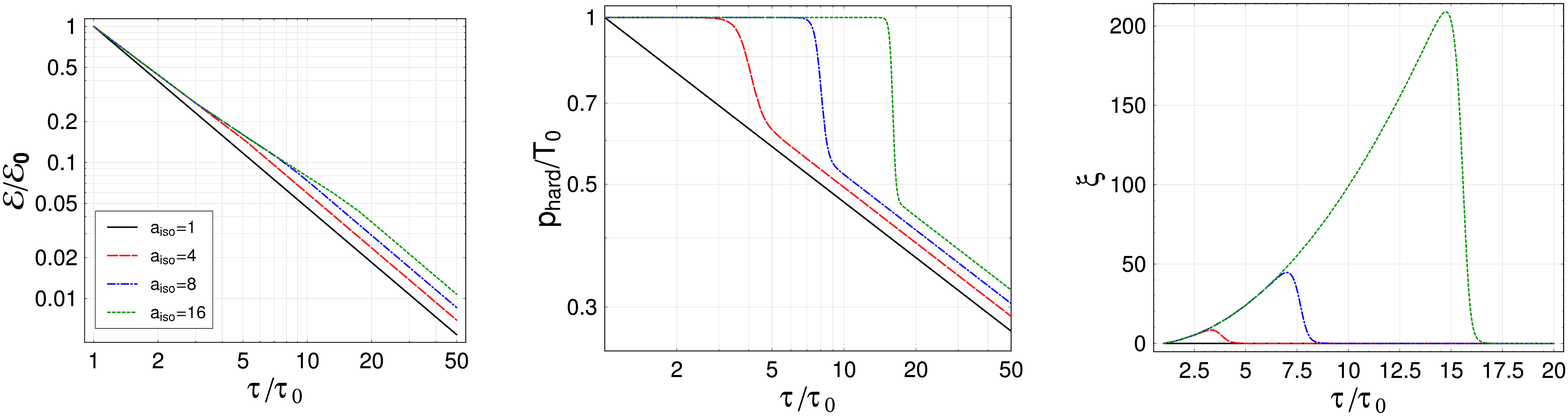}
\vspace{-2mm}
\caption{Model energy density (left), hard momentum scale (middle),
and anisotropy parameter (right) for four
different isotropization times $\tau_{\rm iso} \in \{0.1,0.4,0.8,1.6\}$ fm/c
assuming $\tau_0 =$ 0.1 fm/c.
The transition width is taken to be $\gamma = 2$. Initial energy density and $\tau_0$ are for LHC. \cite{Mauricio:2007vz}}
\label{fig:modelPlot}
\end{figure*}
%\end{center}
%\end{widetext}
%%%%%%%%%%%%%%%%%%%%%%%%%%%%%%%%%%%%%%%%%%%%%%%%%%%%%%%%%%%%%%%%%%%%%%%%%%%%%%
In order to quantify the effect of an evolving anisotropic
distribution function on experimental observables, we apply the
space time model introduced in \cite{Mauricio:2007vz} and \cite{Mauricio:new}. 
This model allows to interpolate
between the hydrodynamical and the free streaming limit, when
performing the integral over the space-time volume
\begin{equation}
 E\frac{dN}{d^3q}=\left.\frac{dN}{d^2q_\perp dy}\right|_{\text{CM}}=\pi R_A^2\int_{\tau_0}^{\tau_f}d\tau \tau\int_{-y_{\text{nucl}}}^{y_{\text{nucl}}}d\eta\left.E \frac{dR}{d^3q}\right|_{\text{LR}}\,,
\end{equation}
where $q_\perp$ is the transverse momentum of the photon,
$y=1/2\,\text{ln}[(E+q_z)/(E-q_z)]$ its rapidity, and $\eta=1/2\,
\text{ln}[(t+z)/(t-z)]$ its space-time rapidity.
$\tau=\sqrt{t^2-z^2}$ is proper time, and $R_A\approx\, 1.2
A^{1/3}$ is the radius of the nucleus in the transverse plane.
$y_{\text{nucl}}=\text{arccos}(\sqrt{s}/(2\,\text{AGeV}))$ is the
center-of-mass rapidity of the projectiles. Note that the final
yield is evaluated in the center of mass (CM) frame, while the
differential rate is calculated in the local rest (LR) frame of
the emitting fluid cell, such that the photon energy in the local
rest-frame is given by $E_{\text{LR}}=q_\perp
\text{cosh}(y-\eta)$, while its longitudinal momentum reads
$q_z=q_\perp \text{sinh}(y-\eta)$. We neglect transverse expansion
of the system, because we are interested in the early time
behavior for which it is negligible compared to the longitudinal
expansion \cite{Ollitrault:2007du}.

One limit that the space-time evolution can have is the
hydrodynamical expansion, for which the system is assumed to be
isotropic already at the partonic formation time
$\tau_0=\tau_\text{iso}$, and stay isotropic throughout the
evolution: $\xi(\tau)=0$. $\tau_0$ is the time after the coherence
effects in the nuclear wave functions can be ignored. Using 1+1
dimensional \textsc{Bjorken} expansion, the temperature drops like
$T(\tau)=T_0(\tau_0/\tau)^{1/3}$, with the initial temperature
$T_0$. Another limit is the free-streaming case, in which the
distribution function is a solution to the collisionless
\textsc{Boltzmann} equation
\begin{equation}
    p\cdot \partial_x f(p,x)=0\,.
\end{equation}
We assume that it is initially isotropic at time $\tau=\tau_0$ and then by longitudinal expansion becomes more and more anisotropic, such that we have
\begin{equation}
    f(p,x)_{\text{f.s.}}=f\left(\frac{p_\perp}{T_0}\sqrt{1+\frac{\tau^2}{\tau_0^2}\text{sinh}^2(y-\eta)}\right)\,,
\end{equation}
where the initial temperature represents the hard momentum scale and is equal to the one used in the hydrodynamic expansion.
The anisotropy parameter $\xi$ can be expressed via the transverse and longitudinal parton momentum by
\begin{equation}
    \xi=\frac{\langle p_\perp^2 \rangle}{2\,\langle p_L^2 \rangle} - 1\,.
\end{equation}
Using the nuclear saturation scale $Q_s$, and $\tau_0\sim
Q_s^{-1}$, $\langle p_\perp^2\rangle\sim Q_s^2$, and $\langle
p_L^2 \rangle\sim \tau^{-2}$, we find for the free streaming limit
\begin{equation}
    \xi(\tau)=\frac{\tau^2}{\tau_0^2} - 1\,.
\end{equation}
Given that we start with an isotropic distribution, this poses an
upper bound on the anisotropy parameter, due to  causality. The
temperature stays constant and isotropization is never reached:
$\tau_{\text{iso}}\rightarrow \infty$.

The interpolating model, described in detail in
\cite{Mauricio:2007vz} and \cite{Mauricio:new},
introduces the following time dependencies of the hard momentum
scale and the anisotropy parameter:
\begin{subequations}
\label{twoparams}
\begin{align}
p_{\rm hard}(\tau) &= \,T_0 \,\left( {\cal R}\!\left(a_{\rm iso}^2-1\right) \right)^{\lambda/4} \left(\frac{a_{\rm iso}}{a}\right)^{\lambda/3}  , \label{pdependence} \\
\xi(\tau) &= a^{2(1-\lambda)} - 1 ,
\label{xidependence}
\end{align}
where $\lambda(\tau,\tau_0,\tau_{\rm iso},\gamma) \equiv
\left\{{\rm tanh}\left[\gamma (\tau-\tau_{\rm iso})/\tau_0
\right]+1\right\}/2$, with the free parameter $\gamma$, which sets
the width of the transition, ${\cal R}(\xi) = \left[ 1/(\xi+1) +
{\rm arctan}\sqrt{\xi}/\sqrt{\xi} \right]/2$, $a \equiv
\tau/\tau_0$ and $a_{\rm iso} \equiv \tau_{\rm  iso}/\tau_0$.
\end{subequations}
Then the energy density behaves as follows
\begin{equation}
{\cal E}(\tau) = {\cal E}_{\rm f.s.}(\tau)\left( {\cal R}\left(a_{\rm iso}^2-1\right)\right)^\lambda\left(\frac{a_{\rm iso}}{a}\right)^{4\,\lambda/3}, \label{endependence}\\
\end{equation}
where ${\cal E}_{\rm f.s.}\sim \tau^{-1}$.
In the limit $\tau\ll \tau_{\text{iso}}$ the system behaves as in the free streaming limit, while for $\tau\gg \tau_{\text{iso}}$ it expands hydrodynamically. Fig. \ref{fig:modelPlot} shows the time dependence of ${\cal E}$,
$p_{\rm hard}$, and $\xi$ assuming $\gamma=2$ for different values of $\tau_{\rm iso}$.
%%%%%%%%%%%%%%%%%%%%%%%%%%%%%%%%%%%%%%%%%%%%%%%%%%%%%%%%%%%%%%%%%%%%%%%%%%%%%%

\begin{figure}[htb]
\begin{center}
\includegraphics[width=10cm]{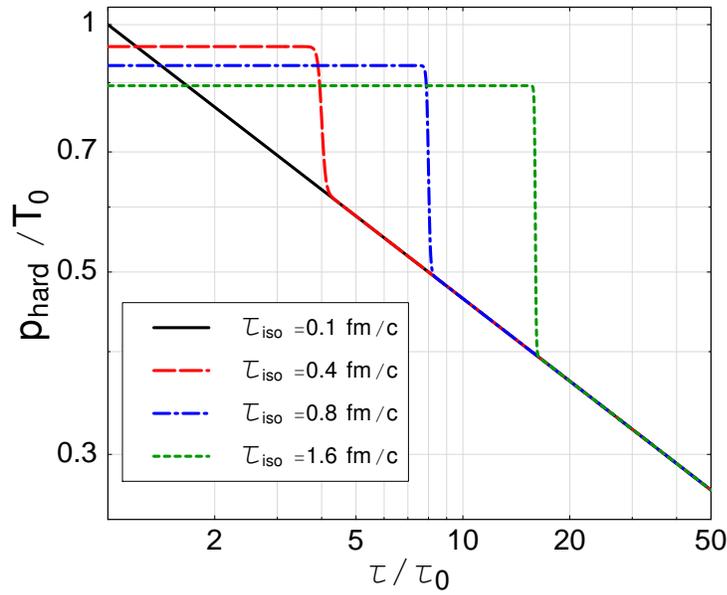}
\vspace{-2mm} \caption{Modification of the time dependence of the
hard momentum scale for four different isotropization times
$\tau_{\rm iso} \in \{0.1,0.4,0.8,1.6\}$ fm/c assuming $\tau_0 =$
0.1 fm/c and transition width $\gamma = 2$. $\tau_0$ is for LHC.}
\label{modifmodel}
\end{center}
\end{figure}

Implementing the model like this leads to different multiplicities
in the hydrodynamical phase for different $\tau_{\text{iso}}$.
However, what is measured in experiments is the multiplicity of
particles, such that we should have this quantity fixed to be able
to compare different model realizations as discussed in \cite{Mauricio:new}. 
To guarantee the same multiplicity
in the hydrodynamical phase for different $\tau_{\text{iso}}$, we
modify the time dependence of the hard momentum scale in the
following way: \bqa \label{phardmodification} p_{\rm hard}(\tau)\,
= \,T_0\, \frac{\left[ {\cal R}\!\left(a_{\rm iso}^2-1\right)
\right]^{\lambda/4} }{\left[ {\cal R}\!\left(a_{\rm
iso}^2-1\right) \right]^{1/4}}\left(\frac{a_{\rm
iso}}{a}\right)^{\lambda/3}\left(\frac{1}{a_{\rm
iso}}\right)^{1/3} \eqa Fig. \ref{modifmodel} shows the modified
ansatz (\ref{phardmodification}), which leads to the same
multiplicity in the hydrodynamical phase, but, as a consequence, a
different initial hard momentum scale and energy density for
different $\tau_{\text{iso}}$.

%%%%%%%%%%%%%%%%%%%%%%%%%%%%%%%%%%%%%%%%%%%%%%%%%%%%%%%%%%%%%%%%%%%%%%%%%%%%%%
%%%%%%%%%%%%%%%%%%%%%%%%%%%%%%%%%%%%%%%%%%%%%%%%%%%%%%%%%%%%%%%%%%%%%%%%%%%%%%

\begin{figure}[htb]
\begin{center}
\includegraphics[angle=270,width=10cm]{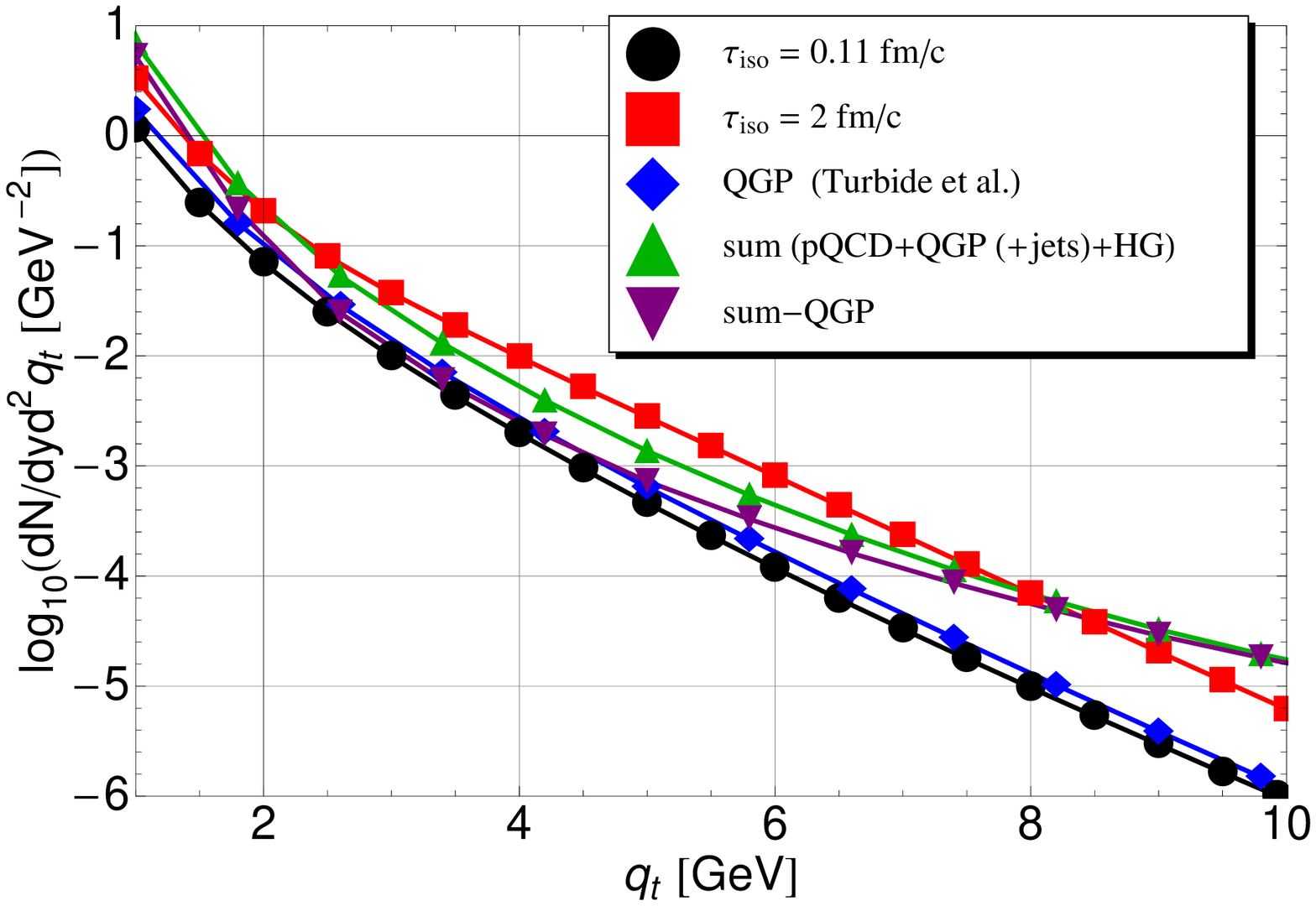}
\end{center}
\vspace{-5mm} \caption{Photon yield for fixed initial conditions.
$\sqrt{s}=5.5$ TeV, $T_0=845$ MeV, $y_\gamma=0$. The calculations
for $\tau_{\text{iso}}=\tau_0=0.11$ fm/c (Hydrodynamic case) and
$\tau_{\text{iso}}=2$ fm/c are compared to the results from
Turbide et al. \cite{Turbide:2005fk} where the thermal QGP result
(also with $\tau_{\text{iso}}=\tau_0=0.11$ fm/c) additionally
contains leading order bremsstrahlung and inelastic pair
annihilation contributions. The sum contains all contributions
from prompt photons (including jet fragmentation), jet
bremsstrahlung, thermal QGP, jet-photon conversion, and the hadron
gas phase. The last line shows the sum without the thermal QGP
contribution.} \label{fig:turbidecomp1}
\end{figure}

We now present results for the expected photon yields at
mid-rapidity for Pb-Pb collisions at full LHC beam energy of
$\sqrt{s}=5.5$ TeV. The critical temperature is chosen to be
$T_{\text{C}}=160$ MeV, which sets $\tau_f$, the time at which all
emission from the QGP-phase stops. The initial time is
$\tau_0=0.11$ fm/c and the initial temperature $T_0=0.845$ GeV.
The same values as used in \cite{Turbide:2003si} and
\cite{Turbide:2005fk} to allow for a comparison. Fig.
\ref{fig:turbidecomp1} shows the results for fixed initial
conditions and parametrization (\ref{twoparams}), with
$\tau_{\text{iso}}=\tau_0=0.11$ fm/c, representing pure
hydrodynamical evolution and $\tau_{\text{iso}}=2$ fm/c including
an evolving anisotropic stage for 1.89 fm/c before the system
becomes isotropic and continues to expand hydrodynamically. We
find that mainly due to the constant initial hard momentum scale
the photon yield increases in the second scenario (red squares in
Fig. \ref{fig:turbidecomp1}). Compared to the pure hydrodynamical
evolution (black circles in Fig. \ref{fig:turbidecomp1}) the
increase is approximately a factor of 6 at $q_t=5$ GeV. The yield
even lies above the sum from all contributions considered by
\cite{Turbide:2005fk}, which apart from the thermal QGP
contributions includes those from jet-photon conversion and the
hadron gas phase. However, it should be compared to the sum of all
contributions without the thermal QGP part (purple triangles in
Fig. \ref{fig:turbidecomp1}, because it replaces the thermal
calculation (neglecting bremsstrahlung as discussed above). This
way we find it to be the dominating contribution in the range $1
{\rm ~GeV}<q_t<8$ GeV. Assuming that the background, mainly coming from
pion decays, can be subtracted efficiently, one can extract
information about the anisotropy of the system from photon yields.

Fixing the initial multiplicity in the hydrodynamical phase leads
to very similar results in the two scenarios of pure
hydrodynamical evolution and for an isotropization time of 2 fm/c.
This is mainly due to the now adjusted initial hard momentum scale
that goes down for later $\tau_{\text{iso}}$ (see Fig.
\ref{modifmodel}). The most realistic scenario lies somewhere
between the two presented parametrizations. The soft particles,
which are responsible for the main part of the entropy production,
isotropize earlier while the hard particles remain anisotropic
longer and dominate the high momentum photon production. Hence,
$\tau_{\text{iso}}(p_{\text{soft}})$ should be used for fixing the
initial multiplicity, while $\tau_{\text{iso}}(p\geq
p_{\text{soft}})>\tau_{\text{iso}}(p_{\text{soft}})$ should be
used in the calculation of the yield.
In addition, modifications to the free streaming 
parametrization of $\xi$ follow from calculations in \cite{Bodeker:2005nv,Arnold:2005qs}.

%%%%%%%%%%%%%%%%%%%%%%%%%%%%%%%%%%%%%%%%%%%%%%%%%%%%%%%%%%%%%%%%%%%%%%%%%%%%%%
%%%%%%%%%%%%%%%%%%%%%%%%%%%%%%%%%%%%%%%%%%%%%%%%%%%%%%%%%%%%%%%%%%%%%%%%%%%%%%

%%%%%%%%%%%%%%%%%%%%%%%%%%%%%%%%%%%%%%%%%%%%%%%%%%%%%%%%%%%%%%%%%%%%%%%%%%%%%%
\begin{figure}[htb]
\begin{center}
\includegraphics[angle=270,width=10cm]{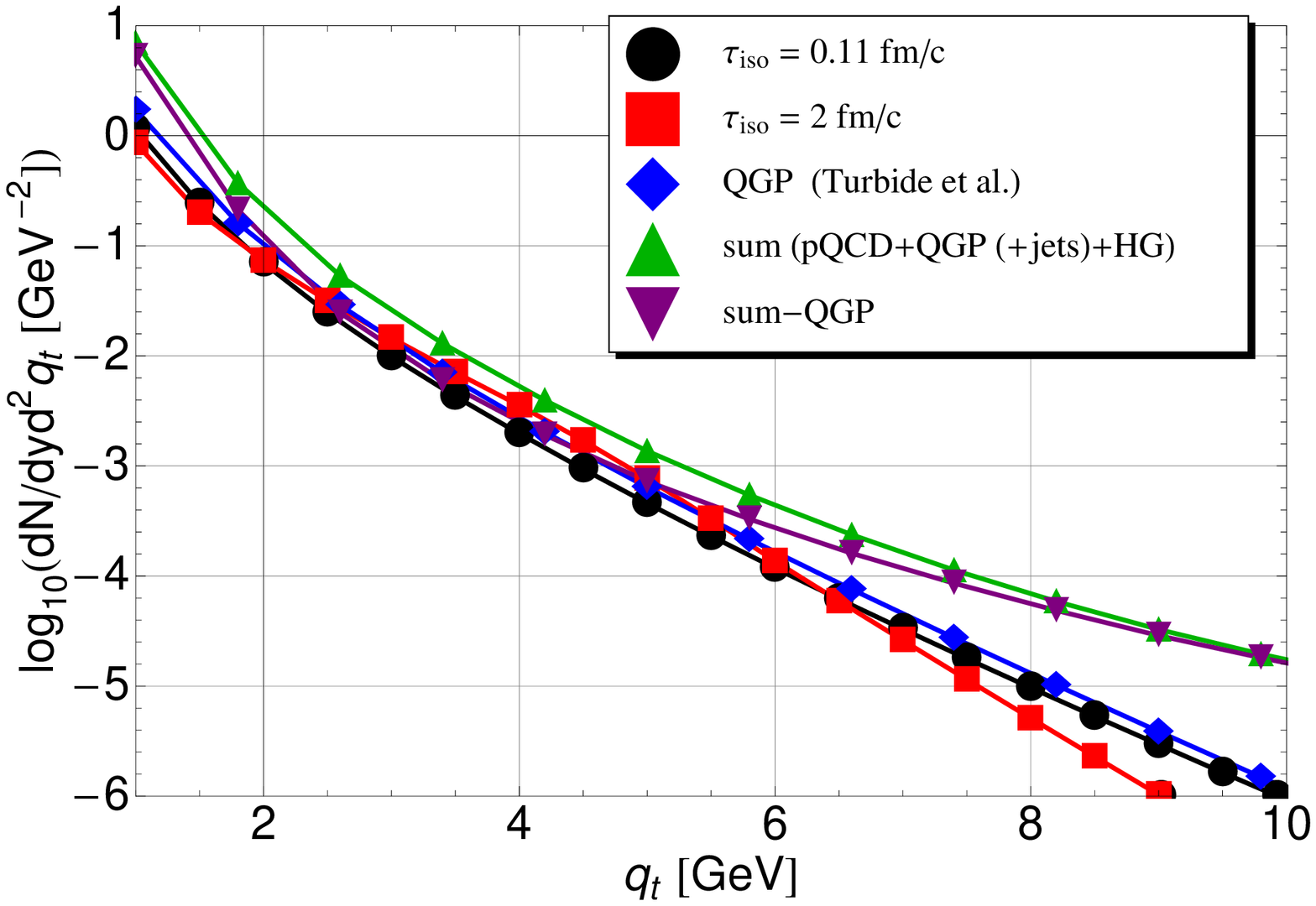}
\end{center}
\vspace{-5mm} \caption{Same as Fig. \ref{fig:turbidecomp1} but for fixed initial multiplicity in the hydrodynamic phase. Again the last three curves are results obtained by Turbide et al. \cite{Turbide:2005fk}.} \label{fig:turbidecomp2}
\end{figure}
%%%%%%%%%%%%%%%%%%%%%%%%%%%%%%%%%%%%%%%%%%%%%%%%%%%%%%%%%%%%%%%%%%%%%%%%%%%%%%

\section{Conclusions}
\label{sec:conclusions} We showed that the differential
high-energy medium photon production rate is sensitive to
quark-gluon plasma momentum-space anisotropies with the
sensitivity increasing with photon energy. In order to translate
this behavior into an experimental observable, we applied a
phenomenological model that takes into account the evolution of
the momentum-space anisotropy of the parton distribution
functions, and calculated the photon yield. For the same initial
hard momentum scale and energy density, the scenario with an
initial anisotropy lead to an approximately 6 times larger yield
above $q_t=4$ GeV, as compared to the scenario with instantaneous
isotropization and a purely hydrodynamical evolution. Fixing the
initial multiplicity of the hydrodynamical phase reduced this
enhancement dramatically, since the initial energy density and
hard momentum scale were reduced for a larger isotropization time.
However, since the isotropization time should be a function of the
particle momenta, the correct answer should lie somewhere between
the two simplified scenarios. In addition, other observables might 
help to remove the ambiguity in the plasma initial conditions at $\tau\sim Q_s^{-1}$.

In summary, we found that although the strong dependence on the
system's anisotropy of the rate is reduced significantly when
integrating over the space-time evolution of the system, there is
still an effect on the observable photon yield. Of course,
extracting those differences of factors of $2-6$ in the photon
yield, is a daunting task experimentally, especially due to the
large background from decay photons that we have not shown here.

      \chapter{Conclusions and Outlook}
\label{summary}
\epigraphwidth 250pt \epigraph{Wie alles sich zum Ganzen webt,\\
Eins in dem andern wirkt und lebt!}{\emph{Faust. Eine Trag\"odie, Nacht. Faust.}\\
Johann Wolfgang von Goethe (1749-1832)}

In this work, we have investigated the collective modes of the
pre-equilibrium quark-gluon plasma, as created in heavy-ion
collisions. We concentrated particularly on the unstable modes, the
so-called chromo-\textsc{Weibel}-instabilities, which could provide an
explanation for the fast isotropization and thermalization of the
QGP. After having derived the framework of kinetic theory for
non-Abelian plasmas, we reviewed the extension to
anisotropic systems, in which plasma instabilities are generically
present.
To answer the question, whether plasma instabilities can actually
contribute significantly to isotropization and equilibration during
the early stage of a heavy-ion collision, we studied the effect 
of collisions among the particles, which are
expected to reduce the growth rate of the unstable modes due to
their randomizing effect.
At infinitely small coupling, instabilities dominate and
collisions among the particles are parametrically suppressed.
However, at couplings present in heavy-ion collision
experiments, they become important. 
Additionally, the equilibration due to instabilities
only happens indirectly, because the instabilities driven
isotropization is a mean-field reversible process, which
does not produce entropy. Collisions, being responsible
for the dissipation are needed to reach the equilibrium state
of maximum entropy.
We investigated the effect of collisions on the unstable modes 
in a model calculation introducing a BGK-collision-term (similar to the
relaxation time approximation) in the hard-loop limit, as well as
in a classical real-time lattice calculation beyond the hard-loop limit,
which allows to simulate the full dynamics of the system. In both
cases it turns out that the randomizing effect of the collisions
slows down the growth of the unstable modes. Furthermore, it was shown
that there exists an upper limit of the collisional frequency beyond which 
the instabilities cease to exist.
However, they are still present for a reasonable estimate of the collision rate
in heavy-ion collisions, and were shown to continue to dominate the isotropization of the system in
the numerical simulation.

The introduction of hard binary collisions to the
colored-particle-in-cell simulation of the
\textsc{Wong-Yang-Mills} system lead to a new kind of
parton-cascade including particle-field interactions in addition
to the usually present collision term. With this advance the simulation
does not have to implicitly rely on assumptions from
equilibrium physics, such as thermal \textsc{Debye}-screening, to
regulate infrared problems encountered in the existing cascade
codes. This way one can study non-equilibrium dynamics, for 
which the screening mass can become imaginary, because the soft
momentum exchanges are mediated by the self-consistently generated
fields and do not have to be screened artificially. This also means 
that the simulations are capable of describing collective phenomena
like plasma instabilities, which is impossible in usual
parton-cascade simulations.

In the simulation, interactions with momentum exchange below the
separation momentum are mediated by the fields, whereas higher
momentum scatterings are described by the collision term. A
physically sensible choice for the separation scale is the
temperature (or hard momentum scale for a non-equilibrium system),
because below the temperature occupation numbers are large and the
description of the degrees of freedom by classical fields becomes
valid. Above the temperature, occupation numbers are small and the
used collision term describes the interactions more appropriately.
We showed that the momentum space diffusion of high transverse
momentum particles is independent of the lattice spacing, and
hence the separation scale between the field and particle
description, when matching the energy densities of the fields and
particles appropriately. This lead to the first cutoff independent estimate 
for the transport coefficient $\hat{q}$.
Another important result is that the
contribution from low momentum exchange scatterings, mediated by
the fields, is not negligible (contributes about 25\%) when the separation
momentum is of the order of the temperature. 
%This may cast some doubt
%on the validity of the usual parton-cascade approach, where these
%scatterings are strongly suppressed by using thermal
%\textsc{Debye}-screening.

Extending the investigation of hard probes to pre-equilibrium
situations, we studied the effect of plasma instabilities
on jet propagation. Interestingly, the domains of growing fields
turn out to bend transversely directed particles stronger along the beam direction than in the transverse
plane. The reason is the creation of large domains of strong chromo-electro-magnetic fields, with $B_\perp>E_\perp$ and $E_z>B_z$, which cause the observed broadening.
This is not obvious for a non-Abelian plasma, because as opposed to
electromagnetic plasmas all field components grow during instability evolution.
This finding provides a possible explanation (or at least a contribution to the full effect)
for the experimental observation that high-energy jets traversing the
plasma perpendicular to the beam axis experience much stronger
broadening in rapidity than in azimuth.

In the future, the extended CPIC-simulation will be used to further
the understanding of energy deposition of jets in the medium and 
study medium reponse to the jets. In addition, one can study the 
shear-viscosity to entropy ratio of the simulated system, a measure 
that is used to determine whether the system is strongly coupled or not.

To complete our investigation of unstable modes in the hard-loop
limit, we extended the exploration of the collective modes of an
anisotropic quark-gluon plasma by studying the quark collective
modes. Specifically, we derived integral expressions for the quark
self-energy for arbitrary anisotropy and evaluated these
numerically. In the direct numerical calculation only real
time-like fermionic modes and no unstable modes were found.
Additionally, using complex contour integration, we have proven
analytically for the cases when the wave vector of the
collective mode is parallel to the anisotropy direction with
arbitrary oblate distributions, and for all angles of propagation
in the limit of an infinitely oblate distribution, that there are no
fermionic instabilities.

Since we want to determine the the role of instabilities
for equilibration in heavy-ion collisions, the knowledge 
of the momentum-space anisotropy is particularly important.
This is due to the observation that the growth rate depends 
strongly on the degree of anisotropy. 
Therefore, we turned to the possible experimental determination of
the system's anisotropy using photon production. 
Electromagnetic probes are sensitive to the early phases of a heavy-ion
collision because of their long mean free path, which allows them to leave
the created medium without further interaction.
This may provide an undistorted signal of the
momentum-space anisotropies in the quark and gluon distribution
functions. 
It turned out that the differential high-energy medium
photon production rate is highly sensitive to quark-gluon plasma
momentum-space anisotropies with the sensitivity increasing with
photon energy. However, the strong dependence on the system's
anisotropy of the rate is reduced significantly when integrating
over the space-time evolution of the system. Principally, the
photon yield could provide a means to extract the system's
anisotropy, however, extracting the differences of factors of
$2-6$ in the yield is a daunting task experimentally, especially
due to the large background. 

The next step towards a complete description
of photon production from an anisotropic quark-gluon plasma should
be the determination of additional photon sources, like that from jet-medium interaction, 
in the presence of a momentum-space anisotropy.

%  \addchap{}
%      \input{}

  \appendix
  \cleardoublepage
  \addappheadtotoc
  \appendixpage
  \chapter{Gauge covariance of the Coulomb type gauge fixing term}
\index{Gauge fixing} \label{appgaugecov} We show that the \textsc{Coulomb}
type gauge fixing term
\begin{equation}\label{agft}
    G^a\equiv\left[D_i[A],a^i\right]^a=\partial^ia_i^a-gf^{abc}A_i^ba^{i\,c}
\end{equation}
transforms covariantly under infinitesimal gauge transformations
of the various gauge fields:
\begin{align}\label{agt}
    A_\mu & \rightarrow h A_\mu h^{\dag}-\frac{i}{g}h\partial_\mu
    h^{\dag}\notag\,,\\
    a_\mu &\rightarrow h a_\mu h^{\dag}\,,
\end{align}
with
\begin{equation}\label{ah}
h(x)=\exp(i\theta^a(x)T^a)=
1+i\theta^a(x)T^a+\mathcal{O}(\theta^2)\,.
\end{equation}
That is,
\begin{align}
    A_i^b&\rightarrow
    A_i^b-\frac{1}{g}\partial_i\theta^b(x)+f^{bde}A_i^d\theta^e(x)\,,\notag\\
    a_i^a&\rightarrow a_i^a-f^{abc}\theta^b(x)a_i^c\,.
\end{align}
Under these transformations the gauge fixing term (\ref{agft})
transforms into
\begin{align}\label{agtlong}
    \partial^ia_i^a-gf^{abc}A_i^ba^{i\,c} &\rightarrow \,\partial^i
    a^a_i-f^{abc}\partial^i\left(\theta^b(x)a_i^c\right)-gf^{abc}A_i^b
    a^{i\,c}+f^{abc}\left(\partial_i\theta^b(x)\right)a^{i\,c}\notag\\
    &~~~~~~-g f^{abc}f^{bde}A_i^d\theta^e(x)a^{i\,c}+gf^{abc}f^{cde}A_i^b\theta^d(x)a^{i\,e}+\mathcal{O}(\theta^2)\notag\\
    &=\,\partial^i a^a_i -gf^{abc}A_i^b
    a^{i\,c}-f^{abc}\theta^b(x)\partial^ia_i^c\notag\\
    &~~~~~~-g
    f^{aec}f^{ebd}A_i^b\theta^d(x)a^{i\,c}+gf^{abe}f^{edc}A_i^b\theta^d(x)a^{i\,c}+\mathcal{O}(\theta^2)\,,
\end{align}
where we renamed color indices in the second step. If this result
is equal to
\begin{equation}\label{ahgh}
    h G h^{\dag}\,,
\end{equation}
we have shown that $G$ transforms covariantly under the gauge
transformations (\ref{agt}). Using (\ref{ah}) we can write
(\ref{ahgh}) as
\begin{align}\label{atrans}
    &\partial^ia_i^a\,T^a+i\theta^b(x)\partial^ia_i^c\,[T^b,T^c]-
    gf^{abc} A_i^ba^{i\,c} \,T^a+i g
    f^{ebc}\theta^d(x)A_i^ba^{i\,c}\,[T^e,T^d]+\mathcal{O}(\theta^2)\notag\\
    &=\partial^ia_i^a\,T^a-g f^{abc} A_i^ba^{i\,c} \,T^a-f^{abc}\theta^b(x)\partial^ia_i^c\,T^a
    -g f^{ebc}f^{aed}\theta^d(x)A_i^ba^{i\,c}\,T^a+\mathcal{O}(\theta^2)\,.
\end{align}
Using the Jacobi identity
\begin{equation}
    f^{bec}f^{ade}+f^{aec}f^{dbe}+f^{dec}f^{bae}=0\,,
\end{equation}
we can rewrite the last term in Eq. (\ref{atrans}) to read
\begin{equation}
    -g f^{ebc}f^{aed}\theta^d(x)A_i^ba^{i\,c}\,T^a=-g
    f^{aec}f^{ebd}A_i^b\theta^d(x)a^{i\,c}\,T^a+gf^{abe}f^{edc}A_i^b\theta^d(x)a^{i\,c}\,T^a\,,
\end{equation}
and have thereby shown that the right hand sides of Eqs.
(\ref{agtlong}) and (\ref{atrans}) are equivalent and hence that
\begin{equation}
    G\rightarrow hGh^{\dag}\,.
\end{equation}

  \chapter{Experimental evidence of the Weibel instability in plasma physics}
\index{Instabilities!Weibel instability} \label{experiment} To
stress that plasma instabilities are not merely theoretical ideas
but can be observed experimentally outside heavy-ion physics, we
quote recent results in plasma physics showing experimental
evidence for the occurance of \textsc{Weibel}-like instabilities.

In measurements of energetic electron beams generated from
ultrahigh intensity laser interactions the beams have been shown
to be unstable to filamentation instabilities in the regime where
the beam density approaches the density of the background plasma.
\cite{Tatarkis:2003,Bret:2005}

Development of laser systems producing pulses focused to extreme
intensities made possible the exploration of relativistic plasma
physics in the laboratory. When such high intensity laser pulses
are focused into a plasma, very energetic electrons, ions, and
gamma rays can be observed. One of the most exciting applications
for these exotic laser-produced plasmas is in the context of fast
ignition \cite{Tabak:1994,Kodama:2001} for inertial confinement
fusion. In this scheme, a high power laser generates a short
pulse, high current electron beam at the edge of a cold, highly
compressed plasma of deuterium and tritium. This electron
beam is then used to spark a burn wave able to propagate
throughout the fuel - consequently generating fusion energy. Since
current laser systems are unable to generate electron beams with
sufficient current density to be useful for fast ignition the
propagation of electron beams produced by state-of-the-art laser
systems in lower density plasmas has been studied
\cite{Tatarkis:2003}.
\begin{figure}[H]
    \begin{center}
        \includegraphics[width=19.5pc]{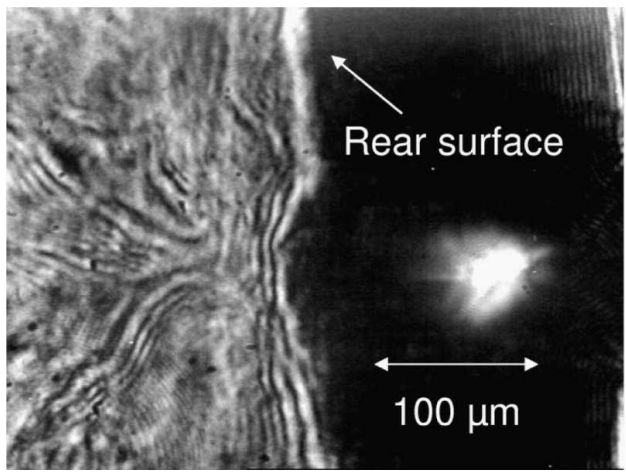}
        \caption{Shadowgraph at later times in helium (300 ps after
the main interaction) Target is 50 $\mu$m thick Mylar. The
filamentation is likely the result of a \textsc{Weibel}-like
instability. Picture taken from \cite{Tatarkis:2003}}
\label{fig:filamentsexperiment}
    \end{center}
\end{figure}
Fig. \ref{fig:filamentsexperiment} shows the propagation path of
an electron beam exiting a 50 $\mu$m Mylar target into a low
density helium plasma. The probe image shown is taken at late time
(300 ps after the pulse). In this experiment the electron beam
remains collimated over about 50 $\mu$m upon entering the gas
plasma after which it filaments. This was observed for
distances of the order of 100 $\mu$m which is about an order of
magnitude larger than the electron plasma wavelength. Clearly such
structures cannot be the result of energy deposition by radiation
or by neutral plasma ``jets'' - but rather of the beam of electrons
produced during the interaction. Also no significant proton or ion
acceleration was observed at the rear of the target during these
experiments so this is unlikely to have contributed to the
observed structure. These data were also taken before shock
breakout at the rear of the target which occurs several
nanoseconds later. Consequently, the observed filamentation in
Fig. \ref{fig:filamentsexperiment} is likely the result of a
\textsc{Weibel}-like instability \cite{Weibel:1959} which is
observed after the beam has propagated some distance in the
plasma. The \textsc{Weibel}- instability is observed as a
filamentation of the electron beam due to variations in the plasma
return current and will grow until the filament charge density
exceeds that of the return current.

This is the most direct observation of the
\textsc{Weibel}-instability, which in a similar form is expected
to be relevant for the evolution of the quark-gluon plasma, as
discussed in Section \ref{instabilities}.

  \chapter{Abelianization}
\label{app:abel}
We explain the phenomenon of ``Abelianization'' as described by \textsc{Arnold} and \textsc{Lenaghan} in \cite{Arnold:2004ih}.
At linear order in the hard-loop approximation unstable growth continues indefinitely. However, it is possible that the self interactions of the soft fields stop the instability growth. To see how this can happen, we analyze the effective potential energy without linearization in the gauge field $A^\mu$.
The hard-loop effective action for anisotropic distribution functions $f$ is given by \cite{Mrowczynski:2004kv}
\begin {equation}
   S_{\rm eff} =
   - \int_x \frac{1}{4} {F^a}_{\mu\nu} F^{a\mu\nu}
   - c g^2 \int_x \int_{\mathbf{p}}
     \frac{f(\mathbf{p})}{p} \, {F^a}_{\alpha\mu}(x)
     \left( \frac{v^\mu v^\nu}{(v\cdot D)^2} \right)_{ab}
     {{F^b}_\nu}^\alpha(x)\,.
\label {eq:Seff1}
\end {equation} 
The \textsc{Penrose} criterion for instabilities is fulfilled when the potential energy defined by the static limit ($\omega \rightarrow 0$) of the effective action has an unstable direction.
When we start with an oblate momentum distribution $f$ for that $p_z\ll p_x,p_y$ then the spectrum of unstable modes typically has $k_z\ll k_\perp$, see \cite{Arnold:2003rq}. That means, they vary more rapidly with $z$ than with $x$ or $y$. Let us therefore consider gauge field configurations that only depend on $z$:
\begin{equation}
 	\mathbf{A}=\mathbf{A}(z)\,,
\end{equation}
for which we will now simplify (\ref{eq:Seff1}).
First rewrite (\ref{eq:Seff1}) as 
\begin {equation}
   S_{\rm eff} =
   - \int_x \frac{1}{4} {F^a}_{\mu\nu} F^{a\mu\nu}
   - c g^2 \int_x \int_{\mathbf{p}}
     \frac{f(\mathbf{p})}{p} \, W^a_{\alpha}W^{a\,\alpha}\,,
\label {eq:Seff2}
\end {equation} 
where $W_\alpha=W_\alpha(x,\mathbf{v})=\frac{v^\mu}{v\cdot D}F_{\mu\alpha}(x)$, using the anti-symmetry of $\mathbf{v}\cdot\mathbf{D}$. For static configurations $\mathbf{A}(\mathbf{x})$ in $A_0=0$ gauge, we obtain the effective potential
\begin {equation}
   V_{\rm eff} =
   \int_{\mathbf{x}} \frac{1}{4} F^a_{ij} F^a_{ij} +  c g^2 \int_{\mathbf{x}} \int_{\mathbf{p}}
   \frac{f(\mathbf{p})}{p}\, W^a_k W^a_k\,,
\end {equation}
where
\begin {equation}
   W_k = W_k(\mathbf{x},\mathbf{v})
   = \frac{v_i}{\mathbf{v}\cdot\mathbf{D}} \, F_{ik}(\mathbf{x}) .
\end {equation}
Let us now specialize to $\mathbf{A}=\mathbf{A}(z)$, and use that
\begin {equation}
   v\cdot D \left(A^\alpha - \frac{n^\alpha v\cdot A}{n\cdot v}\right)
   = v^\nu {F_\nu}^\alpha\,,
\label {eq:BI}
\end {equation}
which holds when $A=A(n\cdot x)$ for some constant four-vector $n$ \cite{Blaizot:1994vs}.
Now apply $(v\cdot D)^{-1}$ to both sides of (\ref{eq:BI}) to get
\begin {equation}
   W^\alpha =
   A^\alpha - \frac{n^\alpha v\cdot A}{n\cdot v}\,,
\label {eq:copout}
\end {equation}
which in our case ($\mathbf{n}=\mathbf{e}_z$) becomes
\begin {equation}
   W_k = A_k - \delta_{kz} \, \frac{\mathbf{v}\cdot\mathbf{A}}{v_z} .
\end {equation}
Substituting this into the effective potential, we find that the second term is quadratic in $\mathbf{A}(z)$, meaning that it must be the same as in the linear approximation. Using the result of the linear theory we have
\begin {equation}
   V_{\rm eff} =
   \int_{\mathbf{x}} \frac{1}{4} F^a_{ij} F^a_{ij} +  \int_{\mathbf{x}} \frac{1}{2} A^a_i \Pi_{ij} A^a_j\,,
\end {equation}
where $\Pi$ is the self energy of the linearized theory.
In $\mathbf{k}$ space the second part can be written as
\begin {equation}
   \int_{\mathbf{k}} \frac{1}{2} \, A^a_i(\mathbf{k})^* \, \Pi_{ij}(0,\hat{\mathbf{k}}) \, A^a_j(\mathbf{k}) .
\end {equation}
Since the Fourier transform of $\mathbf{A} = \mathbf{A}(z)$ has support only for
$\mathbf{k}$'s proportional to $\hat{\mathbf{e}}_z$, we can replace $\Pi^{ij}(0,\hat{\mathbf{k}})$
by the matrix of constants $\Pi_{ij}(0,\hat{\mathbf{e}}_z)$.
The effective potential for $\mathbf{A} = \mathbf{A}(z)$ is then local
in $\mathbf{x}$ and may be written as
\begin {eqnarray}
   V[\mathbf{A}(z)] &=& \int_\mathbf{x} \left[
      \frac{1}{4} F^a_{ij} F^a_{ij}
       + \frac{1}{2} A^a_i \, \Pi_{ij}(0,\hat{\mathbf{e}}_z) \, A^a_j \right]\,.
\label {eq:Vz}
\end {eqnarray}
The first term contains cubic and quartic interactions, while the second term, representing the effects of hard particles, is quadratic in $\mathbf{A}$ as in the linearized theory.
In order to study the stability of the system, we consider the effective potential for the low momentum modes $k\rightarrow 0$:
\begin {eqnarray}
   {\cal V} &=&
       - \frac{1}{4} \, g^2 [A_i,A_j]^a [A_i,A_j]^a
       + \frac{1}{2} A^a_i \, \Pi_{ij}(0,\hat{\mathbf{e}}_z) \, A^a_j ,
\nonumber\\
   &=&
       \frac{1}{4} \, g^2 f^{abc} f^{ade} A^b_i A^c_j A^d_i A^e_j
       + \frac{1}{2} A^a_i \, \Pi_{ij}(0,\hat{\mathbf{e}}_z) \, A^a_j ,
\end {eqnarray}
Let us now assume $f(\mathbf{p})$ to be axially symmetric about the $z$-axis and $\Pi_{ij}(0,\mathbf{e}_z)$ to have a negative eigenvalue, which is the case for oblate distributions.
Using the transversality of the hard-loop self-energy ($K_\mu \Pi^{\mu\nu}=0$) and its symmetry, we find $\Pi^{iz}=\Pi^{zi}=0$. Then we choose the other two eigenvectors to lie along the $x$- and $y$-axis and find
\begin {equation}
   {\cal V} =
       \frac{1}{4} \, g^2 f^{abc} f^{ade} A^b_i A^c_j A^d_i A^e_j
       - \frac{1}{2} \, \mu^2 (A^a_x A^a_x + A^a_y A^a_y) ,
\label {eq:pota}
\end {equation}
where
\begin {equation}
   \mu^2 \equiv -\Pi_{xx}(0,\hat{\mathbf{e}}_z) = -\Pi_{yy}(0,\hat{\mathbf{e}}_z) > 0.
\end {equation}
${\cal V}$ is unbounded below. In particular for an Abelian configuration, which is one for that all components $A_i$ of $\mathbf{A}$ commute (e.g. $A_i^a=A_i\delta^{a1}$), the quartic term vanishes, leaving
${\cal V} = - \frac{1}{2} \, \mu^2 (A^a_x A^a_x + A^a_y A^a_y)$, which goes to $-\infty$ as $A$ increases.
Note that this argument only holds as long as $A$ is small enough such that the assumption $\delta f \ll f$ is fulfilled.
The unstable growth of Abelian configurations should stop when $A$ gets as large as the scale $p_{\text{hard}}/g$. 

To visualize the topography of ${\cal V}$ for non-Abelian configurations we make some simplifying restrictions to $\mathbf{A}$: (i) assume $\mathbf{A}$ to lie in an SU(2) subgroup of color SU(3) and (ii) take $A_z=0$.
With (i) we can replace the $f^{abc}$ by $\varepsilon^{abc}$, use $\varepsilon^{abc}\varepsilon^{ade}=\delta_{bd}\delta_{ce}-\delta_{be}\delta_{cd}$, and get
\begin {equation}
   {\cal V} =
       \frac{1}{4} \, g^2 \left[(A_i^b A_i^b)^2 - A^b_i A^c_j A^c_i A^b_j\right]
       - \frac{1}{2} \, \mu^2 (A^a_x A^a_x + A^a_y A^a_y)\,.
\label {eq:pot}
\end {equation}
To make use of the symmetries we rewrite (\ref{eq:pot}) in the form
\begin{equation}
 	 {\cal V} = \frac{1}{4} g^2 \left\{\left[\text{tr}\left({\cal A}^\top{\cal A}\right)\right]^2-\text{tr}\left[\left({\cal A}^\top{\cal A}\right)^2\right]\right\}-\frac{1}{2} \mu^2\, \text{tr}\left({\cal A}^\top P^{(xy)}{\cal A}\right)\,,
\end{equation}
with
\begin {equation}
   {\cal A} = \begin{pmatrix}
                 A_x^1 & A_x^2 & A_x^3 \\[2pt]
                 A_y^1 & A_y^2 & A_y^3 \\[2pt]
                 A_z^1 & A_z^2 & A_z^3
               \end{pmatrix} \,, ~~~~   P^{(xy)} = \begin{pmatrix} 1 & & \\ & 1 & \\ & & 0 \end{pmatrix}\,,
\end {equation}
where the latter is the projector on the $x$-$y$-plane.
\begin{figure}[htb]
\begin{center}
\includegraphics[width=3.4in]{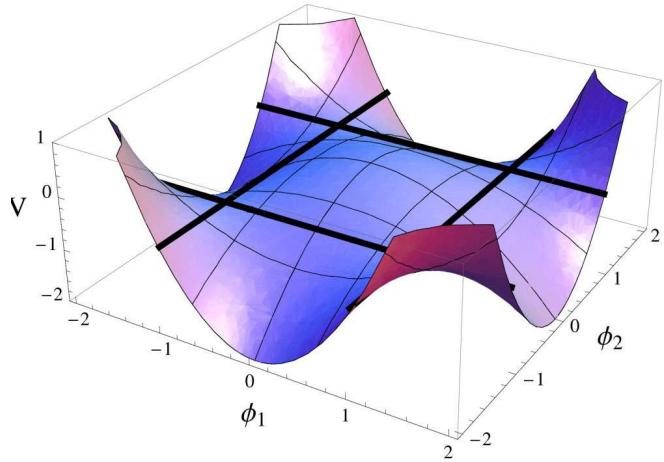}
\caption{Depiction of the potential $V(\phi_1,\phi_2)$
    of (\ref{eq:Vphi}).  The $\phi_1$ and $\phi_2$ axis are in units
    of $\mu/g$, and the values of $V$ are in units of $\mu^4/g^2$.
    There is a local maximum with ${\cal V}=0$ at the origin.
    The four straight lines in the plots correspond to
    the equipotential $V = -\mu^4/2g^2$, and
    the four intersection points of those lines
    are saddle points corresponding
    to static, unstable, non-Abelian configurations.\label{fig:potential}}
\end{center}
\end{figure}
${\cal V}$ is symmetric under spatial rotations in the $x$-$y$-plane and color rotations. Together we can write that
it is symmetric under the rotation ${\cal A} \rightarrow {\cal R A C}^\top$, where ${\cal C}$ is a color rotation for the adjoint representation of SU(2), represented by any $3\times 3$ real, orthogonal matrix with det${\cal C}=1$. ${\cal R}$ is a $3 \times 3$ real, orthogonal matrix, representing spatial rotations in the $x$-$y$-plane.
We can assume without loss of generality that ${\cal A}$ is symmetric, since this can always be achieved by a color rotation.
Restriction to $A_z=0$ makes ${\cal A}_{ia}$ zero except for $i=x,y$ and $a=1,2$. Then we can diagonalize ${\cal A}$ in its $2 \times 2$ subspace by a simultaneous space and color rotation ${\cal A}\rightarrow {\cal R A R}^\top$.
So we have for $k\rightarrow 0$
\begin{equation}	
 	A_i^a=\phi_1\delta_{ix}\delta_{a1}+\phi_2\delta_{iy}\delta_{a2}\,,
\end{equation}
and the potential becomes
\begin{equation}\label{eq:Vphi}
 	{\cal V}(\phi_1,\phi_2)=\frac{1}{2}g^2 \phi_1^2\phi_2^2-\frac{1}{2}\mu^2\left(\phi_1^2+\phi_2^2\right)\,.
\end{equation}
This potential is depicted in Fig. \ref{fig:potential}. The Abelian
configurations correspond to the $\phi_1$ axis ($\phi_2=0$)
and the $\phi_2$ axis ($\phi_1 = 0$).
The static non-Abelian solutions are indicated by
the intersection points of the four straight lines in the figure.
At these points, the amplitude of the gauge field is $A \sim \mu/g$.
Recalling that unstable modes typically have $k_{\text{soft}} \sim \mu$,
this corresponds to the non-Abelian scale $A \sim k_{\text{soft}}/g$.
However, these solutions are unstable to rolling down and
subsequently growing in amplitude along one of the axes.
The picture suggests that, if we start from $A$ near zero, the system
might possibly at first roll toward one of these configurations with
$A \sim k_{\text{soft}}/g$, but its trajectory would eventually roll away, growing along either the $\pm \phi_1$ or
$\pm \phi_2$ axis until the effective action breaks down at
$A \sim p_{\text{hard}}/g$. This effect of unstable growth of the gauge fields towards an Abelian configuration is called ``Abelianization''.

  \chapter{Hamiltonian formulation of lattice gauge theory}
\index{Kogut-Susskind Hamiltonian} \label{kogut} We present the
derivation of the Hamiltonian formulation of lattice gauge theory
based on the work by \textsc{Kogut} and \textsc{Susskind}
\cite{Kogut:1974ag} using simple correspondences to the rigid
rotator in three dimensions.

\section{Fermion fields on a lattice}
Although we restrict our investigations to pure gauge field dynamics, we now start with the lattice formulation for fermion fields for generality.
We denote a link between two lattice sites by $(\mathbf{r},\hat{m})$, where $\mathbf{r}$ denotes the position on the lattice, and $\hat{m}$ points
in one of the 6 directions emanating from that site. Defining the two-component spinor $\Psi(r)$, we can write down the discrete Hamiltonian that reproduces the \textsc{Dirac} theory in the continuum:
\begin{equation}
\label{H1}
H=a^{-1}\sum_{r,n}\Psi^{\dag}(r)\frac{{\boldsymbol\sigma}\cdot\mathbf{n}}{i}\Psi(r+n)+m_0\sum_r(-1)^r\Psi^{\dag}(r)\Psi(r)\,,
\end{equation}
with lattice spacing $a$.\\
The transformation of the fermion field under \emph{global} gauge transformations reads
\begin{equation}
    \Psi'(r)=e^{i{\boldsymbol\tau}\cdot{\boldsymbol\alpha}/2}\Psi(r)\equiv V\Psi(r)\,,
\end{equation}
where ${\boldsymbol\tau}$ is a vector of generators of the group and the parameter ${\boldsymbol\alpha}$ the analog of a vector of rotation angles.
The Hamiltonian (\ref{H1}) is invariant under such transformations. However, under \emph{local} transformations
\begin{equation}
    \Psi'(r)=e^{i{\boldsymbol\tau}\cdot{\boldsymbol\alpha}(r)/2}\Psi(r)\equiv V(r)\Psi(r)\,,
\end{equation}
this is not the case, since $H$ in (\ref{H1}) contains products of fields at separated points.
Note that without loss of generality we concentrate on the invariance under spatially dependent gauge transformations.
Then we can set the time component of the vector potential to zero when the gauge field enters the theory.
To compensate the lack of local invariance, we introduce a gauge field $\tilde{\mathbf{A}}$ and place it on each link.
Furthermore, on each link we define a unitary transformation:
\begin{equation}
\label{ymtrafo}
    U(r,m)=\exp\left(\frac{i}{2}{\boldsymbol\tau}\cdot\tilde{\mathbf{A}}(r,m)\right)\,.
\end{equation}
To place the gauge field on the link is natural because it
transports color information between lattice points. The two
indices of the matrix $U(r,m)^i_{~j}$ can be identified with the
two ends of the link $m$.

A gauge transformation acts on $U$ like
\begin{equation}
    U'(r,m)=V(r)U(r,m)V^{-1}(r+m)\,.
\end{equation}
We can now use $U$ to convert non-gauge-invariant products of spatially separated fields to gauge-invariant products as initially intended. For example the non-gauge-invariant product
\begin{equation}
    \Psi^{\dag}(r)\Psi(r+m) \rightarrow \Psi^{\dag}(r) V^{-1}(r)V(r+m)\Psi(r+m)
\end{equation}
is rewritten as
\begin{align}
    \Psi^{\dag}(r)U(r,m)\Psi(r+m) \rightarrow & \Psi^{\dag}(r) V^{-1}(r)V(r)U(r,m)V^{-1}(r+m)V(r+m)\Psi(r+m)\notag\\
                         =\,& \Psi^{\dag}(r)U(r,m)\Psi(r+m)\,,
\end{align}
which is gauge-invariant. The gauge transformations acting on the
ends of the link undo the gauge transformations of the fermion
fields.

Let us render the Hamiltonian gauge-invariant:
\begin{equation}
    H=a^{-1}\sum_{r,m}\Psi^{\dag}(r)\frac{{\boldsymbol\sigma}\cdot\mathbf{m}}{i}U(r,m)\Psi(r+m)\,,
\end{equation}
where we dropped the mass term, which is not affected by the changes.
To check whether the continuum limit is correctly reproduced we first write
\begin{align}   H=&a^{3}\sum_{r,m}\Psi'^{\dag}(r)\frac{{\boldsymbol\sigma}\cdot\mathbf{m}}{i\,a}e^{i{\boldsymbol\tau}\cdot\tilde{\mathbf{A}}(r,m)/2}\left(1-e^{-i{\boldsymbol\tau}\cdot\tilde{\mathbf{A}}(r,m)/2}\right)\Psi'(r+m)\notag\\
&+a^3\sum_{r,m}\Psi'^{\dag}(r)\frac{{\boldsymbol\sigma}\cdot\mathbf{m}}{i\,a}\Psi'(r+m)\,,
\end{align}
with $\Psi'(r)=a^{-3/2}\Psi(r)$. To take the continuum limit, we
have to assume that
$(1-e^{-i{\boldsymbol\tau}\cdot\tilde{\mathbf{A}}(r,m)/2})$ tends
to 0 as $a\rightarrow 0$. Then we can expand the exponential in
$H$:
\begin{align}   H=&a^{3}\sum_{r,m}\Psi'^{\dag}(r)\frac{{\boldsymbol\sigma}\cdot\mathbf{m}}{i\,a}\left(i\frac{1}{2}{\boldsymbol\tau}\cdot\tilde{\mathbf{A}}(r,m)\right)\Psi'(r+m)\notag\\
&+a^3\sum_{r,m}\Psi'^{\dag}(r)\frac{{\boldsymbol\sigma}\cdot\mathbf{m}}{i\,a}\Psi'(r+m)\,.
\end{align}
For $a\rightarrow 0$,
\begin{equation}
    \Psi'^{\dag}(r)\frac{{\boldsymbol\sigma}\cdot\mathbf{m}}{i\,a}\Psi'(r+m)\rightarrow \overline{\Psi}'(r)\gamma_i\partial_i\Psi'(r)\,,
\end{equation}
the well known kinetic energy term. Reintroducing the mass term,
the full continuum Hamiltonian reads
\begin{equation}
    H=\int_r\left[\overline{\Psi}'(r) i \gamma_i \partial_i \Psi'(r)-\overline{\Psi}'(r)\gamma_i \frac{\tilde{\mathbf{A}}_i(r)}{a}\cdot\frac{{\boldsymbol\tau}}{2}\Psi'(r)+m_0\overline{\Psi}'(r)\Psi'(r)\right]\,,
\end{equation}
which we recognize as the one for usual \textsc{Yang-Mills} gauge theory when identifying
\begin{equation}
    \tilde{\mathbf{A}}_i(r)=ag\mathbf{A}_i(r)\,,
\end{equation}
with the vector potential $\mathbf{A}$ and the coupling constant $g$.

\section{Gauge field on a lattice and analogy to the rigid rotator}
In \textsc{Yang-Mills} theory the local degree of freedom, $U(r,m)$, is an element of the group. The $SU(2)$ algebra is isomorphic to that of $SO(3)$, the group of rotations in three-dimensional space. This is fortunate, since, being a non-Abelian compact group, the topology of the configuration space at a link is closed and nontrivial.
We use this correspondence to the quantum rigid rotator to gain some insight of the nature of the gauge-field degree of freedom.

A configuration of the rigid rotator is specified by a rotation from the space-fixed to the body-fixed axes. It may be
represented in the form
\begin{equation}
\label{rrtrafo}
    U_j=\exp(i\mathbf{T}_j\cdot{\boldsymbol\Omega})
\end{equation}
with $T_{j\,\alpha}, (\alpha=1,2,3)$ being the representation
matrices of the generators of the rotation group for angular
momentum $j$. $j=1/2$ corresponds to $2\times 2$ matrices, the
defining representation of $SU(2)$ \footnote[1]{This
representation is responsible for the name $SU$, an acronym for
"Special Unitary". Exponentiating the generators of the $j=1/2$
representation to get the representation of the finite group
elements gives the most general $2\times 2$ unitary matrices with
determinant 1. The "special" means that the determinant is equal
to 1 instead of being a complex number of absolute value 1.},
$j=1$ to $3\times 3$ matrices, $j=3/2$ to $4\times 4$ matrices and
so on. We define lower indices of matrices and vectors to refer to
space-fixed axes, and upper indices to refer to body-fixed axes.
So $V_i$ are the components of a vector in the space-fixed frame
and $(U_1)^l_{~i}V_i=V^l$ the corresponding body components.
Remembering that we associated the two indices of $U$ in the
\textsc{Yang-Mills} theory with the two ends of a link, we find a
first correspondence: Rotation from a space fixed to a body fixed
coordinate system for the rigid rotator corresponds to a
`rotation' from the beginning of a link to its end. Simultaneous
body and space rotations correspond to global gauge
transformations, while separate body and space rotations
correspond to local gauge transformations. The body (upper) index
corresponds to the final end of a link, the space (lower) index to
the beginning end of a link.
The angular velocity of the rigid
rotator is
\begin{equation}
    {\boldsymbol\omega}=\frac{d}{dt}{\boldsymbol\Omega}\,.
\end{equation}
The angular momentum $\mathbf{J}$, which is the generator of space
rotations, is given by $I{\boldsymbol\omega}$, where $I$ is the
moment of inertia of the rigid body. The Hamiltonian reads
\begin{equation}
\label{H2}
    H=\frac{J^2}{2I}=\frac{1}{2}I\omega^2\,.
\end{equation}
The action of a rotation of the space-axes on $U$ is given by left
multiplication of the appropriate rotation matrix, say $V$.
Similarly, that of a rotation of the body-axes relative to the
body is given by right multiplication. Considering a spherical
rotator (only a spherical rotator has invariance under rotation of
the body-axes), we find that invariance under separate space and
body rotations for the rigid rotator corresponds to local gauge
invariance in the \textsc{Yang-Mills} theory. The Hamiltonian
(\ref{H2}) is invariant under individual body and space rotations.
In fact, with $\boldsymbol{\mathcal{J}}=U_1\mathbf{J}$, the
generator of body rotations, (\ref{H2}) may be written as
\begin{equation}
\label{H2a}
    H=\frac{\mathcal{J}^2}{2I}\,.
\end{equation}
The body- (space-) fixed angular momenta,
$\boldsymbol{\mathcal{J}}$ ($\mathbf{J}$), correspond to the
generators of gauge transformations in the \textsc{Yang-Mills}
theory that rotate only one end of the link. Let us denote these
operators as $Q_-$ and $Q_+$, where the $-$ ($+$) indicates the
beginning (end) of a link. By correspondence to
\begin{equation}
    \mathcal{J}^{\alpha}=(U_1)^\alpha_{~\beta} J_\beta\,,
\end{equation}
we find
\begin{equation}
    Q_+^{\alpha}=(U_1)^\alpha_{~\beta} (Q_-)_\beta\,.
\end{equation}
We can further identify the total color carried by a link with the
difference $Q=Q_+-Q_-$. Another important correspondence can be
drawn from the expressions for rotations (\ref{ymtrafo}) and
(\ref{rrtrafo}):
\begin{align}
    {\boldsymbol\Omega} \rightarrow \tilde{\mathbf{A}}\,,
\end{align}
and hence
\begin{align}
\label{adot}
   \dot{{\boldsymbol\Omega}}= {\boldsymbol\omega} = \mathbf{J}/I \rightarrow
   \dot{\tilde{\mathbf{A}}}\,.
\end{align}

\section{Gauge-invariant space of states}
We have now shown the correspondence between a link in the
\textsc{Yang-Mills} theory with a rigid rotator in three
dimensions. Evidently the space of states of the
\textsc{Yang-Mills} theory is the product of an infinite number of
rigid rotators. The physical states are drawn from the space of
gauge-invariant states.

An arbitrary gauge transformation can be built from individual
gauge transformations at the points of the lattice. Therefore it
suffices to consider only a gauge transformation at a single
lattice site $i$. This gauge transformation affects all of the six
links that emanate from site $i$. Thus, the generator $G$ must be
equal to the sum of the generators $Q_+$ over the six links
\begin{equation}
    G(r)=\sum_m Q_+(r,m)\,.
\end{equation}
In (\ref{adot}) we showed that $\mathbf{J}/I$ corresponds to
$\dot{\tilde{\mathbf{A}}}$. Hence, by the correspondence between
$\mathbf{J}$ and $Q_+$, $Q_+$ is proportional to
$\dot{\tilde{\mathbf{A}}}$. Accordingly, the generator $G(r)$ may
be written as
\begin{equation}
    G(r)=\text{const}\times\sum_m \dot{\tilde{\mathbf{A}}}(r,m)\,.
\end{equation}
The time derivative of the vector potential can be identified with
the component of the non-Abelian electric field at
position $r$ in direction $m$. The sum of electric fields
emanating from a single lattice site is the lattice analog of
$\boldsymbol{\nabla}\cdot\mathbf{E}$ at site $r$. Because
$\mathbf{E}$ varies from $Q_-(r,m)$ to $Q_+(r,m)$ along a link,
there is an additional contribution to the lattice analog of
$\boldsymbol{\nabla}\cdot\mathbf{E}$, which is associated with the
links: $Q_+(r,m) - Q_-(r,m)$, the charge carried by the links. So
\begin{equation}
    G(r)=(\boldsymbol{\nabla}\cdot\mathbf{E})(r)-\frac{1}{2}\sum_m Q(r,m) \,.
\end{equation}
The gauge-invariance of the physical sector is defined by
$G(r)|\Psi\rangle=0$. Identifying $\frac{1}{2}\sum_m Q(r,m)$ as
the local color density $\rho_\text{G}$ (G for gluons), this
constraint becomes the familiar
\begin{equation}
    \boldsymbol{\nabla}\cdot\mathbf{E}=\rho_\text{G}\,.
\end{equation}

\begin{figure}[htb]
    \begin{center}
        \includegraphics[width=10.5pc]{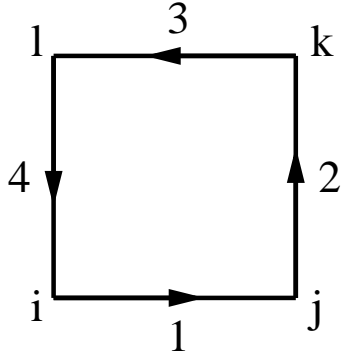}
        \caption{Graphical representation of the gauge-invariant operator $\text{Tr}\,U_{1/2}(1)U_{1/2}(2)U_{1/2}(3)U_{1/2}(4)$.}
\label{fig:plaquette2}
    \end{center}
\end{figure}

Defining $|0\rangle_\text{G}$ as the gauge-invariant state given
by the product over all lattice sites of the individual gauge
field ground states, we can construct the full gauge-invariant
space of states by acting on $|0\rangle_\text{G}$ with any product
of components of $U_{1/2}(r,m)$
\begin{equation}
\label{states}
    \prod_{r,m\in \{s\}} U_{1/2}(r,m)^i_{~l} |0\rangle_\text{G}\,.
\end{equation}
The set $\{s\}$ may contain any link any number of times. Note
that (\ref{states}) describes a gauge-invariant state only if the
color indices at each point are contracted to form a local
singlet. Indices associated with different lattice sites may not
be contracted, because they do not transform identically under
local gauge transformations. Let us give some examples:
\begin{equation}
    U_{1/2}(r,m)^i_{~l}|0\rangle
\end{equation}
is not a gauge-invariant state because it has uncontracted
indices.
\begin{equation}
\label{loop1}
    U_{1/2}(1)^i_{~j} U_{1/2}(2)^j_{~k} U_{1/2}(3)^k_{~l}
    U_{1/2}(4)^l_{~i}|0\rangle\,,
\end{equation}
where the numbers 1, 2, 3 and 4 refer to the links shown in Fig.
\ref{fig:plaquette2}, is gauge-invariant.
However, if the contractions do not involve the same lattice site,
the object would not be gauge-invariant as in this example:
\begin{equation}
    U_{1/2}(1)^i_{~j} U_{1/2}(3)^j_{~k} U_{1/2}(2)^k_{~l}
    U_{1/2}(4)^l_{~i}|0\rangle\,.
\end{equation}
When we include the fermion field $\Psi$ some more gauge-invariant
objects can be constructed. We can construct a gauge-invariant
state by considering the lowest eigenstate of the gauge-invariant
charge-conjugation-invariant operator
\begin{equation}
    \sum_r(-1)^r\Psi^{\dag}(r)\Psi(r)|0\rangle_\text{F}\,,
\end{equation}
where the state $|0\rangle_\text{F}$ is a product of fermion vacua
over all lattice sites. The product state
$|0\rangle=|0\rangle_\text{F} |0\rangle_\text{G}$ is
gauge-invariant. In addition to the gauge-invariant operators
formed from closed paths (e.g. \ref{loop1}) we can now form
gauge-invariant operators from paths with ends. For example,
considering a path $\Gamma$ beginning at site $r$ and ending at
site $s$, we can form the gauge-invariant object
\begin{equation}
    U(\Gamma,\Sigma)=\Psi^{\dag}(r)\Sigma U_{1/2}(r,n)
    U_{1/2}(r+n,m) \cdots U_{1/2}(s-l,l)\Psi(s)\,,
\end{equation}
where $\Sigma$ is any $2\times 2$ spin matrix. Physically the
lines between occupied sites represent electric flux. To see this
recall from above that the operator $Q_+(r,m)$ is proportional to
the electric field at site $r$ and points in direction $m$. On all
links where there is no $U_{1/2}(r,m)$, $Q_+(r,m)$ gives zero,
while on the links through which a line $\Gamma_j$ has passed,
$Q_+^2(r,m)$ gives $j(j+1)$ (again in analogy to the rigid
rotator). Hence, one can think of the lines as containing electric
flux of magnitude $\sqrt{j(j+1)}$.

Now that the fermions have been added again, the generator of
gauge transformations at point $r$ must include the additional
operator $\Psi^{\dag}(r)\frac{1}{2}{\boldsymbol\tau}\Psi(r)$,
which generates color rotations of $\Psi$. The full
gauge-invariance condition on the space of states becomes
\begin{equation}
    \left(\sum_m
    Q_+^\alpha(r,m)-\Psi^{\dag}(r)\frac{1}{2}\tau^\alpha\Psi(r)\right)\Big|\Big\rangle=0\,,
\end{equation}
which is again analogous to the condition
\begin{equation}
     \boldsymbol{\nabla}\cdot\mathbf{E}=\rho_\text{G}+\rho_\text{F}\,,
\end{equation}
where $\rho_\text{G}$ and $\rho_\text{F}$ are the color densities
of the gauge and fermion fields.

\section{The gauge-field Hamiltonian}
To give the field $\tilde{\mathbf{A}}$ some non-trivial dynamics
we must add a pure gauge-field term to the Hamiltonian $H$. This
term must be built from gauge-invariant operators to keep $H$
gauge-invariant. So $H$ may contain objects like $U(\Gamma)$ with
any closed path $\Gamma$. In addition, gauge-invariant operators
can be built from the $Q_{\pm}(r,m)$. Note that in particular
$Q_+^2 (=Q_-^2)$ is the analog of $J^2$ for the rigid rotator.
Since $J^2$ commutes with space and body rotations, $Q_+^2$
commutes with left and right gauge transformations and is
therefore gauge-invariant. Since $Q_+^2$ does not commute with
$U$, its appearance in the Hamiltonian will generate non-trivial
dynamics. In analogy to the Hamiltonian for the rigid rotator let
us include
\begin{equation}
\label{ekin}
    \sum_{r,m} Q_+^2(r,m)/(2I)
\end{equation}
in the Hamiltonian, where $I$ is a constant. (\ref{ekin}) represents
the energy of an assembly of uncoupled rotators. We have already
mentioned that $Q_+$ is proportional to the color electric field.
To get the correct continuum limit, the constant $I$ in (\ref{ekin})
has to be $I=a/g^2$, such that (\ref{ekin}) becomes
\begin{equation}
    \frac{a}{2g^2}\sum_{r,m}\dot{\tilde{A}}^2(r,m)=\frac{a^3}{2}\sum_{r,m}\dot{A}^2(r,m)=\frac{a^3}{2}\sum_{r,m}E^2(r,m)\,,
\end{equation}
where we used that $\dot{\tilde{\mathbf{A}}}=Q_+/I$ and
$\tilde{\mathbf{A}}_i=ag\mathbf{A}_i$. In the continuum limit this
turns into
\begin{equation}
    \frac{1}{2}\int d^3x E^2\,.
\end{equation}

\begin{figure}[H]
    \begin{center}
        \includegraphics[width=10.5pc]{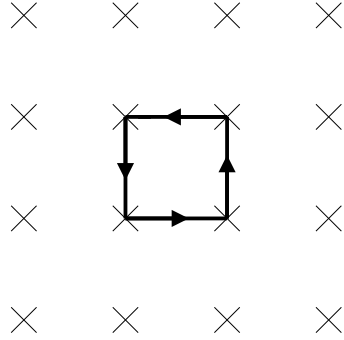}
        \caption{The simplest gauge-invariant closed loop, the so called plaquette $U_\Box$. The crosses denote lattice sites.}
\label{fig:plaquette}
    \end{center}
\end{figure}

To make the pure \textsc{Yang-Mills} theory non-trivial we have to
include terms that couple different links using the operators
$U(\Gamma)$. Obviously there is a great deal of arbitrariness
involved in the choice of the $\Gamma$. Following \textsc{Wilson}
\cite{Wilson:1974sk}, we pick the simplest object
$U_\Box:=U(\Gamma)$ that reproduces the continuum
\textsc{Yang-Mills} theory in the limit $a \to 0$, with $\Gamma$
shown in Fig. \ref{fig:plaquette}. Please note that from this
point on we stick to the conventions of the main text and set the
$\tau^a$ to the \textsc{Pauli} matrices without a factor of two
and absorb another factor of 2 into $A$, such that $U(r,x)$
becomes $e^{iagA_x(r)}=e^{iagA_x^a(r)\sigma^a}$.
\begin{align}
    U_\Box&=U(r,x)U(r+x,y)U^\dag(r+y,x)U^\dag(r,y)\notag\\
          &=e^{iagA_x(r)}e^{iagA_y(r+x)}e^{-iagA_x(r+y)}e^{-iagA_y(r)}\notag\\
          &=e^{iagA_x(r)}e^{iagA_y(r)+ia^2g\partial_xA_y(r)}e^{-iagA_x(r)-ia^2g\partial_yA_x(r)}e^{-iagA_y(r)}
\end{align}
where $A_x=A_x^a \tau^a$ etc.
Let us determine the continuum limit of $U_\Box$. Using
\begin{equation}\label{exprel}
    e^x e^y \simeq e^{x+y+\frac{1}{2}[x,y]}\,,
\end{equation}
we can write
\begin{align}\label{plaquettecont}
        U_\Box&=\exp\left(iagA_x(r)+iagA_y(r)+ia^2g\partial_xA_y(r)-\frac{a^2g^2}{2}[A_x(r),A_y(r)]+\mathcal{O}(a^3)\right)\notag\\
              &~~~\times\exp\left(-iagA_x(r)-iagA_y(r)-ia^2g\partial_yA_x(r)-\frac{a^2g^2}{2}[A_x(r),A_y(r)]+\mathcal{O}(a^3)\right)\notag\\
              &\approx\exp\left[ia^2g\left(\partial_xA_y(r)-\partial_yA_x(r)+ig[A_x(r),A_y(r)]\right)\right]\notag\\
              &=\exp\left(ia^2gF_{x y}(r)\right)\,,
\end{align}
with the spatial components $F_{x y}$ of the field tensor.
Realizing that
\begin{equation}
    \text{Re}\,\text{Tr}\left(U_\Box(r)\right) \xrightarrow{a \to 0}
    \text{Re}\,\text{Tr}\left(1+ia^2g
    F_{xy}-a^4\frac{g^2}{2}F_{xy}^2+\ldots\right)=N_c-\frac{a^4g^2}{2}\text{Re}\,\text{Tr}\left(F_{xy}^2\right)\,,
\end{equation}
we can identify the lattice version of the color magnetic field
contribution to the Hamiltonian to be
\begin{equation}
    \frac{2}{a g^2}\sum_\Box\left(N_c-\text{Re}\text{Tr}U_\Box\right)\,,
\end{equation}
where the sum runs over all plaquettes $\Box$, which includes all
directions of the B-field ($B_x^2$, $B_y^2$ and $B_z^2$, where
$B_z$ is shown in the example path above) at every site. The
prefactor is chosen to reproduce the correct continuum limit,
which reads
\begin{equation}
    \sum_{i\leq j} \frac{2}{ag^2} \frac{a^4g^2}{2} \text{Tr} F_{ij}^2 = 2 a^3 F_{ij}^aF_{ij}^a = \frac{1}{2}\int d^3x B^2\,.
\end{equation}
where you should note that we are sticking to our convention
$[\tau^a,\tau^b]=2\delta^{ab}$ instead of
$[\tau^a,\tau^b]=1/2\,\delta^{ab}$, and that $F_{ij}$ is half the
conventional $F_{ij}$. $B$ is then the usual physical $B$-field.

Adding the two pieces that correspond to the continuum terms
$\frac{1}{2}\int d^3x E^2$ and $\frac{1}{2}\int d^3x B^2$, we find
the lattice version of the Hamiltonian to be
\begin{align}
    \frac{a^3}{2}\sum_{r,m}E^2(r,m)+
    \frac{2}{ag^2}\sum_\Box\left(N_c-\text{Re}\text{Tr}U_\Box\right)\,,
\end{align}
which we can rewrite to lattice units using
$\mathbf{E}^a_L=ga^2/2\, \mathbf{E}^a$
\begin{align}
    H=\frac{2}{ag^2}\sum_{r,m}E_L^2(r,m)+
    \frac{2}{ag^2}\sum_\Box\left(N_c-\text{Re}\text{Tr}U_\Box\right)\,.
\end{align}
This form allows us to rescale and write
\begin{align}
    H_L=\frac{1}{2}\sum_{i}E_L^2(r,m)+
    \frac{1}{2}\sum_\Box\left(N_c-\text{Re}\text{Tr}U_\Box\right)=\frac{g^2a}{4}H\,,
\end{align}
the field part of the \textsc{Kogut-Susskind}-Hamiltonian used in
this work. The first sum goes over all (spatial) links, the second
one over all (spatial) plaquettes. Note again that the lattice is
just a spatial one whereas time is continuous. Of course for
actual calculations, time will be discretized. As opposed to the
usual lattice formulation the Hamiltonian formulation derived
above allows for realtime simulations. We only consider pure gauge
field simulations in this work, but for completeness let us give
the fermionic contribution to the Hamiltonian as well:
\begin{align}
    H_F=a^{-1}\sum\Psi^{\dag}(r)\frac{{\boldsymbol\sigma}\cdot\mathbf{n}}{i}U(r,n)\Psi(r+n)+m_0\sum(-1)^r\Psi^\dag(r)\Psi(r)\,.
\end{align}

  \chapter{The relativistic relative velocity}
\index{Relative velocity} \label{appvrel} 
We derive the relativistic relative velocity of two particles and point out that $\tilde{v}_{\text{rel}}$ in Section \ref{crosssctn} is not the actual relative velocity. The argument is based on \cite{landauv2}.

Let us consider two particles with velocities $\mathbf{v}_1$ and $\mathbf{v}_2$ and go to a system in that $\mathbf{v}_2=0$.
In this system the number of collisions per time $dt$ and volume $dV$ is given by
\begin{equation}
    d\nu=\sigma v_{\text{rel}} n_1 n_2 dV dt\,,
\end{equation}
which is a \textsc{Lorentz} invariant. $\sigma$ is the total cross section for a collision and $n_1$ and $n_2$ are the particle densities. 
The probability for two particles to collide is 
\begin{equation}
    P=\sigma v_{\text{rel}} \frac{dt}{dV}\,,
\end{equation}
such that
\begin{equation}
    d\nu=P n_1 n_2 dV dV=P\Delta N_1 \Delta N_2\,,
\end{equation}
with $N_1$ and $N_2$ the number of particles of species 1 and 2 within the volume $dV$.
Since $d\nu$ is an invariant and so are $n_1 dV$ and $n_2 dV$, $P$ is invariant as well.
We shall now derive a general expression for the invariant $d\nu$, which so far we have only expressed in the system where $\mathbf{v}_2=0$.
In all systems $d\nu$ shall be given by 
\begin{equation}\label{generalnu} 
    d\nu=An_1n_2dVdt\,,
\end{equation}
where $A$ is equal to $\sigma v_{\text{rel}}$ in the system where $\mathbf{v}_2=0$.
Since $dVdt$ is invariant, $A n_1 n_2$ has to be invariant too. Let us rewrite the condition that $A n_1 n_2$ is invariant in terms of the energies of the particles:
$n$ transforms as 
\begin{equation}
    n=\frac{n_0}{\sqrt{1-v^2}}\,,
\end{equation}
because $n dV$ is invariant and $dV$ transforms as $dV=dV_0\sqrt{1-v^2}$, where $n_0$ and $dV_0$ are given in the rest frame of the particle.
Because $E=\gamma m=1/\sqrt{1-v^2}\, m$, we can write
\begin{equation}
    n=n_0\frac{E}{m}\,.
\end{equation}
This means that if $An_1n_2$ is invariant, so is $AE_1E_2$. Now we may as well demand that $AE_1E_2/P_{1\mu}P_2^\mu$ is invariant, since we just devided by another invariant $P_{1\mu}P_2^\mu$. 
In the rest frame of particle 2 we have $E_2=m_2$ and $\mathbf{p}_2=0$, such that
\begin{equation}
    A \frac{E_1 E_2}{P_{1\mu}P_2^\mu}=A\,.
\end{equation}
Furthermore, in this rest frame, $A=\sigma v_{\text{rel}}$, such that in an arbitrary frame it has to hold that
\begin{equation}\label{myA}
    A=\sigma v_{\text{rel}} \frac{P_{1\mu}P_2^\mu}{E_1 E_2}\,.
\end{equation}
Again going to the rest frame of particle 2, we can calculate the invariant
\begin{equation}
    P_{1\mu}P_2^\mu=\frac{m_1}{\sqrt{1-v_{\text{rel}}^2}}m_2\,,
\end{equation}
which leads to
\begin{equation}\label{firstvrel}
    v_{\text{rel}}=\sqrt{1-\frac{m_1^2m_2^2}{(P_{1\mu}P_2^\mu)^2}}=\sqrt{1-\frac{m_1^2m_2^2}{(E_1E_2-\mathbf{p}_1\cdot\mathbf{p}_2)^2}}\,.
\end{equation}
Now using $\mathbf{p}=m\mathbf{v}/\sqrt{1-v^2}$ and $E=m/\sqrt{1-v^2}$, this becomes
\begin{align}
    v_{\text{rel}}&=\sqrt{1-\frac{(1-v_1^2)(1-v_2^2)}{(1-\mathbf{v}_1\cdot\mathbf{v}_2)^2}}=\sqrt{\frac{(1-\mathbf{v}_1\cdot\mathbf{v}_2)^2-(1-v_1^2)(1-v_2^2)}{(1-\mathbf{v}_1\cdot\mathbf{v}_2)^2}}\notag\\
    &=\frac{\sqrt{1-2\mathbf{v}_1\cdot\mathbf{v}_2+(\mathbf{v}_1\cdot\mathbf{v}_2)^2-(1-v_1^2)(1-v_2^2)}}{1-\mathbf{v}_1\cdot\mathbf{v}_2}\notag\\
    &=\frac{\sqrt{-2\mathbf{v}_1\cdot\mathbf{v}_2+(\mathbf{v}_1\cdot\mathbf{v}_2)^2-v_1^2v_2^2+v_1^2+v_2^2}}{1-\mathbf{v}_1\cdot\mathbf{v}_2}\notag\\
    &=\frac{\sqrt{(\mathbf{v}_1-\mathbf{v}_2)^2+(\mathbf{v}_1\cdot\mathbf{v}_2)^2-v_1^2v_2^2}}{1-\mathbf{v}_1\cdot\mathbf{v}_2}
    =\frac{\sqrt{(\mathbf{v}_1-\mathbf{v}_2)^2-(v_1^2v_2^2(1-\cos^2\phi))}}{1-\mathbf{v}_1\cdot\mathbf{v}_2}\notag\\
    &=\frac{\sqrt{(\mathbf{v}_1-\mathbf{v}_2)^2-(v_1^2v_2^2(\sin^2\phi))}}{1-\mathbf{v}_1\cdot\mathbf{v}_2}
    =\frac{\sqrt{(\mathbf{v}_1-\mathbf{v}_2)^2-(\mathbf{v}_1\times\mathbf{v}_2)^2}}{1-\mathbf{v}_1\cdot\mathbf{v}_2}\label{truevrel}\,.
\end{align}
This is the correct expression for the relative velocity, which is bounded from above by 1. To see how on the other hand $\tilde{v}_{\text{rel}}$ is defined, let us plug Eq. (\ref{firstvrel}) into (\ref{myA}) and then $A$ into (\ref{generalnu}). We get
\begin{equation}
    d\nu=\sigma\sqrt{1-\frac{m_1^2m_2^2}{(P_{1\mu}P_2^\mu)^2}} \frac{P_{1\mu}P_2^\mu}{E_1 E_2} n_1n_2dVdt=\sigma\frac{\sqrt{(P_{1\mu}P_2^\mu)^2-m_1^2m_2^2}}{E_1E_2}n_1n_2dVdt\,.
\end{equation}
In the massless limit $m_1$, $m_2\rightarrow 0$, we have 
\begin{equation}
    d\nu=\sigma\frac{P_{1\mu}P_2^\mu}{E_1E_2}n_1n_2dVdt\,,
\end{equation}
which can be expressed as
\begin{equation}
    d\nu=\sigma\frac{s}{2E_1E_2}n_1n_2dVdt\,,
\end{equation}
using the \textsc{Mandelstam} variable 
\begin{align}
    s&=(P_1+P_2)^2=(E_1+E_2)^2-(\mathbf{p}_1+\mathbf{p}_2)^2\notag\\
     &=E_1^2+E_2^2+2E_1E_2-\mathbf{p}_1^2-\mathbf{p}_2^2-2\mathbf{p}_1\cdot\mathbf{p}_2\notag\\
     &=2(E_1E_2-\mathbf{p}_1\cdot\mathbf{p}_2)=2 P_{1\mu}P_2^\mu\,,   
\end{align}
because $\mathbf{p}^2=E^2$.
One may now define $\tilde{v}_{\text{rel}}=\frac{s}{2E_1E_2}$ and write
\begin{equation}
        d\nu=\sigma\tilde{v}_{\text{rel}}n_1n_2dVdt\,,
\end{equation}
however, one should keep in mind that 
\begin{equation}
    \tilde{v}_{\text{rel}} = v_{\text{rel}}\frac{P_{1\mu}P_2^\mu}{E_1E_2} \,,
\end{equation}
is not the true relative velocity, which is given by (\ref{firstvrel}) or (\ref{truevrel}). $\tilde{v}_{\text{rel}}$ is not even less than $1$, i.e., can be larger than the speed of light. The difference is just that we have absorbed the factor $P_{1\mu}P_2^\mu/E_1E_2$ into $\tilde{v}_{\text{rel}}$.

  \chapter{Validity of the equilibrium fluctuation dissipation relation in the hard-loop limit}
\index{Fluctuation dissipation relation} \label{fdrappendix} In
this appendix we argue that Eq. (\ref{fdr}) is valid for a
non-equilibrium system within the hard-loop- (HL-) framework. This is necessary
because \textsc{Greiner} and \textsc{Leupold} showed in
\cite{Greiner:1998ri} and \cite{Greiner:1998vd} that it does not
hold in a general non-equilibrium system. As \textsc{Baier} et al.
did in \cite{Baier:1997xc}, we resum in the following way, using
the self energy calculated from {\it free} propagators):
\begin{equation}
\label{DSeqa} D =  D_0 \, + \, D_0 \Sigma D_0 \, + \, D_0 \Sigma
D_0 \Sigma D_0 \, + \ldots = D_0 \, + \, D_0 \Sigma D    \,.
\end{equation}
With the definitions $\Gamma _0(\vec{ p}, p_0) := \frac{i}{2 p_0
}[\Sigma ^>(\vec{ p},p_0 ) - \Sigma ^<(\vec{ p},p_0 ) ]$ and ${\rm
Re}\Sigma := {\rm Re}\Sigma^{\rm ret}  = {\rm Re}\Sigma^{\rm av} $
we end up with
\begin{eqnarray}
\label{Dret} D^{\rm ret} & = & D_0^{\rm ret} \, + \, D_0^{\rm
ret}\Sigma ^{\rm ret} D^{\rm ret} \, = \, \frac{1}{p^2 - m^2 -
{\rm Re}\Sigma +ip_0 \Gamma _0} \, ,
\\
\label{Dav} D^{\rm av} & = & D_0^{\rm av} \, + \, D_0^{\rm
av}\Sigma ^{\rm av} D^{\rm av} \, = \, \frac{1}{p^2 - m^2 - {\rm
Re}\Sigma - i p_0 \Gamma _0} \, ,
\\
\label{D<} D^{<} & = & D^{\rm ret} \Sigma^< D^{\rm av} \, = \,
(-2i) \frac{p_0 \Gamma _0}{(p^2-m^2-{\rm Re} \Sigma)^2 + p_0^2
\Gamma _0 ^2 } \, \frac{\Sigma^< }{\Sigma^> - \Sigma ^<} \, \, \,
.
\end{eqnarray}
The quantity
\begin{eqnarray}
n_{\Sigma } (\vec{ p}, p_0 )\,  := \, \frac{\Sigma^< }{\Sigma ^> -
\Sigma^<} \label{nsig}
\end{eqnarray}
appearing in the emerging fluctuation dissipation relation has to
be interpreted as the `occupation number' demanded by the self
energy parts \cite{Greiner:1998vd}. If the equilibrium KMS
conditions (\ref{kmseq}) apply for the self energy part, then $
n_{\Sigma } (\vec{ p}, p_0 )$
$\stackrel{\mbox{KMS}}{\longrightarrow }$ $n_B(p_0 ) $ becomes
just the \textsc{Bose} distribution function. In the limit $g\to
0$ this factor is just $1/2$ (cf. Eq.(\ref{kms}) and the
discussion thereafter). Generally $D^<$ has to be proportional to
some non-equilibrium distribution function $f$:
\begin{eqnarray}
\label{Dfull<} D^{<}(\vec{x},t;p) & \approx & (-2i) \frac{p_0
\Gamma (\vec{x},t;p)}{(p^2-m^2(\vec{x},t) - {\rm Re} \Sigma
(\vec{x},t;p) )^2 + p_0^2 \Gamma ^2(\vec{x},t;p)  } \,
f(\vec{x},t;\vec{p}) \, \, \, ,
\end{eqnarray}
which it is not in all cases, when one uses the resummation
(\ref{DSeqa}). \textsc{Greiner} and \textsc{Leupold} argue that
the reason for this is that the self energy parts $\Sigma^<$ and
$\Sigma^>$ should also evolve with time, and do not in this
scheme. They should depend on the {\em evolving} distribution
function and not persistently on the initial one which enters
$\Sigma$ in (\ref{DSeqa}). Thus the resummation of (\ref{DSeqa})
does not cover all relevant contributions. Speaking more
technically, the self energy operators must also be evaluated
consistently by the fully dressed and temporally evolving
one-particle propagators. To get the correct description for every
possible non-equilibrium situation, one would need to apply the
appropriate (quantum) transport equations with the difference to
the resummation of (\ref{DSeqa}) that the propagators entering
into the self-energy operators are then the fully dressed ones.
Unfortunately, the full quantum transport equations are generally
hard to solve and thus are not so much of practical use.

However, when we use squeezed equilibrium distributions as we do
in Chapter \ref{chap:photons}, the same arguments as those for the
equilibrium case hold and the resummation (\ref{DSeqa}) and in
particular Eq. (\ref{D<}) is valid. To make this more explicit, we
can rewrite the expression (\ref{nsig}) using
$\Sigma_{12}=-\Sigma_{21}$, which holds for all self energies in
the HL-limit, now for the case of Fermions:
\begin{equation}
    n_{\Sigma } (\vec{ p}, p_0 )\,  := \,\frac{\Sigma^< }{\Sigma ^< - \Sigma^>} =  \frac{\Sigma^< }{\Sigma ^< + \Sigma^<} =\frac{1}{2}
\end{equation}
This is equivalent to writing $f$ to order $\mathcal{O}(1)$:
\begin{equation}
    f(\mathbf{p},\xi)=\frac{1}{e^{\sqrt{\mathbf{p}^2+\xi(\mathbf{n}\cdot{\mathbf{p})^2}}/p_{\text{hard}}}+1}\rightarrow\frac{1}{2}\,,
\end{equation}
because all components of $\mathbf{p}$ are of order $g
p_{\text{hard}}$. This shows that at leading order in the coupling
constant $g$ (and for $g\to0$), the fluctuation dissipation
relation is valid for the hard loop self energies $\Sigma$ and
hence can be used in the derivation of the soft contribution to
the production rate as done in Section \ref{sec:softpart}.

  \cleardoublepage
  \phantomsection
  \addcontentsline{toc}{chapter}{Bibliography}
  \bibliography{phd}

\providecommand{\href}[2]{#2}\begingroup\raggedright\begin{thebibliography}{10%
0}

\bibitem{Baym:2001in}
G.~Baym, {\it Rhic: From dreams to beams in two decades},  {\em Nucl. Phys.}
  {\bf A698} (2002) XXIII--XXXII,
  [\href{http://xxx.lanl.gov/abs/hep-ph/0104138}{{\tt hep-ph/0104138}}].

\bibitem{Gross:1973id}
D.~J. Gross and F.~Wilczek, {\it Ultraviolet behavior of non-abelian gauge
  theories},  {\em Phys. Rev. Lett.} {\bf 30} (1973) 1343--1346.

\bibitem{Politzer:1973fx}
H.~D. Politzer, {\it Reliable perturbative results for strong interactions?},
  {\em Phys. Rev. Lett.} {\bf 30} (1973) 1346--1349.

\bibitem{Shuryak:1977ut}
E.~V. Shuryak, {\it Theory of hadronic plasma},  {\em Sov. Phys. JETP} {\bf 47}
  (1978) 212--219.

\bibitem{Shuryak:1978ij}
E.~V. Shuryak, {\it Quark-gluon plasma and hadronic production of leptons,
  photons and psions},  {\em Phys. Lett.} {\bf B78} (1978) 150.

\bibitem{Shuryak:1980tp}
E.~V. Shuryak, {\it Quantum chromodynamics and the theory of superdense
  matter},  {\em Phys. Rept.} {\bf 61} (1980) 71--158.

\bibitem{Kalashnikov:1979dp}
O.~K. Kalashnikov and V.~V. Klimov, {\it Phase transition in quark - gluon
  plasma},  {\em Phys. Lett.} {\bf B88} (1979) 328.

\bibitem{Kapusta:1979fh}
J.~I. Kapusta, {\it Quantum chromodynamics at high temperature},  {\em Nucl.
  Phys.} {\bf B148} (1979) 461--498.

\bibitem{Bevalac:2007}
``{Bevalac Synchrotron. The Bevalac started life as the Bevatron in 1954 (the
  ``Bev'' in ``Bevatron'' standing for ``billion electron volts'' - BeV). At
  the time, it was the world's largest particle accelerator, capable of
  accelerating a beam of protons and charging the particles to 6.2 BeV. In
  1974, at the age of 20, the Bevatron was at the end of its useful life for
  research. It was "saved" when a proposal to convert it to a heavy ion
  accelerator was acted on - this would allow it to accelerate all elements on
  the periodic table to relativistic energies. The SuperHILAC linear
  accelerator located 550 ft. up the hill above the Bevatron was linked,
  replacing the Bevatron's own linear accelerator. This linkage renamed the
  Bevatron to the Bevalac. The first beams were produced on August 1, 1974. The
  Bevalac ceased operations when the beam was turned off for the last time on
  February 21, 1993}.'' \url{http://www.uer.ca/locations/show.asp?locid=21423}.

\bibitem{AGS:2007}
``{Alternating Gradient Synchrotron. The name of the AGS, running since 1960,
  is derived from the concept of alternating gradient focusing, in which the
  field gradients of the accelerator's 240 magnets are successively alternated
  inward and outward, permitting particles to be propelled and focused in both
  the horizontal and vertical plane at the same time. Capable of accelerating
  25 trillion protons with every pulse, and heavy ions such as gold and iron,
  the AGS is used by 850 users from 180 institutions from around the world
  annually. The AGS receives protons from Brookhaven's 200 million
  electron-volt (MeV) linear accelerator (LINAC). The AGS Booster, constructed
  in 1991, further augmented the capabilities of the AGS, enabling it to
  accelerate protons and heavy ions to much higher energies than before. Even
  now, the applications for the AGS continue to be expanded with the
  construction of the NASA Space Radiation Laboratory. Among its other duties,
  the AGS is now used as an injector for the Relativistic Heavy Ion Collider
  (RHIC)}.'' \url{http://www.bnl.gov/bnlweb/facilities/AGS.asp}.

\bibitem{SPS:2007}
``{Super Proton Synchrotron. The Super Proton Synchrotron is the second largest
  machine in CERN's accelerator complex. Measuring nearly 7 km in
  circumference, it takes particles from the Proton Synchrotron (PS) and
  accelerates them to provide beams for the Large Hadron Collider (LHC), the
  COMPASS experiment and the CNGS project. When it switched on in 1976, the SPS
  became the workhorse of CERN's particle physics programme. Research using SPS
  beams has probed the inner structure of protons, investigated nature's
  preference for matter over antimatter, looked for matter as it might have
  been in the first instants of the Universe and searched for exotic forms of
  matter. A major highlight came in 1983 with the Nobel-prize-winning discovery
  of W and Z particles made with the SPS running as a proton-antiproton
  collider. The SPS has 1317 conventional (room temperature) electromagnets,
  including 744 dipoles to bend the beams round the ring, and it operates at up
  to 450 GeV. It has handled many different kinds of particles - sulphur and
  oxygen nuclei, electrons, positrons, protons and antiprotons}.''
  \url{http://info.web.cern.ch/public/en/Research/SPS-en.html}.

\bibitem{RHIC:2007}
``{Relativistic Heavy Ion Collider. The Relativistic Heavy Ion Collider (RHIC)
  achieved its first successful operation in the summer of 2000, capping ten
  years of development. However, the history of RHIC stretches back more than
  30 years beginning with an idea for a machine called the Intersecting Storage
  Accelerator (ISA). When single ring accelerators like the Alternating
  Gradient Synchrotron propel protons against a stationary target to
  investigate nuclear properties, much of the usable energy in the reaction is
  lost, carried away in the forward motion of the incident proton. This limits
  the energy which can be practically achieved by a single ring accelerator. It
  was recognized by early accelerator physicists that much higher reaction
  energies could be achieved by colliding two accelerated beams head-on. Using
  this technique, the energy released can equal the sum total carried in each
  beam. The idea of using storage rings for a colliding beam accelerator was
  considered at a summer study held at Brookhaven in 1963. Such technology was
  considered feasible at that time, though it was decided that no construction
  would take place since it was felt that storage rings lacked the versatility
  of a single proton accelerator of the same equivalent energy. The idea was
  revived in 1970 by John Blewett and this time, it was greeted with
  enthusiasm. A group called the Fitch committee recommended that Brookhaven
  apply its pioneering development work in superconducting magnets to build two
  proton intersecting storage rings. This was the beginning of the ISABELLE
  project. In 1981 technical problems were encountered in the fabrication of
  the superconducting magnets which would power the machine, causing Brookhaven
  to replace the ISABELLE proposal with a different machine design called the
  Colliding Beam Accelerator (CBA). Due to the change in machine plans, the
  HEPAP voted in 1983 to discontinue construction of the CBA in favor of a new
  machine to be built in Texas called the Superconducting Supercollider (SSC)
  (which would ironically be abandoned itself a decade later). Physicists at
  Brookhaven persevered, continuing to push for an advanced accelerator design.
  In 1984, the first proposal was submitted for the machine now known as RHIC.
  RHIC's main function is to search for a state of matter called the
  quark-gluon plasma. It was considered very cost-effective to build RHIC at
  Brookhaven because of the existing accelerator infrastructure which could be
  used to inject protons and heavy ions into the machine as well as the fact
  that a tunnel (originally excavated for ISABELLE/CBA) was already completed.
  Brookhaven received funding to proceed with the construction of RHIC in
  1991}.'' \url{http://www.bnl.gov/bnlweb/history/RHIC_history.asp}.

\bibitem{LHC:2007}
``{Large Hadron Collider. The Large Hadron Collider (LHC) is being built in a
  circular tunnel 27 km in circumference. The tunnel is buried around 50 to 175
  m underground. It straddles the Swiss and French borders on the outskirts of
  Geneva. It is planned to circulate the first beams in May 2008. First
  collisions at high energy are expected mid-2008 with the first results from
  the experiments soon after. The LHC is designed to collide two counter
  rotating beams of protons or heavy ions. Proton-proton collisions are
  foreseen at an energy of 7 TeV per beam. The beams move around the LHC ring
  inside a continuous vacuum guided by magnets. The magnets are superconducting
  and are cooled by a huge cryogenics system. The cables conduct current
  without resistance in their superconducting state. The beams will be stored
  at high energy for hours. During this time collisions take place inside the
  four main LHC experiments}.'' \url{http://lhc.web.cern.ch/lhc/}.

\bibitem{Turbide:2005fk}
S.~Turbide, C.~Gale, S.~Jeon, and G.~D. Moore, {\it Energy loss of leading
  hadrons and direct photon production in evolving quark-gluon plasma},  {\em
  Phys. Rev.} {\bf C72} (2005) 014906,
  [\href{http://xxx.lanl.gov/abs/hep-ph/0502248}{{\tt hep-ph/0502248}}].

\bibitem{Lee:1974ma}
T.~D. Lee and G.~C. Wick, {\it Vacuum stability and vacuum excitation in a spin
  0 field theory},  {\em Phys. Rev.} {\bf D9} (1974) 2291.

\bibitem{Collins:1974ky}
J.~C. Collins and M.~J. Perry, {\it Superdense matter: Neutrons or
  asymptotically free quarks?},  {\em Phys. Rev. Lett.} {\bf 34} (1975) 1353.

\bibitem{Baym:1976yu}
G.~Baym and S.~A. Chin, {\it Can a neutron star be a giant mit bag?},  {\em
  Phys. Lett.} {\bf B62} (1976) 241--244.

\bibitem{Freedman:1976ub}
B.~A. Freedman and L.~D. McLerran, {\it Fermions and gauge vector mesons at
  finite temperature and density. 3. the ground state energy of a relativistic
  quark gas},  {\em Phys. Rev.} {\bf D16} (1977) 1169.

\bibitem{Chapline:1976gy}
G.~Chapline and M.~Nauenberg, {\it Asymptotic freedom and the baryon-quark
  phase transition},  {\em Phys. Rev.} {\bf D16} (1977) 450.

\bibitem{Nobel:2004}
``{The Nobel Prize in Physics 2004}.''
  \url{http://nobelprize.org/nobel_prizes/physics/laureates/2004/}.

\bibitem{Bethke:2006ac}
S.~Bethke, {\it {Experimental tests of asymptotic freedom}},  {\em Prog. Part.
  Nucl. Phys.} {\bf 58} (2007) 351--386,
  [\href{http://xxx.lanl.gov/abs/hep-ex/0606035}{{\tt hep-ex/0606035}}].

\bibitem{Yao:2006px}
{\bf Particle Data Group} Collaboration, W.~M. Yao {\em et~al.}, {\it {Review
  of particle physics}},  {\em J. Phys.} {\bf G33} (2006) 1--1232.

\bibitem{Nasa:2007}
``{The sun fact sheet}.''
  \url{http://nssdc.gsfc.nasa.gov/planetary/factsheet/sunfact.html}.

\bibitem{FAIR:2007}
``{The proposed project FAIR (Facility for Antiproton and Ion Research) is an
  international accelerator facility of the next generation. It builds on the
  experience and technological developments already made at the existing GSI
  facility, and incorporates new technological concepts. At its heart is a
  double ring facility with a circumference of of 1100 meters. A system of
  cooler-storage rings for effective beam cooling at high energies and various
  experimental halls will be connected to the facility }.''
  \url{http://www.gsi.de/fair/}.

\bibitem{RHICdata:2007}
{For a complete list of RHIC publications see Spires and use ``FIND CN BRAHMS
  OR CN PHOBOS OR CN STAR OR CN PHENIX AND PS PUBLISHED''}.
  \url{http://www.slac.stanford.edu/spires/}.

\bibitem{Allton:2003vx}
C.~R. Allton {\em et~al.}, {\it The equation of state for two flavor qcd at
  non-zero chemical potential},  {\em Phys. Rev.} {\bf D68} (2003) 014507,
  [\href{http://xxx.lanl.gov/abs/hep-lat/0305007}{{\tt hep-lat/0305007}}].

\bibitem{Karsch:2000ps}
F.~Karsch, E.~Laermann, and A.~Peikert, {\it The pressure in 2, 2+1 and 3
  flavour qcd},  {\em Phys. Lett.} {\bf B478} (2000) 447--455,
  [\href{http://xxx.lanl.gov/abs/hep-lat/0002003}{{\tt hep-lat/0002003}}].

\bibitem{Karsch:2001cy}
F.~Karsch, {\it Lattice qcd at high temperature and density},  {\em Lect. Notes
  Phys.} {\bf 583} (2002) 209--249,
  [\href{http://xxx.lanl.gov/abs/hep-lat/0106019}{{\tt hep-lat/0106019}}].

\bibitem{Bernard:1996cs}
{\bf MILC} Collaboration, C.~W. Bernard {\em et~al.}, {\it The equation of
  state for two flavor qcd at n(t) = 6},  {\em Phys. Rev.} {\bf D55} (1997)
  6861--6869, [\href{http://xxx.lanl.gov/abs/hep-lat/9612025}{{\tt
  hep-lat/9612025}}].

\bibitem{Gupta:2003be}
S.~Gupta, {\it The quark gluon plasma: Lattice computations put to experimental
  test},  {\em Pramana} {\bf 61} (2003) 877--888,
  [\href{http://xxx.lanl.gov/abs/hep-ph/0303072}{{\tt hep-ph/0303072}}].

\bibitem{Fodor:2004nz}
Z.~Fodor and S.~D. Katz, {\it Critical point of qcd at finite t and mu, lattice
  results for physical quark masses},  {\em JHEP} {\bf 04} (2004) 050,
  [\href{http://xxx.lanl.gov/abs/hep-lat/0402006}{{\tt hep-lat/0402006}}].

\bibitem{Csikor:2004me}
F.~Csikor {\em et~al.}, {\it The qcd equation of state at finite t/mu on the
  lattice},  {\em Prog. Theor. Phys. Suppl.} {\bf 153} (2004) 93--105,
  [\href{http://xxx.lanl.gov/abs/hep-lat/0401022}{{\tt hep-lat/0401022}}].

\bibitem{Aoki:2006br}
Y.~Aoki, Z.~Fodor, S.~D. Katz, and K.~K. Szabo, {\it The qcd transition
  temperature: Results with physical masses in the continuum limit},  {\em
  Phys. Lett.} {\bf B643} (2006) 46--54,
  [\href{http://xxx.lanl.gov/abs/hep-lat/0609068}{{\tt hep-lat/0609068}}].

\bibitem{Cheng:2006qk}
M.~Cheng {\em et~al.}, {\it The transition temperature in qcd},  {\em Phys.
  Rev.} {\bf D74} (2006) 054507,
  [\href{http://xxx.lanl.gov/abs/hep-lat/0608013}{{\tt hep-lat/0608013}}].

\bibitem{Gyulassy:2004zy}
M.~Gyulassy and L.~McLerran, {\it New forms of qcd matter discovered at rhic},
  {\em Nucl. Phys.} {\bf A750} (2005) 30--63,
  [\href{http://xxx.lanl.gov/abs/nucl-th/0405013}{{\tt nucl-th/0405013}}].

\bibitem{Fodor:2001au}
Z.~Fodor and S.~D. Katz, {\it {A new method to study lattice QCD at finite
  temperature and chemical potential}},  {\em Phys. Lett.} {\bf B534} (2002)
  87--92, [\href{http://xxx.lanl.gov/abs/hep-lat/0104001}{{\tt
  hep-lat/0104001}}].

\bibitem{deForcrand:2002ci}
P.~de~Forcrand and O.~Philipsen, {\it {The QCD phase diagram for small
  densities from imaginary chemical potential}},  {\em Nucl. Phys.} {\bf B642}
  (2002) 290--306, [\href{http://xxx.lanl.gov/abs/hep-lat/0205016}{{\tt
  hep-lat/0205016}}].

\bibitem{BraunMunzinger:1996mq}
P.~Braun-Munzinger and J.~Stachel, {\it {Probing the phase boundary between
  hadronic matter and the quark-gluon-plasma in relativistic heavy ion
  collisions}},  {\em Nucl. Phys.} {\bf A606} (1996) 320--328,
  [\href{http://xxx.lanl.gov/abs/nucl-th/9606017}{{\tt nucl-th/9606017}}].

\bibitem{Philipsen:2007rm}
O.~Philipsen, {\it {Exploring the QCD phase diagram}},
  \href{http://xxx.lanl.gov/abs/0710.1217}{{\tt 0710.1217}}.

\bibitem{Stephanov:2007fk}
M.~A. Stephanov, {\it {QCD phase diagram: An overview}},  {\em PoS} {\bf
  LAT2006} (2006) 024, [\href{http://xxx.lanl.gov/abs/hep-lat/0701002}{{\tt
  hep-lat/0701002}}].

\bibitem{Alford:2006wn}
M.~G. Alford, {\it {Color superconductivity in ultra-dense quark matter}},
  {\em PoS} {\bf LAT2006} (2006) 001,
  [\href{http://xxx.lanl.gov/abs/hep-lat/0610046}{{\tt hep-lat/0610046}}].

\bibitem{BraunMunzinger:2008tz}
P.~Braun-Munzinger and J.~Wambach, {\it {The Phase Diagram of
  Strongly-Interacting Matter}},  \href{http://xxx.lanl.gov/abs/0801.4256}{{\tt
  0801.4256}}.

\bibitem{Arsene:2004fa}
{\bf BRAHMS} Collaboration, I.~Arsene {\em et~al.}, {\it Quark gluon plasma and
  color glass condensate at rhic? the perspective from the brahms experiment},
  {\em Nucl. Phys.} {\bf A757} (2005) 1--27,
  [\href{http://xxx.lanl.gov/abs/nucl-ex/0410020}{{\tt nucl-ex/0410020}}].

\bibitem{Back:2004je}
B.~B. Back {\em et~al.}, {\it The phobos perspective on discoveries at rhic},
  {\em Nucl. Phys.} {\bf A757} (2005) 28--101,
  [\href{http://xxx.lanl.gov/abs/nucl-ex/0410022}{{\tt nucl-ex/0410022}}].

\bibitem{Adams:2005dq}
{\bf STAR} Collaboration, J.~Adams {\em et~al.}, {\it Experimental and
  theoretical challenges in the search for the quark gluon plasma: The star
  collaboration's critical assessment of the evidence from rhic collisions},
  {\em Nucl. Phys.} {\bf A757} (2005) 102--183,
  [\href{http://xxx.lanl.gov/abs/nucl-ex/0501009}{{\tt nucl-ex/0501009}}].

\bibitem{Adcox:2004mh}
{\bf PHENIX} Collaboration, K.~Adcox {\em et~al.}, {\it Formation of dense
  partonic matter in relativistic nucleus nucleus collisions at rhic:
  Experimental evaluation by the phenix collaboration},  {\em Nucl. Phys.} {\bf
  A757} (2005) 184--283, [\href{http://xxx.lanl.gov/abs/nucl-ex/0410003}{{\tt
  nucl-ex/0410003}}].

\bibitem{Muller:2006ee}
B.~Muller and J.~L. Nagle, {\it Results from the relativistic heavy ion
  collider},  {\em Ann. Rev. Nucl. Part. Sci.} {\bf 56} (2006) 93--135,
  [\href{http://xxx.lanl.gov/abs/nucl-th/0602029}{{\tt nucl-th/0602029}}].

\bibitem{Muller:2007rs}
B.~Muller, {\it {From Quark-Gluon Plasma to the Perfect Liquid}},  {\em Acta
  Phys. Polon.} {\bf B38} (2007) 3705--3730,
  [\href{http://xxx.lanl.gov/abs/0710.3366}{{\tt 0710.3366}}].

\bibitem{Heinz:2005ja}
U.~W. Heinz, {\it Equation of state and collective dynamics},  {\em J. Phys.
  Conf. Ser.} {\bf 50} (2006) 230--237,
  [\href{http://xxx.lanl.gov/abs/nucl-th/0504011}{{\tt nucl-th/0504011}}].

\bibitem{Retiere:2004wa}
F.~Retiere, {\it Flow in ultra-relativistic heavy ion collisions},  {\em J.
  Phys.} {\bf G30} (2004) S827--S834,
  [\href{http://xxx.lanl.gov/abs/nucl-ex/0405024}{{\tt nucl-ex/0405024}}].

\bibitem{Poskanzer:1998yz}
A.~M. Poskanzer and S.~A. Voloshin, {\it Methods for analyzing anisotropic flow
  in relativistic nuclear collisions},  {\em Phys. Rev.} {\bf C58} (1998)
  1671--1678, [\href{http://xxx.lanl.gov/abs/nucl-ex/9805001}{{\tt
  nucl-ex/9805001}}].

\bibitem{Heinz:2004pj}
U.~W. Heinz, {\it Thermalization at rhic},  {\em AIP Conf. Proc.} {\bf 739}
  (2005) 163--180, [\href{http://xxx.lanl.gov/abs/nucl-th/0407067}{{\tt
  nucl-th/0407067}}].

\bibitem{Arnold:2004ti}
P.~Arnold, J.~Lenaghan, G.~D. Moore, and L.~G. Yaffe, {\it Apparent
  thermalization due to plasma instabilities in quark gluon plasma},  {\em
  Phys. Rev. Lett.} {\bf 94} (2005) 072302,
  [\href{http://xxx.lanl.gov/abs/nucl-th/0409068}{{\tt nucl-th/0409068}}].

\bibitem{Shuryak:2004kh}
E.~Shuryak, {\it A strongly coupled quark-gluon plasma},  {\em J. Phys.} {\bf
  G30} (2004) S1221--S1224.

\bibitem{Sorge:1998mk}
H.~Sorge, {\it Highly sensitive centrality dependence of elliptic flow: A novel
  signature of the phase transition in qcd},  {\em Phys. Rev. Lett.} {\bf 82}
  (1999) 2048--2051, [\href{http://xxx.lanl.gov/abs/nucl-th/9812057}{{\tt
  nucl-th/9812057}}].

\bibitem{Baier:2000sb}
R.~Baier, A.~H. Mueller, D.~Schiff, and D.~T. Son, {\it 'bottom-up'
  thermalization in heavy ion collisions},  {\em Phys. Lett.} {\bf B502} (2001)
  51--58, [\href{http://xxx.lanl.gov/abs/hep-ph/0009237}{{\tt
  hep-ph/0009237}}].

\bibitem{Baier:2002bt}
R.~Baier, A.~H. Mueller, D.~Schiff, and D.~T. Son, {\it Does parton saturation
  at high density explain hadron multiplicities at rhic?},  {\em Phys. Lett.}
  {\bf B539} (2002) 46--52, [\href{http://xxx.lanl.gov/abs/hep-ph/0204211}{{\tt
  hep-ph/0204211}}].

\bibitem{Xu:2004mz}
Z.~Xu and C.~Greiner, {\it Thermalization of gluons in ultrarelativistic heavy
  ion collisions by including three-body interactions in a parton cascade},
  {\em Phys. Rev.} {\bf C71} (2005) 064901,
  [\href{http://xxx.lanl.gov/abs/hep-ph/0406278}{{\tt hep-ph/0406278}}].

\bibitem{Arnold:1998cy}
P.~Arnold, D.~T. Son, and L.~G. Yaffe, {\it Hot b violation, color
  conductivity, and log(1/alpha) effects},  {\em Phys. Rev.} {\bf D59} (1999)
  105020, [\href{http://xxx.lanl.gov/abs/hep-ph/9810216}{{\tt
  hep-ph/9810216}}].

\bibitem{Mrowczynski:1993qm}
S.~Mrowczynski, {\it Plasma instability at the initial stage of
  ultrarelativistic heavy ion collisions},  {\em Phys. Lett.} {\bf B314} (1993)
  118--121.

\bibitem{Mrowczynski:1994xv}
S.~Mrowczynski, {\it Color collective effects at the early stage of
  ultrarelativistic heavy ion collisions},  {\em Phys. Rev.} {\bf C49} (1994)
  2191--2197.

\bibitem{Randrup:2003cw}
J.~Randrup and S.~Mrowczynski, {\it Chromodynamic weibel instabilities in
  relativistic nuclear collisions},  {\em Phys. Rev.} {\bf C68} (2003) 034909,
  [\href{http://xxx.lanl.gov/abs/nucl-th/0303021}{{\tt nucl-th/0303021}}].

\bibitem{Romatschke:2003ms}
P.~Romatschke and M.~Strickland, {\it Collective modes of an anisotropic quark
  gluon plasma},  {\em Phys. Rev.} {\bf D68} (2003) 036004,
  [\href{http://xxx.lanl.gov/abs/hep-ph/0304092}{{\tt hep-ph/0304092}}].

\bibitem{Arnold:2003rq}
P.~Arnold, J.~Lenaghan, and G.~D. Moore, {\it Qcd plasma instabilities and
  bottom-up thermalization},  {\em JHEP} {\bf 08} (2003) 002,
  [\href{http://xxx.lanl.gov/abs/hep-ph/0307325}{{\tt hep-ph/0307325}}].

\bibitem{Rebhan:2004ur}
A.~Rebhan, P.~Romatschke, and M.~Strickland, {\it Hard-loop dynamics of
  non-abelian plasma instabilities},  {\em Phys. Rev. Lett.} {\bf 94} (2005)
  102303, [\href{http://xxx.lanl.gov/abs/hep-ph/0412016}{{\tt
  hep-ph/0412016}}].

\bibitem{Dumitru:2005gp}
A.~Dumitru and Y.~Nara, {\it Qcd plasma instabilities and isotropization},
  {\em Phys. Lett.} {\bf B621} (2005) 89,
  [\href{http://xxx.lanl.gov/abs/hep-ph/0503121}{{\tt hep-ph/0503121}}].

\bibitem{Csanad:2007fj}
{\bf PHENIX} Collaboration, M.~Csanad, {\it {Milestones of the PHENIX
  experiment at RHIC}},  \href{http://xxx.lanl.gov/abs/0712.1435}{{\tt
  0712.1435}}.

\bibitem{Adler:2005ig}
{\bf PHENIX} Collaboration, S.~S. Adler {\em et~al.}, {\it {Centrality
  dependence of direct photon production in s(NN)**(1/2) = 200-GeV Au + Au
  collisions}},  {\em Phys. Rev. Lett.} {\bf 94} (2005) 232301,
  [\href{http://xxx.lanl.gov/abs/nucl-ex/0503003}{{\tt nucl-ex/0503003}}].

\bibitem{Adams:2003kv}
{\bf STAR} Collaboration, J.~Adams {\em et~al.}, {\it {Transverse momentum and
  collision energy dependence of high p(T) hadron suppression in Au + Au
  collisions at ultrarelativistic energies}},  {\em Phys. Rev. Lett.} {\bf 91}
  (2003) 172302, [\href{http://xxx.lanl.gov/abs/nucl-ex/0305015}{{\tt
  nucl-ex/0305015}}].

\bibitem{Busching:2005yy}
{\bf PHENIX} Collaboration, H.~Busching {\em et~al.}, {\it {Medium effects on
  high particle production measured with the PHENIX experiment}},  {\em Eur.
  Phys. J.} {\bf C43} (2005) 303--310.

\bibitem{Vitev:2002pf}
I.~Vitev and M.~Gyulassy, {\it {High-p(T) tomography of d + Au and Au + Au at
  SPS, RHIC, and LHC}},  {\em Phys. Rev. Lett.} {\bf 89} (2002) 252301,
  [\href{http://xxx.lanl.gov/abs/hep-ph/0209161}{{\tt hep-ph/0209161}}].

\bibitem{d'Enterria:2005cs}
D.~d'Enterria, {\it {High p(T) leading hadron suppression in nuclear collisions
  at s(NN)**(1/2) = 20-GeV - 200-GeV: Data versus parton energy loss models}},
  {\em Eur. Phys. J.} {\bf C43} (2005) 295--302,
  [\href{http://xxx.lanl.gov/abs/nucl-ex/0504001}{{\tt nucl-ex/0504001}}].

\bibitem{Adler:2002tq}
{\bf STAR} Collaboration, C.~Adler {\em et~al.}, {\it Disappearance of
  back-to-back high p(t) hadron correlations in central au + au collisions at
  s(nn)**(1/2) = 200-gev},  {\em Phys. Rev. Lett.} {\bf 90} (2003) 082302,
  [\href{http://xxx.lanl.gov/abs/nucl-ex/0210033}{{\tt nucl-ex/0210033}}].

\bibitem{Adler:2005ee}
{\bf PHENIX} Collaboration, S.~S. Adler {\em et~al.}, {\it Modifications to
  di-jet hadron pair correlations in au + au collisions at s(nn)**(1/2) =
  200-gev},  {\em Phys. Rev. Lett.} {\bf 97} (2006) 052301,
  [\href{http://xxx.lanl.gov/abs/nucl-ex/0507004}{{\tt nucl-ex/0507004}}].

\bibitem{Baumgardt:1975qv}
H.~G. Baumgardt {\em et~al.}, {\it Shock waves and mach cones in fast
  nucleus-nucleus collisions},  {\em Z. Phys.} {\bf A273} (1975) 359--371.

\bibitem{Stoecker:2007su}
H.~Stoecker, B.~Betz, and P.~Rau, {\it Hydrodynamic flow and jet induced mach
  shocks at rhic and lhc},  {\em PoS} {\bf CPOD2006} (2006) 029,
  [\href{http://xxx.lanl.gov/abs/nucl-th/0703054}{{\tt nucl-th/0703054}}].

\bibitem{Baeuchle:2007qw}
B.~Baeuchle, L.~P. Csernai, and H.~Stoecker, {\it Mace -- mach cones in heavy
  ion collisions},  \href{http://xxx.lanl.gov/abs/0710.1476}{{\tt 0710.1476}}.

\bibitem{Putschke:2007mi}
J.~Putschke, {\it Intra-jet correlations of high-$p_t$ hadrons from star},
  {\em J. Phys.} {\bf G34} (2007) S679--684,
  [\href{http://xxx.lanl.gov/abs/nucl-ex/0701074}{{\tt nucl-ex/0701074}}].

\bibitem{Adams:2004pa}
{\bf STAR} Collaboration, J.~Adams {\em et~al.}, {\it Minijet deformation and
  charge-independent angular correlations on momentum subspace (eta, phi) in
  au-au collisions at s(nn)**(1/2) = 130-gev},  {\em Phys. Rev.} {\bf C73}
  (2006) 064907, [\href{http://xxx.lanl.gov/abs/nucl-ex/0411003}{{\tt
  nucl-ex/0411003}}].

\bibitem{Kajantie:1981wg}
K.~Kajantie and H.~I. Miettinen, {\it Temperature measurement of quark - gluon
  plasma formed in high-energy nucleus-nucleus collisions},  {\em Zeit. Phys.}
  {\bf C9} (1981) 341.

\bibitem{Halzen:1981kz}
F.~Halzen and H.~C. Liu, {\it Experimental signatures of phase transition to
  quark matter in high-energy collisions of nuclei},  {\em Phys. Rev.} {\bf
  D25} (1982) 1842.

\bibitem{Kajantie:1982nj}
K.~Kajantie and P.~V. Ruuskanen, {\it Shielding of quark mass singularities in
  photon emission from hot quark - gluon plasma},  {\em Phys. Lett.} {\bf B121}
  (1983) 352.

\bibitem{Sinha:1983jm}
B.~Sinha, {\it Universal signals of quark - gluon plasma},  {\em Phys. Lett.}
  {\bf B128} (1983) 91--94.

\bibitem{Hwa:1985xg}
R.~C. Hwa and K.~Kajantie, {\it Diagnosing quark matter by measuring the total
  entropy and the photon or dilepton emission rates},  {\em Phys. Rev.} {\bf
  D32} (1985) 1109.

\bibitem{Staadt:1985uc}
G.~Staadt, W.~Greiner, and J.~Rafelski, {\it Photons from strange quark
  annihilation in quark - gluon plasma},  {\em Phys. Rev.} {\bf D33} (1986) 66.

\bibitem{Neubert:1989hu}
M.~Neubert, {\it Photon production in ultrarelativistic heavy ion collisions at
  200-gev/u},  {\em Z. Phys.} {\bf C42} (1989) 231--242.

\bibitem{Kapusta:1991qp}
J.~I. Kapusta, P.~Lichard, and D.~Seibert, {\it High-energy photons from quark
  - gluon plasma versus hot hadronic gas},  {\em Phys. Rev.} {\bf D44} (1991)
  2774--2788.

\bibitem{Redlich:1992fr}
K.~Redlich, R.~Baier, H.~Nakkagawa, and A.~Niegawa, {\it Dynamical screening
  and real photon production in a hot quark gluon plasma},  {\em Nucl. Phys.}
  {\bf A544} (1992) 511--512.

\bibitem{Aurenche:1998nw}
P.~Aurenche, F.~Gelis, R.~Kobes, and H.~Zaraket, {\it Bremsstrahlung and photon
  production in thermal {QCD}},  {\em Phys. Rev.} {\bf D58} (1998) 085003,
  [\href{http://xxx.lanl.gov/abs/hep-ph/9804224}{{\tt hep-ph/9804224}}].

\bibitem{Aurenche:1999tq}
P.~Aurenche, F.~Gelis, and H.~Zaraket, {\it Kln theorem, magnetic mass, and
  thermal photon production},  {\em Phys. Rev.} {\bf D61} (2000) 116001,
  [\href{http://xxx.lanl.gov/abs/hep-ph/9911367}{{\tt hep-ph/9911367}}].

\bibitem{Arnold:2001ba}
P.~Arnold, G.~D. Moore, and L.~G. Yaffe, {\it Photon emission from
  ultrarelativistic plasmas},  {\em JHEP} {\bf 11} (2001) 057,
  [\href{http://xxx.lanl.gov/abs/hep-ph/0109064}{{\tt hep-ph/0109064}}].

\bibitem{Arnold:2001ms}
P.~Arnold, G.~D. Moore, and L.~G. Yaffe, {\it Photon emission from quark gluon
  plasma: Complete leading order results},  {\em JHEP} {\bf 12} (2001) 009,
  [\href{http://xxx.lanl.gov/abs/hep-ph/0111107}{{\tt hep-ph/0111107}}].

\bibitem{Turbide:2003si}
S.~Turbide, R.~Rapp, and C.~Gale, {\it Hadronic production of thermal photons},
   {\em Phys. Rev.} {\bf C69} (2004) 014903,
  [\href{http://xxx.lanl.gov/abs/hep-ph/0308085}{{\tt hep-ph/0308085}}].

\bibitem{we90}
H.~A. Weldon, {\it Reformulation of finite temperature dilepton production},
  {\em Phys. Rev.} {\bf D42} (1990) 2384--2387.

\bibitem{gk91}
C.~Gale and J.~I. Kapusta, {\it Vector dominance model at finite temperature},
  {\em Nucl. Phys.} {\bf B357} (1991) 65--89.

\bibitem{ra96}
R.~Rapp, G.~Chanfray, and J.~Wambach, {\it Medium modifications of the rho
  meson at cern sps energies},  {\em Phys. Rev. Lett.} {\bf 76} (1996)
  368--371, [\href{http://xxx.lanl.gov/abs/hep-ph/9508353}{{\tt
  hep-ph/9508353}}].

\bibitem{br97}
E.~L. Bratkovskaya and W.~Cassing, {\it Dilepton production from ags to sps
  energies within a relativistic transport approach},  {\em Nucl. Phys.} {\bf
  A619} (1997) 413--446, [\href{http://xxx.lanl.gov/abs/nucl-th/9611042}{{\tt
  nucl-th/9611042}}].

\bibitem{le98}
S.~Leupold, W.~Peters, and U.~Mosel, {\it What qcd sum rules tell about the rho
  meson},  {\em Nucl. Phys.} {\bf A628} (1998) 311--324,
  [\href{http://xxx.lanl.gov/abs/nucl-th/9708016}{{\tt nucl-th/9708016}}].

\bibitem{Cooper:1998hu}
F.~Cooper, {\it Inclusive dilepton production at rhic: A field theory approach
  based on a non-equilibrium chiral phase transition},  {\em Phys. Rept.} {\bf
  315} (1999) 59--81, [\href{http://xxx.lanl.gov/abs/hep-ph/9811246}{{\tt
  hep-ph/9811246}}].

\bibitem{Rapp:1999ej}
R.~Rapp and J.~Wambach, {\it Chiral symmetry restoration and dileptons in
  relativistic heavy-ion collisions},  {\em Adv. Nucl. Phys.} {\bf 25} (2000)
  1, [\href{http://xxx.lanl.gov/abs/hep-ph/9909229}{{\tt hep-ph/9909229}}].

\bibitem{Renk:2002md}
T.~Renk, R.~A. Schneider, and W.~Weise, {\it Phases of qcd, thermal
  quasiparticles and dilepton radiation from a fireball},  {\em Phys. Rev.}
  {\bf C66} (2002) 014902, [\href{http://xxx.lanl.gov/abs/hep-ph/0201048}{{\tt
  hep-ph/0201048}}].

\bibitem{Ruppert:2005id}
J.~Ruppert, T.~Renk, and B.~Muller, {\it {Mass and width of the rho meson in a
  nuclear medium from Brown-Rho scaling and QCD sum rules}},  {\em Phys. Rev.}
  {\bf C73} (2006) 034907, [\href{http://xxx.lanl.gov/abs/hep-ph/0509134}{{\tt
  hep-ph/0509134}}].

\bibitem{Schenke:2005ry}
B.~Schenke and C.~Greiner, {\it {Dilepton production from hot hadronic matter
  in nonequilibrium}},  {\em Phys. Rev.} {\bf C73} (2006) 034909,
  [\href{http://xxx.lanl.gov/abs/hep-ph/0509026}{{\tt hep-ph/0509026}}].

\bibitem{Schenke:2006uh}
B.~Schenke and C.~Greiner, {\it {Dilepton yields from Brown-Rho scaled vector
  mesons including memory effects}},  {\em Phys. Rev. Lett.} {\bf 98} (2007)
  022301, [\href{http://xxx.lanl.gov/abs/hep-ph/0608032}{{\tt
  hep-ph/0608032}}].

\bibitem{Brown:1991kk}
G.~E. Brown and M.~Rho, {\it {Scaling effective Lagrangians in a dense
  medium}},  {\em Phys. Rev. Lett.} {\bf 66} (1991) 2720--2723.

\bibitem{Hatsuda:1991ez}
T.~Hatsuda and S.~H. Lee, {\it {QCD sum rules for vector mesons in nuclear
  medium}},  {\em Phys. Rev.} {\bf C46} (1992) 34--38.

\bibitem{ce95}
{\bf CERES} Collaboration, G.~Agakichiev {\em et~al.} {\em Phys. Rev. Lett.}
  {\bf 75} (1995) 1272.

\bibitem{Arnaldi:2006jq}
{\bf NA60} Collaboration, R.~Arnaldi {\em et~al.}, {\it {First measurement of
  the rho spectral function in high- energy nuclear collisions}},  {\em Phys.
  Rev. Lett.} {\bf 96} (2006) 162302,
  [\href{http://xxx.lanl.gov/abs/nucl-ex/0605007}{{\tt nucl-ex/0605007}}].

\bibitem{Elze:1989un}
H.-T. Elze and U.~W. Heinz, {\it Quark - gluon transport theory},  {\em Phys.
  Rept.} {\bf 183} (1989) 81--135.

\bibitem{Blaizot:1993zk}
J.~P. Blaizot and E.~Iancu, {\it Kinetic equations for long wavelength
  excitations of the quark - gluon plasma},  {\em Phys. Rev. Lett.} {\bf 70}
  (1993) 3376--3379, [\href{http://xxx.lanl.gov/abs/hep-ph/9301236}{{\tt
  hep-ph/9301236}}].

\bibitem{Blaizot:1993be}
J.~P. Blaizot and E.~Iancu, {\it Soft collective excitations in hot gauge
  theories},  {\em Nucl. Phys.} {\bf B417} (1994) 608--673,
  [\href{http://xxx.lanl.gov/abs/hep-ph/9306294}{{\tt hep-ph/9306294}}].

\bibitem{Kelly:1994dh}
P.~F. Kelly, Q.~Liu, C.~Lucchesi, and C.~Manuel, {\it Classical transport
  theory and hard thermal loops in the quark - gluon plasma},  {\em Phys. Rev.}
  {\bf D50} (1994) 4209--4218,
  [\href{http://xxx.lanl.gov/abs/hep-ph/9406285}{{\tt hep-ph/9406285}}].

\bibitem{Blaizot:2001nr}
J.-P. Blaizot and E.~Iancu, {\it The quark-gluon plasma: Collective dynamics
  and hard thermal loops},  {\em Phys. Rept.} {\bf 359} (2002) 355--528,
  [\href{http://xxx.lanl.gov/abs/hep-ph/0101103}{{\tt hep-ph/0101103}}].

\bibitem{Lifshitz81}
E.~Lifshitz and L.~Pitaevskii, {\it Physical kinetics},  {\em Pergamon Press,
  Oxford} (1981).

\bibitem{Linde:1980ts}
A.~D. Linde, {\it {Infrared Problem in Thermodynamics of the Yang-Mills Gas}},
  {\em Phys. Lett.} {\bf B96} (1980) 289.

\bibitem{Pisarski:1988vd}
R.~D. Pisarski, {\it {Scattering Amplitudes in Hot Gauge Theories}},  {\em
  Phys. Rev. Lett.} {\bf 63} (1989) 1129.

\bibitem{Braaten:1989mz}
E.~Braaten and R.~D. Pisarski, {\it Soft amplitudes in hot gauge theories: A
  general analysis},  {\em Nucl. Phys.} {\bf B337} (1990) 569.

\bibitem{DeWitt:1967}
B.~D. Witt {\em Phys. Rev.} {\bf 162} (1967) 1195.

\bibitem{DeWitt:1975ys}
B.~S. DeWitt, {\it Quantum field theory in curved space-time},  {\em Phys.
  Rept.} {\bf 19} (1975) 295--357.

\bibitem{Abbott:1980hw}
L.~F. Abbott, {\it The background field method beyond one loop},  {\em Nucl.
  Phys.} {\bf B185} (1981) 189.

\bibitem{Hansson:1987um}
T.~H. Hansson and I.~Zahed, {\it Electric and magnetic properties of hot
  gluons},  {\em Phys. Rev. Lett.} {\bf 58} (1987) 2397.

\bibitem{Heinz:1984yq}
U.~W. Heinz, {\it Quark - gluon transport theory. part 1. the classical
  theory},  {\em Ann. Phys.} {\bf 161} (1985) 48.

\bibitem{Elze:1986qd}
H.~T. Elze, M.~Gyulassy, and D.~Vasak, {\it Transport equations for the qcd
  quark wigner operator},  {\em Nucl. Phys.} {\bf B276} (1986) 706--728.

\bibitem{Gyulassy:1986da}
M.~Gyulassy, H.~T. Elze, A.~Iwazaki, and D.~Vasak, {\it Introduction to quantum
  chromo transport theory for quark - gluon plasmas}, . LBL-22072.

\bibitem{Mrowczynski:2000ed}
S.~Mrowczynski and M.~H. Thoma, {\it Hard loop approach to anisotropic
  systems},  {\em Phys. Rev.} {\bf D62} (2000) 036011,
  [\href{http://xxx.lanl.gov/abs/hep-ph/0001164}{{\tt hep-ph/0001164}}].

\bibitem{Weldon:1982aq}
H.~A. Weldon, {\it Covariant calculations at finite temperature: The
  relativistic plasma},  {\em Phys. Rev.} {\bf D26} (1982) 1394.

\bibitem{Gribov:1984tu}
L.~V. Gribov, E.~M. Levin, and M.~G. Ryskin, {\it {Semihard Processes in QCD}},
   {\em Phys. Rept.} {\bf 100} (1983) 1--150.

\bibitem{Mueller:1985wy}
A.~H. Mueller and J.-w. Qiu, {\it {Gluon Recombination and Shadowing at Small
  Values of x}},  {\em Nucl. Phys.} {\bf B268} (1986) 427.

\bibitem{Blaizot:1987nc}
J.~P. Blaizot and A.~H. Mueller, {\it {The Early Stage of Ultrarelativistic
  Heavy Ion Collisions}},  {\em Nucl. Phys.} {\bf B289} (1987) 847.

\bibitem{McLerran:1993ni}
L.~D. McLerran and R.~Venugopalan, {\it {Computing quark and gluon distribution
  functions for very large nuclei}},  {\em Phys. Rev.} {\bf D49} (1994)
  2233--2241, [\href{http://xxx.lanl.gov/abs/hep-ph/9309289}{{\tt
  hep-ph/9309289}}].

\bibitem{McLerran:1993ka}
L.~D. McLerran and R.~Venugopalan, {\it {Gluon distribution functions for very
  large nuclei at small transverse momentum}},  {\em Phys. Rev.} {\bf D49}
  (1994) 3352--3355, [\href{http://xxx.lanl.gov/abs/hep-ph/9311205}{{\tt
  hep-ph/9311205}}].

\bibitem{McLerran:1994vd}
L.~D. McLerran and R.~Venugopalan, {\it {Green's functions in the color field
  of a large nucleus}},  {\em Phys. Rev.} {\bf D50} (1994) 2225--2233,
  [\href{http://xxx.lanl.gov/abs/hep-ph/9402335}{{\tt hep-ph/9402335}}].

\bibitem{JalilianMarian:1996xn}
J.~Jalilian-Marian, A.~Kovner, L.~D. McLerran, and H.~Weigert, {\it {The
  intrinsic glue distribution at very small x}},  {\em Phys. Rev.} {\bf D55}
  (1997) 5414--5428, [\href{http://xxx.lanl.gov/abs/hep-ph/9606337}{{\tt
  hep-ph/9606337}}].

\bibitem{Kovchegov:1996ty}
Y.~V. Kovchegov, {\it {Non-Abelian Weizsaecker-Williams field and a two-
  dimensional effective color charge density for a very large nucleus}},  {\em
  Phys. Rev.} {\bf D54} (1996) 5463--5469,
  [\href{http://xxx.lanl.gov/abs/hep-ph/9605446}{{\tt hep-ph/9605446}}].

\bibitem{JalilianMarian:1997jx}
J.~Jalilian-Marian, A.~Kovner, A.~Leonidov, and H.~Weigert, {\it {The BFKL
  equation from the Wilson renormalization group}},  {\em Nucl. Phys.} {\bf
  B504} (1997) 415--431, [\href{http://xxx.lanl.gov/abs/hep-ph/9701284}{{\tt
  hep-ph/9701284}}].

\bibitem{JalilianMarian:1997gr}
J.~Jalilian-Marian, A.~Kovner, A.~Leonidov, and H.~Weigert, {\it {The Wilson
  renormalization group for low x physics: Towards the high density regime}},
  {\em Phys. Rev.} {\bf D59} (1999) 014014,
  [\href{http://xxx.lanl.gov/abs/hep-ph/9706377}{{\tt hep-ph/9706377}}].

\bibitem{McLerran:1998nk}
L.~D. McLerran and R.~Venugopalan, {\it {Fock space distributions, structure
  functions, higher twists and small x}},  {\em Phys. Rev.} {\bf D59} (1999)
  094002, [\href{http://xxx.lanl.gov/abs/hep-ph/9809427}{{\tt
  hep-ph/9809427}}].

\bibitem{Kovner:1999bj}
A.~Kovner and J.~G. Milhano, {\it {Vector potential versus colour charge
  density in low-x evolution}},  {\em Phys. Rev.} {\bf D61} (2000) 014012,
  [\href{http://xxx.lanl.gov/abs/hep-ph/9904420}{{\tt hep-ph/9904420}}].

\bibitem{Iancu:2000hn}
E.~Iancu, A.~Leonidov, and L.~D. McLerran, {\it {Nonlinear gluon evolution in
  the color glass condensate. I}},  {\em Nucl. Phys.} {\bf A692} (2001)
  583--645, [\href{http://xxx.lanl.gov/abs/hep-ph/0011241}{{\tt
  hep-ph/0011241}}].

\bibitem{Iancu:2001ad}
E.~Iancu, A.~Leonidov, and L.~D. McLerran, {\it {The renormalization group
  equation for the color glass condensate}},  {\em Phys. Lett.} {\bf B510}
  (2001) 133--144, [\href{http://xxx.lanl.gov/abs/hep-ph/0102009}{{\tt
  hep-ph/0102009}}].

\bibitem{Gavai:1996vu}
R.~V. Gavai and R.~Venugopalan, {\it {Lattice computations of small-x parton
  distributions in a model of parton densities in very large nuclei}},  {\em
  Phys. Rev.} {\bf D54} (1996) 5795--5803,
  [\href{http://xxx.lanl.gov/abs/hep-ph/9605327}{{\tt hep-ph/9605327}}].

\bibitem{Romatschke:2004jh}
P.~Romatschke and M.~Strickland, {\it Collective modes of an anisotropic
  quark-gluon plasma. ii},  {\em Phys. Rev.} {\bf D70} (2004) 116006,
  [\href{http://xxx.lanl.gov/abs/hep-ph/0406188}{{\tt hep-ph/0406188}}].

\bibitem{Dumitru:2007hy}
A.~Dumitru, Y.~Guo, and M.~Strickland, {\it {The heavy-quark potential in an
  anisotropic (viscous) plasma}},
  \href{http://xxx.lanl.gov/abs/0711.4722}{{\tt 0711.4722}}.

\bibitem{Kra73}
N.~Krall and A.~Trivelpiece, {\it Principles of plasma physics},  {\em
  McGraw-Hill, New York} (1973).

\bibitem{Heinz:1985vf}
U.~W. Heinz, {\it Quark - gluon transport theory},  {\em Nucl. Phys.} {\bf
  A418} (1984) 603c--612c.

\bibitem{Pokrovsky:1988bm}
Y.~E. Pokrovsky and A.~V. Selikhov, {\it Filamentation in a quark - gluon
  plasma},  {\em JETP Lett.} {\bf 47} (1988) 12--14.

\bibitem{Pokrovsky:1990sz}
Y.~E. Pokrovsky and A.~V. Selikhov, {\it Filamentation in quark plasma at
  finite temperatures},  {\em Sov. J. Nucl. Phys.} {\bf 52} (1990) 146--152.

\bibitem{Pokrovsky:1990uh}
Y.~E. Pokrovsky and A.~V. Selikhov, {\it Filamentation in the quark-gluon
  plasma at finite temperatures},  {\em Sov. J. Nucl. Phys.} {\bf 52} (1990)
  385--387.

\bibitem{Mrowczynski:1988dz}
S.~Mrowczynski, {\it Stream instabilities of the quark - gluon plasma},  {\em
  Phys. Lett.} {\bf B214} (1988) 587.

\bibitem{Pavlenko:1990as}
O.~P. Pavlenko, {\it Dynamical instabilities in quark - gluon plasma with hard
  jet},  {\em Sov. J. Nucl. Phys.} {\bf 54} (1991) 884--886.

\bibitem{Pavlenko:1991ih}
O.~P. Pavlenko, {\it Filamentation instability of hot quark - gluon plasma with
  hard jet},  {\em Sov. J. Nucl. Phys.} {\bf 55} (1992) 1243--1245.

\bibitem{Weibel:1959}
E.~Weibel, {\it Spontaneously growing transverse waves in a plasma due to an
  anisotropic velocity distribution},  {\em Phys. Rev. Lett.} {\bf 2} (1959)
  83--84.

\bibitem{Birse:2003qp}
M.~C. Birse, C.-W. Kao, and G.~C. Nayak, {\it Magnetic screening effects in
  anisotropic qed and qcd plasmas},  {\em Phys. Lett.} {\bf B570} (2003)
  171--179, [\href{http://xxx.lanl.gov/abs/hep-ph/0304209}{{\tt
  hep-ph/0304209}}].

\bibitem{Mrowczynski:2004kv}
S.~Mrowczynski, A.~Rebhan, and M.~Strickland, {\it Hard-loop effective action
  for anisotropic plasmas},  {\em Phys. Rev.} {\bf D70} (2004) 025004,
  [\href{http://xxx.lanl.gov/abs/hep-ph/0403256}{{\tt hep-ph/0403256}}].

\bibitem{Arnold:2005vb}
P.~Arnold, G.~D. Moore, and L.~G. Yaffe, {\it The fate of non-abelian plasma
  instabilities in 3+1 dimensions},  {\em Phys. Rev.} {\bf D72} (2005) 054003,
  [\href{http://xxx.lanl.gov/abs/hep-ph/0505212}{{\tt hep-ph/0505212}}].

\bibitem{Rebhan:2005re}
A.~Rebhan, P.~Romatschke, and M.~Strickland, {\it Dynamics of quark-gluon
  plasma instabilities in discretized hard-loop approximation},  {\em JHEP}
  {\bf 09} (2005) 041, [\href{http://xxx.lanl.gov/abs/hep-ph/0505261}{{\tt
  hep-ph/0505261}}].

\bibitem{Mrowczynski:1996vh}
S.~Mrowczynski, {\it Color filamentation in ultrarelativistic heavy-ion
  collisions},  {\em Phys. Lett.} {\bf B393} (1997) 26--30,
  [\href{http://xxx.lanl.gov/abs/hep-ph/9606442}{{\tt hep-ph/9606442}}].

\bibitem{Kato:2005wv}
T.~N. Kato, {\it Saturation mechanism of the weibel instability in weakly
  magnetized plasmas},  {\em Phys. Plasmas} {\bf 12} (2005) 080705,
  [\href{http://xxx.lanl.gov/abs/physics/0501110}{{\tt physics/0501110}}].

\bibitem{Arnold:2004ih}
P.~Arnold and J.~Lenaghan, {\it The abelianization of qcd plasma
  instabilities},  {\em Phys. Rev.} {\bf D70} (2004) 114007,
  [\href{http://xxx.lanl.gov/abs/hep-ph/0408052}{{\tt hep-ph/0408052}}].

\bibitem{Arnold:2005ef}
P.~Arnold and G.~D. Moore, {\it Qcd plasma instabilities: The nonabelian
  cascade},  {\em Phys. Rev.} {\bf D73} (2006) 025006,
  [\href{http://xxx.lanl.gov/abs/hep-ph/0509206}{{\tt hep-ph/0509206}}].

\bibitem{Dumitru:2005hj}
A.~Dumitru and Y.~Nara, {\it Numerical simulation of non-abelian particle-field
  dynamics},  {\em Eur. Phys. J.} {\bf A29} (2006) 65--69,
  [\href{http://xxx.lanl.gov/abs/hep-ph/0511242}{{\tt hep-ph/0511242}}].

\bibitem{Dumitru:2006pz}
A.~Dumitru, Y.~Nara, and M.~Strickland, {\it Ultraviolet avalanche in
  anisotropic non-abelian plasmas},  {\em Phys. Rev.} {\bf D75} (2007) 025016,
  [\href{http://xxx.lanl.gov/abs/hep-ph/0604149}{{\tt hep-ph/0604149}}].

\bibitem{Romatschke:2006wg}
P.~Romatschke and A.~Rebhan, {\it Plasma instabilities in an anisotropically
  expanding geometry},  {\em Phys. Rev. Lett.} {\bf 97} (2006) 252301,
  [\href{http://xxx.lanl.gov/abs/hep-ph/0605064}{{\tt hep-ph/0605064}}].

\bibitem{Rebhan:2008uj}
A.~Rebhan, M.~Strickland, and M.~Attems, {\it {Instabilities of an
  anisotropically expanding non-Abelian plasma: 1D+3V discretized hard-loop
  simulations}},  \href{http://xxx.lanl.gov/abs/0802.1714}{{\tt 0802.1714}}.

\bibitem{Romatschke:2005pm}
P.~Romatschke and R.~Venugopalan, {\it Collective non-abelian instabilities in
  a melting color glass condensate},  {\em Phys. Rev. Lett.} {\bf 96} (2006)
  062302, [\href{http://xxx.lanl.gov/abs/hep-ph/0510121}{{\tt
  hep-ph/0510121}}].

\bibitem{Romatschke:2005ag}
P.~Romatschke and R.~Venugopalan, {\it Signals of a weibel instability in the
  melting color glass condensate},  {\em Eur. Phys. J.} {\bf A29} (2006)
  71--75, [\href{http://xxx.lanl.gov/abs/hep-ph/0510292}{{\tt
  hep-ph/0510292}}].

\bibitem{Romatschke:2006nk}
P.~Romatschke and R.~Venugopalan, {\it The unstable glasma},  {\em Phys. Rev.}
  {\bf D74} (2006) 045011, [\href{http://xxx.lanl.gov/abs/hep-ph/0605045}{{\tt
  hep-ph/0605045}}].

\bibitem{Iancu:2003xm}
E.~Iancu and R.~Venugopalan, {\it The color glass condensate and high energy
  scattering in qcd},  \href{http://xxx.lanl.gov/abs/hep-ph/0303204}{{\tt
  hep-ph/0303204}}.

\bibitem{Bhatnagar:1954}
P.~L. Bhatnagar, E.~P. Gross, and M.~Krook, {\it A model for collision
  processes in gases. i. small amplitude processes in charged and neutral
  one-component systems},  {\em Phys. Rev.} {\bf 94} (1954) 511--525.

\bibitem{Schenke:2006xu}
B.~Schenke, M.~Strickland, C.~Greiner, and M.~H. Thoma, {\it A model of the
  effect of collisions on qcd plasma instabilities},  {\em Phys. Rev.} {\bf
  D73} (2006) 125004, [\href{http://xxx.lanl.gov/abs/hep-ph/0603029}{{\tt
  hep-ph/0603029}}].

\bibitem{Manuel:2004gk}
C.~Manuel and S.~Mrowczynski, {\it Whitening of the quark-gluon plasma},  {\em
  Phys. Rev.} {\bf D70} (2004) 094019,
  [\href{http://xxx.lanl.gov/abs/hep-ph/0403024}{{\tt hep-ph/0403024}}].

\bibitem{Carrington:2003je}
M.~E. Carrington, T.~Fugleberg, D.~Pickering, and M.~H. Thoma, {\it Dielectric
  functions and dispersion relations of ultra- relativistic plasmas with
  collisions},  {\em Can. J. Phys.} {\bf 82} (2004) 671--678,
  [\href{http://xxx.lanl.gov/abs/hep-ph/0312103}{{\tt hep-ph/0312103}}].

\bibitem{Selikhov:1993ns}
A.~Selikhov and M.~Gyulassy, {\it Color diffusion and conductivity in a quark -
  gluon plasma},  {\em Phys. Lett.} {\bf B316} (1993) 373--380,
  [\href{http://xxx.lanl.gov/abs/nucl-th/9307007}{{\tt nucl-th/9307007}}].

\bibitem{Bodeker:1998hm}
D.~Bodeker, {\it On the effective dynamics of soft non-abelian gauge fields at
  finite temperature},  {\em Phys. Lett.} {\bf B426} (1998) 351--360,
  [\href{http://xxx.lanl.gov/abs/hep-ph/9801430}{{\tt hep-ph/9801430}}].

\bibitem{Bodeker:1999ey}
D.~Bodeker, {\it From hard thermal loops to langevin dynamics},  {\em Nucl.
  Phys.} {\bf B559} (1999) 502--538,
  [\href{http://xxx.lanl.gov/abs/hep-ph/9905239}{{\tt hep-ph/9905239}}].

\bibitem{Arnold:2002zm}
P.~Arnold, G.~D. Moore, and L.~G. Yaffe, {\it Effective kinetic theory for high
  temperature gauge theories},  {\em JHEP} {\bf 01} (2003) 030,
  [\href{http://xxx.lanl.gov/abs/hep-ph/0209353}{{\tt hep-ph/0209353}}].

\bibitem{Thoma:1993vs}
M.~H. Thoma, {\it Parton interaction rates in the quark - gluon plasma},  {\em
  Phys. Rev.} {\bf D49} (1994) 451--459,
  [\href{http://xxx.lanl.gov/abs/hep-ph/9308257}{{\tt hep-ph/9308257}}].

\bibitem{Braaten:1991we}
E.~Braaten and M.~H. Thoma, {\it Energy loss of a heavy quark in the quark -
  gluon plasma},  {\em Phys. Rev.} {\bf D44} (1991) 2625--2630.

\bibitem{Romatschke:2003vc}
P.~Romatschke and M.~Strickland, {\it Energy loss of a heavy fermion in an
  anisotropic qed plasma},  {\em Phys. Rev.} {\bf D69} (2004) 065005,
  [\href{http://xxx.lanl.gov/abs/hep-ph/0309093}{{\tt hep-ph/0309093}}].

\bibitem{Romatschke:2004au}
P.~Romatschke and M.~Strickland, {\it Collisional energy loss of a heavy quark
  in an anisotropic quark-gluon plasma},  {\em Phys. Rev.} {\bf D71} (2005)
  125008, [\href{http://xxx.lanl.gov/abs/hep-ph/0408275}{{\tt
  hep-ph/0408275}}].

\bibitem{Peshier:2004bv}
A.~Peshier, {\it Hard gluon damping in hot qcd},  {\em Phys. Rev.} {\bf D70}
  (2004) 034016, [\href{http://xxx.lanl.gov/abs/hep-ph/0403225}{{\tt
  hep-ph/0403225}}].

\bibitem{Peshier:2005pp}
A.~Peshier and W.~Cassing, {\it The hot non-perturbative gluon plasma is an
  almost ideal colored liquid},  {\em Phys. Rev. Lett.} {\bf 94} (2005) 172301,
  [\href{http://xxx.lanl.gov/abs/hep-ph/0502138}{{\tt hep-ph/0502138}}].

\bibitem{Krasnitz:1998ns}
A.~Krasnitz and R.~Venugopalan, {\it Non-perturbative computation of gluon
  mini-jet production in nuclear collisions at very high energies},  {\em Nucl.
  Phys.} {\bf B557} (1999) 237,
  [\href{http://xxx.lanl.gov/abs/hep-ph/9809433}{{\tt hep-ph/9809433}}].

\bibitem{Krasnitz:1999wc}
A.~Krasnitz and R.~Venugopalan, {\it The initial energy density of gluons
  produced in very high energy nuclear collisions},  {\em Phys. Rev. Lett.}
  {\bf 84} (2000) 4309--4312,
  [\href{http://xxx.lanl.gov/abs/hep-ph/9909203}{{\tt hep-ph/9909203}}].

\bibitem{Krasnitz:2001qu}
A.~Krasnitz, Y.~Nara, and R.~Venugopalan, {\it Coherent gluon production in
  very high energy heavy ion collisions},  {\em Phys. Rev. Lett.} {\bf 87}
  (2001) 192302, [\href{http://xxx.lanl.gov/abs/hep-ph/0108092}{{\tt
  hep-ph/0108092}}].

\bibitem{Krasnitz:2002mn}
A.~Krasnitz, Y.~Nara, and R.~Venugopalan, {\it Gluon production in the color
  glass condensate model of collisions of ultrarelativistic finite nuclei},
  {\em Nucl. Phys.} {\bf A717} (2003) 268--290,
  [\href{http://xxx.lanl.gov/abs/hep-ph/0209269}{{\tt hep-ph/0209269}}].

\bibitem{Krasnitz:2003jw}
A.~Krasnitz, Y.~Nara, and R.~Venugopalan, {\it Classical gluodynamics of high
  energy nuclear collisions: An erratum and an update},  {\em Nucl. Phys.} {\bf
  A727} (2003) 427--436, [\href{http://xxx.lanl.gov/abs/hep-ph/0305112}{{\tt
  hep-ph/0305112}}].

\bibitem{Krasnitz:2002ng}
A.~Krasnitz, Y.~Nara, and R.~Venugopalan, {\it Elliptic flow of colored glass
  in high energy heavy ion collisions},  {\em Phys. Lett.} {\bf B554} (2003)
  21--27, [\href{http://xxx.lanl.gov/abs/hep-ph/0204361}{{\tt
  hep-ph/0204361}}].

\bibitem{Heinz:1983nx}
U.~W. Heinz, {\it Kinetic theory for nonabelian plasmas},  {\em Phys. Rev.
  Lett.} {\bf 51} (1983) 351.

\bibitem{Heinz:1985qe}
U.~W. Heinz, {\it Quark - gluon transport theory. part 2. color response and
  color correlations in a quark - gluon plasma},  {\em Ann. Phys.} {\bf 168}
  (1986) 148.

\bibitem{Wong:1970fu}
S.~K. Wong, {\it Field and particle equations for the classical yang-mills
  field and particles with isotopic spin},  {\em Nuovo Cim.} {\bf A65S10}
  (1970) 689--694.

\bibitem{Litim:1999id}
D.~F. Litim and C.~Manuel, {\it {Effective transport equations for non-Abelian
  plasmas}},  {\em Nucl. Phys.} {\bf B562} (1999) 237--274,
  [\href{http://xxx.lanl.gov/abs/hep-ph/9906210}{{\tt hep-ph/9906210}}].

\bibitem{Litim:1999ns}
D.~F. Litim and C.~Manuel, {\it {Mean field dynamics in non-Abelian plasmas
  from classical transport theory}},  {\em Phys. Rev. Lett.} {\bf 82} (1999)
  4981--4984, [\href{http://xxx.lanl.gov/abs/hep-ph/9902430}{{\tt
  hep-ph/9902430}}].

\bibitem{Bertsch:1988ik}
G.~F. Bertsch and S.~Das~Gupta, {\it A guide to microscopic models for
  intermediate-energy heavy ion collisions},  {\em Phys. Rept.} {\bf 160}
  (1988) 189--233.

\bibitem{Ambjorn:1990pu}
J.~Ambjorn, T.~Askgaard, H.~Porter, and M.~E. Shaposhnikov, {\it Sphaleron
  transitions and baryon asymmetry: A numerical real time analysis},  {\em
  Nucl. Phys.} {\bf B353} (1991) 346--378.

\bibitem{Hu:1996sf}
C.~R. Hu and B.~Muller, {\it Classical lattice gauge field with hard thermal
  loops},  {\em Phys. Lett.} {\bf B409} (1997) 377--381,
  [\href{http://xxx.lanl.gov/abs/hep-ph/9611292}{{\tt hep-ph/9611292}}].

\bibitem{Moore:1997sn}
G.~D. Moore, C.-r. Hu, and B.~Muller, {\it Chern-simons number diffusion with
  hard thermal loops},  {\em Phys. Rev.} {\bf D58} (1998) 045001,
  [\href{http://xxx.lanl.gov/abs/hep-ph/9710436}{{\tt hep-ph/9710436}}].

\bibitem{Hockney:1981}
R.~W. Hockney and J.~Eastwood, {\em Computer Simulation using Particles}.
\newblock MacGraw-Hill, New York, 1981.

\bibitem{Birdsall:1985}
C.~Birdsall and A.~Langdon, {\em Plasma Physics via Computer Simulation}.
\newblock MacGraw-Hill, New York, 1985.

\bibitem{Eastwood:1991Co}
J.~Eastwood, {\it The virtual particle electromagnetic particle-mesh method},
  {\em Computer Physics Communications} {\bf 64} (1991) 252--266.

\bibitem{Umeda:2003}
T.~Umeda, Y.~Omura, T.~Tominaga, and H.~Matsumoto, {\it A new charge
  conservation method in electromagnetic particle-in-cell simulations},  {\em
  Comput. Phys. Comm.} {\bf 156} (2003) 73.

\bibitem{PhysRevC.29.2146}
T.~Kodama, S.~B. Duarte, K.~C. Chung, R.~Donangelo, and R.~A. M.~S. Nazareth,
  {\it Causality and relativistic effects in intranuclear cascade
  calculations},  {\em Phys. Rev. C} {\bf 29} (Jun, 1984) 2146--2152.

\bibitem{Kortemeyer:1995di}
G.~Kortemeyer, W.~Bauer, K.~Haglin, J.~Murray, and S.~Pratt, {\it Causality
  violations in cascade models of nuclear collisions},  {\em Phys. Rev.} {\bf
  C52} (1995) 2714--2724, [\href{http://xxx.lanl.gov/abs/nucl-th/9509013}{{\tt
  nucl-th/9509013}}].

\bibitem{Zhang:1998tj}
B.~Zhang, M.~Gyulassy, and Y.~Pang, {\it Equation of state and collision rate
  tests of parton cascade models},  {\em Phys. Rev.} {\bf C58} (1998)
  1175--1182, [\href{http://xxx.lanl.gov/abs/nucl-th/9801037}{{\tt
  nucl-th/9801037}}].

\bibitem{Cheng:2001dz}
S.~Cheng {\em et~al.}, {\it The effect of finite-range interactions in
  classical transport theory},  {\em Phys. Rev.} {\bf C65} (2002) 024901,
  [\href{http://xxx.lanl.gov/abs/nucl-th/0107001}{{\tt nucl-th/0107001}}].

\bibitem{PhysRevC.40.2611}
G.~Welke, R.~Malfliet, C.~Gr\'egoire, M.~Prakash, and E.~Suraud, {\it
  Collisional relaxation in simulations of heavy-ion collisions using
  boltzmann-type equations},  {\em Phys. Rev. C} {\bf 40} (Dec, 1989)
  2611--2620.

\bibitem{Molnar:2001ux}
D.~Molnar and M.~Gyulassy, {\it Saturation of elliptic flow at rhic: Results
  from the covariant elastic parton cascade model mpc},  {\em Nucl. Phys.} {\bf
  A697} (2002) 495--520, [\href{http://xxx.lanl.gov/abs/nucl-th/0104073}{{\tt
  nucl-th/0104073}}].

\bibitem{Xu:2004gw}
X.-M. Xu, Y.~Sun, A.-Q. Chen, and L.~Zheng, {\it Triple-gluon scatterings and
  early thermalization},  {\em Nucl. Phys.} {\bf A744} (2004) 347--377.

\bibitem{Xu:2007aa}
Z.~Xu and C.~Greiner, {\it {Transport rates and momentum isotropization of
  gluon matter in ultrarelativistic heavy-ion collisions}},  {\em Phys. Rev.}
  {\bf C76} (2007) 024911, [\href{http://xxx.lanl.gov/abs/hep-ph/0703233}{{\tt
  hep-ph/0703233}}].

\bibitem{Danielewicz:1991dh}
P.~Danielewicz and G.~F. Bertsch, {\it Production of deuterons and pions in a
  transport model of energetic heavy ion reactions},  {\em Nucl. Phys.} {\bf
  A533} (1991) 712--748.

\bibitem{Lang:1993dh}
A.~Lang, H.~Babovsky, W.~Cassing, U.~Mosel, H.~Reusch, and K.~Weber, {\it A new
  treatment of boltzmann-like collision integrals in nuclear kinetic
  equations},  {\em J. Comp. Phys.} {\bf 106} (1993) 391--396.

\bibitem{Groot:1980}
S.~de~Groot, W.~van Leeuwen, and C.~van Weert, {\em Relativistic Kinetic
  Theory: Principles and Applications}.
\newblock North Holland, Amsterdam, 1980.

\bibitem{Combridge:1977dm}
B.~L. Combridge, J.~Kripfganz, and J.~Ranft, {\it Hadron production at large
  transverse momentum and qcd},  {\em Phys. Lett.} {\bf B70} (1977) 234.

\bibitem{Owens:1977sj}
J.~F. Owens, E.~Reya, and M.~Gluck, {\it Detailed quantum chromodynamic
  predictions for high p(t) processes},  {\em Phys. Rev.} {\bf D18} (1978)
  1501.

\bibitem{Bern:2002tk}
Z.~Bern, A.~De~Freitas, and L.~J. Dixon, {\it Two-loop helicity amplitudes for
  gluon gluon scattering in qcd and supersymmetric yang-mills theory},  {\em
  JHEP} {\bf 03} (2002) 018,
  [\href{http://xxx.lanl.gov/abs/hep-ph/0201161}{{\tt hep-ph/0201161}}].

\bibitem{Press:1992}
W.~Press, S.~Teukolsky, W.~Vetterling, and B.~Flannery, {\em Numerical Recipes
  in C}.
\newblock Cambridge University Press, 1992.

\bibitem{Baier:1996sk}
R.~Baier, Y.~L. Dokshitzer, A.~H. Mueller, S.~Peigne, and D.~Schiff, {\it
  Radiative energy loss and p(t)-broadening of high energy partons in nuclei},
  {\em Nucl. Phys.} {\bf B484} (1997) 265--282,
  [\href{http://xxx.lanl.gov/abs/hep-ph/9608322}{{\tt hep-ph/9608322}}].

\bibitem{Majumder:2007iu}
A.~Majumder, {\it A comparative study of jet-quenching schemes},  {\em J.
  Phys.} {\bf G34} (2007) S377--388,
  [\href{http://xxx.lanl.gov/abs/nucl-th/0702066}{{\tt nucl-th/0702066}}].

\bibitem{Dumitru:2007rp}
A.~Dumitru, Y.~Nara, B.~Schenke, and M.~Strickland, {\it Jet broadening in
  unstable non-abelian plasmas},  \href{http://xxx.lanl.gov/abs/0710.1223}{{\tt
  0710.1223}}.

\bibitem{Nara:2005fr}
Y.~Nara, {\it Isotropization by qcd plasma instabilities},  {\em Nucl. Phys.}
  {\bf A774} (2006) 783--786,
  [\href{http://xxx.lanl.gov/abs/nucl-th/0509052}{{\tt nucl-th/0509052}}].

\bibitem{Arnold:2005qs}
P.~Arnold and G.~D. Moore, {\it The turbulent spectrum created by non-abelian
  plasma instabilities},  {\em Phys. Rev.} {\bf D73} (2006) 025013,
  [\href{http://xxx.lanl.gov/abs/hep-ph/0509226}{{\tt hep-ph/0509226}}].

\bibitem{Moore:2004tg}
G.~D. Moore and D.~Teaney, {\it How much do heavy quarks thermalize in a heavy
  ion collision?},  {\em Phys. Rev.} {\bf C71} (2005) 064904,
  [\href{http://xxx.lanl.gov/abs/hep-ph/0412346}{{\tt hep-ph/0412346}}].

\bibitem{Romatschke:2006bb}
P.~Romatschke, {\it Momentum broadening in an anisotropic plasma},  {\em Phys.
  Rev.} {\bf C75} (2007) 014901,
  [\href{http://xxx.lanl.gov/abs/hep-ph/0607327}{{\tt hep-ph/0607327}}].

\bibitem{Jacobs:2005pk}
P.~Jacobs, {\it Jets in nuclear collisions: Status and perspective},  {\em Eur.
  Phys. J.} {\bf C43} (2005) 467--473,
  [\href{http://xxx.lanl.gov/abs/nucl-ex/0503022}{{\tt nucl-ex/0503022}}].

\bibitem{Majumder:2006wi}
A.~Majumder, B.~Muller, and S.~A. Bass, {\it Longitudinal broadening of
  quenched jets in turbulent color fields},  {\em Phys. Rev. Lett.} {\bf 99}
  (2007) 042301, [\href{http://xxx.lanl.gov/abs/hep-ph/0611135}{{\tt
  hep-ph/0611135}}].

\bibitem{Armesto:2004vz}
N.~Armesto, C.~A. Salgado, and U.~A. Wiedemann, {\it Low-p(t) collective flow
  induces high-p(t) jet quenching},  {\em Phys. Rev.} {\bf C72} (2005) 064910,
  [\href{http://xxx.lanl.gov/abs/hep-ph/0411341}{{\tt hep-ph/0411341}}].

\bibitem{Mrowczynski:2001az}
S.~Mrowczynski, {\it Quasiquarks in two stream system},  {\em Phys. Rev.} {\bf
  D65} (2002) 117501, [\href{http://xxx.lanl.gov/abs/hep-ph/0112100}{{\tt
  hep-ph/0112100}}].

\bibitem{fp1a}
H.~Stoof {\em et~al.} {\em Phys. Rev. Lett.} {\bf 76} (1996) 10.

\bibitem{fp1b}
M.~Houbiers {\em et~al.} {\em Phys. Rev.} {\bf A56} (1997) 4864.

\bibitem{fp2a}
M.~Holland {\em et~al.} {\em Phys. Rev. Lett.} {\bf 87} (2001) 120406.

\bibitem{fp2b}
E.~Timmermans {\em et~al.} {\em Phys. Lett.} {\bf A285} (2001) 228.

\bibitem{fp2c}
Y.~Ohashi and A.~Griffin {\em Phys. Rev. Lett.} {\bf 89} (2002) 130402.

\bibitem{fp3}
J.~Kinast, S.~Hemmer, M.~Gehm, A.~Turlapov, and J.~Thomas {\em Phys. Rev.
  Lett.} {\bf 92} (2004) 150402.

\bibitem{Schenke:2006fz}
B.~Schenke and M.~Strickland, {\it Fermionic collective modes of an anisotropic
  quark-gluon plasma},  {\em Phys. Rev.} {\bf D74} (2006) 065004,
  [\href{http://xxx.lanl.gov/abs/hep-ph/0606160}{{\tt hep-ph/0606160}}].

\bibitem{Romatschke:2003yc}
P.~Romatschke, {\it Quasiparticle description of the hot and dense quark gluon
  plasma},  \href{http://xxx.lanl.gov/abs/hep-ph/0312152}{{\tt
  hep-ph/0312152}}.

\bibitem{Schenke:2006yp}
B.~Schenke and M.~Strickland, {\it Photon production from an anisotropic
  quark-gluon plasma},  {\em Phys. Rev.} {\bf D76} (2007) 025023,
  [\href{http://xxx.lanl.gov/abs/hep-ph/0611332}{{\tt hep-ph/0611332}}].

\bibitem{Weldon:1983jn}
H.~A. Weldon, {\it Simple rules for discontinuities in finite temperature field
  theory},  {\em Phys. Rev.} {\bf D28} (1983) 2007.

\bibitem{McLerran:1984ay}
L.~D. McLerran and T.~Toimela, {\it Photon and dilepton emission from the quark
  - gluon plasma: Some general considerations},  {\em Phys. Rev.} {\bf D31}
  (1985) 545.

\bibitem{Ke64}
L.~Keldysh {\em Zh. Eks. Teor. Fiz.} {\bf 47} (1964) 1515.

\bibitem{Ke65}
L.~Keldysh {\em Sov. Phys. JETP} {\bf 20} (1965) 1018.

\bibitem{Chou:1984es}
K.-c. Chou, Z.-b. Su, B.-l. Hao, and L.~Yu, {\it Equilibrium and nonequilibrium
  formalisms made unified},  {\em Phys. Rept.} {\bf 118} (1985) 1.

\bibitem{Mrowczynski:1992hq}
S.~Mrowczynski and U.~W. Heinz, {\it Towards a relativistic transport theory of
  nuclear matter},  {\em Ann. Phys.} {\bf 229} (1994) 1--54.

\bibitem{Calzetta:1986cq}
E.~Calzetta and B.~L. Hu, {\it Nonequilibrium quantum fields: Closed time path
  effective action, wigner function and boltzmann equation},  {\em Phys. Rev.}
  {\bf D37} (1988) 2878.

\bibitem{Baier:1997xc}
R.~Baier, M.~Dirks, K.~Redlich, and D.~Schiff, {\it Thermal photon production
  rate from non-equilibrium quantum field theory},  {\em Phys. Rev.} {\bf D56}
  (1997) 2548--2554, [\href{http://xxx.lanl.gov/abs/hep-ph/9704262}{{\tt
  hep-ph/9704262}}].

\bibitem{hm84}
F.~Halzen and A.~Martin, {\em Quarks \& Leptons}.
\newblock John Wiley \& Sons, New York, 1984.

\bibitem{PS}
M.~Peskin and D.~Schroeder, {\em An Introduction to Quantum Field Theory}.
\newblock Westview Press, 1995.

\bibitem{Braaten:1991dd}
E.~Braaten and T.~C. Yuan, {\it Calculation of screening in a hot plasma},
  {\em Phys. Rev. Lett.} {\bf 66} (1991) 2183--2186.

\bibitem{Baier:1991em}
R.~Baier, H.~Nakkagawa, A.~Niegawa, and K.~Redlich, {\it Production rate of
  hard thermal photons and screening of quark mass singularity},  {\em Z.
  Phys.} {\bf C53} (1992) 433--438.

\bibitem{Greiner:1998vd}
C.~Greiner and S.~Leupold, {\it Stochastic interpretation of kadanoff-baym
  equations and their relation to langevin processes},  {\em Annals Phys.} {\bf
  270} (1998) 328--390, [\href{http://xxx.lanl.gov/abs/hep-ph/9802312}{{\tt
  hep-ph/9802312}}].

\bibitem{Ipp:2007ng}
A.~Ipp, A.~Di~Piazza, J.~Evers, and C.~H. Keitel, {\it {Photon polarization as
  a probe for quark-gluon plasma dynamics}},
  \href{http://xxx.lanl.gov/abs/0710.5700}{{\tt 0710.5700}}.

\bibitem{Mauricio:2007vz}
M.~Martinez and M.~Strickland, {\it {Measuring QGP thermalization time with
  dileptons}},  {\em Phys. Rev. Lett.} {\bf 100} (2008) 102301,
  [\href{http://xxx.lanl.gov/abs/0709.3576}{{\tt 0709.3576}}].

\bibitem{Mauricio:new}
M.~Martinez and M.~Strickland {\em forthcoming}.

\bibitem{Ollitrault:2007du}
J.-Y. Ollitrault, {\it Relativistic hydrodynamics},
  \href{http://xxx.lanl.gov/abs/0708.2433}{{\tt 0708.2433}}.

\bibitem{Bodeker:2005nv}
D.~Bodeker, {\it {The impact of QCD plasma instabilities on bottom-up
  thermalization}},  {\em JHEP} {\bf 10} (2005) 092,
  [\href{http://xxx.lanl.gov/abs/hep-ph/0508223}{{\tt hep-ph/0508223}}].

\bibitem{Tatarkis:2003}
M.~Tatarakis, F.~N. Beg, E.~L. Clark, A.~E. Dangor, R.~D. Edwards, R.~G. Evans,
  T.~J. Goldsack, K.~W.~D. Ledingham, P.~A. Norreys, M.~A. Sinclair, M.-S. Wei,
  M.~Zepf, and K.~Krushelnick, {\it Propagation instabilities of high-intensity
  laser-produced electron beams},  {\em Phys. Rev. Lett.} {\bf 90} (2003)
  175001.

\bibitem{Bret:2005}
A.~Bret, M.-C. Firpo, and C.~Deutsch, {\it Characterization of the initial
  filamentation of a relativistic electron beam passing through a plasma},
  {\em Phys. Rev. Lett.} {\bf 94} (2005) 115002.

\bibitem{Tabak:1994}
M.~Tabak {\em et~al.}, {\it Ignition and high gain with ultrapowerful lasers},
  {\em Phys. Plasmas} {\bf 1} (1994) 1626.

\bibitem{Kodama:2001}
R.~Kodama {\em et~al.} {\em Nature} {\bf 412} (2001) 798.

\bibitem{Blaizot:1994vs}
J.-P. Blaizot and E.~Iancu, {\it Nonabelian plane waves in the quark - gluon
  plasma},  {\em Phys. Lett.} {\bf B326} (1994) 138--144,
  [\href{http://xxx.lanl.gov/abs/hep-ph/9401323}{{\tt hep-ph/9401323}}].

\bibitem{Kogut:1974ag}
J.~B. Kogut and L.~Susskind, {\it Hamiltonian formulation of wilson's lattice
  gauge theories},  {\em Phys. Rev.} {\bf D11} (1975) 395.

\bibitem{Wilson:1974sk}
K.~G. Wilson, {\it Confinement of quarks},  {\em Phys. Rev.} {\bf D10} (1974)
  2445--2459.

\bibitem{landauv2}
L.~D. Landau and E.~M. Lifschitz, {\em Lehrbuch der Theoretischen Physik,
  Klassische Feldtheorie}.
\newblock Harri Deutsch, Thun und Frankfurt am Main, 1997.

\bibitem{Greiner:1998ri}
C.~Greiner and S.~Leupold, {\it Interpretation and resolution of pinch
  singularities in non-equilibrium quantum field theory},  {\em Eur. Phys. J.}
  {\bf C8} (1999) 517--522, [\href{http://xxx.lanl.gov/abs/hep-ph/9804239}{{\tt
  hep-ph/9804239}}].

\end{thebibliography}\endgroup
  \cleardoublepage
  \chapter*{Danksagung\markboth{Danksagung}{Danksagung}}
\label{danksagung}
Zuallererst m\"ochte ich mich bei meinen Eltern f\"ur Ihre Unterst\"utzung bedanken,
die mir das Studium der Physik und die Erstellung dieser Arbeit erst erm\"oglicht hat.

Mein Dank gilt auch besonders Prof.\,Dr.\,Carsten Greiner f\"ur die Themenstellung und die Betreuung dieser Arbeit.
Mit gro\ss em Interesse hat er sie begleitet und ihre Entwicklung verfolgt, mir aber auch in vielen Punkten die n\"otige Freiheit gelassen.

Ebenso danke ich Dr.\,Michael Strickland f\"ur die tolle Zusammenarbeit, seine Hilfe und Unterst\"utzung
bei Problemen, viele erleuchtende Diskussionen und auch f\"ur das Korrekturlesen dieser Arbeit.

Ich danke JProf.\,Dr.\,Adrian Dumitru f\"ur die gute
Zusammenarbeit, Hilfsbereitschaft und viele hilfreiche
Diskussionen.

Dr.\,Yasushi Nara danke ich f\"ur die Einf\"uhrung in die numerische Simulation und seine Geduld, mit der er alle meine Fragen beantwortet hat.

Ich danke Dr.\,Zhe Xu, Oliver Fochler und Dr.\,Paul Romatschke f\"ur viele hilfreiche Diskussionen und Erl\"auterungen ihrer Arbeit.

Zudem danke ich meinem Zimmerkollegen Oliver Fochler f\"ur viele
auch \"uber die Physik hinausf\"uhrende Gespr\"ache, die gro\ss
artige Arbeitsatmosph\"are und eine tolle Zeit an der Uni.

Ich danke Barbara Betz, JProf.\,Dr.\,Marcus Bleicher, Andreas Ipp, Mauricio
Martinez, Hannah Petersen und Sascha Vogel f\"ur einige hilfreiche
Diskussionen und/oder Erkl\"arungen.

Auch danke ich Prof.\,Dr.\,Horst St\"ocker f\"ur sein Interesse an meiner Arbeit.

Ich danke Joe Laperal-Gomez, Denise Meixler, Gabriela Meyer,
Veronica Palade, Daniela Radulescu und Astrid Steidl f\"ur ihre
Hilfe in vielen Angelegenheiten.

Alexander Achenbach, Thilo Kalkbrenner und Oliver Fochler danke ich f\"ur ihre schnelle Hilfe bei all meinen Computerproblemen.

Ebenso danke ich dem Team des Center for Scientific Computing
f\"ur die Bereitstellung von enorm viel Rechenleistung und ihre
Hilfe im Falle von Problemen.

Ich danke dem Institute for Nuclear Theory an der University of
Washington, an dem ein Teil dieser Arbeit entstanden ist, f\"ur seine
Gastfreundschaft und Unterst\"utzung.

Diese Arbeit wurde von der Deutschen Forschungsgemeinschaft unterst\"utzt (DFG Grant GR 1536/6-1).

\section*{Acknowledgments}
First I want to thank my parents for their support, which made my studies and this work possible.

I thank my supervisor Prof.\,Dr.\,Carsten Greiner for his advise and help, for his continued interest in my work and for giving me a lot of freedom, which is so important in research.

I thank Dr.\,Michael Strickland for great collaboration, his help and advise, many enlightening discussions and for proof reading this work.

I thank JProf.\,Dr.\,Adrian Dumitru for his collaboration, his
constant will to help, and for many useful discussions.

I thank Dr.\,Yasushi Nara for introducing me to the numerical simulation and for his patience in answering all my questions.

I thank Dr.\,Zhe Xu, Oliver Fochler, and Dr.\,Paul Romatschke for many helpful discussions and explanations of their work.

In addition I thank my office-mate Oliver Fochler for many additional discussions beyond physics, a fun work environment and a great time at the university.

I thank Barbara Betz, JProf.\,Dr.\,Marcus Bleicher, Andreas Ipp, Mauricio
Martinez, Hannah Petersen and Sascha Vogel for some helpful
discussions and/or explanations.

I also thank Prof.\,Dr.\,Horst St\"ocker for his interest in my work.

I thank Joe Laperal-Gomez, Denise Meixler, Gabriela Meyer, Veronica Palade, Daniela Radulescu and Astrid Steidl for
their help with many issues.

I thank Alexander Achenbach, Thilo Kalkbrenner and Oliver Fochler for their help with all my computer problems.

I also thank the team of the Center for Scientific Computing for providing a huge amount of computer power and their help in case of problems.

I thank the Institute for Nuclear Theory at the University of Washington, where part of this work was done, for its hospitality and support.

This work was supported by the Deutsche Forschungsgemeinschaft (DFG Grant GR 1536/6-1).

  \cleardoublepage
  \phantomsection
  \addcontentsline{toc}{chapter}{Index}
  \printindex

\end{document}